\documentclass[a4paper,11pt,oneside]{Thesis}
\usepackage{
  amsmath,
  slashed,
  multirow,
  axodraw,
  bm,
  array,
  moreverb,
  booktabs,
  subfigure,
  appendix,
  amsfonts,
  cite}
\date{14 August 2008}
\newcommand\ie{\textit{i.e.}}  \newcommand\eg{\textit{e.g.}}
\newcommand\cf{\textit{cf.}}  \newcommand{\etal}{\textit{et al.}}
 \newcommand{\etc}{\textit{etc.}}
\newcommand{\viz}{\textit{viz.}}  
\newcommand{\mhvbar}{$\overline{\text{MHV}}$}
\DeclareMathOperator{\tr}{tr} \renewcommand{\vec}[1]{{\mathbf{#1}}}
\DeclareMathOperator*{\Res}{Res}
\newcommand\mdot{\!\cdot\!}
\newcommand\gA{{\cal A}} \newcommand\gB{{\cal B}}
\newcommand\gD{{\cal D}} \newcommand\gF{{\cal F}}

\newcommand\Oomega{\omega} \newcommand\Ozeta{\zeta}
\newcommand\mpp{\text{$-$$+$$+$}} \newcommand\ppm{\text{$+$$+$$-$}}
 \newcommand\mmp{\text{$-$$-$$+$}}
\newcommand\mmpp{\text{$-$$-$$+$$+$}}

\newcommand\fourplus{\text{$+$$+$$+$$+$}}
\newcommand\mmmm{\text{$-$$-$$-$$-$}}

\newcommand\NC{N_{\text{C}}}

\newcommand{\ud}[2]{_{i_{#1}}{}^{\bar\imath_{#2}}}
\newcommand{\di}[2]{\delta_{i_{#1}}^{\bar\imath_{#2}}}

\begin{document}
\title{MHV Lagrangians for Yang--Mills and QCD}
\supervisor{Prof.~Tim~R.~Morris}
\authors{James Ettle}
\date{\today}
\addresses{\groupname\\\deptname\\\univname}

\maketitle

\abstract{%
Over the past few decades, it has been realised that gauge theory
scattering amplitudes have structures much simpler than the
traditional Feynman graph driven approach would suggest. In
particular, Parke and Taylor found a particularly simple expression
for the tree-level amplitudes with two gluons of different helicity
than the others (the so-called MHV amplitudes).
Cachazo, Svr\v cek and Witten (CSW) devised rules for constructing 
tree-level amplitudes by sewing lower-valence MHV amplitudes together
with scalar propagators.
It was shown by Mansfield in 2005 that a canonical change of the field
variables could be constructed that resulted in a lagrangian whose
vertices were proportional to MHV amplitudes, continued off-shell by
CSW's prescription, the so-called Canonical MHV Lagrangian.
We derive the explicit form of this
transformation and use this to show that the vertices are indeed the
Parke--Taylor amplitudes for up to five gluons. Noting that CSW's MHV
rules cannot be used to construct the tree-level $(\mpp)$ or
one-loop $(\fourplus)$
amplitudes, we extend our work to augment the MHV rules with so-called
completion vertices. These permit construction of these missing
amplitudes by means of evasion of the $S$-matrix equivalence
theorem. Indeed, together they reconstruct off-shell light-cone Yang-Mills
amplitudes algebraically. We also give a prescription for dimensional
regularisation of the Canonical MHV Lagrangian. Finally, we construct
a canonical MHV lagrangian with massless fermions in the fundamental
representation using a similar methodology.
}

\tableofcontents
\listoffigures

\Declaration{
  I, James Ettle, declare that this thesis titled, `MHV Lagrangians
  for Yang--Mills and QCD' and the work presented in it are my own. I
  confirm that:

  \begin{itemize}
  \item This work was done wholly or mainly while in candidature for a
    research degree at this University.
  \item Where any part of this thesis has previously been submitted
    for a degree or any other qualification at this University or any
    other institution, this has been clearly stated.
  \item Where I have consulted the published work of others, this is
    always clearly attributed.
  \item Where I have quoted from the work of others, the source is
    always given. With the exception of such quotations, this thesis
    is entirely my own work.
  \item I have acknowledged all main sources of help.
  \item Where the
    thesis is based on work done by myself jointly with others, I have
    made clear exactly what was done by others and what I have
    contributed myself.
  \end{itemize}
 
  Signed:\\
 
  Date:\\
}
\clearpage 

\acknowledgements{
I would like to start by thanking my supervisor, Tim Morris, for his
support and wisdom throughout this project.  It has also been an
absolute pleasure to work in collaboration with Chih-Hao Fu, Jonathan
Fudger, Zhiguang Xiao, and of course Paul Mansfield and I would like
to extend the greatest gratitude to them all.

I would like to thank the following individuals, who have provided
interesting discussions on the subject and/or helped in an
indirect manner: Rutger Boels, Andreas Brandhuber, Jacques Distler
(for putting MHV lagrangians on the Blog-o-Sphere),
Nick Evans,
Nigel Glover, Simon McNamara, Doug Ross,
Bill Spence, Gabriele Travaglini and
Konstantinos Zubos.  I also thank the STFC (n\'e PPARC) for financial
support throughout.

The many years I have spent at the University of Southampton would not
have been so joyful were it not for the help, support, encouragement,
good humour, and out-and-out tolerance of the staff, post-docs and my
fellow students. In particular, I'd like to thank Michael Donnellan,
Jonathan Flynn, Chris Sachrajda, Andrew Tedder, and Martin Wiebusch.

Finally, I would like to thank my family for their patience and
support.
}

\chapter{Introduction}
\label{cha:intro}

The Standard Model\footnote{The purpose of this discussion is to give
  an overview of the Standard Model that provides sufficient context
  for the rest of the thesis. Further details can be found in many
  textbooks, \eg\ that of ref.\ \cite{Peskin:1995ev}} of Particle
Physics is a quantum field theory with an ${\rm SU}(3)_C \times {\rm
  SU}(2)_I \times {\rm U}(1)_Y$ gauge symmetry which gives rise to
three of the four fundamental forces: electromagnetism, the Weak
interaction and the Strong force.  The quantisation of gravity is, at
time of writing, an open question and it is omitted from the Standard
Model framework.

The ${\rm SU}(2)_I \times {\rm U}(1)_Y$ factor of the Standard Model
gauge symmetry provides the \emph{electroweak} interaction, a union of
electromagnetism and the Weak interaction. The ${\rm SU}(2)_I$
component acts on representations classified by \emph{weak isospin}
$I$, and the ${\rm U}(1)$ on \emph{weak hypercharge} $Y$.  Below
energies of around $100\:{\rm GeV}$ this is spontaneously broken to
the Weak interaction and electromagnetism via the Higgs
mechanism. This endows the Weak interaction gauge bosons ${\rm W}^\pm$
and ${\rm Z}^0$ (discovered in 1983 at the UA1 and UA2 experiments at
CERN) with masses $80\:{\rm GeV}$ and $91\:{\rm GeV}$, respectively.

The Strong force is governed by the ${\rm SU}(3)_C$ component of the
gauge group. It acts only on quark matter, generations of which live
in its fundamental representation called \emph{colour}, hence the
choice of name `Quantum Chromodynamics' (QCD) for the underlying
theory. This force is seen to exhibit two properties of particular
phenomenological interest. The first is \emph{confinement}: physical
states are only ever in the colour singlet representation, and free
individual quarks and gluons are never observed. This property has
never been proved analytically for QCD\footnote{Studies with lattice
  gauge theory have yielded an analytical proof for confinement and
  the linear form of quark-antiquark potential \cite{DeGrand:2006zz},
  but it is not known how this behaves in the continuum limit.}. The
second interesting property of QCD is that it demonstrates
\emph{asymptotic freedom} \cite{Gross:1973id, Politzer:1973fx} in the
ultraviolet: its coupling $\alpha_{\rm s}$ runs with energy scale $Q$
according to (see \eg\ section 17.2 of \cite{Peskin:1995ev})
\[
\alpha_{\rm s}(Q) = \frac{2\pi}{(11-2 n_{\rm f}/3) \ln
  (Q/\Lambda_{\text QCD})}
\]
to one-loop order in perturbation theory
for $n_{\rm f}$ quark flavours. Clearly the coupling becomes smaller
as the energy scale increases. $\Lambda_{\text{QCD}}$ characterises
the energy scale around which $\alpha_{\rm s}$ transitions between
strong and weak, and is determined experimentally to be around
$200\:{\rm MeV}$.  At energies below $\Lambda_{\text{QCD}}$ the Strong
force coupling is larger than unity so we cannot expect perturbation
theory to work. Instead one must resort to lattice techniques (see
\eg\ \cite{Wilson:1977nj, Montvay:1994cy}) or strong-weak dualities
such as the AdS/CFT correspondence \cite{Maldacena:1997re}. But at
energy scales further above $\Lambda_{\text{QCD}}$ (of the order
$1\:{\rm GeV}$ or more), the running coupling is less than unity, so
we may attempt to use perturbation theory with increasing confidence
as the energy scale rises.  For high-energy processes, QCD exhibits a
factorisation property which allows us to split cross-sections into a
non-perturbative hadronic part and a perturbative partonic piece. This
is a core concept in perturbative
QCD phenomenology, a rich subject in itself and
we refer the interested reader to \eg\ refs.\ \cite{Sterman:1995fz,
  Tung:2001cv}, but the important idea in this is that long- and
short-distance physics can be decoupled. The perturbative piece is
computed as if the partons were free physical states, `weighted' by
form factors computed from hadronic physics.

The Standard Model has been tested extensively and found to be a
remarkably robust theory. Nevertheless, it suffers from a number of
limitations. First, there is the Higgs mechanism.  The ${\rm W}^\pm$
and ${\rm Z}^0$ particles acquire mass through \emph{spontaneous
  symmetry breaking} --- the scalar Higgs field acquires a non-zero
vacuum expectation value. Yet its particle the Higgs boson, believed
to have a mass somewhere between $115\:{\rm GeV}$ and $180\:{\rm GeV}$
(see \eg\ `Higgs Boson Theory and Searches' in \cite{Yao:2006px}), has
yet to be observed. Furthermore, there are indications of the presence
of physics beyond the Standard Model arising from questions concerning
the origin of neutrino masses, the observed matter/antimatter
asymmetry in the universe, the nature of dark matter, and the
so-called `hierarchy problem' of the vastly different energy scales in
physics.

It is hoped that the Large Hadron Collider (LHC) at CERN will help
resolve (at least some of) these issues from the experimental
end. This machine is a $14\:{\rm TeV}$ proton-proton collider and,
assuming that the current bounds on the Higgs mass are correct, Higgs
production should be well within its reach. Furthermore it is believed
that it is at the ${\rm TeV}$ energy scales at which the first signals
of trans-Standard Model `New Physics' should emerge. There are a
number of candidate models for this, amongst which the most
phenomenologically promising are rooted in supersymmetric\footnote{For
  an introduction, see \eg\ \cite{Martin:1997ns}; a more technical
  review can be found in the `Supersymmetry' sections of
  \cite{Yao:2006px}.} and/or extra dimensional\footnote{Such as those
  of the section `Extra Dimensions' of ref. \cite{Yao:2006px}.}
extensions of the Standard Model.  But being a hadronic collider, the
signal from any collision event it generates will be dominated by the
QCD background --- processes involving quarks and gluons, often with
many being involved in any given process. Furthermore,
experimentalists have published a next-to-leading order `LHC
``priority'' wish list' \cite{Buttar:2006zd} of processes important to
both Higgs production and New Physics discovery, some of which involve
sub-processes with potentially large numbers of gluons and
quark-antiquark pairs. These background effects are all Standard Model
physics, and we know how to compute perturbative quantities with the
Standard Model using Feynman graphs. Putting this into practice is
where we come unstuck: computing amplitudes for many-particle
processes quickly becomes prohibitively complicated; for example, even
at tree-level, a six gluon amplitude has 216 Feynman diagrams, a
figure which grows faster than the factorial of the number of gluons
involved.

Thankfully, a number of methods for quickly computing tree-level
amplitudes have been developed. First is the realisation that much of
the complexity associated with non-abelian gauge theories, the colour
structure, can be separated from the kinematical part of the dynamics
leaving one to consider just `partial' amplitudes. Famously, Parke and
Taylor in \cite{Parke:1986gb} conjectured a compact formula for the
tree-level maximally helicity-violating (MHV) partial amplitudes
containing only gluons, with no more than two having a different
helicity than the others:
\begin{align*}
  A(1^+, \dots, n^+) &= 0, \\
  A(1^-, 2^+, \dots, n^+) &= 0, \\
  A(1^-, 2^+, \dots, j^-, \dots, n^+) &= \frac{\langle 1\:j\rangle^4}
  {\langle 1\: 2 \rangle \langle 2\:3 \rangle \cdots \langle n-1,n
    \rangle \langle n\: 1 \rangle}.
\end{align*}
(Note that in the last line only gluons $1$ and $j$ have negative
helicity; the notation is explained in section
\ref{sec:background-spinorhel}.)  This was later proved by Berends and
Giele \cite{Berends:1988zn}.  More recently, Cachazo, Svr\v cek and
Witten developed the CSW rules for connecting MHV vertices with scalar
propagators (using a particular prescription for off-shell spinors)
\cite{Cachazo:2004kj, Risager:2005vk} that can be used to compute
tree-level amplitudes with arbitrary helicity configurations, using a
\emph{polynomial} number of diagrams. We will consider these in some
detail. Additionally, Britto, Cachazo and Feng discovered (in ref.\
\cite{Britto:2004ap}) the BCF recursion relations between tree-level
amplitudes, which were proved with the help of Witten in ref.\
\cite{Britto:2004ap, Britto:2005fq}. These arise from general
considerations on the pole structure and analytic properties of
tree-level scattering amplitudes, rather than relying on particular
details of the theory under consideration; correspondingly, recursion
relations have been found for theories involving non-gluonic matter
\cite{Ozeren:2005mp, Luo:2005rx, Ozeren:2006ft, Badger:2005zh,
  Badger:2005jv} and even for perturbative gravity
\cite{Bedford:2005yy, Cachazo:2005ca, Benincasa:2007qj}. The CSW rules
have been extended likewise \cite{Georgiou:2004wu, Wu:2004jxa,
  Ozeren:2005mp}, and indeed it has come to be understood that the CSW
rules arise from a particular adaptation of the BCF relations.

Work has also been done to search for labour-saving devices for the
one-loop contributions, although no satisfactorily complete systematic
method has been found to date. A variety of techniques have been shown
to reproduce known results both in QCD and Supersymmetric Yang-Mills
(SYM) theories, and have even been used to predict results for all
helicity configurations for up to six gluons and particular
configurations at higher orders. These methods include:
\begin{itemize}
\item the twistor-string inspired holomorphic anomaly
  \cite{Cachazo:2004zb, Cachazo:2004by, Cachazo:2004dr,
    Britto:2004nj};
\item (generalised) unitarity \cite{Eden:1966aa, Bern:1994cg,
    Brandhuber:2005jw, Britto:2004nc},
  whereby cuts in the complex plane of the momentum invariants are
  matched to those of functions known to arise in one-loop amplitudes,
  thereby recovering the \emph{cut constructible} parts. The ${\cal N}
  = 1$ and ${\cal N} = 4$ supersymmetric Yang--Mills theories are
  completely cut constructible since their rational terms are
  inextricably linked to the cut-containing pieces. In QCD, however,
  the rational terms cannot be correctly recovered by cut construction
  (indeed, some classes of loop amplitudes are purely rational) and
  these must be determined \eg\ by extracting the cut-free parts from
  the Feynman integrals \cite{Xiao:2006vr,Su:2006vs,Xiao:2006vt} or
  using:
\item `bootstrap' methods \cite{Bern:2005hs, Bern:2005ji,
    Bern:2005cq}, which exploit recursion relations between amplitudes
  at one-loop to extract the rational parts given the
  cut-constructible pieces; and
\item one-loop MHV diagrams \cite{Brandhuber:2004yw, Bedford:2004nh,
    Brandhuber:2005kd}, essentially the application of the CSW rules
  at one-loop to recover the cut-constructible parts of QCD
  amplitudes.
\end{itemize}

A feature common to all these techniques is that they have taken place
outside the usual Lagrangian framework, drawing inspiration from the
twistor-string dual theories in the cases of the CSW rules and
holomorphic anomaly computations, or studies of the analytic structure
of scattering amplitudes.  The Lagrangian formulation underpins a
large body of our understanding of quantum field theory, in particular
making manifest the symmetries and guiding us towards a systematic
application of regularisation structure, something which these
techniques currently lack. This brings us neatly to the motivation of
the work documented in this thesis: to formulate the modern
techniques, in particular the CSW rules, from a field theory
viewpoint. We construct the rules in terms of the Lagrangian
formalism, and begin a study of the consequences thereof.

\section{Outline}
The rest of this thesis is organised as follows.  In chapter
\ref{cha:background}, we establish the necessary background materials
and preliminaries from perturbative gauge theory that provide the
tools and the context for the work presented in chapters
\ref{cha:mhvym}--\ref{cha:mhvqcd}. In particular, we review gauge
theory itself; the important techniques of colour-ordered
decomposition, the spinor helicity formalism, the Parke--Taylor MHV
amplitudes, and the use of supersymmetry to extract information about
scattering amplitudes. We describe the BCFW construction of recurrence
relations between on-shell amplitudes, and the CSW rules at
tree-level. Although loop-level techniques are not the main focus of
this thesis, we will also review the most successful developments in
this area, in particular unitarity and generalised unitarity.

Chapter \ref{cha:mhvym} introduces and develops the Canonical MHV
Lagrangian. In ref.\ \cite{Mansfield:2005yd}\footnote{We note that a
  similar transformation was proposed by Gorsky and Rosly in ref.\
  \cite{Gorsky:2005sf}.}, Mansfield gave a canonical (in the sense of
classical mechanics) field transformation that re-wrote the lagrangian
of light-cone gauge Yang--Mills theory as one in terms of an infinite
series of Parke--Taylor MHV vertices connected by scalar propagators
and continued off-shell in a manner that follows the CSW
rules. Following the construction set out in \cite{Mansfield:2005yd},
we proceed to solve for this transformation explicitly as a power
series. We demonstrate explicitly that the transformation results in
vertices of the Parke--Taylor form for up to five gluons. This work
was originally published in ref.\ \cite{Ettle:2006bw}.

Next, chapter \ref{cha:etv} addresses some of the issues identified at
the end of the previous chapter. We follow the research published in
ref.\ \cite{Ettle:2007qc}, where we use the transformation discovered
previously to define \emph{MHV completion} vertices. We use these to
reconstruct amplitudes which could not be obtained from MHV vertices
alone, such as $(\mpp)$ at tree-level and $(\fourplus)$ at one
loop. We apply the transformation to dimensionally regulated
light-cone Yang--Mills and obtain $D$-dimensional MHV vertices. We
find strong evidence that MHV vertices and completion vertices
together algebraically reconstruct the original LCYM theory, even off
shell.

Chapter \ref{cha:mhvqcd} extends the construction of chapter
\ref{cha:mhvym} to include massless quarks in the fundamental
representation, and follows the work published in ref.\
\cite{Ettle:2008ey}.  By considering a similarly specified canonical
transformation which results in a lagrangian consisting of an infinite
series of vertices with an MHV helicity content, we construct a series
solution from which we define completion vertices and test the
resulting lagrangian vertices against known QCD MHV amplitudes.

Finally, in chapter \ref{cha:conclusion}, we review this work and
consider it in the context laid out in this introduction, considering
its value as a computational tool and as a means of gaining insight
into the structure of gauge theories.  We also outline what is left to
be understood in the techniques developed herein, and what routes for
further research this implies. Finally, we take the opportunity to
consider similar developments.

\chapter{Modern Perturbative Gauge Theory}
\label{cha:background}

The Strong Interaction is described by QCD, a non-Abelian gauge theory
with the following action:
\begin{equation}
  \label{eq:textbook-qcd-action}
  S_{\text{QCD}} = \int d^4x\: \left\{ \bar\psi(i \slashed D - m)\psi
    - \frac14 \tr F_{\mu\nu} F^{\mu\nu} \right\}.
\end{equation}
The phenomenologically relevant implementation of QCD has an ${\rm
  SU}(3)$ gauge group with the quark field $\psi$ transforming in the
fundamental representation. It also features (from QCD's point of
view) six different flavours of quark, each with a different
mass. However, for the purposes of the studies conducted in this
thesis, we will be concerned with the case of just one flavour with
$m=0$ when we come to consider quarks. We will also generalise the
gauge group slightly and work with ${\rm SU}(\NC)$ (and sometimes
${\rm U}(\NC)$ where convenient) where $\NC$ is the number of colours
of quark (\ie\ the dimension of the fundamental representation).

Let us make \eqref{eq:textbook-qcd-action} more concrete by making
some further definitions. First, the gauge covariant derivative is
\begin{equation}
  \label{eq:textbook-cov-der}
  D_\mu = \partial_\mu - \frac{ig}{\sqrt 2} A_\mu,
\end{equation}
where $A_\mu = A^a_\mu T^a$ is the gluon field, living in the adjoint
representation of the gauge group. The $\NC^2 - 1$ generators of the
$\mathfrak{su}(\NC)$ Lie algebra are normalised according to $\tr (T^a
T^b) = \delta^{ab}$, giving the factor of $2^{-1/2}$ in
\eqref{eq:textbook-cov-der}\footnote{This factor would be absent had
  we chosen the `textbook' normalisation $\tr (T^a T^b) =
  \frac12\delta^{ab}$, but this would lead to a proliferation of
  factors of $\sqrt 2$ elsewhere.}, and the structure constants
$f^{abc}$ defined according to
\begin{equation}
  \label{eq:structureconstants}
  [T^a, T^b] = i \sqrt 2 f^{abc} T^c.
\end{equation}
The field strength $F_{\mu\nu}$ is an algebra-valued 2-form, defined
here by
\begin{equation}
  \label{eq:textbook-F}
  F_{\mu\nu} = [D_\mu, D_\nu] = \partial_\mu A_\nu - \partial_\nu A_\mu
  + [A_\mu, A_\nu]. 
\end{equation}
Note that in this thesis, we will use the term `QCD' to refer to the
massless, quark-containing ${\rm SU}(\NC)$ gauge theory. In absence of
quarks, the non-Abelian nature of the gauge group (\ie\ $f^{abc} \ne
0$) means that the gauge field still interacts with itself, and so
what is left over is still interesting. We will refer to this theory
as `(pure) Yang--Mills'.

As noted in chapter \ref{cha:intro}, QCD is asymptotically free, and
its running coupling constant decreases with increasing energy scale
$Q$ like $(\ln Q)^{-1}$. From a phenomenological point of view, it
will be the perturbatively amenable, high-energy regime on which we
focus our studies. Alternatively, noting that in the absence of an
experimental impetus the theory described continues to exhibit UV
asymptotic freedom, we can simply \emph{presume} to work at
\emph{some} energy scale at which the coupling constant is much
smaller than $1$.

\begin{figure}[h]
\centering\begin{tabular}{ccc}
\(\begin{matrix}\begin{picture}(50,20)
\SetOffset(25,8)
\ArrowLine(-15,0)(15,0)
\Text(-17,0)[cr]{$i$}
\Text(17,0)[cl]{$\bar\jmath$}
\Text(0,3)[cb]{$p$}
\end{picture}\end{matrix}\)
& $\displaystyle \frac i{\slashed p - m} \delta_i^{\bar\jmath} $
& $\displaystyle \frac i{\slashed p - m} $
\\[4ex]
\(\begin{matrix}\begin{picture}(72,30)
\SetOffset(36,13)
\Gluon(-15,0)(15,0){2}{5}
\Text(-17,0)[cr]{$\mu,a$}
\Text(17,0)[cl]{$\nu,b$}
\LongArrow(-7,6)(7,6)
\Text(0,9)[cb]{$p$}
\end{picture}\end{matrix}\)
& $\displaystyle -\frac i{p^2} g^{\mu\nu}\delta^{ab} $
& $\displaystyle -\frac i{p^2} g^{\mu\nu} $
\\[4ex]
\(\begin{matrix}\begin{picture}(60,48)
\SetOffset(30,18)
\Gluon(0,0)(0,20){2}{4}
\ArrowLine(0,0)(17.3205,-10)
\ArrowLine(-17.3205,-10)(0,0)
\Vertex(0,0){1}
\Text(0,22)[cb]{$\mu,a$}
\Text(-19,-12)[cr]{$i$}
\Text(19,-12)[cl]{$\bar\jmath$}
\end{picture}\end{matrix}\)
& $\displaystyle \frac{ig}{\sqrt 2} \gamma^\mu (T^a)_i{}^{\bar\jmath} $
& $\displaystyle \frac{ig}{\sqrt 2} \gamma^\mu $
\\[4ex]
\(\begin{matrix}\begin{picture}(80,54)
\SetOffset(40,20)
\Gluon(0,0)(0,24){2}{4}
\Gluon(0,0)(20.7846,-12){2}{4}
\Gluon(-20.7846,-12)(0,0){-2}{4}
\Vertex(0,0){1}
\LongArrow(6,4)(6,20)
\LongArrow(2,-7)(15.9,-15)
\LongArrow(-6,2)(-19.9,-6)
\Text(9,12)[cl]{$k$}
\Text(8,-12)[tr]{$p$}
\Text(-14,-1)[br]{$q$}
\Text(0,26)[cb]{$\mu,a$}
\Text(-23,-14)[cr]{$\rho,c$}
\Text(23,-14)[cl]{$\nu,b$}
\end{picture}\end{matrix}\)
& \parbox{5cm}{
  \[\begin{split}
    &g\, f^{abc}[g^{\mu\nu}(p - k)^\rho \\
    &\hphantom{g\, f^{abc}[} + g^{\nu\rho} (q - p)^\mu \\
    &\hphantom{g\, f^{abc}[} + g^{\rho\mu}(k - q)^\nu]
  \end{split}\]
}
& \parbox{5cm}{
  \[\begin{split}
    & -\frac{ig}{\sqrt2} [g^{\mu\nu}(p - k)^\rho \\
    &\hphantom{-\frac{ig}{\sqrt2} [} + g^{\nu\rho} (q - p)^\mu \\
    &\hphantom{-\frac{ig}{\sqrt2} [} + g^{\rho\mu}(k - q)^\nu]
  \end{split}\]
}
\\[4ex]
\(\begin{matrix}\begin{picture}(78,50)
\SetOffset(39,25)
\Gluon(0,0)(17,17){2}{4}
\Gluon(0,0)(17,-17){2}{4}
\Gluon(0,0)(-17,-17){2}{4}
\Gluon(0,0)(-17,17){2}{4}
\Vertex(0,0){1}
\Text(-20,17)[cr]{$\mu,a$}
\Text(20,17)[cl]{$\nu,b$}
\Text(20,-17)[cl]{$\rho,c$}
\Text(-20,-17)[cr]{$\sigma,d$}
\end{picture}\end{matrix}\)
& \parbox{6cm}{
\[\begin{split}
& -ig^2[ f^{abe} f^{cde} (g^{\mu\rho} g^{\nu\sigma} - g^{\mu\sigma}
g^{\nu\rho}) \\
& \hphantom{-ig^2[}
+ f^{ade} f^{bce} (g^{\mu\nu} g^{\rho\sigma} - g^{\mu\rho} g^{\nu\sigma})
\\
& \hphantom{-ig^2[}
+ f^{ace} f^{bde} (g^{\mu\nu} g^{\rho\sigma} - g^{\mu\sigma}
g^{\nu\rho}) ]
\end{split}\]
}
& \parbox{5cm}{
\begin{multline*}
  ig^2(g^{\mu\rho} g^{\nu\sigma} - \tfrac12 [g^{\mu\sigma} g^{\nu\rho}
  + g^{\mu\nu} g^{\sigma\rho}])
\end{multline*}
}
\end{tabular}
\caption{Feynman rules for QCD, excluding Faddeev--Popov ghosts.
  The expressions in the second column
  are for the traditional formulation that includes colour
  explicitly. The third column gives the colour-ordered Feynman
  rules. All momenta are out-going.}
\label{fig:qcd-feynrules}
\end{figure}

The textbook treatment of the theory of \eqref{eq:textbook-qcd-action}
is to expand the expression for the action in terms of its component
fields, and then gauge fix in order to define the path integral. This
is conventionally carried out by the Faddeev--Popov procedure working
in Feynman gauge. The procedure introduces extra `ghost' fields which
may be understood as `anti-degrees-of-freedom' that eliminate the
non-physical degrees of freedom that arise from gauge
invariance. Ghosts couple to gluons but are not physical particles
(which may be understood by considering the BRST transformation
properties of the theory \cite{Weinberg:1996kr}), so they
only show up in loop diagrams. Nevertheless, the Feynman gauge is not
the only possible gauge and a judicious choice (such as an axial
gauge, $n \cdot A^a = 0$) decouples the ghosts entirely (albeit often
at the cost of increased computational difficulties).

With a gauge-fixed action, we can construct Feynman rules and thence
Feynman graphs. The Feynman rules (minus ghost vertices) for the
theory \eqref{eq:textbook-qcd-action} in Feynman gauge are shown in
fig.~\ref{fig:qcd-feynrules}. Calculating perturbative gauge theory
amplitudes is from then on a straightforward procedure: one draws all
the relevant diagrams contributing to an amplitude at a desired order
in $g$, contracts with external polarisation vectors and spinors and
adds them up.

Unfortunately, in practise this turns out to be a computationally
intensive task as the number of particles in the amplitude
grows. Consider the multi-gluon amplitudes, which have a
phenomenological relevance to the computation of the LHC background
(amongst other things): just at tree level, the number of diagrams and
the complexity of their expressions grows incredibly rapidly. Indeed,
for $n$ gluons the number of Feynman graphs increases faster than $n!$
\cite{Kleiss:1988ne}: 220 diagrams for 6 gluons, increasing (from 7
gluons) as $2~485, 34~300, 589~405, 10~525~900, \dots$, and any
attempt to calculate these amplitudes soon encounters the limitations
of time or memory. It is, therefore, surprising to find that the final
expressions for these amplitudes often take a very simple form (when
expressed in the right framework), even at the loop
level. Computational techniques have also been discovered that have a
great computational advantage over Feynman graphs. All this hints that
gauge theories have a structure much simpler than we would otherwise
expect.

The remainder of this chapter aims to establish some of the formalism
that makes these simplifications manifest, and illustrate some of the
modern techniques that exploit this. Sections
\ref{sec:background-spinorhel}, \ref{sec:background-colourorder} and
\ref{sec:background-susy} introduce the spinor-helicity formalism, the
concept of colour ordered amplitudes and the Parke--Taylor MHV
amplitude, and explain the role of supersymmetry in dealing with a
non-supersymmetric theory. Sections \ref{sec:background-csw} and
\ref{sec:background-bcf} discuss two important modern tools for
computing tree-level amplitudes, the CSW rules and BCF recursion
relations, and finally in section \ref{sec:background-loops}, we
review the techniques that have been applied at the loop level.

\section{The spinor formalism}
\label{sec:background-spinorhel}

Much development in modern perturbative gauge theory uses the spinor
framework to represent momenta and helicity. Ref.\
\cite{Witten:2003nn} has a particularly clear exposition of this, and
we will follow closely the formalism therein.

It is well-known that the complexified Lorentz group ${\rm
  SO}(1,3;\mathbb{C})$ is locally isomorphic to ${\rm SL}(2;
\mathbb{C}) \times {\rm SL}(2; \mathbb{C})$, so its representations
may be classified as $(m,n)$ where $m$ and $n$ are half-integers
giving the spin of the representation. $(\tfrac12,0)$ and
$(0,\tfrac12)$ are the left- and right-handed chiral Weyl spinors,
respectively, whose direct sum gives the Dirac spinor. $(\tfrac12,
\tfrac12)$ is the 4-vector, so it can be represented as a bispinor
$p_{\alpha\dot\alpha}$, and we can map from Minkowski co-ordinates
$(p^t, p^1, p^2, p^3)$ into this representation using
\begin{equation}
  \label{eq:bispinor-mink}
  p_{\alpha\dot\alpha} = p_\mu (\bar\sigma^\mu)_{\dot\alpha\alpha}
  = \begin{pmatrix}
    p^t + p^3 & p^1 - ip^2 \\
    p^1 + ip^2 & p^t - p^3
  \end{pmatrix},
\end{equation}
where the 4-vectors of Pauli matrices are
\[
(\sigma^\mu)^{\dot\alpha\alpha} = (1, {\pmb\sigma})
\quad\text{and}\quad (\bar\sigma^\mu)_{\alpha\dot\alpha} = (1,
-{\pmb\sigma})
\]
and ${\pmb\sigma}$ is the usual 3-vector of Pauli matrices. (The
spinor indices are printed here to show their `natural' position, \ie\
as they dereference the elements of the matrices on the RHS.)  From
this, it is clear that $\det p = p^2$. If $p$ is null then
$p_{\alpha\dot\alpha}$ may be factorised into two Weyl spinors as
\[
p_{\alpha\dot\alpha} = \lambda_\alpha \tilde\lambda_{\dot\alpha}.
\]
$\lambda_\alpha$ is commonly termed the \emph{holomorphic} spinor,
transforming in the $(0,\tfrac12)$ representation, and
$\tilde\lambda_{\dot\alpha}$ the \emph{antiholomorphic} spinor in the
$(\tfrac12,0)$ representation\footnote{While we are following the
  discussion of \cite{Witten:2003nn}, this is closer to the convention
  of \cite{Cachazo:2004kj}.}.

We define raising (and hence lowering) of spinor indices as follows:
\begin{equation}
  \label{eq:spinorraiselower}
  \lambda^\alpha = \epsilon^{\alpha\beta} \lambda_\beta
  \quad\Leftrightarrow\quad
  \lambda_\alpha = \epsilon_{\alpha\beta} \lambda^\beta,
\end{equation}
and likewise with dotted indices. In our convention, the invariant
antisymmetric bispinor has $\epsilon^{12} = \epsilon^{\dot1\dot2} =
1$. We define the downstairs-indexed $\epsilon$ to be just the inverse
of the one with indices raised:
$\epsilon^{\alpha\beta}\epsilon_{\beta\gamma} = \delta^\alpha_\gamma$.
Thus,
\begin{equation}
  \label{eq:bispinor-dot}
  p^{\alpha\dot\alpha} q_{\alpha\dot\alpha} = 2 p \cdot q,
\end{equation}
which can be seen either by direct application of
\eqref{eq:bispinor-mink} and \eqref{eq:spinorraiselower}, or by
noticing that $(\sigma^\mu)^{\dot\alpha\alpha} =
(\bar\sigma^\mu)^{\alpha\dot\alpha}$, and hence
\[
(\bar\sigma^\mu)_{\alpha\dot\alpha}
(\bar\sigma^\nu)^{\alpha\dot\alpha} = \tr \bar\sigma^\mu \sigma^\nu =
2 g^{\mu\nu}.
\]
Since the $\epsilon$ bispinor is an ${\rm SL}(2; \mathbb{C})$
invariant, the following products of spinors are Lorentz invariants:
\begin{equation}
  \label{eq:spinorbrackets}
  \langle \lambda \: \mu \rangle := \epsilon^{\alpha\beta}
  \lambda_\alpha \mu_\beta
  \quad\text{and}\quad
  [ \lambda \: \mu ] := \epsilon^{\dot\alpha\dot\beta}
  \tilde\lambda_{\dot\alpha} \tilde\mu_{\dot\beta}
\end{equation}
In what follows, for null momenta we will often simply use that
momentum's symbol or its number directly in the $\langle\:\rangle$ and
$[\:]$ brackets, using its (anti)holomorphic spinor as the context
above suggests. Clearly these brackets are antisymmetric in their
arguments. Combining \eqref{eq:spinorbrackets} with
\eqref{eq:bispinor-dot}, we have for any two null momenta $p =
\lambda\tilde\lambda$ and $q = \mu\tilde\mu$
\begin{equation}
  \label{eq:dot-brackets}
  2 p\cdot q = \langle \lambda\:\mu \rangle [ \lambda\:\mu ].
\end{equation}

\subsection{Spinor helicity}

For massless vector particles, we choose the following polarisation
vectors, written as bispinors \cite{Witten:2003nn}:
\begin{equation}
  \label{eq:gluon-poln}
  \epsilon^+_{\alpha\dot\alpha}(p,\mu) =
  \sqrt 2 \frac{\nu_\alpha \tilde\lambda_{\dot\alpha}}
  {\langle \nu\:\lambda \rangle}
  \quad\text{and}\quad
  \epsilon^-_{\alpha\dot\alpha}(p, \mu) =
  \sqrt 2 \frac{\lambda_\alpha \tilde\nu_{\dot\alpha}}{[ \nu\:\lambda ]}
\end{equation}
for a gluon with momentum $p_{\alpha\dot\alpha} = \lambda_\alpha
\tilde\lambda_{\dot\alpha}$.  The spinors $\nu$ and $\tilde\nu$
correspond to a null \emph{reference vector} $\mu_{\alpha\dot\alpha} =
\nu_\alpha \tilde\nu_{\dot\alpha}$. It is easy to verify (\eg\ using
\eqref{eq:bispinor-dot} and the antisymmetry of
$\epsilon^{\alpha\beta}$) that these polarisation vectors satisfy the
following:
\begin{align}
  p \cdot \epsilon^\pm(p, \mu) &= p^{\alpha\dot\alpha}
  \epsilon^+_{\alpha\dot\alpha}(p,\mu) = 0, \nonumber \\
  \epsilon^+(p,\mu)\cdot\epsilon^+(q,\mu) &= 0,
  \label{eq:poln-ident-1} \\
  \epsilon^+(p,q)\cdot\epsilon^-(q,\mu) &= 0.
  \label{eq:poln-ident-2}
\end{align}
The choice of $\mu$ is arbitrary (so long as its respective spinors
are not proportional to those of $p$), and this is due to residual
gauge invariance. We can add to $\epsilon^\pm$ any multiple of $p$ and
still preserve the Lorentz gauge condition, and clearly any change of
$\nu$ and $\tilde\nu$ effects precisely that: since they are not
parallel, $\nu$ and $\lambda$ (and their conjugates) form a basis for
the Weyl spinors. Any change along the $\nu$ direction cancels in
\eqref{eq:gluon-poln}, whereas the component of $\lambda$ leaves a
piece proportional to $\lambda\tilde\lambda$ --- that is, $p$.

Last but not least, we state the Schouten identity,
\begin{equation}
  \label{eq:schouten}
  \langle i\:j \rangle \langle k\:l \rangle = \langle i\:k \rangle
  \langle j\:l \rangle + \langle i\:l \rangle \langle k\:j \rangle
\end{equation}
and its conjugate under $\langle\:\rangle \leftrightarrow [\:]$. This
may be derived by noting Fierz-type identity for Weyl spinors
$\epsilon_{\alpha\beta} \epsilon_{\gamma\delta} =
\epsilon_{\alpha\gamma} \epsilon_{\beta\delta} +
\epsilon_{\alpha\delta} \epsilon_{\gamma\beta}$.
 \section{Colour ordered decomposition}
\label{sec:background-colourorder}

The first significant simplification of the process of computing QCD
scattering amplitudes comes from the observation that we can separate
the management of the colour information from the kinematics, and this
gives rise to \emph{colour ordering} and \emph{colour ordered partial
  amplitudes}.

Now consider the Feynman rules of the Yang--Mills theory. We see that
the vertices have group theory factors that are either linear or
quadratic in the gauge theory structure constants $f^{abc}$. We can
re-write these as traces:
\[
f^{abc} = \frac i{\sqrt 2} \tr (T^a T^b T^c - T^c T^b T^a).
\]
Internal ${\rm SU}(\NC)$ colour lines in the adjoint representation
make contractions between indices of these structure constants, but by
definition, $i \sqrt2 T^c f^{cde} = [T^d, T^e]$, so inserting this
into the above simply unrolls the further permutations of gauge
generator matrices. Thus for a purely gluonic tree-level amplitude,
the entire colour content can be moved out into a trace. A $n$-gluon
tree-level amplitude with all particles out-going,
where the gluon label $i$ subsumes a momentum
$p_i$ and helicity state $h_i$, and $a_i$ is an adjoint index, can be
decomposed as\footnote{Note that it will be our convention to absorb
  the factors of $i$ and $g$ into the partial amplitude.}:
\begin{equation}
  \label{eq:colourorder-decomp-tree}
  {\cal A}(1\cdots n) = (2\pi)^4 \delta({\textstyle \sum_{i=1}^n p_i})
  \sum_{\sigma \in S_n/\mathbb{Z}_n} \tr (T^{a_{\sigma(1)}} \cdots
  T^{a_{\sigma(n)}} ) \: A(\sigma(1) \cdots \sigma(n)),
\end{equation}
where $S_n/\mathbb{Z}_n$ is the group of all permutations of $n$
objects, modulo cycles (under which the trace is invariant).  The
colour-stripped object $A(1\cdots n)$ is a \emph{partial amplitude}.

When a single quark-antiquark pair is present in an amplitude,
carrying fundamental indices $\bar\imath$ and $j$, respectively, we
obtain colour structures with exposed fundamental indices of the form
\[
(X_0 T^{a_1} T^{a_2} Y_0)_i{}^{\bar\jmath} \tr(X_1 T^{a_1} Y_1)
\tr(X_2 T^{a_2} Y_2) \cdots
\]
where the ellipsis on the extreme right denotes further omitted
traces, and $X_i$ and $Y_i$ are products of gauge generators.  We can
use the ${\rm SU}(\NC)$ Fierz identity
\begin{equation}
  \label{eq:sun-fierz}
  (T^a)_i{}^{\bar\jmath} (T^a)_k{}^{\bar l} = \delta_i^{\bar l}
  \delta_k^{\bar\jmath} - \frac1{N_{\rm C}}
  \delta_i^{\bar\jmath} \delta_k^{\bar l}
\end{equation}
to break up these products into a product of all the gauge generators,
plus a product of traces suppressed by $1/\NC$ of the form
\[
(X_0 Y_1 X_1 Y_2 X_2 \cdots Y_0)_i{}^{\bar\jmath} + \frac1\NC
(\text{products of traces}) \delta_i^{\bar\jmath}.
\]
Now for one quark line, we can ignore the ${\cal O}(1/\NC)$ piece: to
see this, consider that it is just the statement that the
$\mathfrak{su}(\NC)$ generators are traceless. Had we considered the
${\rm U}(\NC)$ gauge group instead, the ${\cal O}(1/\NC)$ term would
be absent from \eqref{eq:sun-fierz}. Now, ${\rm U}(\NC) = {\rm
  SU}(\NC) \times {\rm U}(1)$ so in this case the theory has an extra
gauge boson (a photon); its corresponding gauge generator is
proportional to the identity and so commutes with the ${\rm SU}(\NC)$
generators and decouples from the gluons. Thus, \emph{with one quark
  line at tree level}, the ${\rm U}(\NC)$ and ${\rm SU}(\NC)$
amplitudes are the same and we may discard terms ${\cal
  O}(1/\NC)$. The colour decomposition for an amplitude with one
quark-antiquark pair is therefore
\begin{multline}
  \label{eq:colourstruct-2q}
  {\cal A}(1_{{\rm q}},2,\dots, n-1,n_{\bar{\rm q}}) = (2\pi)^4
  \delta({\textstyle \sum_{i=1}^n p_i}) \sum_{\sigma\in S_{n-2}}
  (T^{a_{\sigma(2)}} \cdots
  T^{a_{\sigma(n-1)}})_{i_1}{}^{\bar\imath_n} \\ \times A(1_{{\rm
      q}},\sigma(2),\dots,\sigma(n-1),n_{\bar{\rm q}}).
\end{multline}

With more than one quark line, the extra photon from the ${\rm
  U}(\NC)$ theory can couple between the quark lines. Thus, in this
situation we do have to consider terms ${\cal O}(1/\NC)$ in the
decomposition of the colour structures. This is explained in detail in
\cite{Mangano:1990by, Wu:2004jxa}.
We will not reproduce the full construction here for the purposes of
this thesis, since we will only encounter this in section
\ref{ssec:four-quarks} for four-quark amplitudes with no gluons; these
are simple enough to permit us to work out the colour structures
on-the fly.  However, we note below that for an amplitude with two
quark-antiquark pairs and an arbitrary number of positive-helicity
gluons, the \emph{leading} partial amplitude is
\begin{equation}
  \label{eq:colourorder-4q}
  A(1^{h_1}_{{\rm q}},2,\dots,j\!-\!2,(j\!-\!1)^{-h_j}_{\bar{\rm q}};
  j^{h_j}_{{\rm q}},j\!+\!1,\dots,n\!-\!1,n^{-h_1}_{\bar{\rm q}}),
\end{equation}
which is associated with the colour factor
\begin{equation}
  \label{eq:colourstruct-4q}
  (T^{a_{2}} \cdots
  T^{a_{j-2}})_{i_1}{}^{\bar\imath_{j-1}}
  (T^{a_{j+1}} \cdots
  T^{a_{n-1}})_{i_j}{}^{\bar\imath_{n}}.
\end{equation}
There are also sub-leading partial amplitudes associated with the
colour structure
\begin{equation}
  \label{eq:colourstruct-4q-sub}
  - \frac1{\NC}
  (T^{a_{2}} \cdots
  T^{a_{j-2}})_{i_1}{}^{\bar\imath_{n}}
  (T^{a_{j+1}} \cdots
  T^{a_{n-1}})_{i_j}{}^{\bar\imath_{j-1}}.
\end{equation}
The full amplitude is obtained by summing over the permutations of the
gluon labels $\{2,\dots,j-2,j+1,\dots,n-1$\}, over permutations of the
quarks $\{1,j\}$ and antiquarks $\{j-1,n\}$ so as not to over-count
and remembering a Fermi statistics factor of $-1$ for an odd
permutation; and finally over $j$, interpreted as the \emph{position}
of the second quark in \eqref{eq:colourstruct-4q} and
\eqref{eq:colourstruct-4q-sub} (or equivalently as a parametrisation
of the length of $T$-strings).

\subsubsection*{'t~Hooft double-line diagrams and colour-ordered
  Feynman rules}

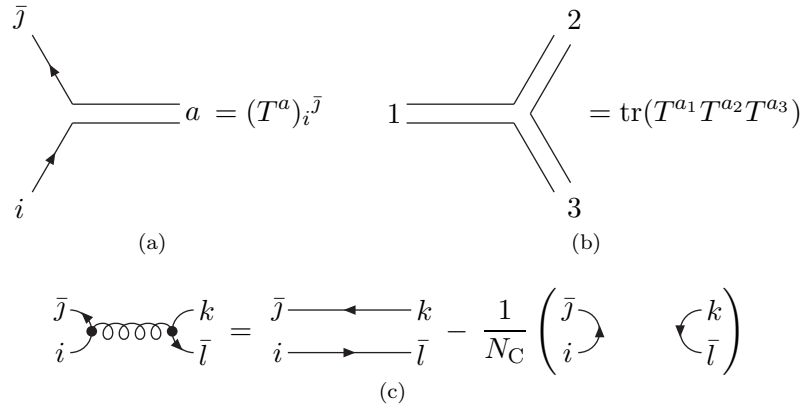
\begin{figure}[h]
\centering
\subfigure[]{
$\displaystyle
\begin{matrix}\begin{picture}(80,84)
\SetOffset(50,40)
\ArrowLine(-20,3.5356)(-35,29.5164)
\Line(-20,3.5356)(20,3.5356)
\Line(20,-3.5356)(-20,-3.5356)
\ArrowLine(-35,-29.5164)(-20,-3.5356)
\Text(-38,32)[br]{$\bar\jmath$}
\Text(22,0)[cl]{$a$}
\Text(-38,-32)[tr]{$i$}
\end{picture}\end{matrix}
= (T^a)_i{}^{\bar\jmath}
$
\label{fig:thooft-qqg}
}
\quad
\subfigure[]{
$\displaystyle
\begin{matrix}\begin{picture}(75,84)
\SetOffset(30,40)
\Line(-20,3.5356)(20,3.5356)
\Line(20,3.5356)(35,29.5164)
\Line(41.1237,25.9808)(26.1237,0)
\Line(26.1237,0)(41.1237,-25.9808)
\Line(35,-29.5164)(20,-3.5356)
\Line(20,-3.5356)(-20,-3.5356)
\Text(-22,0)[cr]{$1$}
\Text(40,32)[bl]{$2$}
\Text(40,-32)[tl]{$3$}
\end{picture}\end{matrix}
= \tr(T^{a_1} T^{a_2} T^{a_3})
$
\label{fig:thooft-v3}
}
\\
\subfigure[]{
$\displaystyle 
\begin{matrix}\begin{picture}(62,32)
\SetOffset(31,14)
\Gluon(-15,0)(15,0){3}{4}
\ArrowArc(-23,0)(8,0,90)
\CArc(-23,0)(8,270,360)
\Vertex(-15,0){2}
\ArrowArc(23,0)(8,180,270)
\CArc(23,0)(8,90,180)
\Vertex(15,0){2}
\Text(-25,8)[cr]{$\bar\jmath$}
\Text(-25,-8)[cr]{$i$}
\Text(25,8)[cl]{$k$}
\Text(25,-8)[cl]{$\bar l$}
\end{picture}\end{matrix}
\:=\:
\begin{matrix}\begin{picture}(62,32)
\SetOffset(31,14)
\ArrowLine(23,8)(-23,8)
\ArrowLine(-23,-8)(23,-8)
\Text(-25,8)[cr]{$\bar\jmath$}
\Text(-25,-8)[cr]{$i$}
\Text(25,8)[cl]{$k$}
\Text(25,-8)[cl]{$\bar l$}
\end{picture}\end{matrix}
\:-\: \frac1{\NC} \left(
\begin{matrix}\begin{picture}(62,32)
\SetOffset(31,14)
\ArrowArc(-23,0)(8,-90,90)
\ArrowArc(23,0)(8,90,270)
\Text(-25,8)[cr]{$\bar\jmath$}
\Text(-25,-8)[cr]{$i$}
\Text(25,8)[cl]{$k$}
\Text(25,-8)[cl]{$\bar l$}
\end{picture}\end{matrix} \right)
$
\label{fig:thooft-fierz}
}
\caption{
  't~Hooft double line diagrams for (a) the quark-gluon vertex, (b) a
  three-gluon vertex, and (c) the ${\rm SU}(\NC)$ Fierz identity \eqref{eq:sun-fierz}.}
\label{fig:tHooft-gluon}
\end{figure}

In \cite{tHooft:1973jz}, 't Hooft described a novel way of analysing
the colour structure of an amplitude. One assigns a directed line to
each line in the graph that carries a charge in the fundamental
representation. In our convention where all particles are out-going,
lines start on fundamental indices $i$ and end on antifundamental
indices $\bar\jmath$, such as with the gluon vertex of
fig.~\ref{fig:thooft-qqg}.  A trace of gauge matrices, such as at a
three-gluon vertex, joins up fundamental lines cyclically as shown in
fig.~\ref{fig:thooft-v3}.  Gluon lines contract fundamental indices,
and may be dealt with using the ${\rm SU}(\NC)$ Fierz identity. This
manifests itself as two possible ways of connecting lines to the
fundamental indices of the matrices on which the gluons end, and so
connections between vertices in double-line diagrams are made using
the right-hand side of in fig.~\ref{fig:thooft-fierz}.

The third column of fig.~\ref{fig:qcd-feynrules} shows the
\emph{colour ordered Feynman rules} for QCD, which are derived from
the normal Feynman rules by taking their colour factors and writing
them in terms of [traces of] the various $T^a$ matrices, and one then
simply extracts the term bearing the leading colour order (\eg\
$\tr(T^aT^bT^cT^d)$ for the four gluon vertex).  These rules are used
to compute colour-ordered partial amplitudes, which we obtain by
summing all planar graphs whose external legs' labels (into which we
remind the reader we have subsumed both colour and helicity
information) are ordered so as to match the colour factor with which
the partial amplitude is associated. That these graphs must be planar
and ordered in this fashion is clear upon considering the associated
't~Hooft diagrams: non-planar graphs contribute to other colour
orders' partial amplitudes.

\subsubsection*{Symmetries of gluonic partial amplitudes}

The pure-gluon partial amplitude is gauge invariant (inasmuch as it
does not change under redefinitions of polarisation $\epsilon_i
\rightarrow \epsilon_i + \alpha p_i$), and satisfies a number of
useful symmetry properties.
\begin{itemize}
\item The argument list is only defined up to cycles, so $A(1\cdots n)
  = A(2\cdots n1) = \cdots = A(n,1,\dots,n-1)$.
\item Under complete reversal of the arguments, $A(1\cdots n) = (-1)^n
  A(n\cdots 1)$. To see this, note that under \emph{complete reversal}
  of a planar diagram, the three-point Feynman rule changes sign but
  the four point rule does not. An $n$-gluon tree with $m$ 4-point
  vertices has $n-2-2m$ 3-point vertices, so the overall sign change
  is $(-1)^{n-2-2m} = (-1)^n$.
\item The \emph{dual Ward identity} links cycles of $n-1$ consecutive
  arguments:
  \[
  A(1\cdots n) + A(23\cdots1n) + A(34\cdots12n) \cdots +
  A(n-1,\dots,n-2,n) = 0.\] This can be seen from the construction of
  the amplitudes in terms of colour-ordered Feynman rules: one can
  always pair up each graph from one term with another graph in
  another term of the opposite sign. Alternatively, one works with the
  the gauge group ${\rm U}(\NC)$ and lets particle $n$ be the photon
  in \eqref{eq:colourorder-decomp-tree}. Carry out the sum noting now
  that $T^{a_n}$ is proportional to the identity, and by collecting
  terms of the same trace structure we arrive at the expression above.
\end{itemize}

Already, we can see that these symmetries cut down on the amount of
work we have to do to assemble amplitudes. For example, for five
gluons, we first note that parity invariance means that we only need
to consider partial amplitudes with up to two negative helicity
gluons, since for more we can flip the helicities by complex
conjugation. Furthermore, we note (as seen in section
\ref{ssec:parke-taylor}), the partial amplitudes with fewer than two
negative helicity gluons vanish. Thus, applying the rotational
symmetry, we are left with just two independent partial amplitudes:
\[
A(1^-2^-3^+4^+5^+) \quad\text{and}\quad A(1^-2^+3^-4^+5^+).
\]

\subsection{The Parke--Taylor amplitudes}
\label{ssec:parke-taylor}

Some of the most important colour-ordered amplitudes were first given
in ref.\ \cite{Parke:1986gb} by Parke and Taylor. The following
expressions are for tree-level colour-ordered gluon amplitudes with an
arbitrary number of particles with positive helicity, and up to two
with negative helicity:
\begin{align}
  A(1^+\cdots n^+) &= 0,
  \label{eq:allplus-tree}\\
  A(1^- 2^+ \cdots n^+) &= 0,
  \label{eq:oneminus-tree}\\
  A(1^- 2^+ \cdots j^- \cdots n^+) &= ig^{n-2} \frac{\langle 1\: j
    \rangle^4}{\langle 1\:2 \rangle \langle 2\:3 \rangle \cdots
    \langle n-1,n \rangle \langle n\:1\rangle}.
  \label{eq:parke-taylor}
\end{align}
(The amplitudes with the opposite helicity follow by parity symmetry
upon exchanging $\langle\:\rangle \leftrightarrow [\:]$.) The last
amplitude above is the first one which is non-vanishing at tree-level
for the largest difference between the number of positive- and
negative-helicity gluons, and hence is referred to as \emph{maximally
  helicity-violating}, or MHV for short. When the helicities are
flipped, we refer to this as an \mhvbar\ amplitude. An amplitude with
three negative-helicity gluons is termed a next-to-MHV or NMHV
amplitude, with four a NNMHV and so-on so that an N${}^n$MHV amplitude
has $n+2$ negative-helicity gluons.

It is fairly straightforward to prove the first two expressions
\eqref{eq:allplus-tree} and \eqref{eq:oneminus-tree} by considering
the (colour-ordered) Feynman graphs that contribute to them. For the
all-$+$ amplitude with $n$ gluons, first consider the gluon vertices
of fig.~\ref{fig:qcd-feynrules}: each one can contribute at the most
one external momentum $p^\mu_i$, and there are at the most $n-2$
vertices. Now contract the polarisation vectors $\epsilon_i^+ =
\epsilon^+_\mu(p_i,k_i)$ into this expression.  There are $n$ of
these, so each term must contain a factor of the form $\epsilon^+_i
\cdot \epsilon^+_j$, which vanishes by \eqref{eq:poln-ident-1} if we
choose all the reference momenta to be the same; the amplitude
therefore vanishes by gauge invariance. When one gluon has negative
helicity, the same thing happens: we obtain factors $\epsilon^-_i
\cdot \epsilon^+_j$, $j\ge2$, which vanish by \eqref{eq:poln-ident-2}
if we pick reference momenta $k_j = p_1$ for $j\ge2$.

Berends and Giele proved \eqref{eq:parke-taylor} in
\cite{Berends:1987me} using a recursive technique that connected
together off-shell gluon currents. We will not reproduce this here;
instead, we refer the reader to the proof obtained by Britto, Cachazo
and Feng in \cite{Britto:2004ap}, and shown in section
\ref{ssec:background-bcf-mhv}, using on-shell recursion relations.

Now, it is clear these are remarkably simple expressions: even for
tree-level amplitudes, given what we remarked earlier about the growth
in the complexity of the expressions arising in conventional
perturbation theory. What might not be so clear is precisely why.
They hint that gauge theories have a much simpler structure than
implied by traditional Feynman diagrams --- and as such we might well
ask how this simplicity extends to higher order computations, and
moreover wherein lie its origins.

\subsection{Colour ordering at one loop}
\label{ssec:background-colourorder-1loop}

At one loop, the colour ordered decomposition is more complicated due
to the presence of non-planar diagrams. For the planar pieces, the
leading colour structure is still of the form $\tr(T^{a_1} \cdots
T^{a_n})$, but is amplified by a factor of $\NC$ arising from the
trace of the identity of the ${\rm SU}(\NC)$'s fundamental
representation. The non-planar graphs contribute structures of the
form $\tr(T^{a_1} \cdots T^{a_{c-1}}) \tr(T^{a_c} \cdots T^{a_n})$,
which are correspondingly suppressed relative to the planar terms. The
one-loop colour ordered decomposition for gluon amplitudes was
obtained in \cite{Bern:1990ux} as
\begin{multline} {\cal A}^{\text{1 loop}}(1\cdots n) = \sum_{\sigma\in
    S_n/\mathbb{Z}_n} \NC
  \tr(T^{a_1} \cdots T^{a_n}) A_{n;1}(\sigma(1) \cdots \sigma(n)) \\
  + \sum_{c=2}^{\lfloor n/2 \rfloor + 1} \sum_{\sigma\in S_n/S_{n;c}}
  \tr(T^{a_{\sigma(1)}} \cdots T^{a_{\sigma(c-1)}})
  \tr(T^{a_{\sigma(c-1)}} \cdots T^{a_{\sigma(n)}}) A_{n;c}(\sigma(1)
  \cdots \sigma(n)).
\end{multline}
$S_{n;c}$ is the subset of $S_n$ that leaves the traces invariant.
Amazingly, it turns out \cite{Bern:1994zx} that the non-planar partial
amplitudes $A_{n;c}(\sigma(1) \cdots \sigma(n))$, $c>1$, can be
computed as sums of permutations of the \emph{primitive} amplitudes
$A_{n;1}(\sigma(1) \cdots \sigma(n))$, which are obtained from planar
graphs. As such, the majority of the literature is devoted to the
computation of the latter.

\section{Techniques from Supersymmetry}
\label{sec:background-susy}

While QCD is not itself a supersymmetric theory, we can use
prototypical unbroken supersymmetric theories as elements in the
calculation of QCD amplitudes, and thereby leverage the cancellations
and simplifications the supersymmetry produces.

\subsection{Supersymmetric Ward identities}
\label{sec:swis}

SUSY transformations mix fermions and bosons in a supermultiplet, and
so construct relationships between pure-gluon amplitudes and those
also containing gluinos. This is useful for QCD since \emph{at tree
  level} the theory is effectively supersymmetric (once we have
removed the colour information) by virtue of sharing its kinematic
structure with unbroken SUSY theories. (Because SUSY multiplets have a
different field content to QCD, this breaks at loop level; however,
this turns out to be a blessing in disguise as discussed in section
\ref{ssec:background-susy-loops}.)  We can map the gluon--gluino
amplitudes back to gluon--quark amplitudes (albeit with a few caveats
for certain configurations of pure-quark amplitudes).

Consider a `parallel' theory with ${\cal N}=1$ supersymmetry and a
vector supermultiplet consisting of a gauge field $A_\mu$, with two
bosonic physical degrees of freedom, and the Majorana spinor $\Lambda$
(also in the adjoint representation) with its two fermionic degrees of
freedom. The theory has the lagrangian density
\begin{equation}
  \label{eq:superqcd}
  {\cal L} = \tr \left( -\frac14 F^{\mu\nu} F_{\mu\nu} + i
    \bar\Lambda \slashed D \Lambda + \frac{D^2}2 \right),
\end{equation}
where $D$ is the auxiliary field.  Now let $q$ be a null vector with
corresponding chiral Weyl polarisation spinors $\varphi(q)$ and
$\bar\varphi(q)$ (defined, \eg\ in \eqref{eq:lc-spinors} upon
identifying $\varphi = \lambda$, $\bar\varphi = \tilde\lambda$), from
which we define $\eta_\alpha(q) = \theta \varphi_\alpha(q) / \sqrt2$
and $\bar\eta^{\dot\alpha}(q) = \theta \bar\varphi^{\dot\alpha}(q) /
\sqrt2$ where $\theta$ is a Grassman number.  Then for on-shell fields
in the asymptotic (free) limit, one can show that the commutators of
the SUSY generator $Q(\eta) = \eta^\alpha Q_\alpha +
\bar\eta_{\dot\alpha} \bar Q^{\dot\alpha}$ with the gluon and gluino
helicity--momentum annihilation operators, $A^\pm(k)$ and
$\Lambda^\pm(k)$ respectively, are as follows\footnote{These may be
  obtained directly by considering the effect of SUSY transformations
  acting on the annihilation operators expressed in terms of the
  asymptotic (free) fields \cite{Grisaru:1976vm}. We note that the
  convention used here differs slightly from that found in
  \cite{Dixon:1996wi}, which results in a sign difference for every
  fermion-antifermion pair in the amplitudes.}:
\begin{align}
  \label{eq:susy-comm-A}
  [Q(\eta), A^\pm(k)] &= i \Gamma^\pm(k, \eta) \Lambda^\pm(k), \\
  \label{eq:susy-comm-L}
  [Q(\eta), \Lambda^\pm(k)] &= i \Gamma^\mp(k, \eta) A^\pm(k),
\end{align}
with
\begin{equation}
  \label{eq:susy-comm-G}
  \Gamma^+(k,\eta) = -\theta [k\:q] \quad\text{and}\quad
  \Gamma^-(k,\eta) = \theta \langle k\:q \rangle.
\end{equation}
Applying the commutation relations
\eqref{eq:susy-comm-A}--\eqref{eq:susy-comm-G} to strings of
annihilation operators gives relationships known as
\emph{supersymmetric Ward identities} \cite{Grisaru:1976vm,
  Grisaru:1977px} that link different amplitudes. Let us study some
examples of this in action to see how it is useful for obtaining QCD
amplitudes.

First, let us use these ideas to derive the gluon--gluino MHV
amplitude.  We define $\lvert \text{vac} \rangle = S \lvert 0
\rangle$, where the $S$-matrix $S$ evolves asymptotic `in'
states (defined in the infinite past) to `out' states (defined in the
infinite future), and $\lvert 0 \rangle$ is the free vacuum state.
$Q(\eta)$ annihilates the vacuum, so
\[
0 = \langle 0 \rvert [Q(\eta), A^-_1 A^+_2 \cdots A_j^- \cdots
A^+_{n-1} \Lambda_n^+] \lvert \text{vac} \rangle .
\]
We expand the commutator, and choose $\eta$, $\bar\eta$ to be the
spinors associated with the null momentum $p_j$ to obtain (noting, of
course, that the terms with two $\Lambda^+$ operators vanish by
helicity conservation)
\begin{align}
  \langle 0 \rvert \Lambda_1^- A_2^+ \cdots A_j^- \cdots A_{n-1}^+
  \Lambda_n^+ \lvert \text{vac} \rangle &= -\frac{\langle
    n\:j\rangle}{\langle 1 \:j\rangle} \langle 0 \rvert
  A_1^-A_2^+\cdots A_j^-\cdots A_n^+ \lvert \text{vac}
  \rangle \notag \\
  \Rightarrow A(\Lambda_1^- A_2^+ \cdots A_j^- \cdots A_{n-1}^+
  \Lambda_n^+) &= ig^{n-2} \frac{ \langle 1\:j \rangle^3 \langle j\:n
    \rangle}{\langle 1\:2 \rangle \langle 2\:3 \rangle \cdots \langle
    n-1, n\rangle \langle n\:1 \rangle}.
\end{align}
Similarly,
\begin{align}
  \langle 0 \rvert \Lambda_1^+A_2^+\cdots A_j^-\cdots
  A_{n-1}^+\Lambda_n^- \lvert \text{vac} \rangle &= \frac{\langle 1\:j
    \rangle}{\langle n \:j \rangle} \langle 0 \rvert A_1^+ A_2^+
  \cdots
  A_j^- \cdots A_{n-1}^+ A_n^- \lvert \text{vac} \rangle \notag \\
  \label{eq:swi-qg3q}
  \Rightarrow A(\Lambda_1^+ A_2^+ \cdots A_j^- \cdots A_{n-1}^+
  \Lambda_n^-) &= ig^{n-2} \frac{\langle 1\:j \rangle \langle n\:j
    \rangle^3}{\langle 1\:2 \rangle \langle 2\:3 \rangle \cdots
    \langle n-1,n \rangle \langle n\:1 \rangle}.
\end{align}

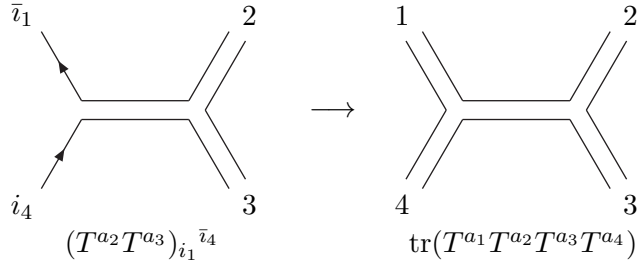
\begin{figure}[h]
\[\begin{matrix}
\begin{matrix}\begin{picture}(100,84)
\SetOffset(50,40)
\ArrowLine(-20,3.5356)(-35,29.5164)
\Line(-20,3.5356)(20,3.5356)
\Line(20,3.5356)(35,29.5164)
\Line(41.1237,25.9808)(26.1237,0)
\Line(26.1237,0)(41.1237,-25.9808)
\Line(35,-29.5164)(20,-3.5356)
\Line(20,-3.5356)(-20,-3.5356)
\ArrowLine(-35,-29.5164)(-20,-3.5356)
\Text(-38,32)[br]{$\bar\imath_1$}
\Text(40,32)[bl]{$2$}
\Text(40,-32)[tl]{$3$}
\Text(-38,-32)[tr]{$i_4$}
\end{picture}\end{matrix}
& \longrightarrow &
\begin{matrix}\begin{picture}(100,84)
\SetOffset(50,40)
\Line(-20,3.5356)(-35,29.5164)
\Line(-20,3.5356)(20,3.5356)
\Line(20,3.5356)(35,29.5164)
\Line(41.1237,25.9808)(26.1237,0)
\Line(26.1237,0)(41.1237,-25.9808)
\Line(35,-29.5164)(20,-3.5356)
\Line(20,-3.5356)(-20,-3.5356)
\Line(-35,-29.5164)(-20,-3.5356)
\Line(-41.1237,25.9808)(-26.1237,0)
\Line(-26.1237,0)(-41.1237,-25.9808)
\Text(-40,32)[br]{$1$}
\Text(40,32)[bl]{$2$}
\Text(40,-32)[tl]{$3$}
\Text(-40,-32)[tr]{$4$}
\end{picture}\end{matrix} \\
(T^{a_2} T^{a_3})_{i_1}{}^{\bar\imath_4}
&&
\tr(T^{a_1} T^{a_2} T^{a_3} T^{a_4})
\end{matrix}\]
\caption{'t~Hooft double-line diagrams relating QCD to SUSY colour
  structures with one quark-antiquark pair. On the left is the QCD
  structure with the fundamental indices, the quark line running on
  the outside. It is lifted to the colour trace on the right by adding
  an extra line.}
\label{fig:thooft-1q}
\end{figure}

Extracting a QCD amplitude from this is straightforward. For example,
when $n=4$,
\begin{equation}
  \label{eq:example-swi}
  \langle 0 \rvert \Lambda_1^- A_2^+ A_3^- \Lambda_4^+ \lvert \text{vac}
  \rangle
  = - \frac{\langle 4\:3 \rangle}{\langle 1\:3 \rangle} \langle 0 \rvert A_1^- A_2^+ A_3^- A_4^+ \lvert \text{vac}
  \rangle.
\end{equation}
Now since everything is in the adjoint representation, we can
interpret both sides as \emph{partial} amplitudes associated with $\tr
(T^{a_1} T^{a_2} T^{a_3} T^{a_4})$.  Conversely, we can interpret the
LHS of \eqref{eq:example-swi} as $A(1^-_{\rm q} 2^+ 3^- 4^+_{\bar{\rm
    q}})$, to be associated with the $(T^{a_2} T^{a_3})\ud14$ colour
structure, as follows. We draw the 't~Hooft double-line diagram for
the QCD colour structure such that the quark lines run on the outside,
shown for this case in fig.~\ref{fig:thooft-1q}.  To lift to the SUSY
side, we add an extra line alongside each quark line. Note that in the
case of up to one quark-antiquark pair, there is a unique way to do
this, hence we can map these QCD partial amplitudes directly onto SUSY
partial amplitudes with the associated gluinos \emph{adjacent} to each
other.

Correspondingly, we can define the MHV amplitudes with a
quark-antiquark pair and a single gluon of negative helicity by
\begin{equation}
  \label{eq:swi1}
  A(1_{\rm q}^- 2^+ \cdots j^- \cdots (n\!-\!1)^+ n_{\bar{\rm q}}^+)
  =ig^{n-2} \frac{ \langle 1\:j \rangle^3 \langle j\:n
    \rangle}{\langle 1\:2 \rangle \langle 2\:3 \rangle \cdots \langle
    n-1, n\rangle \langle n\:1 \rangle}
\end{equation}
and
\begin{equation}
  \label{eq:swi2}
  A(1_{\rm q}^+ 2^+ \cdots j^- \cdots (n\!-\!1)^+ n_{\bar{\rm q}}^-)
  =ig^{n-2} \frac{\langle 1\:j \rangle \langle n\:j
    \rangle^3}{\langle 1\:2 \rangle \langle 2\:3 \rangle \cdots
    \langle n-1,n \rangle \langle n\:1 \rangle},
\end{equation}
both of which are associated with the $(T^{a_2} \cdots
T^{a_{n-1}})_{i_1}{}^{\bar\imath_n}$ colour structure.

To add more gluino pairs, we continue the process. For example, to
obtain four gluino amplitudes, start from
\begin{equation}
  \begin{split}
    0 &= \langle0\rvert [Q(\eta),\Lambda_1^-\Lambda_2^+A_3^-
    \Lambda_4^+] \lvert \text{vac} \rangle
    \\
    &=\langle0\rvert \Gamma^+(1,\eta)A_1^-\Lambda_2^+A_3^-\Lambda_4^+-\Gamma^-(2,\eta)\Lambda_1^-A_2^+A_3^-\Lambda_4^+\\
    &\hphantom{=\langle0\rvert}
    +\Gamma^-(3,\eta)\Lambda_1^-\Lambda_2^+\Lambda_3^-\Lambda_4^+
    +\Gamma^-(4,\eta)\Lambda_1^-\Lambda_2^+A_3^-A_4^+
    \lvert\text{vac}\rangle.
  \end{split}
\end{equation}
If we choose $\eta$ and $\bar\eta$ to be the spinors associated with
the momentum $p_4$, then
\begin{equation}
  A(\Lambda_1^-\Lambda_2^+\Lambda_3^-\Lambda_4^+)
  =-\frac{\langle 4\:2\rangle}{\langle 4\:3\rangle} A(\Lambda_1^-A_2^+A_3^-\Lambda_4^+ )
  =ig^2 \frac{\langle2\:4\rangle\langle 1\:3\rangle^3}
  {\langle1\:2\rangle\langle2\:3\rangle\langle3\:4\rangle\langle 4\:1\rangle}.
\end{equation}
Similarly
\begin{equation}
  A(\Lambda_1^+\Lambda_2^-\Lambda_3^+\Lambda_4^-) =
  \frac{\langle 3\:1\rangle}{\langle 4\:1\rangle} A(\Lambda_2^-A_3^+A_4^-\Lambda_1^+ )
  =ig^2 \frac{\langle2\:4\rangle^3\langle 1\:3\rangle}{\langle1\:2\rangle\langle2\:3\rangle\langle3\:4\rangle\langle 4\:1\rangle},
\end{equation} 
and
\begin{equation}
  A(\Lambda_1^+\Lambda_2^+\Lambda_3^-\Lambda_4^-)
  = -\frac{\langle 1\:2\rangle}{\langle 2\:4\rangle} A
  (\Lambda_3^-A_4^-A_1^+\Lambda_2^+ )
  =-ig^2 \frac{\langle3\:4\rangle^2}{\langle2\:3\rangle\langle 4\:1\rangle}.
\end{equation}

\begin{figure}[h]
\[\begin{matrix}
\begin{matrix}\begin{picture}(90,94)
\SetOffset(45,45)
\Line(30,35)(10,15)
\ArrowLine(10,15)(-10,15)
\Line(-10,15)(-30,35)
\Line(-30,-35)(-10,-15)
\ArrowLine(-10,-15)(10,-15)
\Line(10,-15)(30,-35)
\Text(-35, 35)[br]{$\bar\imath_1$}
\Text( 35, 35)[bl]{$i_2$}
\Text( 35,-35)[tl]{$\bar\imath_3$}
\Text(-35,-35)[tr]{$i_4$}
\end{picture}\end{matrix} 
& + &
\begin{matrix}\begin{picture}(90,94)
\SetOffset(45,45)
\Line(-35,30)(-15,10)
\ArrowLine(-15,10)(-15,-10)
\Line(-15,-10)(-35,-30)
\Line(35,-30)(15,-10)
\ArrowLine(15,-10)(15,10)
\Line(15,10)(35,30)
\Text(-35, 35)[br]{$i_1$}
\Text( 35, 35)[bl]{$\bar\imath_2$}
\Text( 35,-35)[tl]{$i_3$}
\Text(-35,-35)[tr]{$\bar\imath_4$}
\end{picture}\end{matrix}
& \longrightarrow &
\begin{matrix}\begin{picture}(90,94)
\SetOffset(45,45)
\Line(30,35)(10,15)
\ArrowLine(10,15)(-10,15)
\Line(-10,15)(-30,35)
\Line(-30,-35)(-10,-15)
\ArrowLine(-10,-15)(10,-15)
\Line(10,-15)(30,-35)
\Line(-35,30)(-15,10)
\ArrowLine(-15,10)(-15,-10)
\Line(-15,-10)(-35,-30)
\Line(35,-30)(15,-10)
\ArrowLine(15,-10)(15,10)
\Line(15,10)(35,30)
\Text(-35, 35)[br]{$1$}
\Text( 35, 35)[bl]{$2$}
\Text( 35,-35)[tl]{$3$}
\Text(-35,-35)[tr]{$4$}
\end{picture}\end{matrix} \\
\delta_{i_1}^{\bar\imath_2} \delta_{i_3}^{\bar\imath_4}
&&
\delta_{i_4}^{\bar\imath_1} \delta_{i_2}^{\bar\imath_3}
&&
\tr(T^{a_1} T^{a_2} T^{a_3} T^{a_4})
\end{matrix}\]
\caption{Certain amplitudes with two or more quark-antiquark pairs in
  QCD have multiple colour structures lifting to the same trace on the
  SUSY side.}
\label{fig:thooft-2q}
\end{figure}
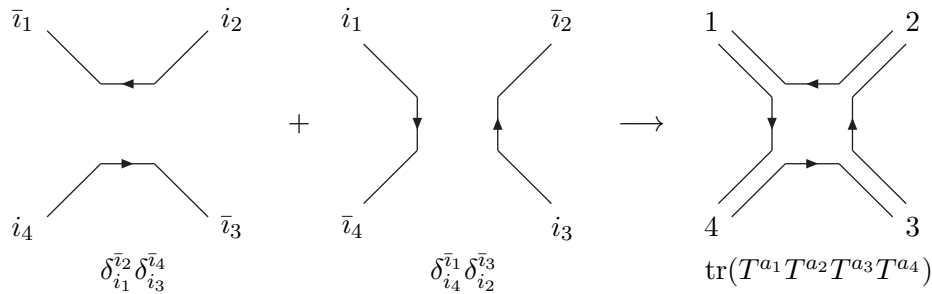

There is one caveat to note when mapping between QCD and the parallel
SUSY theory. Certain pure quark amplitudes with the alternating
$+-+-\dots$ helicity configuration contain colour structures whereby
the (anti)fundamental indices of adjacent quarks are connected by
Kronecker $\delta$s, such as shown for $n=4$ in
fig.~\ref{fig:thooft-2q}.  In this particular example, we see that two
separate QCD amplitudes bear colour structures that lift to the same
trace on the SUSY side upon adding additional lines to the 't~Hooft
diagram. Indeed, the quarks' charges are flipped between the left and
centre diagrams, necessary because the gluinos are in the adjoint
representation.  As such, we expect both QCD partial amplitudes to
contribute to the SUSY partial amplitude (something which we
demonstrate explicitly in section \ref{sec:mhvqcd-4qproblem}).

We end this section by noting that this programme can be extended to
include scalar particles by considering a ${\cal N}=2$ extended
supersymmetric theory with a vector supermultiplet (which has four
bosonic degrees of freedom, from the two gluon states, the scalar and
anti-scalar, and four fermionic degrees of freedom). From this we
obtain the result
\[
A(1^-2^-_P3^+_P4^+\cdots n^+) = \left(\frac{\langle 1\:2
    \rangle}{[1\:2]} \right)^{2s_P} A(1^-2^-_\phi3^+_\phi4^+\cdots
n^+)
\]
where $P$ is one of $\{A,\Lambda,\phi\}$ for gluon, gluino and scalar,
respectively. $s_P$ is the spin of particle species $P$.  (For a
scalar, $+$ and $-$ `helicity' refer to particle and antiparticle,
respectively). Whence,
\[
A(1^\pm2^+\cdots n^+)=0
\]
for any spin content.

It is important to point out here that \emph{for supersymmetric
  theories}, the SWIs were derived without resorting to any particular
perturbative expansion. Therefore they hold to all orders of
perturbation theory.

\subsection{Supersymmetric decomposition of loop amplitudes}
\label{ssec:background-susy-loops}

The properties (such as those above) endowed upon SUSY theories by
their extra symmetries can be exploited at the loop level to make
Yang--Mills and QCD calculations easier. By counting the respective
degrees of freedom, a gluon amplitude with a gluon running around the
loop
\begin{equation}
  \label{eq:susy-decomp-g}
  A^{\text{gluon}} = A^{{\cal N}=4} - 4 A^{{\cal N}=1,\chi} + A^{\text{scalar}}.
\end{equation}
Here,
\begin{itemize}
\item $A^{{\cal N}=4}$ is the same amplitude but with a ${\cal N}=4$
  supermultiplet running around the loop, which contains a gluon (one
  bosonic d.o.f.\ for each helicity), four gluinos (each contributing
  one fermionic d.o.f.\ for each helicity) and six complex scalars
  (two bosonic d.o.f.s each);
\item $A^{{\cal N}=1,\chi}$ has an ${\cal N}=1$ supermultiplet in its
  loop, consisting of a Weyl fermion and a complex scalar; and
\item $A^{\text{scalar}}$ has just a complex scalar in its loop.
\end{itemize}
Similarly for a fermion in the loop,
\begin{equation}
  \label{eq:susy-decomp-f}
  A^{\text{fermion}} = A^{{\cal N}=1,\chi} - A^{\text{scalar}}.
\end{equation}

Why is this helpful? Well firstly, we saw that the supersymmetric Ward
identities tell us that a significant number of amplitudes vanish
exactly. This means that gluon amplitudes with a $(\pm+\cdots+)$
helicity content receive contributions only from the scalar
loop. These have an algebraically simpler form, and as a result much
of the recent effort in computing multi-gluon loop amplitudes has
focused on obtaining scalar loop amplitudes.  For cases where the
supersymmetric contributions do not vanish, the supersymmetry provides
useful cancellations that simplify the calculation.  In particular,
the SUSY amplitudes are `cut-constructible' (see section
\ref{ssec:background-loops-cutcon}) --- a knowledge of its unitarity
cuts is sufficient to reconstruct the amplitude itself. At one-loop
level, these cuts can be computed from the tree graphs, which are much
easier to obtain.
 \section{The CSW construction}
\label{sec:background-csw}

Motivated by the observation \cite{Witten:2003nn} that MHV amplitudes
\eqref{eq:parke-taylor}
localise on complex lines in twistor space\footnote{Specifically,
  copies of $\mathbb{CP}^1$ embedded in $\mathbb{CP}^3$.}, and that
lines in twistor space correspond to points in space-time, Cachazo,
Svr\v cek and Witten proposed \cite{Cachazo:2004kj} a framework for
computing tree-level colour-ordered partial amplitudes for gluons by
sewing together MHV amplitudes according to the following rules:
\begin{enumerate}
\item Use MHV amplitudes as vertices.
\item Join the vertices together, helicities $+$ to $-$, using a
  scalar propagator $i/P^2$, where $P$ is the momentum flowing between
  the vertices.
\item For each leg of a vertex that joins to a propagator carrying
  momentum $P$, define its corresponding holomorphic spinor as
  $(\lambda_P)_\alpha = P_{\alpha\dot\alpha} \tilde\eta^{\dot\alpha}$,
  where $\tilde\eta$ is some arbitrary spinor.
\end{enumerate}
On the last point, we note that if we supplement $\tilde\eta$ with a
holomorphic spinor $\eta$ and use it define a null vector $\mu =
\eta\tilde\eta$, then we can construct the null projection of an
arbitrary four vector $p$ by
\begin{equation}
\label{eq:nullproj}
p_{\text{null}} = p - \frac{p^2}{2 \mu \cdot p} \mu.
\end{equation}
Then if $(p_{\text{null}})_{\alpha\dot\alpha} = \lambda_\alpha
\tilde\lambda_{\dot\alpha}$, $\lambda_\alpha = p_{\alpha\dot\alpha}
\tilde\eta^{\dot\alpha} / [\lambda\:\eta]$. This is coincident with
point 3 above: a propagator links a $+$ line on one vertex to a $-$
line on another, so
under a scaling $\lambda \rightarrow c \lambda$ the diagram will scale
like $c^{-2} \cdot (c^4 \cdot c^{-2}) = 1$.  Hence we can discard the
$[\lambda\:\eta]$ denominator.

Adding all the so-called CSW diagrams constructed this way returns the
desired scattering amplitude. Given the unbounded positive-helicity
valence of the vertices, an amplitude with $n_-$ negative helicity
gluons has $n_--1$ vertices; for $n$ gluons overall, the number of
diagrams grows no faster than $n^2$, so this method clearly presents a
significant computational advantage over Feynman diagrams.

\subsection{An example: $A(1^+2^-3^-4^-)$}
In \cite{Cachazo:2004kj}, the authors give examples of calculations
for the tree-level amplitudes $A(1^+2^-3^-4^-)$ as well as for the
five-gluon amplitudes. They also note verification of the technique
for certain configurations of higher-valence tree amplitudes, and even
apply it to obtain the $n$-gluon amplitude with three consecutive
gluons of negative helicity. We refer the interested reader to the
literature for these examples; here, we will repeat the calculation of
vanishing amplitude $A(1^+2^-3^-4^-)$, since this not only illustrates
the construction in action, but also forms a consistency check.

\begin{figure}[h]
\centering\subfigure[]{
\begin{picture}(100,100) \SetOffset(50,50)
      \ArrowLine(0,-17.6777)(0,17.6777) \Line(30.6186,35.3553)(0,17.6777)
      \Line(0,17.6777)(-30.6186,35.3553)
      \Line(-30.6186,-35.3553)(0,-17.6777)
      \Line(0,-17.6777)(30.6186,-35.3553) \Vertex(0,-17.6777){2}
      \Vertex(0,17.6777){2} \Text(-32,-37)[tr]{$4^-$}
      \Text(32,-37)[tl]{$3^-$} \Text(32,37)[bl]{$2^-$}
      \Text(-32,37)[br]{$1^+$} \Text(-3,-15.6)[br]{$+$}
      \Text(-3,15.6)[tr]{$-$} \Text(3,0)[cl]{$P$}
    \end{picture}
\label{fig:csw-mppp-a}
}\quad\subfigure[]{
    \begin{picture}(100,100) \SetOffset(50,50)
      \ArrowLine(17.6777,0)(-17.6777,0) \Line(35.3553,30.6186)(17.6777,0)
      \Line(17.6777,0)(35.3553,-30.6186)
      \Line(-35.3553,-30.6186)(-17.6777,0)
      \Line(-17.6777,0)(-35.3553,30.6186) \Vertex(-17.6777,0){2}
      \Vertex(17.6777,0){2} \Text(-37,-32)[tr]{$4^-$}
      \Text(-37,32)[br]{$1^+$} \Text(37,32)[bl]{$2^-$}
      \Text(37,-32)[tl]{$3^-$} \Text(-15.6,-3)[tl]{$-$}
      \Text(15.6,-3)[tr]{$+$} \Text(0,3)[bc]{$Q$}
    \end{picture}
\label{fig:csw-mppp-b}
}
\caption{CSW diagrams contributing to the $A(1^+2^-3^-4^-)$ amplitude.}
\label{fig:csw-mppp}
\end{figure}
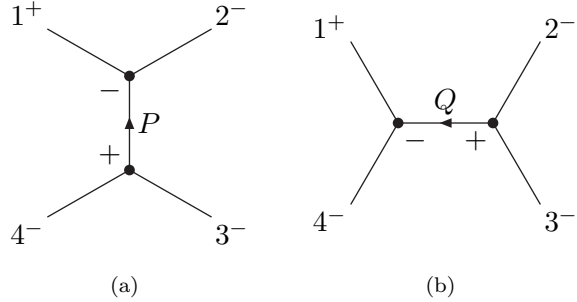

The contributing CSW rules diagrams are shown in
fig.~\ref{fig:csw-mppp}. Since the final result vanishes anyway, we
will drop any phase and coupling factors from propagators and the
formula \eqref{eq:parke-taylor} for the MHV amplitudes. The first
diagram, fig.~\ref{fig:csw-mppp-a}, has a contribution
\begin{equation}
  \label{eq:csw-example-1}
  \frac{\langle 2\:\lambda_P\rangle^3}{\langle 1\:3 \rangle \langle
    \lambda_P\:1 \rangle} \frac1{P^2} \frac{\langle 3\:4
    \rangle^3}{\langle 4\:\lambda_{-P} \rangle \langle \lambda_{-P}\:3 \rangle}  
\end{equation}
where $P = p_1 + p_2$. We pick a reference spinor $\tilde\eta$ (whose
specific value will not be needed). Then according to the
CSW prescription,
\begin{align*}
  (\lambda_P)_\alpha &= [(\lambda_1)_\alpha
  (\tilde\lambda_1)_{\dot\alpha} + (\lambda_2)_\alpha
  (\tilde\lambda_2)_{\dot\alpha}] \tilde\eta^{\dot\alpha}, \\
  (\lambda_{-P})_\alpha &= [(\lambda_3)_\alpha
  (\tilde\lambda_3)_{\dot\alpha} + (\lambda_4)_\alpha
  (\tilde\lambda_4)_{\dot\alpha}] \tilde\eta^{\dot\alpha},
\end{align*}
so that $\langle 2\:\lambda_P \rangle = \langle 2\:1 \rangle
\alpha_1$, $\langle \lambda_P\:1 \rangle = \langle 2\:1 \rangle
\alpha_2$, $\langle 4\:\lambda_{-P} \rangle = \langle 4\:3 \rangle
\alpha_3$ and $\langle \lambda_{-P}\:3 \rangle = \langle 4\:3 \rangle
\alpha_4$, where $\alpha_i := [i\:\eta]$. Substituting, and using
\eqref{eq:dot-brackets} to deal with the propagator,
\eqref{eq:csw-example-1} becomes
\begin{equation}
  \label{eq:csw-example-2}
  \frac{\langle 3\:4 \rangle}{[1\:2]} \frac{\alpha_1^3}{\alpha_2
    \alpha_3 \alpha_4}.
\end{equation}
Repeating the exercise for the graph of fig.~\ref{fig:csw-mppp-b}
gives, similarly,
\[
\frac{\langle 3\:2 \rangle}{[1\:4]} \frac{\alpha_1^3}{\alpha_2
  \alpha_3 \alpha_4}.
\]
Adding this to \eqref{eq:csw-example-2} results in an expression with
a numerator proportional to
\[
\langle 3\:4 \rangle [4\:1] + \langle 3\:2 \rangle [2\:1] =
\sum_{i=1}^4 \langle 3\:i \rangle [i\:1] = 0
\]
by conservation of momentum. Thus, the amplitude vanishes.

\subsection{Proving the CSW rules}
In the latter part of ref.\ \cite{Britto:2005fq}, the authors gave a
neat proof of the validity of the CSW construction by considering
Lorentz invariance and the singularity structure of tree-level
scattering amplitudes. It proceeds as follows.

It turns out that upon analytic continuation to complex momenta, a
tree amplitude can be determined entirely by its physical
singularities --- something which gives rise to recursion relations
between on-shell amplitudes (see section
\ref{sec:background-bcf}). Now suppose we calculate a particular
amplitude in two ways: by colour-ordered Feynman graphs, to give
$A_{\text{Feyn}}$, and $A_{\text{CSW}}$ by the CSW
construction. Both are clearly rational functions of the external
momenta's spinors. In ref. \cite{Cachazo:2004kj}, the authors argue
that $A_{\text{CSW}}$ contains the same physical singularities as
$A_{\text{Feyn}}$ (individual graphs in $A_{\text{CSW}}$ might
contain outstanding singularities of the form $1/\lambda^\alpha
P_{\alpha \dot\alpha} \tilde\eta^{\dot\alpha}$, but in their sum these
must
vanish by Lorentz invariance) so $A_{\text{CSW}} - A_{\text{Feyn}}$
must be a polynomial. For $n$ gluons, we can check directly that this
polynomial vanishes when $n=3,4$, and for $n>4$, such a polynomial
cannot exist on dimensional grounds. Therefore $A_{\text{CSW}} =
A_{\text{Feyn}}$.

\subsection{CSW rules for fermions}
One extension of the CSW rules that will have relevance to the work in
chapter \ref{cha:mhvqcd} is that of ref.\ \cite{Wu:2004jxa} which adds
massless quarks in the fundamental representation. This extension
works as one might expect: the propagator and off-shell prescription
are as in the purely gluonic case above, but one adds two new classes
of vertices (using the expressions for the corresponding amplitudes
found in ref.\ \cite{Mangano:1990by}) to the fold.

The first has either the quark or antiquark carrying a negative
helicity (by conservation of helicity) and one gluon of negative
helicity. Thus these vertices are
\begin{align}
  \label{eq:csw-ferm-q-gq+}
  \begin{split}
    \begin{matrix}
      \begin{picture}(112,50) \SetOffset(45,5.5) \ArrowLine(30,0)(0,0)
        \ArrowLine(0,0)(-30,0) \Gluon(0,0)(-25.9807,15){2}{5}
        \Gluon(0,0)(0,30){2}{5} \Gluon(0,0)(25.9807,15){2}{5}
        \DashCArc(0,0)(15,35,85){1} \DashCArc(0,0)(15,95,145){1}
        \Vertex(0,0){1.5} \Text(-32,0)[cr]{$1^-$}
        \Text(32,0)[cl]{$n^+$} \Text(0,32)[bc]{$j^-$}
        \Text(27,13)[bl]{$(n\!-\!1)^+$} \Text(-27,13)[br]{$2^+$}
      \end{picture}
    \end{matrix} &= A(1_{{\rm q}}^-,2^+,\dots,
    j^-,\dots,(n-1)^+,n_{\bar{\rm q}}^+) \\
    &= ig^{n-2} \frac{ \langle 1\:j \rangle^3 \langle j\:n
      \rangle}{\langle 1\:2 \rangle \langle 2\:3 \rangle \cdots
      \langle
      n-1, n\rangle \langle n\:1 \rangle}, \end{split} \\
  \label{eq:csw-ferm-q+gq-}
  \begin{split}
    \begin{matrix}
      \begin{picture}(112,50) \SetOffset(45,5.5) \ArrowLine(30,0)(0,0)
        \ArrowLine(0,0)(-30,0) \Gluon(0,0)(-25.9807,15){2}{5}
        \Gluon(0,0)(0,30){2}{5} \Gluon(0,0)(25.9807,15){2}{5}
        \DashCArc(0,0)(15,35,85){1} \DashCArc(0,0)(15,95,145){1}
        \Vertex(0,0){1.5} \Text(-32,0)[cr]{$1^+$}
        \Text(32,0)[cl]{$n^-$} \Text(0,32)[bc]{$j^-$}
        \Text(27,13)[bl]{$(n\!-\!1)^+$} \Text(-27,13)[br]{$2^+$}
      \end{picture}
    \end{matrix} &= A(1_{{\rm q}}^+,2^+,\dots,
    j^-,\dots,(n-1)^+,n_{\bar{\rm q}}^-) \\
    &= ig^{n-2} \frac{\langle 1\:j \rangle \langle n\:j
      \rangle^3}{\langle 1\:2 \rangle \langle 2\:3 \rangle \cdots
      \langle n-1,n \rangle \langle n\:1 \rangle}, \end{split}
\end{align}
associated with the colour structure of
\eqref{eq:colourstruct-2q}.\footnote{We note that our expressions have
  a sign difference with those of \cite{Wu:2004jxa} due to a different
  choice of external state fermion ordering.}  The second has two
quark-antiquark pairs. For this to be an MHV vertex, all the gluons
have positive helicity, and it is
\begin{multline}
  \label{eq:csw-ferm-qgqqgq}
    \begin{matrix}
      \begin{picture}(130,84) \SetOffset(54,40)
        \ArrowLine(28.5317,9.2705)(0,0)
        \ArrowLine(0,0)(28.5317,-9.2705)
        \ArrowLine(-28.5317,-9.2705)(0,0)
        \ArrowLine(0,0)(-28.5317,9.2705)
        \Gluon(0,0)(17.6336,24.2705){2}{5}
        \Gluon(0,0)(-17.6336,24.2705){2}{5}
        \Gluon(0,0)(-17.6336,-24.2705){2}{5}
        \Gluon(0,0)(17.6336,-24.2705){2}{5}
        \DashCArc(0,0)(15,59,121){1} \DashCArc(0,0)(15,239,301){1}
        \Vertex(0,0){1.5} \Text(-30,10)[cr]{$1^{h_1}$}
        \Text(30,10)[cl]{$(j\!-\!1)^{-h_j}$}
        \Text(-30,-10)[cr]{$n^{-h_1}$} \Text(30,-10)[cl]{$j^{h_j}$}
        \Text(-10,27)[br]{$2^+$} \Text(10,27)[bl]{$(j\!-\!2)^+$}
        \Text(-10,-27)[tr]{$(n\!-\!1)^+$}
        \Text(10,-27)[tl]{$(j\!+\!1)^+$}
      \end{picture}
    \end{matrix}
    = A\!\begin{aligned}[t] (&1^{h_1}_{{\rm q}}, 2^+, \dots ,(j\!-\!2)^+,
    (j\!-\!1)^{-h_j}_{\bar{\rm q}}; \\
    &j^{h_j}_{{\rm q}},(j\!+\!1)^+,\dots,(n\!-\!1)^+,n^{-h_1}_{\bar{\rm
        q}})
\end{aligned} \\
    = ig^{n-2}\frac{F(h_1, h_j)}{\langle 1\:n\rangle\langle
      j,j\!-\!1\rangle} \frac{\langle
      1,j\!-\!1\rangle}{\langle1\:2\rangle \cdots \langle
      j\!-\!2,j\!-\!1\rangle} \frac{\langle j\:n\rangle}{\langle
      j,j\!+\!1 \rangle \cdots \langle n\!-\!1,n\rangle}
\end{multline}
with
\begin{align*}
  F(+,+) &= \langle n , j\!-\!1 \rangle^2 &
  F(+,-) &= -\langle n \: j \rangle^2, \\
  F(-,+) &= -\langle 1 , j\!-\!1 \rangle^2, &
  F(-,-) &= \langle 1\:j \rangle^2, \\
\end{align*}
and associated with the colour structure of
\eqref{eq:colourstruct-4q}. The sub-leading amplitude associated with
\eqref{eq:colourstruct-4q-sub} is
\begin{multline}
  \label{eq:csw-ferm-qgqqgq-sub}
    \begin{matrix}
      \begin{picture}(130,84)
        \SetOffset(54,40)
        \ArrowLine(28.5317,9.2705)(0,0) \ArrowLine(0,0)(28.5317,-9.2705)
        \ArrowLine(-28.5317,-9.2705)(0,0)
        \ArrowLine(0,0)(-28.5317,9.2705)
        \Gluon(0,0)(17.6336,24.2705){2}{5}
        \Gluon(0,0)(-17.6336,24.2705){2}{5}
        \Gluon(0,0)(-17.6336,-24.2705){2}{5}
        \Gluon(0,0)(17.6336,-24.2705){2}{5} \DashCArc(0,0)(15,59,121){1}
        \DashCArc(0,0)(15,239,301){1} \GBoxc(0,0)(3,3){0}
\Text(-30,10)[cr]{$1^{h_1}$}
        \Text(30,10)[cl]{$(j\!-\!1)^{-h_j}$}
        \Text(-30,-10)[cr]{$n^{-h_1}$} \Text(30,-10)[cl]{$j^{h_j}$}
        \Text(-10,27)[br]{$2^+$} \Text(10,27)[bl]{$(j\!-\!2)^+$}
        \Text(-10,-27)[tr]{$(n\!-\!1)^+$}
        \Text(10,-27)[tl]{$(j\!+\!1)^+$}
      \end{picture}
    \end{matrix}
    = A_{(1)}\!\begin{aligned}[t](&1^{h_1}_{{\rm q}}, 2^+, \dots ,(j\!-\!2)^+,
    n^{-h_1}_{\bar{\rm q}};\\&
    j^{h_j}_{{\rm q}},(j\!+\!1)^+,\dots,(n\!-\!1)^+,
    (j\!-\!1)^{-h_j}_{\bar{\rm q}} )\end{aligned}
\\
    = ig^{n-2}\frac{F(h_1, h_j)}{\langle 1\:n\rangle\langle
      j,j\!-\!1\rangle} \frac{\langle 1\:n\rangle}{\langle1\:2\rangle
      \cdots \langle j\!-\!3, j\!-\!2 \rangle \langle j\!-\!2, n
      \rangle} \frac{\langle
      j,j\!-\!1\rangle}{\langle j,j\!+\!1 \rangle \cdots \langle
      n\!-\!2,n\!-\!1 \rangle \langle n\!-\!1,j\!-\!1 \rangle}.
\end{multline}
There are no MHV vertices with more quark-antiquark lines.  We notice
that in both cases the gluons lie to the \emph{right} of the quark
lines as one travels along them in the direction of the arrows. This
can be seen from the 't~Hooft diagram construction, where in planar
graphs the external fermion lines can always be arranged to run on the
outside.


 \section{BCF recursion relations}
\label{sec:background-bcf}

Britto, Cachazo and Feng discovered a recursion relation that shows
one how to construct tree-level amplitudes in terms of lower-valence,
on-shell amplitudes and a scalar propagator. This works by shifting
the external momenta by an amount parametrised by a complex number
$z$; the amplitude sought is therefore a rational function of $z$, and
we can use knowledge of its pole structure to reconstruct
it. Ultimately, every amplitude reduces to sums of products of MHV and
\mhvbar\ amplitudes.

Concretely, suppose we are to compute the amplitude $A(1\cdots n)$
for some helicity configuration. We
pick external lines $k$ and $l$ which without loss of generality we
may assume to have helicities $-,+$, $-,-$ or $+,+$,
respectively\footnote{The $+,-$ case can be dealt with by re-labelling
  the external momenta using the cyclic symmetry of partial
  amplitudes.}. We then shift these null momenta $p_k = \lambda_k
\tilde\lambda_k \rightarrow \hat p_k(z)$ and $p_l = \lambda_l
\tilde\lambda_l \rightarrow \hat p_l(z)$ by shifting their spinors according to
\begin{equation}
\label{eq:bcf-shift}
\tilde\lambda_k \rightarrow \tilde\lambda_k - z \tilde\lambda_l,
\quad
\lambda_l \rightarrow \lambda_l + z \lambda_k,
\end{equation}
which defines the amplitude $A(z) \equiv A(1,\dots,\hat
k(z),\dots,\hat l(z),\dots,n)$.
Now let us choose a partition\footnote{We defined `partition' here and
  hereafter to mean a choice of contiguous, non-intersecting,
  sequential subsets.} of the external lines, $i,\dots,j$ such that
this range includes $l$. Since we are dealing with tree diagrams,
there is (for general momenta) a unique propagator carrying momentum
$P_{ij} := p_i + \dots + p_j$, and all propagators can be found this
way by choosing an appropriate range. $P_{ij}$ becomes shifted to
\begin{equation}
\label{eq:bcf-p-shift}
\hat P_{ij}(z) = P_{ij} + z \lambda_k \tilde\lambda_l.
\end{equation}
Now $A(z)$ will have poles whenever the $\hat P_{ij}(z)^2 = 0$, \ie\ when
the shifted propagators go on shell. As
\[\begin{split}
\hat P_{ij}(z)^2 &= \tfrac12 \hat P_{ij}(z)_{\alpha \dot\alpha}
\hat P_{ij}(z)^{\alpha \dot\alpha} \\
&= \tfrac12 \{ (P_{ij})_{\alpha \dot\alpha} (P_{ij})^{\alpha
  \dot\alpha} + 2z (P_{ij})_{\alpha \dot\alpha} \lambda_k^\alpha
\tilde\lambda_l^{\dot\alpha} \} \\
&= P_{ij}^2 - z \langle k \rvert P_{ij} \lvert l ],
\end{split}\]
these poles are located at
\begin{equation}
\label{eq:bcf-poles}
z = z_{ij} := \frac{P^2}{\langle k \rvert P_{ij} \lvert l ]},
\end{equation}
using the notation
\begin{equation}
\label{eq:bcf-anglesq-not}
\langle k \rvert P_{ij} \lvert l ] \equiv - \lambda^\alpha_k P_{\alpha
  \dot\alpha} \tilde\lambda^{\dot\alpha}_l.
\end{equation}
Since $A(z)$ is a rational function, these are its only
poles\footnote{There may be poles in $z$ arising from the action of the
  shifts \eqref{eq:bcf-shift} on the denominators of the polarisation
  vectors \eqref{eq:gluon-poln}, but these may always be removed by
  an appropriate choice of reference spinor and as such are gauge
  artifacts that do not contribute to the final amplitude.}.

Let us now assume that $A(z) \rightarrow 0$ as $|z| \rightarrow
\infty$, so that
\begin{equation}
\label{eq:bcf-poleint}
0 = \frac1{2\pi i} \oint_C \frac{A(z)}z 
= A(0) + \sum_{z\in \{z_{ij}\}} \frac1{z_{ij}} \Res_{z=z_{ij}} A(z),
\end{equation}
where $C$ is a contour at infinity. Since the residues in the sum
above are taken at points where we know propagators to go on shell,
each term in the sum must be product of the amplitudes to either side
of that propagator multiplied by the residue of the shifted
propagator, so
\[\begin{split}
\frac1{z_{ij}} \Res_{z=z_{ij}} A(z)
&= \frac{\langle k \rvert P_{ij} \lvert n ]}{P_{ij}^2} \sum_\pm
 A^\pm_{\text{L}}(z_{ij}) \Res_{z=z_{ij}} \frac1{P_{ij}^2 - z \langle k
   \rvert P_{ij} \lvert n ]} A^\mp_{\text{R}}(z_{ij}) \\
&= -\sum_\pm A^\pm_{\text{L}}(z_{ij}) \frac1{P_{ij}^2} A^\mp_{\text{R}}(z_{ij}),
\end{split}\]
where the sub-amplitudes
\begin{align}
A^\pm_{\text{L}}(z_{ij}) &=
A(\hat P_{ij}^\pm(z_{ij}),j+1,\dots,k(z_{ij}),\dots,i-1), \\
A^\mp_{\text{R}}(z_{ij}) &=
A(-\hat P_{ij}^\mp(z_{ij}),i,\dots,l(z_{ij}),\dots,j),
\end{align}
and all index arithmetic should be carried out cyclically.
Thus we obtain the recursion relation
\begin{multline}
\label{eq:bcf-rr}
A(1\cdots n) = \sum_{(i,j)\in{\cal P}} \sum_\pm
A(\hat P_{ij}^\pm(z_{ij}),j+1,\dots,k(z_{ij}),\dots,i-1) \\ \times
\frac1{P_{ij}^2}
A(-\hat P_{ij}^\pm(z_{ij}),i,\dots,l(z_{ij}),\dots,j),
\end{multline}
where $\cal P$ is the set of all partitions into two of ranges of external lines
that include line $l$. This may be described schematically by the
diagram of fig.~\ref{fig:bcf-rr-diag}.

\begin{figure}[h]
\[
A(1\cdots n) = \sum_{(i,j)\in{\cal P}} \sum_\pm \begin{matrix}
\begin{picture}(145,74)
\SetOffset(75,35)
\ArrowLine(-30,0)(30,0)
\Line(30,0)(45,25.9808)
\Line(30,0)(45,-25.9808)
\Line(-30,0)(-45,25.9808)
\Line(-30,0)(-45,-25.9808)
\Line(-30,0)(-60,0)
\Line(30,0)(60,0)
\BCirc(30,0){9}
\BCirc(-30,0){9}
\Text(-30,0)[cc]{$A_{\text{L}}$}
\Text(30,0)[cc]{$A_{\text{R}}$}
\DashCArc(-30,0)(20,125,175){1}
\DashCArc(-30,0)(20,185,235){1}
\DashCArc(30,0)(20,5,55){1}
\DashCArc(30,0)(20,305,355){1}
\Text(0,-4)[tc]{$\hat P_{ij}(z_{ij})$}
\Text(20,2)[br]{$\mp$}
\Text(-20,2)[bl]{$\pm$}
\Text(47,28)[cl]{$i$}
\Text(47,-26)[cl]{$j$}
\Text(-47,28)[cr]{$i-1$}
\Text(-47,-26)[cr]{$j+1$}
\Text(-62,0)[cr]{$\hat k$}
\Text(62,0)[cl]{$\hat l$}
\end{picture}
\end{matrix}
\times \frac1{P_{ij}^2}
\]
\caption{Mnemonic diagram illustrating the terms of the BCF construction.
 Index arithmetic should be carried out cyclically.
  All external momenta are out-going.}
\label{fig:bcf-rr-diag}
\end{figure}

What remains is to prove that for Yang--Mills theory $A(z)$ vanishes
as $|z| \rightarrow 0$. In ref.\ \cite{Britto:2005fq}, the authors use the CSW
construction (discussed in section \ref{sec:background-csw}) to prove
this, and we reproduce this reasoning here. First, we recall the
possible choices of helicity for the gluons $k$ and $l$ as outline
above. We can restrict the analysis to the case where gluon $l$ has
$+$ helicity; the case where gluon $k$ is of $-$ helicity can be
treated by the same reasoning but using the conjugate CSW construction
with \mhvbar\ vertices.

Now a general CSW tree graph consists of a number of MHV vertices
continued off shell by the CSW prescription, multiplied by a number of
propagators. Under the momentum shifts, a subset of these propagators
may be shifted such that they vanish as $|z| \rightarrow \infty$. Now
in the CSW prescription, one defines the holomorphic spinor of a
propagator's momentum by $(\lambda_P)_\alpha = (\hat
P_{ij})_{\alpha\dot\alpha} \tilde\eta^{\dot\alpha}$,
where $\tilde\eta$ is some arbitrary spinor. But
we know that $\hat P_{ij} = P_{ij} + z \lambda_k \tilde\lambda_l$, so
if we pick $\tilde\eta = \tilde\lambda_l$,  $\lambda_P$ will be
independent of $z$. Thus, all MHV vertices that connect to it will be
$z$ independent, \emph{except} for the vertex connected to gluon
$l$. By postulation, $l$ has $+$ helicity, so its shifted spinor
$\lambda_l + z \lambda_k$ shows up in the vertex's denominator. Hence,
the contribution for the vertices vanishes as $|z| \rightarrow
\infty$, and the rest of the amplitude along with it.

\subsection{An example: the Parke--Taylor MHV amplitude}
\label{ssec:background-bcf-mhv}
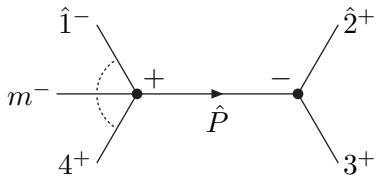
\begin{figure}[h]
\centering\begin{picture}(160,70)
\SetOffset(80,35)
\ArrowLine(-30,0)(30,0)
\Line(30,0)(45,25.9808)
\Line(30,0)(45,-25.9808)
\Line(-30,0)(-45,25.9808)
\Line(-30,0)(-45,-25.9808)
\Line(-30,0)(-60,0)
\Vertex(30,0){2}
\Vertex(-30,0){2}
\DashCArc(-30,0)(15,125,175){1}
\DashCArc(-30,0)(15,185,235){1}
\Text(0,-4)[tc]{$\hat P$}
\Text(28,2)[br]{$-$}
\Text(-28,2)[bl]{$+$}
\Text(47,28)[cl]{$\hat2^+$}
\Text(47,-26)[cl]{$3^+$}
\Text(-47,28)[cr]{$\hat1^-$}
\Text(-47,-26)[cr]{$4^+$}
\Text(-62,0)[cr]{$m^-$}
\end{picture}
\caption{The only term contributing to the BCF recursion computation
  of the $A(1^-2^+\cdots m^- \cdots n^+)$ MHV amplitude. The vertex on
  the left is an MHV amplitude, obtained by a lower-valence
  Parke--Taylor expression. The vertex on the right is a three-gluon
  \mhvbar\ amplitude. Missing lines are for $+$-helicity gluons.
  All external momenta are out-going.}
\label{fig:bcf-pt-diag}
\end{figure}

The BCF recursion relations provide a very neat, quick proof of the
expression \eqref{eq:parke-taylor}
\[
A(1^-2^+\cdots m^- \cdots n^+)
= ig^{n-2} \frac{\langle 1\:m \rangle^4}{\langle 1\:2 \rangle \cdots
  \langle n-1,n \rangle \langle n\:1\rangle}
\]
for the MHV amplitude. We choose to
shift momenta $1$ and $2$ according to
\[
\tilde\lambda_1 \rightarrow \tilde\lambda_1 - z \tilde\lambda_2,
\quad
\lambda_2 \rightarrow \lambda_2 + z \lambda_1
\]
For the purposes of this example, we will take $m>3$; the cases of
$m\le 3$ may be treated similarly. In this case, there is only one
choice of partition and intermediate helicity for the propagator,
specifically $i=2$ and $j=3$, for which neither of
the amplitudes in \eqref{eq:bcf-rr} vanish. This is illustrated in
fig.~\ref{fig:bcf-pt-diag}. The momentum flowing across this split is
$P_{23} = p_2 + p_3$, so under the shift $\hat P_{23}(z)^2$ vanishes when
$z=P_{23}^2 / \langle 1 | P | 2 ]$. Let us denote by $\hat P$ the
value $\hat P_{23}(z)$ takes for this choice of $z$.

Assume that the Parke--Taylor expression is valid for MHV amplitudes
with up to $n-1$ gluons.
If we drop the coupling constant factors, the BCF recursion relations
give
\begin{equation}
\label{eq:bcf-pt-1}
\frac{\langle \hat1 \: m \rangle^4}{\langle 4\:5 \rangle \cdots
    \langle n\:\hat1 \rangle \langle \hat1\:\hat P \rangle \langle
    \hat P\:4 \rangle}
\frac1{P_{23}^2}
\frac{[\hat 2\:3]^3}{[3\:\hat P][\hat P\:2]},
\end{equation}
where we remind the reader that the hats above the numbers denote a
shifted momentum. First, $P_{23}^2 = 2 p_2 \cdot p_3 = \langle 2\:3
\rangle [2\:3]$. Next, we remove the shifted pieces of the
momenta. Since we have only shifted the antiholomorphic part of $p_1$,
we can drop the hat from any $1$ that occurs in a $\langle \:
\rangle$, and likewise for any $2$ in a pair of $[\:]$
brackets. Next, we note that since $\hat P_{ij}(z) = P_{ij} + z
\lambda_k \tilde\lambda_l$, for general spinors $x$ and $y$
\begin{align}
\langle x\:\hat P_{ij}(z) \rangle [ \hat P_{ij}(z)\:l]
&= - \langle x \rvert \hat P_{ij}(z) \lvert l ]
= - \langle x \rvert P_{ij} \lvert l ], \\
\langle k\:\hat P_{ij}(z) \rangle [ \hat P_{ij}(z)\:y]
&= - \langle k \rvert \hat P_{ij}(z) \lvert y]
= - \langle k \rvert P_{ij} \lvert y ],\\
\end{align}
whence
\[
\langle \hat1\:\hat P \rangle = -\frac{\langle 1 | 2+3 |
  2]}{[\hat P\: 2]} = \frac{\langle 1\:3 \rangle [3\:2]}{[\hat P \: 2]},
\]
and similarly
\[
\langle \hat P \: 4 \rangle = -\frac{\langle 4\:3 \rangle
  [3\:2]}{[\hat P\:2]}, \quad
[3\:\hat P] = \frac{\langle 1\:2 \rangle [3\:2]}{\langle 1\:\hat P
  \rangle}, \quad\text{and}\quad
[\hat P\:2] = \frac{\langle 1\:3 \rangle [3\:2]}{\langle 1\:\hat P
  \rangle}.
\]
Inserting into \eqref{eq:bcf-pt-1}, we find we readily obtain the
Parke--Taylor expression for the MHV amplitude.

\subsection{A direct proof of the CSW rules}
By a modification of the BCF recursion relations (studied in section
\ref{sec:background-bcf}), Risager obtained a direct proof of the CSW
rules of section \ref{sec:background-csw}, and we refer the interested
reader to ref.\ \cite{Risager:2005vk} wherein is documented the exact
reasoning. The essence of Risager's proof is to construct a shift of
the momenta such that each propagator that occurs in a CSW rules graph
is shifted. This will occur if the antiholomorphic spinor of the
momentum of \emph{every} negative helicity gluon is shifted and the
shift is crafted such that no proper subset of them vanishes. With
this shift, any possible \mhvbar\ sub-amplitude vanishes, and so one
reasons inductively that the CSW construction must give the correct
amplitude if the lower-valence sub-amplitudes to either side of
partition do as well.

\subsection{Application to other field theories}
Although the BCF recursion relations were introduced in the context of
Yang--Mills theory and multi-gluon scattering, their derivation only
makes use of the fact that shifted amplitudes are vanishing in the
complex infinity limit, and that at tree-level they contain only
simple, physical poles. From a knowledge of these poles, the BCF
recursion relations provide a mechanism to determine tree-level
amplitudes.

There is much literature on the application of BCF recursion relations
to amplitudes of other particle content.  CSW rules and BCF recursion
relations for QED were derived in \cite{Ozeren:2005mp}. Ref.\
\cite{Luo:2005rx} adds massless fermions to the QCD relations, with
massive fermions in \cite{Ozeren:2006ft}. Recursion relations for
amplitudes involving massive propagating scalars can be found in
\cite{Badger:2005zh} (which is of particular use in the cut
construction of loop amplitudes, \cf\ \eqref{eq:susy-decomp-g} and
section \ref{sec:background-loops}), and with massive vector
bosons and fermions in ref.\ \cite{Badger:2005jv}.  BCF relations for
gravity were found in \cite{Bedford:2005yy, Cachazo:2005ca} with the
proof that the amplitudes vanish as $|z| \rightarrow \infty$ appearing
in ref.\ \cite{Benincasa:2007qj}.

\section{Loop level techniques}
\label{sec:background-loops}

In this section we discuss two techniques that are important for the
calculation of one-loop amplitudes. The first is (generalised)
unitarity and cut construction, and the second is the application of
the CSW rules in one-loop diagrams.  Of course, these are not the only
techniques on the market for loop amplitudes. Two others worth
mentioning are the twistor-space studies of Cachazo \etal\
\cite{Cachazo:2004zb, Cachazo:2004by, Cachazo:2004dr, Britto:2004nj};
and so-called `bootstrapping' approaches \cite{Bern:2005hs,
  Bern:2005cq, Bern:2005ji, Berger:2006uc}, which combine unitarity
with on-shell recursion relations to obtain the rational parts of
non-cut-constructible amplitudes. We refer the interested reader to the
literature for further details on these approaches.

\subsection{Unitarity, generalised unitarity,
and cut construction}
\label{ssec:background-loops-cutcon}
The so-called `cut construction' of loop amplitudes stems from the
observation that, on general grounds, a gauge theory loop amplitude
can be written as a rational function of kinematical invariants, plus
some linear combination with rational coefficients over a basis of
loop integrals that contain branch cuts in their kinematic
invariants. By studying an amplitude's cuts, one can (in certain
circumstances) deduce the loop integrals' coefficients; the advantage
of this is that, at one loop, these discontinuities can be computed
from a knowledge of on-shell tree-level amplitudes.

Although \emph{in general} the rational pieces cannot be determined
this way, in a number of theories \emph{all} the rational terms are
associated to functions that do contain cuts. Such theories are termed
\emph{cut-constructible}, and in \cite{Bern:1994cg} Bern, Dixon,
Dunbar and Kosower found a power-counting criterion that is satisfied
by these theories.  For an amplitude in a massless theory, this
criterion is that its $n$-point loop integrals have in their
numerators not more than one power of the loop momentum for $n=2$, and
not more than $n-2$ powers of the loop momentum for $n>2$, \ie\ given
the $n$-point loop integral with momentum $p_i$ flowing out of the
$i^\text{th}$ point
\begin{equation}
  \label{eq:genericloop}
  I_n[P] := \int
  \frac{d^{4-2\epsilon}L}{(2\pi)^{4-2\epsilon}} \frac{P(L)}{L^2 (L-p_1)^2
    (L-p_1-p_2)^2 \cdots (L+p_n)^2}
\end{equation}
where $P$ is a polynomial, then
\begin{equation}
  \label{eq:bdk-criterion}
  \deg P(L) \le \begin{cases} 1 & (n=2) \\ n-2 & (n>2) \end{cases}.
\end{equation}
A key result of \cite{Bern:1994cg} was showing that
supersymmetric Yang--Mills amplitudes satisfy this
criterion.

If the external momenta are all four dimensional, integrals satisfying
this criterion can be processed using a variety of reduction
techniques \cite{Brown:1952eu, Passarino:1978jh, 'tHooft:1978xw,
  Stuart:1987tt, Stuart:1989de, vanNeerven:1983vr, Melrose:1965kb,
  vanOldenborgh:1989wn, vanOldenborgh:1990fy, Aeppli:1992aa} through
which it can be shown \cite{Bern:1994cg} that they can be reduced to
linear combinations of scalar box, triangle and bubble functions (\ie\
those of the form $I_4[1]$ $I_3[1]$ and $I_2[1]$ in
\eqref{eq:genericloop}). These are all one-loop integrals for diagrams
with a massless scalar running between each of the external points;
the external momenta are partitioned amongst these points, the
momentum flowing from each point being the sum of those in its
partition.  For $n\ge4$ gluons, this set is listed with diagrams in
section \ref{sec:loopbasis} and summarised here as:
\begin{itemize}
\item the box functions $I_{4:jkl;i}^\text{4m}$,
  $I_{4:jk,i}^\text{3m}$, $I_{4:j;i}^\text{2me}$,
  $I_{4:j;i}^\text{2mh}$, $I_{4:i}^\text{1m}$ and $I_{4:i}^\text{0m}$;
\item the triangle functions $I_{3:jk;i}^\text{3m}$,
  $I_{3:j;i}^\text{2m}$, $I_{3:i}^\text{1m}$; and
\item the bubble function $I_{2:i}^\text{2m}$.
\end{itemize}
The subscripts denote how the external momenta are partitioned,
starting at $i$ and put into partitions of length $j$, $k$ and $l$.
The external momenta are taken to be on-shell, but when there is more
than one in a particular partition, their sum will in general not be
null. A $m$-mass function (noted in the superscripts of the functions
above) has $m$ such points. For box functions when $m=2$, there are
two possible (cyclically) distinct configurations: when the two null
legs are opposing, the so-called `two-mass easy' function
$I_{4:j;i}^\text{2me}$ (shown in fig.~\ref{fig:I42me}); and the
`two-mass hard' function $I_{4:j;i}^\text{2mh}$ when they are adjacent
(as in fig.~\ref{fig:I42mh}).

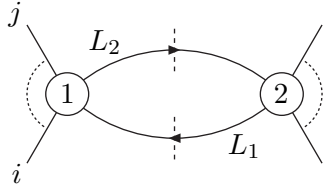
\begin{figure}[h]
  \centering \begin{picture}(220,80)
    \SetOffset(110,40)
    \ArrowArcn(0,-40)(56.5684,-225,-315)
    \ArrowArcn(0,40)(56.5864,315,225)
    \Line(-40,0)(-55,25.9808)
    \Line(-40,0)(-55,-25.9808)
    \Line(40,0)(55,25.9808)
    \Line(40,0)(55,-25.9808)
    \DashCArc(40,0)(15,305,55){1}
    \DashCArc(-40,0)(15,125,235){1}
    \DashLine(0,8.5685)(0,24.5685){2}
    \DashLine(0,-8.5685)(0,-24.5685){2}
    \BCirc(40,0){8}
    \BCirc(-40,0){8}
    \Text(-57,30)[cr]{$j$}
    \Text(-57,-30)[cr]{$i$}
    \Text(-20,15)[rb]{$L_2$}
    \Text(20,-15)[lt]{$L_1$}
    \Text(-40,0)[cc]{1}
    \Text(40,0)[cc]{2}
  \end{picture}
  \caption{Schematic diagram of the two-particle cut in the plane of
    the $(p_i + \cdots + p_j)^2$ invariant. The blobs represent the tree
    amplitudes $A^\text{tree}_1$ and $A^\text{tree}_2$ in the text.}
  \label{fig:bdk-1loop-cut}
\end{figure}

The procedure used by BDDK to extract the rational coefficients may be
summarised as follows:
\begin{enumerate}
\item For the amplitude being constructed, consider a particular
  kinematical channel (characterised by a partition into two of the
  external momenta), and draw all planar cut diagrams that contribute
  in this channel.  (Remember from section
  \ref{ssec:background-colourorder-1loop} that we only need the
  leading one-loop partial amplitude.)  These are constructed by
  connecting tree amplitudes either side of the cut, $A^\text{tree}_1$
  and $A^\text{tree}_2$, together with propagators as shown in
  fig.~\ref{fig:bdk-1loop-cut}. The propagators connect to adjacent
  lines in the amplitudes, whose remaining lines connect to external
  states according to colour order and helicity configuration.  The
  discontinuity in the branch cut of this channel's kinematic
  invariant is obtained by integrating over the two-particle
  \emph{phase space} of the internal particles instead of over the
  propagators \viz\ the Cutkosky rules \cite{Landau:1959fi,
    Mandelstam:1959bc, Cutkosky:1960sp}:
  \[
  \sum_\text{spin} \int d^4L \: \delta^{+4}(L_1^2) \delta^{+4}(L_2^2)
  A^\text{tree}_1 A^\text{tree}_2,
  \]
  where $L_1\equiv L$ and $L_2 = L_1 - p_i - \cdots - p_j$, and
  $\delta^{+4}(p^2) \equiv \theta(p^t) \delta^4(p^2)$. The sum
  here is over the admissible spin configurations of the internal
  particles, as determined by the helicity configuration of the
  particles on either side of the cut.
\item Reconstruct the Feynman integral. To do this, promote the
  $\delta^{+4}(p^2)$ functions to propagators $i/p^2$.
  \[
  \sum_\text{spin} \int d^4L \: \frac{i}{L_1^2} \frac{i}{L_2^2}
  A^\text{tree}_1 A^\text{tree}_2.
  \]
  The resulting Feynman integral has the same cut in the channel under
  consideration as the amplitude. One may use $L_1^2 = L_2^2 = 0$ in
  the numerator above. Now the Passarino--Veltman procedure is applied
  to the reconstructed integral to express it in terms of the basis of
  scalar integrals with rational coefficients.
\item By considering the reconstructed integrals of the cuts in all
  kinematic channels without over-counting each scalar integral's
  contribution, one deduces the rational coefficients and obtains the
  full amplitude.
\end{enumerate}
Note that by reconstructing the integral algebraically in step 2, one
avoids having to explicitly calculate the channel's discontinuity.

QCD is a theory that does not satisfy the BDDK criterion
\eqref{eq:bdk-criterion} and so cannot be reconstructed from knowledge
of the four-dimensional cuts alone. However, we have the
supersymmetric decompositions \eqref{eq:susy-decomp-g} and
\eqref{eq:susy-decomp-f}, so many terms in QCD amplitudes \emph{are}
cut-constructible. This leaves the scalar loops as the source of the
BDDK violation. (Indeed, in situations where \eqref{eq:bdk-criterion}
is not satisfied, the reduction procedure can result in additional
tensor functions (\ie\ those of the form $I_n[L^\mu L^\nu \cdots]$ in
\eqref{eq:genericloop}). In this case, as \cite{Bern:1994cg} notes,
one can construct linear combinations of these and the scalar
functions with rational coefficients that are rational functions in
four dimensions.) Nevertheless, the cut-containing terms in the scalar
loop amplitudes can be obtained through unitarity. The rational pieces
left over must be obtained by other means, such as on-shell recursion
relations \cite{Bern:2005hs, Bern:2005cq, Bern:2005ji, Berger:2006uc}
which use the cut-constructible pieces as inputs, or more direct
methods that extract the rational terms from the original Feynman
integrals \cite{Xiao:2006vr, Su:2006vs, Xiao:2006vt}.

\subsubsection*{Quadruple cuts}
${\cal N}=4$ amplitudes have a particularly special decomposition:
they can be constructed purely from scalar box integrals
\cite{Bern:1994cg} \ie
\begin{equation}
  \label{eq:N=4-boxes}
  A_{n;1}^{{\cal N}=4} = \sum_i b_i I^{1\rm m}_{4:i} + \sum_{ij}
  (c_{ij} I^{2\rm me}_{4:j;i} + d_{ij} I^{2\rm mh}_{4:j;i}) +
  \sum_{ijk} g_{ijk} I^{3\rm m}_{4:jk;i} + \sum_{ijkl} f_{ijkl}
  I^{4\rm m}_{4:jkl;i}.
\end{equation}
$b$, $c$, $d$, $g$ and $f$ are rational coefficients and in
\cite{Britto:2004nc}, Britto, Cachazo and Feng use \emph{generalised
  unitarity} to deduce the rational coefficients of these
functions. In this approach, one cuts two \emph{or more} lines by
replacing their propagators $i/p^2 \rightarrow \delta^{+4}(p^2)$. By
choosing different sets of lines to cut, we can isolate the
discontinuities of different cuts in the kinematic
variables.\footnote{Generalised unitarity is a rich topic, and we
  refer the reader to chapter 2 of \cite{Eden:1966aa} for a thorough
  treatment.}

\begin{figure}[h]
\centering
\begin{picture}(120,96)
\SetOffset(60,48)
\ArrowLine(-20,20)(20,20)
\ArrowLine(20,20)(20,-20)
\ArrowLine(20,-20)(-20,-20)
\ArrowLine(-20,-20)(-20,20)
\Line(20,20)(39.3185,25.1764)
\Line(20,20)(25.1764,39.3185)
\Line(20,-20)(39.3185,-25.1764)
\Line(20,-20)(25.1764,-39.3185)
\Line(-20,20)(-39.3185,25.1764)
\Line(-20,20)(-25.1764,39.3185)
\Line(-20,-20)(-39.3185,-25.1764)
\Line(-20,-20)(-25.1764,-39.3185)
\DashCArc(20,20)(15,20,70){1}
\DashCArc(20,-20)(15,290,340){1}
\DashCArc(-20,-20)(15,200,250){1}
\DashCArc(-20,20)(15,110,160){1}
\DashLine(14,0)(26,0){3}
\DashLine(-14,0)(-26,0){3}
\DashLine(0,14)(0,26){3}
\DashLine(0,-14)(0,-26){3}
\BCirc(20,20){8}
\BCirc(20,-20){8}
\BCirc(-20,-20){8}
\BCirc(-20,20){8}
\Text(-20,-20)[cc]{1}
\Text(-20,20)[cc]{2}
\Text(20,20)[cc]{3}
\Text(20,-20)[cc]{4}
\Text(0,-28)[tc]{$L_1$}
\Text(-28,0)[rc]{$L_2$}
\Text(0,28)[bc]{$L_3$}
\Text(28,0)[lc]{$L_4$}
\Text(-25.2,-41)[tc]{${}_{i}$}
\Text(-41,25.2)[cr]{${}_{i+j}$}
\Text(25.2,41)[bc]{${}_{i+j+k}$}
\Text(30,-20)[bl]{${}_{i+j+k+l}$}
\end{picture}
\caption{Quadruple cut of an amplitude that contributes the
  coefficient of the box function $I_{4:jkl;i}^\text{4m}$ (see
  fig.~\ref{fig:I44m}). The circular blobs represent the tree-level
  amplitudes $A^\text{tree}_{1,2,3,4}$ in the text.}
\label{fig:quadruplecut}
\end{figure}
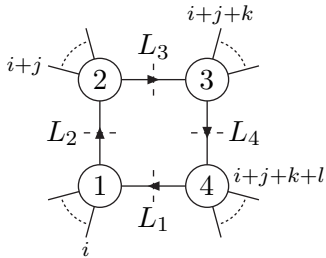

Britto, Cachazo and Feng compute the quadruple cuts of the LHS of
\eqref{eq:N=4-boxes}, and there are a number of such cuts
parametrised by the partitioning of the gluon labels into four.  An
example is shown in fig.~\ref{fig:quadruplecut}: this particular cut
is given by an expression proportional to
\begin{equation}
  \label{eq:N=4-cutintegral}
  \sum_{\text{spin}}
  \int d^4L \: \delta^{+4}(L_1) \delta^{+4}(L_2)
  \delta^{+4}(L_3) \delta^{+4}(L_4)
  \: A^\text{tree}_1 A^\text{tree}_2 A^\text{tree}_3 A^\text{tree}_4
\end{equation}
where $L_1 \equiv L$, $L_2 = L_i - p_i - \cdots - p_{j-1}$, \etc\ and
the sum is over the permissible spin configurations of the internal
particles.  (Since all internal lines are put on
shell by the cut, one must take care to deal with the vanishing of the
three-particle tree amplitudes for partitions containing just one external
gluon. In \cite{Britto:2004nc} this is
handled by working in $(2,2)$ signature; alternatively, one can use
complex momenta.) In four dimensions a quadruple cut freezes the
momentum integral so that $L$ has discrete set of solutions ${\cal S}$
to leave us with an expression proportional to
\[
\sum_{\text{spin},{\cal S}} \: A^\text{tree}_1 A^\text{tree}_2
A^\text{tree}_3 A^\text{tree}_4.
\]
On the other hand, the quadruple cut of a box function is unique to
that box function; that is, there is only one box function on the RHS
of \eqref{eq:N=4-boxes} with the cut computed in
\eqref{eq:N=4-cutintegral} --- namely, the one with the corresponding
external momentum configuration. As such, its coefficient must be
proportional to the product of tree amplitudes computed as given
above.

\subsubsection*{(Generalised) unitarity in
$4-2\epsilon$ dimensions}
While only the supersymmetric Yang--Mills theories are
cut-constructible in four dimensions, by analytically continuing to
$4-2\epsilon$ dimensions BDDK-violating theories also become so
\cite{vanNeerven:1985xr}. This can be understood on dimensional
grounds: terms in the amplitudes must contain factors of the form
$(-K^2)^{-\epsilon}$ for some kinematic invariant $K^2$ and so have
have cuts in the complex plane that allow us to deduce the presence of
terms that are rational in the four-dimensional limit.  This was used
in \cite{Bern:1995db} to obtain the $(\fourplus)$ amplitude with a
scalar in the loop by considering unitarity cuts,
and later in \cite{Brandhuber:2005jw} with generalised
unitarity (using triple and quadruple cuts) to re-derive the one-loop
$(\fourplus)$, $(\text{$-$$+$$+$$+$})$, four-gluon MHV and
$(\text{$+$$+$$+$$+$$+$})$ amplitudes in QCD. These amplitudes vanish
for supersymmetric theories, and so are equal to those with just a
complex scalar in the loop --- thereby avoiding the complications of
$4-2\epsilon$-dimensional polarisations in the internal states.  In
particular, we reproduce here the result \cite{Bern:1995db}
\begin{equation}
  \label{eq:fourplus-oneloop}
  A_{4;1}(1^+2^+3^+4^+) = A_{4;1}^\text{scalar}(1^+2^+3^+4^+) =
  \frac{2ig^4}{(4\pi)^{2-\epsilon}}
  K_4 \frac{[1\:2][3\:4]}{\langle 1\:2 \rangle \langle
    3\:4 \rangle}
\end{equation}
where $K_4$ is the $4-2\epsilon$-dimensional box function,
\begin{equation}
  \label{eq:K4}
  K_4 = -\frac{i}{(4\pi)^{\epsilon-2}}
  \int \frac{d^{4-2\epsilon}L}{(2\pi)^{4-2\epsilon}}
  \frac{\nu^4}{L^2 (L-p_1)^2 (L-p_1-p_2)^2 (L +p_4)^2}.
\end{equation}
As $\epsilon \rightarrow 0$, $K_4 \rightarrow -1/6$, leaving a
rational expression.

The scalar's $D$-dimensional loop momentum $L_D$ is split into two
parts, $L$ in four dimensions and $\nu$ in a $-2\epsilon$-dimensional
orthogonal spacelike subspace, and so the loop integration can be
expressed as two over these subspaces:
\begin{equation}
  \label{eq:Ddim-split}
  \int \frac{d^DL_D}{(2\pi)^D} = \int \frac{d^4L}{(2\pi)^4}
  \frac{d^{-2\epsilon}\nu}{(2\pi)^{-2\epsilon}}.
\end{equation}
Since $L_D$ is the only $4-2\epsilon$-dimensional quantity present
(all external momenta are kept four-dimensional), the
$-2\epsilon$-dimensional parts only show up as $\nu^2$ the
integrand\footnote{We take $\nu^2$ to be the scalar product with
  respect to a $(+\dots +)$-signature metric on the subspace.}  so the
integration over $\nu$ above can be traded for one over $\nu^2$ when
it comes to actually evaluating or analysing the integral. Since
$L_D^2 = L^2 - \nu^2$, the \emph{massless} scalar in $D$ dimensions
may be treated in the four dimensional tree amplitudes used to obtain
unitarity cuts as a \emph{massive} scalar of mass $\nu$.

\subsection{MHV amplitudes and loops}
\label{ssec:background-csw-loop}
Following the success of the CSW rules at computing tree-level gluon
scattering, one might consider applying them at one loop (and beyond).
Brandhuber, Spence and Travaglini (BST) developed a method for
applying the CSW rules at one loop, initially in the context of ${\cal
  N}=4$ scattering amplitudes in \cite{Brandhuber:2004yw}. In that
paper, their procedure was used to obtain an expression for
$A_{n;1}^\text{MHV}$ (the leading one-loop partial amplitude), in
exact agreement with the result obtained by cut construction in ref.\
\cite{Bern:1994zx}. To compute partial amplitudes using the BST
construction:
\begin{enumerate}
\item Draw all planar graphs with a scalar loop and the external lines
  matching the specified configuration. Use the MHV amplitudes as
  vertices, and connect legs of opposing helicities with scalar
  propagators.
\item Continue the internal lines off-shell as per the CSW
  prescription using a null reference momentum $\mu=\eta\tilde\eta$.
\item Integrate over the loop momentum using a measure that splits the
  loop integral into one over phase space and a dispersive integral.
\end{enumerate}
The momenta $L_i$ carried on each internal line in the loop can be
written using \eqref{eq:nullproj} as
\[
L_i = l_i + z_i \mu,
\]
where $z_i = L^2/2l_i\cdot \mu$, and $l_i$ is the null momentum whose
spinors feature in the MHV vertices. The integral over $d^4L_i$ is
traded for one over phase-space and $z_i$ by using
\cite{Brandhuber:2004yw, Nair:1988bq}
\[
\frac{d^4L_i}{L_i^2+i\epsilon} = \frac{dz_i}{z_i+i\epsilon}
\frac{d^3l_i}{2l_i^0},
\]
where to regulate the IR divergences, the phase-space integral over
$l$ is continued to $4-2\epsilon$ dimensions. However, without a full
extension of the vertices to $4-2\epsilon$ dimensions, this process
can only be expected to return the correct cut-constructible parts of
non-supersymmetric amplitudes \cite{Bedford:2004nh}.

\begin{figure}[h]
  \centering \begin{picture}(220,80)
    \SetOffset(110,40)
    \DashArrowArcn(0,-69.2820)(80,-240,-300){2}
    \DashArrowArcn(0,69.2820)(80,300,240){2}
    \Line(-40,0)(-55,25.9808)
    \Line(-40,0)(-70,0)
    \Line(-40,0)(-55,-25.9808)
    \Line(40,0)(55,25.9808)
    \Line(40,0)(70,0)
    \Line(40,0)(55,-25.9808)
    \Vertex(40,0){2}
    \Vertex(-40,0){2}
    \DashCArc(40,0)(15,5,55){1}
    \DashCArc(40,0)(15,305,355){1}
    \DashCArc(-40,0)(15,125,175){1}
    \DashCArc(-40,0)(15,185,235){1}
    \Text(72,0)[cl]{$j^-$}
    \Text(-72,0)[cr]{$i^-$}
    \Text(-57,-27)[tc]{$(b+1)^+$}
    \Text(-57,30)[cr]{$a^+$}
    \Text(57,30)[cl]{$(a+1)^+$}
    \Text(57,-30)[cl]{$b^+$}
    \Text(0,15)[cb]{$L_2$}
    \Text(0,-16)[ct]{$L_1$}
    \Text(-38,7)[bl]{$\pm$}
    \Text(38,7)[br]{$\mp$}
    \Text(-38,-7)[tl]{$\mp$}
    \Text(38,-7)[tr]{$\pm$}
    \Text(-82,0)[cr]{$\displaystyle P_L\left\{ \begin{matrix}\vspace{60pt}\end{matrix} \right.$}
  \end{picture}
  \caption{Contribution to the multi-gluon one-loop MHV amplitude with
    a scalar running in the loop, constructed using MHV diagrams.}
  \label{fig:mhv-1loop-scalar}
\end{figure}
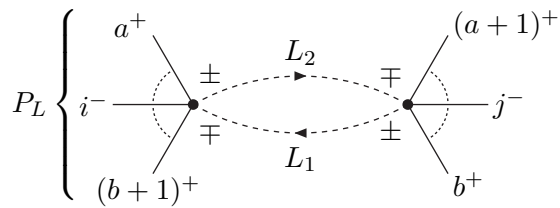

With this in mind, in \cite{Bedford:2004nh} the technique was applied
to the scalar loop contributions to the pure Yang--Mills MHV amplitude
with the two negative-helicity gluons adjacent, and for five gluons in
both of the two independent helicity configurations. The generic graph
for the MHV amplitude, with gluons $i$ and $j$ of negative helicity,
is shown in fig.~\ref{fig:mhv-1loop-scalar}; note that for complex
scalar, helicity should be understood as charge. The measure for this,
incorporating propagators and momentum-conserving $\delta$ function,
is
\[
\frac{d^4L_1}{L_1^2+i\epsilon} \frac{d^4L_2}{L_2^2+i\epsilon}
\delta^4(L_1 - L_2 - P_L).
\]
Now define $P_{L;z}$ such that $L_1 - L_2 - P_L = l_1 - l_2 -
P_{L;z}$, so the full integral is re-written as one over $P_{L;z}$
\cite{Brandhuber:2004yw}:
\[
2\pi i \frac{dP_{L;z}^2}{P_{L;z}^2 - P_L^2 - i\epsilon}
\theta(P^2_{L;z}) d{\rm LIPS}(l_2,-l_1;P_{L;z}),
\]
where the last differential is the two-particle phase space measure
that enforces $l_1 - l_2 = P_{L;z}$.  Note that the phase space
integral computes the unitarity cut in the $P_{L;z}^2$ channel, and
the dispersion integral reconstructs the terms in the amplitude that
have that cut (see \eg\ \cite{Eden:1966aa}); the sum over all diagrams
of fig.~\ref{fig:mhv-1loop-scalar} computes the sum of these integrals
over all kinematic channels.  The phase-space integration is extended
to $4-2\epsilon$ dimensions and both integrals are then evaluated, the
details of which may be found in \cite{Bedford:2004nh}, wherein the
cut-constructible parts of the MHV loop amplitudes specified above are
recovered.

The articles of refs.\ \cite{Quigley:2004pw, Bedford:2004py} apply
this method to ${\cal N}=1$ supersymmetric Yang--Mills theory to
calculate its one-loop MHV amplitude.  In ref.\
\cite{Brandhuber:2005kd}, the authors make arguments for the validity
of their method (for supersymmetric theories), demonstrating its
covariance by using the Feynman Tree Theorem \cite{Feynman:1972mt},
and that all the discontinuities and soft and collinear singularities
are the same as those computed using traditional methods.

\subsubsection*{Yang--Mills inspired approach}
In \cite{Brandhuber:2006bf}, the authors take a different approach to
integrating over the loop momentum which in many ways makes contact
with the field theory-driven approach to the CSW rules we will explore
in detail in chapters \ref{cha:mhvym} and \ref{cha:etv}
.  Again, they compute the scalar
loop part of non-supersymmetric amplitudes using the integral measure
of \eqref{eq:Ddim-split} and the protocol described thereunder to
split the loop momentum across the four- and $-2\epsilon$-dimensional
subspaces.
The vertices used are taken from the light-cone gauge Yang--Mills
theory coupled to a complex scalar, and they depend on the null vector
$\mu=\eta\tilde\eta$ that defines the gauge. They also contract the
$D$-dimensional loop momentum with 4-dimensional spinors corresponding
to an external momentum and $\tilde\eta$, and this has two important
effects. First, it means that $-2\epsilon$-dimensional part of the
loop momentum can be discarded in the vertices. Secondly, the scalar's
off-shell momentum is projected down to a null vector in the vertex
according to \eqref{eq:nullproj}. As a result, one obtains precisely
the scalar-gluon vertex one would have by applying the CSW
prescription with $\eta$ as the reference spinor to the associated
amplitude.  This is something we will come across in more detail in
our studies in section \ref{ssec:mhvym-amps-verts}.

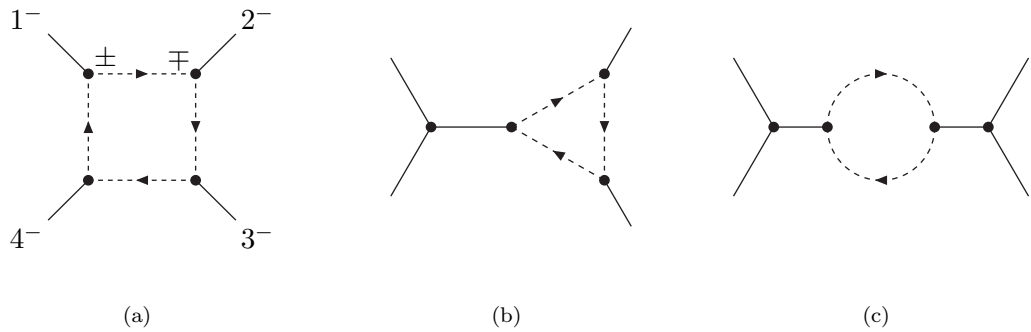
\begin{figure}[h]
  \centering \subfigure[]{
    \begin{picture}(120,120)
      \SetOffset(60,60)
      \DashArrowLine(20,20)(20,-20){2}
      \DashArrowLine(20,-20)(-20,-20){2}
      \DashArrowLine(-20,-20)(-20,20){2}
      \DashArrowLine(-20,20)(20,20){2}
      \Line(20,20)(35,35)
      \Line(20,-20)(35,-35)
      \Line(-20,20)(-35,35)
      \Line(-20,-20)(-35,-35)
      \Vertex(20,20){2}
      \Vertex(20,-20){2}
      \Vertex(-20,20){2}
      \Vertex(-20,-20){2}
      \Text(-37,37)[br]{$1^-$}
      \Text(37,37)[bl]{$2^-$}
      \Text(37,-37)[tl]{$3^-$}
      \Text(-37,-37)[tr]{$4^-$}
      \Text(-18,22)[bl]{$\pm$} \Text(18,22)[br]{$\mp$}
    \end{picture}
    \label{fig:mhv-1loop-4--box}
  }\quad\subfigure[]{
    \begin{picture}(120,120) \SetOffset(60,60)
      \Line(-30,0)(-45,25.9808) \Line(-30,0)(-45,-25.9808)
      \Line(-30,0)(0,0) \DashArrowLine(0,0)(34.6410,20){2}
      \DashArrowLine(34.6410,20)(34.6410,-20){2}
      \DashArrowLine(34.6410,-20)(0,0){2}
      \Line(34.6410,20)(44.6410,37.3205)
      \Line(34.6410,-20)(44.6410,-37.3205) \Vertex(0,0){2}
      \Vertex(-30,0){2} \Vertex(34.6410,20){2} \Vertex(34.6410,-20){2}
    \end{picture}
    \label{fig:mhv-1loop-4--triangle}
  }\quad\subfigure[]{
    \begin{picture}(120,120) \SetOffset(60,60)
      \Line(-20,0)(-40,0) \Line(20,0)(40,0) \Line(40,0)(55,25.9808)
      \Line(40,0)(55,-25.9808) \Line(-40,0)(-55,25.9808)
      \Line(-40,0)(-55,-25.9808) \DashArrowArcn(0,0)(20,180,0){2}
      \DashArrowArcn(0,0)(20,360,180){2} \Vertex(20,0){2}
      \Vertex(-20,0){2} \Vertex(40,0){2} \Vertex(-40,0){2}
      \label{fig:mhv-1loop-4--bubble}
    \end{picture}
  }
  \caption{The box and typical triangle and bubble contributions
    to the one-loop $(\mmmm)$ amplitude with a scalar
    running in the loop, constructed using MHV diagrams. Dashed lines
    are scalars, solid lines gluons, and all momenta are out-going.}
  \label{fig:mhv-1loop-4-}
\end{figure}
    

Using this method, Brandhuber \etal\ were able to obtain the
correct expression for the one-loop $(\mmmm)$ amplitude. This
amplitude vanishes in supersymmetric theories (see \eg\ supersymmetric
Ward identities, section \ref{sec:swis}) and being rational in four
dimensions is not cut constructible.  By considering the diagrams of
fig.~\ref{fig:mhv-1loop-4-} and their rotations, it was found that
this method leads, after performing Passarino--Veltman reduction --- a
purely \emph{algebraic} manipulation --- to an expression in terms of
$4-2\epsilon$-dimensional box, triangle and bubble integrals. The
triangle and bubble functions' contributions cancel to leave the known
answer
\begin{equation}
  \frac{2ig^4}{(4\pi)^{2-\epsilon}}
  K_4 \frac{\langle 1\:2 \rangle \langle
    3\:4 \rangle}{[1\:2][3\:4]}.
\end{equation}
As the authors of \cite{Brandhuber:2006bf} note, the survival of this
piece arises from a cancellation between the $\epsilon^{-1}$ supplied
by the underlying $4-2\epsilon$-dimensional integration over the
propagators, and the factor of $\epsilon$ provided by keeping track of
the factors of $\nu^2$ in the numerator.
 \section{A summary of the state of the art up to one loop}
\label{sec:background-sota}
The methods described in sections \ref{sec:background-csw} and
\ref{sec:background-bcf} (the latter in particular) for the analytic
computation of tree-level amplitudes involving many external gluons
have lead to considerable advancement in our ability to compute new
results.  From an automation standpoint, the discovery of algorithms
with complexity growing no faster than the square of the number of
external partons has lead many to consider the tree-level problem
`solved'.  Numerical techniques useful to builders of Monte Carlo
simulations should not be overlooked, either.  A recent analysis in
\cite{Dinsdale:2006sq} concluded that for numerical calculations of
tree-level amplitudes, the Berends--Giele \emph{off}-shell recursion
relations \cite{Berends:1987me} were fastest for nine or more external
partons, whereas on-shell recursion relations offered better
performance in cases with fewer.

Table \ref{tbl:loop-sota} shows the current state-of-the-art for
one-loop gluon amplitudes. As can be seen, analytic expressions for
gluon loop amplitudes have been obtained via a number of methods and
there is extensive literature on the subject.  The highest valence with
analytic results for all helicity configurations is six gluons.
The higher-point amplitudes are, at time of writing, still
awaiting the various other components required for assembly. On the
other hand, numerical programmes have been forging ahead and for the
more demanding phenomenologist, results in ref.\ \cite{Giele:2008bc}
have been published for up to twenty gluons evaluated using
\textbf{Rocket}, a program which implements the algorithms and
techniques given in refs.\ \cite{Ellis:2006ss, Giele:2008ve}.

\begin{table}[h]
  \centering\begin{tabular}{p{0.3\textwidth}p{0.6\textwidth}}
    \toprule
    \textbf{Amplitude(s)} & \textbf{Methodology} \\ \midrule
    5 gluons & String based calculation \cite{Bern:1993mq} \\ \midrule
    6 gluons & ${\cal N}=4$ part via unitarity
    \cite{Bern:1994cg}; ${\cal N}=1$ MHV in \cite{Bern:1994cg},
    NMHV in \cite{Bidder:2004tx, Britto:2005ha};
    scalar by unitarity \cite{Britto:2006sj}, rational terms by
    Feynman integrals \cite{Xiao:2006vt} \\ \midrule
    $(--+++++)$ &  Bootstrap \cite{Bern:2005cq} \\ \midrule
    Up to 20 gluons & Numerical, selected regions of phase
    space \cite{Giele:2008bc} \\ \midrule
    ${\cal N}=1$ $n$ gluons MHV & Unitarity \cite{Bern:1994cg} \\ \midrule
    ${\cal N}=4$ $n$ gluons MHV & Unitarity \cite{Bern:1994zx} \\ \midrule
    ${\cal N}=4$ $n$ gluons NMHV & Unitarity \cite{Bern:2004bt} \\ \midrule
    $(\pm+\cdots+)$ & Recursive techniques \cite{Mahlon:1993si}  \\ \midrule
    $n$ gluon MHV & Cuts of scalar pieces by MHV diagrams in
    \cite{Bedford:2004nh},
    rational terms by on-shell recursion relations in \cite{Berger:2006vq}
    \\ \midrule
    $(---+\cdots+)$ & Unitarity \cite{Britto:2005ha} \\ \midrule
    $(----+\cdots+)$ & Bootstrap leading to
    recursive solution in \cite{Berger:2006ci} (with
    numerical results given for 7
    and 8 gluon cases) \\
    \bottomrule
  \end{tabular}
  \label{tbl:loop-sota}
  \caption{Summary of the current state-of-the-art in gluon loop
    amplitudes and when and how they were first derived.}
\end{table}

\section{Closing statement: From the CSW rules 
to field theory}
\label{sec:background-closing}

We are now in a position to make more concrete the context and spirit
of the remainder of this thesis.  The CSW rules described in section
\ref{sec:background-csw} were originally derived based on observations
in twistor string theory. As the foregoing bears witness, a lot of
work has been undertaken to develop these ideas at tree and loop
level, powered mostly by applied understanding of the analytic
structure of scattering amplitudes and twistor-space geometry.

While highly inspired, this is perhaps not so satisfying from a field
theorist's point of view.  The CSW rules look \emph{qualitatively}
like the end product of quantising some kind of field theory,
albeit one with an infinite series of vertices of increasing valence
joined together by a scalar propagator.  One might therefore ask: Is
it possible to construct the \emph{field theory} QCD in such a way
that the CSW rules are made manifest? In other words, can we write
down an action for QCD with a scalar propagator and an infinite series
of vertices that take the form of MHV amplitudes?  Such a formulation
would make the construction accessible to the well-established framework
of quantum field theory, as well as indicate how to incorporate (the
as-yet missing) regulation structure needed for quantum corrections.
For a start, the formalisms we have seen so far are fundamentally tied
to four dimensions. Might deriving the CSW rules from an action
viewpoint lead to a dimensionally regulated version of the
construction?\footnote{Here we are talking about a \emph{full}
  dimensional regularisation of amplitudes consisting of purely
  gluonic components, \ie\ accounting for the changing number of
  degrees of freedom this introduces. This is in contrast to the
  calculations mentioned in the foregoing (such as MHV amplitudes in
  loop diagrams and $4-2\epsilon$-dimensional cut construction) with a
  complex \emph{scalar} in the loop.}${}^,$\footnote{We note that
  there are other regulators that work within four dimensions, such as
  that of ref.\ \cite{Thorn:2005ak}, but thanks largely to its Lorentz
  and gauge invariance (ignoring chiral anomalies) dimensional
  regularisation has emerged as the phenomenologists' favourite. }

It turns out that these suspicions are well-founded.  Without wishing
to give away too much at this stage one can construct a transformation
of light-cone gauge Yang--Mills that re-writes the usual action in
precisely the `MHV lagrangian' form we desire. The bulk of this
thesis, chapters \ref{cha:mhvym}--\ref{cha:mhvqcd}, is devoted to
obtaining this transformation and studying the lagrangians it
constructs. As well as deriving the CSW rules at the action level from
a field theoretic point of view, we will see that this approach
provides extra structure that can answer certain questions concerning
the apparent incompleteness of the CSW rules. In particular, we note
here that with MHV vertices alone, one simply cannot assemble a graph
for the one-loop $(+ \cdots +)$ amplitudes.

\begin{subappendices}

\section{4D cut-constructible loop integrals}
\label{sec:loopbasis}
Here we list the set of functions that can appear in cut-constructible
amplitudes satisfying the BDDK criteria with four or more external
particles. These are also given in Figures 1--3 of \cite{Bern:1994cg},
along with explicit expressions in Appendix I of the same reference.

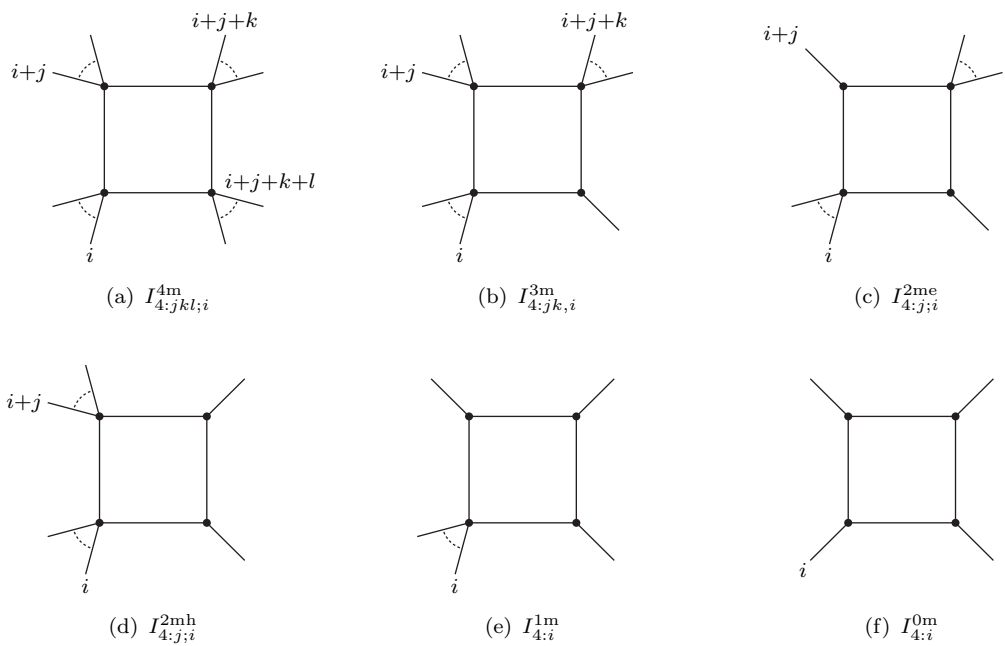
\begin{figure}[h]
  \centering\subfigure[$I_{4:jkl;i}^\text{4m}$]{
    \begin{picture}(120,96)
      \SetOffset(60,48)
      \EBox(-20,-20)(20,20)
      \Line(20,20)(39.3185,25.1764)
      \Line(20,20)(25.1764,39.3185)
      \Line(20,-20)(39.3185,-25.1764)
      \Line(20,-20)(25.1764,-39.3185)
      \Line(-20,20)(-39.3185,25.1764)
      \Line(-20,20)(-25.1764,39.3185)
      \Line(-20,-20)(-39.3185,-25.1764)
      \Line(-20,-20)(-25.1764,-39.3185)
      \DashCArc(20,20)(10,20,70){1}
      \DashCArc(20,-20)(10,290,340){1}
      \DashCArc(-20,-20)(10,200,250){1}
      \DashCArc(-20,20)(10,110,160){1}
      \Vertex(20,20){1.5}
      \Vertex(20,-20){1.5}
      \Vertex(-20,-20){1.5}
      \Vertex(-20,20){1.5}
      \Text(-25.2,-41)[tc]{${}_{i}$}
      \Text(-41,25.2)[cr]{${}_{i+j}$}
      \Text(25.2,41)[bc]{${}_{i+j+k}$}
      \Text(25,-20)[bl]{${}_{i+j+k+l}$}
    \end{picture}
    \label{fig:I44m}
  }\quad\subfigure[$I_{4:jk,i}^\text{3m}$]{
    \begin{picture}(120,96)
      \SetOffset(60,48)
      \EBox(-20,-20)(20,20)
      \Line(20,20)(39.3185,25.1764)
      \Line(20,20)(25.1764,39.3185)
      \Line(20,-20)(34.1421,-34.1421)
      \Line(-20,20)(-39.3185,25.1764)
      \Line(-20,20)(-25.1764,39.3185)
      \Line(-20,-20)(-39.3185,-25.1764)
      \Line(-20,-20)(-25.1764,-39.3185)
      \DashCArc(20,20)(10,20,70){1}
      \DashCArc(-20,-20)(10,200,250){1}
      \DashCArc(-20,20)(10,110,160){1}
      \Vertex(20,20){1.5}
      \Vertex(20,-20){1.5}
      \Vertex(-20,-20){1.5}
      \Vertex(-20,20){1.5}
      \Text(-25.2,-41)[tc]{${}_{i}$}
      \Text(-41,25.2)[cr]{${}_{i+j}$}
      \Text(25.2,41)[bc]{${}_{i+j+k}$}
    \end{picture}
    \label{fig:I43m}
  }\quad\subfigure[$I_{4:j;i}^\text{2me}$]{
    \begin{picture}(120,96)
      \SetOffset(60,48)
      \EBox(-20,-20)(20,20)
      \Line(20,20)(39.3185,25.1764)
      \Line(20,20)(25.1764,39.3185)
      \Line(20,-20)(34.1421,-34.1421)
      \Line(-20,20)(-34.1421,34.1421)
      \Line(-20,-20)(-39.3185,-25.1764)
      \Line(-20,-20)(-25.1764,-39.3185)
      \DashCArc(20,20)(10,20,70){1}
      \DashCArc(-20,-20)(10,200,250){1}
      \Vertex(20,20){1.5}
      \Vertex(20,-20){1.5}
      \Vertex(-20,-20){1.5}
      \Vertex(-20,20){1.5}
      \Text(-25.2,-41)[tc]{${}_{i}$}
      \Text(-36,36)[br]{${}_{i+j}$}
    \end{picture}
    \label{fig:I42me}
  }\\\subfigure[$I_{4:j;i}^\text{2mh}$]{
    \begin{picture}(120,96)
      \SetOffset(60,48)
      \EBox(-20,-20)(20,20)
      \Line(20,-20)(34.1421,-34.1421)
      \Line(20,20)(34.1421,34.1421)
      \Line(-20,20)(-39.3185,25.1764)
      \Line(-20,20)(-25.1764,39.3185)
      \Line(-20,-20)(-39.3185,-25.1764)
      \Line(-20,-20)(-25.1764,-39.3185)
      \DashCArc(-20,-20)(10,200,250){1}
      \DashCArc(-20,20)(10,110,160){1}
      \Vertex(20,20){1.5}
      \Vertex(20,-20){1.5}
      \Vertex(-20,-20){1.5}
      \Vertex(-20,20){1.5}
      \Text(-25.2,-41)[tc]{${}_{i}$}
      \Text(-41,25.2)[cr]{${}_{i+j}$}
    \end{picture}
    \label{fig:I42mh}
  }\quad\subfigure[$I_{4:i}^\text{1m}$]{
    \begin{picture}(120,96)
      \SetOffset(60,48)
      \EBox(-20,-20)(20,20)
      \Line(20,-20)(34.1421,-34.1421)
      \Line(20,20)(34.1421,34.1421)
      \Line(-20,20)(-34.1421,34.1421)
      \Line(-20,-20)(-39.3185,-25.1764)
      \Line(-20,-20)(-25.1764,-39.3185)
      \DashCArc(-20,-20)(10,200,250){1}
      \Vertex(20,20){1.5}
      \Vertex(20,-20){1.5}
      \Vertex(-20,-20){1.5}
      \Vertex(-20,20){1.5}
      \Text(-25.2,-41)[tc]{${}_{i}$}
    \end{picture}
  } \quad\subfigure[$I_{4:i}^\text{0m}$]{
    \begin{picture}(120,96)
      \SetOffset(60,48)
      \EBox(-20,-20)(20,20)
      \Line(20,-20)(34.1421,-34.1421)
      \Line(20,20)(34.1421,34.1421)
      \Line(-20,20)(-34.1421,34.1421)
      \Line(-20,-20)(-34.1421,-34.1421)
      \Vertex(20,20){1.5}
      \Vertex(20,-20){1.5}
      \Vertex(-20,-20){1.5}
      \Vertex(-20,20){1.5}
      \Text(-35,-35)[tr]{${}_{i}$}
    \end{picture}
  }
  \caption{Cut-constructible box integral contributions, $n\ge4$.}
  \label{fig:bdk-boxes}
\end{figure}

\begin{figure}[h]
  \centering\subfigure[$I_{3:jk;i}^\text{3m}$]{
    \begin{picture}(120,96)
      \SetOffset(40,44)
      \Line(-17.3205,-10)(51.9615,30)
      \Line(-17.3205,10)(51.9615,-30)
      \Line(34.6410,40)(34.6410,-40)
      \DashCArc(0,0)(10,155,205){1}
      \DashCArc(34.6410,20)(10,35,85){1}
      \DashCArc(34.6410,-20)(10,275,325){1}
      \Vertex(0,0){1.5}
      \Vertex(34.6410,20){1.5}
      \Vertex(34.6410,-20){1.5}
      \Text(-19,-11)[tr]{${}_{i}$}
      \Text(34.6,42)[bc]{${}_{i+j}$}
      \Text(54,-30)[lc]{${}_{i+j+k}$}
    \end{picture}
  }\quad\subfigure[$I_{3:j;i}^\text{2m}$]{
    \begin{picture}(120,96)
      \SetOffset(40,44)
      \Line(-17.3205,-10)(51.9615,30)
      \Line(-17.3205,10)(34.6410,-20)
      \Line(34.6410,40)(34.6410,-20)
      \Line(34.6410,-20)(44.6410,-37.3205)
      \DashCArc(0,0)(10,155,205){1}
      \DashCArc(34.6410,20)(10,35,85){1}
      \Vertex(0,0){1.5}
      \Vertex(34.6410,20){1.5}
      \Vertex(34.6410,-20){1.5}
      \Text(-19,-11)[tr]{${}_{i}$}
      \Text(34.6,42)[bc]{${}_{i+j}$}
    \end{picture}
  }\quad\subfigure[$I_{3:i}^\text{1m}$]{
    \begin{picture}(120,96)
      \SetOffset(40,44)
      \Line(-17.3205,-10)(34.6410,20)
      \Line(-17.3205,10)(34.6410,-20)
      \Line(34.6410,20)(34.6410,-20)
      \Line(34.6410,-20)(44.6410,-37.3205)
      \Line(34.6410,20)(44.6410,37.3205)
      \DashCArc(0,0)(10,155,205){1}
      \Vertex(0,0){1.5}
      \Vertex(34.6410,20){1.5}
      \Vertex(34.6410,-20){1.5}
      \Text(-19,-11)[tr]{${}_{i}$}
    \end{picture}
  }
  \caption{Cut-constructible triangle integral contributions,
    $n\ge4$.}
  \label{fig:bdk-triangles}
\end{figure}
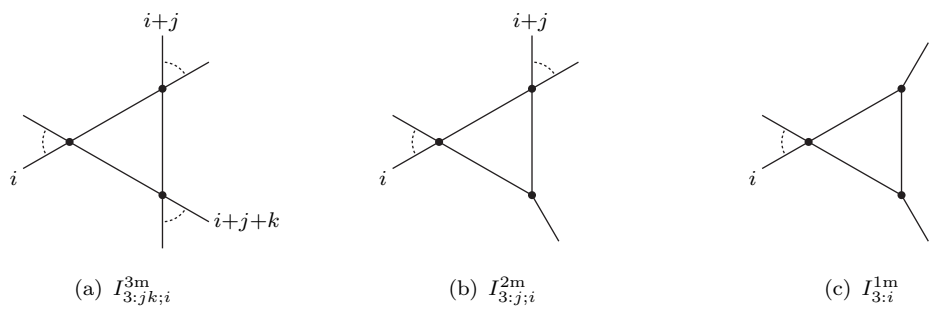

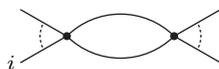
\begin{figure}[h]
  \centering
  \begin{picture}(90,28)
    \SetOffset(45,14)
    \Line(20,0)(37.3205,10)
    \Line(20,0)(37.3205,-10)
    \Line(-20,0)(-37.3205,10)
    \Line(-20,0)(-37.3205,-10)
    \CArc(0,-20)(28.2843,45,135)
    \CArc(0,20)(28.2843,225,315)
    \DashCArc(-20,0)(10,155,205){1}
    \DashCArc(20,0)(10,335,25){1}
    \Vertex(20,0){1.5}
    \Vertex(-20,0){1.5}
    \Text(-39,-10)[cr]{${}_{i}$}
  \end{picture}
  \caption{Cut-constructible bubble integral contribution
    $I_{2:i}^\text{2m}$.}
  \label{fig:bdk-bubbles}
\end{figure}
\end{subappendices}

\newcommand\hatfrac[2]{\frac{\hat{#1}}{\hat{#2}}}

\chapter{An MHV Lagrangian for Pure Yang--Mills}
\label{cha:mhvym}

In the previous chapter, we studied the MHV rules of Cachazo, Svr\v cek
and Witten and how they are applied to the calculation of multi-gluonic
scattering amplitudes. There, we hinted that the construction is
reminiscent of a field theory with an infinite set of MHV vertices
connected by scalar propagators. However, the CSW construction lies
outside the framework of Lagrangian mechanics as usually applied to
quantum field theory and the well understood machinery thereof.

In this chapter, we will show how to put the MHV rules within the
Lagrangian framework by means of a canonical transformation of the
field variables of light-cone gauge Yang--Mills theory.
The result is the Canonical MHV Lagrangian,
consisting of an infinite series of MHV-like vertices in the new field
variables. The structure of the rest of this chapter is as follows. In
section \ref{sec:mhvym-lightcone}, we derive the action for light-cone
Yang--Mills, and in sections
\ref{sec:mhvym-struct}--\ref{sec:mhvym-transf}
we specify and solve the
field transformation that gives us an MHV lagrangian. Next, in section
\ref{sec:mhvym-examples} we compute explicitly terms in our MHV
lagrangian for up to $5$ gluons and demonstrate that (up to
polarisation factors) these are the Parke--Taylor amplitudes of
\eqref{eq:parke-taylor}. Finally we draw our conclusions on this work
in section \ref{sec:mhvym-conclusion}.

This work was published in \cite{Ettle:2006bw}.


\section{Light-cone gauge Yang--Mills Theory}
\label{sec:mhvym-lightcone}

In this section, we detail the preliminaries underlying the
construction of the Canonical MHV Lagrangian, specifically our choice
of co-ordinates and normalisation, and derive the Yang--Mills action
fixed to the light-cone gauge.

\subsection{Light-cone co-ordinates}
\label{ssec:lightcone-coords}

The forthcoming analysis is adapted to a light-cone co-ordinate
system, which is defined by
\begin{equation}
\label{eq:lightcone-coords}
       x^0 = \tfrac1{\sqrt 2} (t - x^3), \quad
x^{\bar 0} = \tfrac1{\sqrt 2} (t + x^3), \quad
         z = \tfrac1{\sqrt 2} (x^1 + i x^2), \quad
    \bar z = \tfrac1{\sqrt 2} (x^1 - i x^2).
\end{equation}
Here, $t$ and $x^i$ are the usual Minkowski co-ordinates. Note here
the presence of the $1/\sqrt 2$ factors that preserve the
normalisation of the volume form. Since we will frequently be dealing with
specific components of momenta, we
will make use of the short-hand $(p_0, p_{\bar 0}, p_z, p_{\bar z})
\equiv (\check p, \hat p, p, \bar p)$ for the components of
$1$-forms. For the $n^{\text{th}}$ external momentum, we write the
number with the embellishments so that the momentum's components are
$(\check n, \hat n, \tilde n, \bar n)$.
In these co-ordinates, the Lorentz invariant is
\begin{equation}
A \cdot B = \check A \hat B + \hat A \check B - A \bar B - \bar A B.
\label{eq:lc-invariant}
\end{equation}

It will also turn out to be useful to define the following quantities which
are bilinear in their momentum arguments:
\begin{equation}
  \label{eq:lc-bilinears}
  (1\:2) := \hat 1 \tilde 2 - \hat 2 \tilde 1, \quad
  \{1\:2\} := \hat 1 \bar 2 - \hat 2 \bar 1.
\end{equation}
These quantities have a simple relationship with the $\langle \cdots
\rangle$ and $[ \cdots ]$ spinor brackets of
\eqref{eq:spinorbrackets}. We will not make much use of the latter in
the forthcoming, but it will be necessary to understand this
relationship in order to compare the techniques we have developed with
established results. First, we begin by writing the bispinor
representation of a $4$-vector $p$ of \eqref{eq:bispinor-mink} in
light-cone co-ordinates:
\begin{equation}
\label{eq:bispinor}
  p \cdot \bar\sigma = \sqrt 2 \begin{pmatrix} \check p & -p \\
    - \bar p & \hat p \end{pmatrix}.
\end{equation}
If $p$ is null, then $p \bar p = \check p \hat p$ and this matrix
factorises as $(p \cdot \bar\sigma)_{\alpha\dot\alpha} =
\lambda_\alpha \tilde\lambda_{\dot\alpha}$, where we can choose
\begin{equation}
  \label{eq:lc-spinors}
  \lambda_\alpha = 2^{1/4} \begin{pmatrix} -p/\sqrt{\hat p} \\ \sqrt{\hat p}
  \end{pmatrix}
  \quad\text{and}\quad
  \tilde\lambda_{\dot\alpha} = 2^{1/4} \begin{pmatrix} -\bar p/\sqrt{\hat p} \\ \sqrt{\hat p}
  \end{pmatrix}.
\end{equation}
Hence the spinor brackets can be expressed as
\begin{equation}
  \label{eq:lc-spinorbrackets}
  \langle 1\:2 \rangle = \epsilon^{\alpha\beta}
  \lambda_{1\alpha} \lambda_{2\beta}
  = \sqrt{2} \frac{(1\:2)}{\sqrt{\hat 1 \hat 2}}
  \quad\text{and}\quad [ 1\:2 ] = \epsilon^{\dot\alpha\dot\beta}
  \lambda_{1\dot\alpha} \lambda_{2\dot\beta}
  = \sqrt{2} \frac{\{1\:2\}}{\sqrt{\hat 1 \hat 2}}.
\end{equation}

\subsection{Gauge-fixing the action}
\label{ssec:mhvym-gaugefix}
Mansfield's programme for the construction of the Canonical MHV
Lagrangian begins with light-cone gauge Yang--Mills theory, which we
will derive in this section. We start with non-gauge-fixed Yang--Mills action
\begin{equation}
\label{eq:ym-action}
S = \frac1{2g^2} \int d^4x \: \tr \gF^{\mu\nu} \gF_{\mu\nu},
\end{equation}
where
$d^4x$ is the usual Minkowski $4$-volume element, and our field
strength tensor, covariant derivative and gauge potential are defined
according to
\begin{equation}
  \label{eq:gauge-conv}
  \gF_{\mu\nu} = [\gD_\mu, \gD_\nu], \quad
  \gD_\mu = \partial_\mu + \gA_\mu, \quad
  \gA_\mu = - \frac{ig}{\sqrt 2} A^a_\mu T^a.
\end{equation}
Note that here, and throughout,
the ${\rm SU}(N_{\rm C})$ generators $T^a$ are normalised as in
chapter \ref{cha:background}, \ie\
\begin{equation}
[T^a, T^b] = i \sqrt 2 f^{abc} T^c, \quad \tr (T^a
  T^b) = \delta^{ab}.
\end{equation}

We will quantise the theory on a null $3$-surface $\Sigma$ of constant
$x^0$. This has a normal vector $\mu = (1, 0, 0, 1)/\sqrt 2$ in
Minkowski co-ordinates. Light-cone gauge imposes the condition $\mu
\cdot \gA \equiv \hat \gA = 0$
for which the Faddeev--Popov
 ghosts decouple (contributing an
overall infinite constant factor to the path integral, which we
discard). Now, with this condition in force, we find that
\eqref{eq:ym-action} becomes
\begin{equation}
\label{eq:ym-action-1}
S = \frac1{2g^2} \int d^4x \: \tr ({\cal L}_2 + {\cal L}_3 + {\cal L}_4)
\end{equation}
where (after integrating by parts)
\begin{align}
\label{eq:ym-l2}
{\cal L}_2 &= 2 (\gA [\check\partial \hat\partial - \partial
\bar\partial] \bar \gA 
 + \check\gA \: \hat\partial^2 \check\gA
 - \gA \: \hat\partial \bar\partial \check\gA - 2 \bar\gA \:
 \hat\partial \partial \check\gA 
+\gA \: \bar\partial^2 \gA + \bar\gA \: \partial^2
\bar\gA + \gA \: \partial \bar\partial \bar\gA ), \\
 \label{eq:ym-l3}
{\cal L}_3 &= 4 (\partial \bar\gA [\bar\gA, \gA] + \bar\partial \gA
[\gA, \bar\gA] - \hat\partial \gA [\check\gA, \bar\gA] - \hat\partial
\bar\gA [\check\gA, \gA]), \\
\label{eq:ym-l4}
{\cal L}_4 &= 2 ([\gA, \bar\gA] [\bar\gA, \gA]).
\end{align}
Importantly, we notice here that the lagrangian density
\eqref{eq:ym-l2}--\eqref{eq:ym-l4} is quadratic in
$\check\gA$. Furthermore, it is non-dynamical with respect to
light-cone time (\ie\ it there are no $\check\partial$ operators
acting on it). Therefore we will integrate it out of the action.

To see how this is done, consider briefly a toy model field theory
with a hermitian, algebra-valued field $\phi$, and a set of other independent
fields which we will label $\Psi_i$. Let $K(\Psi)$ be a hermitian, 
algebra-valued function of the $\Psi_i$. We can write this theory's partition
function as
\begin{equation}
\int {\cal D}\phi \: {\cal D}\Psi \exp
    \int d^4x \: ik \left\{ \tr [\tfrac12 \phi \Delta^{-1} \phi + K(\Psi) \phi] + L(\Psi) \right\}, 
\end{equation}
where $k$ is a constant, $\Delta^{-1}$ is a differential operator,
and $L(\Psi)$ is a real function of the $\Psi_i$ only. Now make the
following change of variables:
\begin{equation}
\phi \rightarrow \phi' = \phi + \Delta K(\Psi),
\end{equation}
where we assume a regularisation prescription that
allows $\Delta$ to exist. Since this is a shift, its jacobian is unity
and the path integral becomes 
\begin{equation}
\label{eq:toymodel-shift}
\int {\cal D}\phi' \: {\cal D}\Psi \exp
    \int d^4x \: ik \left\{ \tr [ \tfrac12 \phi' \Delta^{-1} \phi' - \tfrac12 K(\Psi)
    \Delta K(\Psi)] + L(\Psi) \right\}.
\end{equation}
The first term above calculates $\det(-k \Delta^{-1})^{-1/2}$. 
In the case that $\Delta^{-1}$ is field-independent, this is a
divergent but physically irrelevant factor in the path integral. We
may therefore discard the $\phi'$ terms in the action of
\eqref{eq:toymodel-shift}.

Now for \eqref{eq:ym-l4}, the analogue of $K(\Psi)$ is the
coefficient of $\check\gA$ found the trace therein, specifically
\begin{equation}
\label{eq:Kym}
K_{\text{YM}} = 4 ([\hat\partial \gA, \bar\gA] + [\hat\partial
\bar\gA, \gA] - \partial \hat\partial \bar\gA - \bar\partial
\hat\partial \gA ).
\end{equation}
The operator $\tfrac 12 \Delta^{-1} = 2 \hat\partial^2$, and so by
\eqref{eq:toymodel-shift}, we drop all terms in \eqref{eq:ym-action-1}
that contain $\check\gA$ and replace them with
\begin{equation}
\label{eq:intout}
-\frac1{2g^2} \int d^4x\: \tr \tfrac18 K_{\text{YM}} \hat\partial^{-2} K_{\text{YM}}.
\end{equation}
We plug \eqref{eq:Kym} into \eqref{eq:intout} and after some algebra
(where we take derivatives and inverse derivatives to commute) and
integration by parts, we arrive at the light-cone Yang--Mills
action:
\begin{equation}
\label{eq:lcym-action}
S_{\text{LCYM}} = \frac 4{g^2}\int dx^0 ( L^{-+} + L^\mpp + L^\mmp
    + L^{--++})
\end{equation}
where
\begin{align}
  \label{eq:lcym-mp}
  L^{-+} &= \phantom{-}{\rm tr}\int_\Sigma d^3{\bf x}\: {\bar
    \gA}(\check\partial\hat\partial-
  \partial\bar\partial)\gA, \\
  \label{eq:lcym-mpp} {L}^{\mpp}&=-{\rm tr}\int_\Sigma d^3{\bf x}\:
  ({\bar\partial}{\hat\partial}^{-1} { \gA})\:
  [{  \gA},{\hat\partial} {\bar\gA}], \\
  \label{eq:lcym-mmp} {L}^{\mmp}&=-{\rm tr}\int_\Sigma d^3{\bf x}\:
  [{\bar \gA},{\hat\partial} { \gA}]\:
  ({  \partial}{\hat\partial}^{-1} {\bar \gA}), \\
  \label{eq:lcym-mmpp} {L}^{--++}&=-{\rm tr}\int_\Sigma d^3{\bf x}\:
  [{\bar \gA },{\hat\partial} { \gA }]\:{\hat\partial}^{-2}\: [{ \gA
  },{\hat\partial} {\bar \gA }].
\end{align}

The result, as seen above, is a relatively simple form for the
Yang--Mills action in terms of the physical degrees of freedom
alone.\footnote{We note here that there is still some residual gauge freedom
  left over in \eqref{eq:lcym-action}, something to which we will turn
  our attention later.} This is the motivation behind this choice,
not least because it allows us to identify $\gA$ and $\bar\gA$ with
positive and negative helicity, respectively. To see this, consider
the standard polarisation vectors \eqref{eq:gluon-poln} for a massless on-shell
vector boson of momentum $p = \lambda \tilde\lambda$, which we reprint here:
\[
  (E_+)_{\alpha\dot\alpha} =
  \sqrt 2 \frac{\nu_\alpha \tilde\lambda_{\dot\alpha}}
  {\langle \nu\:\lambda \rangle}
  \quad\text{and}\quad
  (E_-)_{\alpha\dot\alpha} =
  \sqrt 2 \frac{\lambda_\alpha \tilde\nu_{\dot\alpha}}{[ \nu\:\lambda ]}.
\]
Let us choose the reference vector to be $\nu \tilde\nu = \mu$,
the null vector normal to the
quantisation surface $\Sigma$. Since \eqref{eq:lc-spinors} is singular for
$\mu$, instead we choose
\begin{equation}
\label{eq:mu-spinors}
\nu_\alpha = \tilde\nu_{\dot\alpha}
= 2^{1/4} \begin{pmatrix} 1 \\ 0 \end{pmatrix}.
\end{equation}
Substituting into \eqref{eq:gluon-poln} and comparing the bispinor
components with \eqref{eq:bispinor}, we find that the polarisation vectors'
non-zero components are $E_+ = \bar E_- = -1$ and $\check E_\pm$.
Scattering amplitudes are formed by the application of
LSZ reduction to correlation functions of the $\gA$
fields. Schematically, 
for outgoing momenta
$\{p_i\}$ and helicities $\{h_i\}$,
\begin{equation}
\label{eq:mhvym-lsz}
\langle p_1^{h_1},\dots,p_n^{h_n} \rvert S \lvert 0 \rangle
    = (-i)^n \:
      \lim_{p_i^2 \rightarrow 0}
      p_1^2 \cdots p_n^2 \: \langle
          {E^{\mu_1}_{h_1}} \gA_{\mu_1} \cdots
      {E^{\mu_n}_{h_n}} \gA_{\mu_n} \rangle.
\end{equation}
(up to normalisation). By considering the invariant
\eqref{eq:lc-invariant}, we see that each $+$ ($-$) external
state gluon on the LHS above is associated
with an insertion of $\bar\gA$ ($\gA$) in the correlation function on the
RHS. The $\check E_\pm$ component couples to $\hat\gA$, which vanishes
by gauge choice. The  scalar propagator of \eqref{eq:lcym-mp} connects
$\gA$ to $\bar\gA$ and \textit{vice versa}, so when the LSZ procedure
`amputates' these propagators, we come to identify the $\gA$ and
$\bar\gA$ as the positive and negative helicity fields, respectively.

\section{Structure of the MHV lagrangian}
\label{sec:mhvym-struct}

Let us now begin to construct and explore the field transformation
that takes us from \eqref{eq:lcym-action} to a lagrangian that makes
manifest the CSW construction.

\subsection{Form of the transformation and lagrangian}
\label{ssec:mhvym-form}

In \cite{Mansfield:2005yd}, Mansfield defines the field transformation
that gives the new gauge fields $\gB$ and $\bar\gB$ as
\begin{equation}
  \label{eq:mhvym-transf}
  L^{-+}[\gA,\bar\gA] + L^{\mpp}[\gA,\bar\gA] = L^{-+}[\gB,\bar\gB].
\end{equation}
This transformation is to be performed entirely in the quantisation
surface $\Sigma$, so that all the fields involved have the same $x^0$
`time' dependence (which we will henceforth suppress unless clarity
dictates otherwise).  At first, it might seem rather odd to absorb the
interacting part of one field theory into the kinetic term of another.
However, we recognise the LHS of \eqref{eq:mhvym-transf} as the
Chalmers--Siegel self-dual truncation of Yang--Mills
theory \cite{Chalmers:1996rq}; this is a classically free theory\footnote{Inasmuch as all of its tree-level amplitudes are of
  the form $(-+\cdots+)$ and therefore vanish on shell.}
\cite{Dixon:1996wi} which we are mapping onto another free theory.

The momentum conjugate to $\gA^a(x)$ is
\[
\Pi^a(x) = \frac{\delta L^{-+}}{\delta \check\partial \gA^a(x)} = -
\hat\partial \gA^a(x),
\]
and as such the path integral measure
\[ {\cal D}\gA {\cal D}\bar\gA \equiv \prod_{x,a} \gA^a(x)
\bar\gA^a(x)
\]
is proportional to the phase space measure. In the interest of
simplicity, we would like this measure to be preserved, and this will
be so if the field transformation is \emph{canonical} (in the sense of
Hamiltonian mechanics). Section \ref{sec:canon} explains canonical
transformations and their properties in more detail, but
what we need to know here is that
we can construct one by starting with a transformation of the
canonical co-ordinates alone --- in the case here by postulating that
$\gA$ be a functional of just $\gB$.  One important outcome of this
choice is \eqref{eq:canon-mom}, which links the new and old momenta,
and since \eqref{eq:mhvym-transf} is defined on a surface of constant
$x^0$ and preserves the form of the kinetic part, $- \hat\partial
\gB^a(x)$ is the momentum conjugate to $\gB^a(x)$ so
\begin{equation}
  \label{eq:mhvym-canon-mom}
  \hat\partial \bar\gB^a(\vec y) = \int_\Sigma d^3\vec x \: \frac{\delta
    \gA^b(\vec x)}{\delta \gB^a(\vec y)} \hat\partial \bar\gA^b(\vec x).
\end{equation}
Let us write out \eqref{eq:mhvym-transf} explicitly:
\begin{equation}
  \label{eq:mhvym-transf-1}
  \tr \int_\Sigma d^3\vec x \bigl\{
  \check\partial \gA \hat\partial \bar\gA - \Oomega\gA \hat\partial
  \bar\gA + [\Ozeta\gA, \gA] \hat\partial \bar\gA
  \bigr\} = \tr \int_\Sigma d^3\vec x \bigl\{ \check\partial \gB
  \hat\partial \bar\gB - \Oomega\gB \hat\partial \bar\gB \bigr\}
\end{equation}
where we have introduced the differential operators $\Oomega
:= \partial \bar\partial / \hat\partial$ and $\Ozeta := \bar\partial /
\hat\partial$. Substituting with \eqref{eq:mhvym-canon-mom} into its
LHS, we can eliminate the leading terms on either side and obtain
\begin{equation}
  \label{eq:mhvym-transf-2}
  \tr \int_\Sigma d^3\vec x \bigl\{
  \omega\gA \hat\partial
  \bar\gA - [\zeta\gA, \gA] \hat\partial \bar\gA
  \bigr\} = \tr \int_\Sigma d^3\vec x \: \omega\gB \hat\partial \bar\gB.
\end{equation}
We can again use \eqref{eq:mhvym-canon-mom} to eliminate $\hat\partial
\bar\gB$ in favour of $\hat\partial \bar\gA$, and noting that
$\bar\gA$ is arbitrary here, it follows that $\gA$ has a power series
expansion in $\gB$ of the form
\begin{equation}
  \label{eq:A-series-form}
  \gA = \gB + {\cal O}(\gB^2),
\end{equation}
and from \eqref{eq:mhvym-canon-mom} that $\bar\gA$ has an expansion in
$\gB$ and a single power of $\bar\gB$ in each term, \ie\ of the form
\begin{equation}
  \label{eq:Abar-series-form}
  \bar\gA = \bar\gB + {\cal O}(\gB \bar\gB).
\end{equation}

The remaining part of the Yang--Mills lagrangian is $L^\mmp +
L^\mmpp$, and it follows from \eqref{eq:A-series-form} and
\eqref{eq:Abar-series-form} that upon substitution for $\gA$ and
$\bar\gA$ in terms of $\gB$ and $\bar\gB$ that this will generate an
infinite series of terms with the helicity content required to be
identified as a MHV lagrangian: terms with an increasing number $n_+ \ge
1$ of $\gB$ fields, but only ever two $\bar\gB$ fields. These
interaction terms, together with the $\gB$ kinetic term, form what we
refer to as the Canonical MHV Lagrangian.  In more detail, inspection
of \eqref{eq:lcym-mmp} and \eqref{eq:lcym-mmpp} shows that the fields
are within traces and all possible helicity arrangements thereunder
occur.  If we now work in momentum space on the quantisation surface,
we can write the general $n$-gluon term of the MHV lagrangian as
\begin{equation}
  \label{eq:mhvym-Vseries}
  L^{--\overbrace{+ \cdots +}^{n-2}} =
  \frac12 \sum_{s=2}^n \int_{1\cdots n} V^s_{1\cdots n}
  \tr(\bar\gB_{\bar 1} \gB_{\bar 2} \cdots \bar\gB_{\bar s} \cdots
  \gB_{\bar n}).
\end{equation}
Note that by the cyclicity of the trace, we can always arrange for the
first field in \eqref{eq:mhvym-Vseries} to be a $\bar\gB$. We have
introduced a number of new notations here: the subscripts are the
$3$-momentum arguments, with a bar denoting negation, \eg\ $\gB_{\bar
  1} = \gB(-\vec p_1)$, and the momentum integral short-hand is
defined via
\begin{equation}
  \label{eq:integ-sh}
  \int_{1\cdots n} \equiv \prod_{k=1}^n \int_\Sigma \frac{d\hat k\:
    dk \: d\bar k}{(2\pi)^3} 
\end{equation}
Now since \eqref{eq:mhvym-Vseries} must conserve momentum, it will
often be convenient to factor off the implicit $\delta^3$ function as
\begin{equation}
\label{eq:V-delta}
V^s_{1\cdots n} = (2\pi)^2 \delta^3({\textstyle \sum_{i=1}^n \vec
  p_i}) \: V^s(1\cdots n).
\end{equation}
The presence of the $\delta^3$ function above should be taken as an
indication that the vertices $V^s$ are only defined when the sum of
their arguments is zero.  (We could have restricted the upper limit of
the sum in \eqref{eq:mhvym-Vseries} to $\lfloor n/2+1 \rfloor$ and
obtained all possible helicity arrangements, but by writing it this
way we get the same thing except for in the $s=n/2+1$ term when $n$ is
even, which is is accompanied by a factor of $\frac12$. We will defer
discussing the significance of this momentarily.)

\subsection{Amplitudes, vertices and the CSW rules}
\label{ssec:mhvym-amps-verts}

Let us show that this construction coincides with the CSW rules
\cite{Cachazo:2004kj}, as described in section \ref{sec:background-csw}.  First,
inspection of \eqref{eq:mhvym-transf} tells us that the propagator
associated with the new fields is
\begin{equation}
  \label{eq:propagators-mhvym}
  \langle \gB \bar\gB \rangle = - \frac{ig^2}{2p^2}
  \quad \Leftrightarrow \quad
  \langle B \bar B \rangle = \frac{i}{p^2}
\end{equation}
using \eqref{eq:gauge-conv}, \ie\ a scalar propagator.

We must next show that the vertices of the Canonical MHV Lagrangian
constructed in this manner coincide with the Parke--Taylor amplitudes
and that we recover the CSW off-shell prescription.  First, let us address
the off-shell continuation of the Weyl spinors.  The choice
\eqref{eq:lc-spinors} depend only on three of the four momentum
components, and makes sense even for off-shell momenta. For
$\lambda$ and $\tilde\lambda$ obtained from a $4$-momentum $p$, their
product corresponds to the null vector given by
\begin{equation}
  \label{eq:pnull-bispinor}
  (p_{\text{null}})_{\alpha \dot\alpha}
  = \lambda_\alpha \tilde\lambda_{\dot\alpha}
  = \sqrt2 \begin{pmatrix} p \bar p/\hat p & -p \\
    -\bar p & \hat p \end{pmatrix}.
\end{equation}
It is easy to see, \eg\ by subtracting \eqref{eq:pnull-bispinor} from
\eqref{eq:bispinor} that $p_{\text{null}} = p + a \mu$ where $a= {p
  \bar p}/{\hat p} - \check p= -p^2/2\mu\cdot p$ and $\mu =
\nu\tilde\nu$ is the null vector normal to the quantisation surface,
its spinors given by \eqref{eq:mu-spinors}.  But this is just the CSW
prescription of \eqref{eq:nullproj}, coinciding with
\eqref{eq:lc-spinors} by choosing the arbitrary spinor $\tilde\eta
\propto \tilde\nu$.

Turning to the vertices themselves, we first note that the Canonical
MHV Lagrangian is defined in terms of the new $\gB$ fields. We would
expect to formulate $S$-matrix elements by applying the LSZ theorem to
correlation function in $\gA$ and $\bar\gA$ as in
\eqref{eq:mhvym-lsz}, but by the equivalence theorem (explained in
section \ref{sec:equivalencetheorem}), at tree level for generic
momenta we can equally well use $\gB$ and $\bar\gB$ in their places
when we attach external lines to the interaction vertices in the
Feynman graphs. This is because higher order terms in the expansion of
$\gA$ (and likewise $\bar\gA$) introduce products of multiple $\gB$
fields, whose momenta must sum to that of the original
$\gA$. Generically, their propagators cannot generate the poles
required for such terms to survive as the LSZ procedure takes the
external momenta on shell\footnote{This is assuming that the
  transformation does not generate a pole as an external momentum is
  taken on shell in such a way that higher-order terms survive the LSZ
  reduction.  We note here that there exist certain conditions within
  the context of the MHV lagrangian where the equivalence theorem is
  evaded.  Specifically, these include when off-shell at tree-level,
  in certain non-physical on-shell situations, and at
  loop-level. However, this is outside the sequence in which the topic
  was developed and is addressed in more detail in chapter
  \ref{cha:etv}.}.  Now consider formulating an on-shell partial
amplitude using \eqref{eq:mhvym-Vseries}: since such an object is
defined as the term in an $S$-matrix element that multiplies a given
colour trace, it is therefore clear that to extract a colour-ordered
MHV amplitude using \eqref{eq:mhvym-Vseries}, one simply contracts the
external momentum and gauge indices into the trace as appropriate and
multiplies by the relevant polarisation and quantum mechanical
factors.  Thus contracting an external MHV state into the MHV lagrangian
to obtain a partial amplitude amounts to pulling out the term with the
relevant helicity content and colour ordering and replacing
\[
\gB(x^0, \vec p_j) , \bar\gB(x^0, \vec p_j) \rightarrow -1 \times
-\frac{ig}{\sqrt2} \exp(i \check p_j x^0) \: T^{a_j}
\]
in \eqref{eq:mhvym-Vseries} and symmetrising as appropriate.  (The
significance of the factor of $\frac12$ for the $s=n/2+1$ term in
\eqref{eq:mhvym-Vseries} when $n$ is even is now apparent: this term
has a symmetry under the index shift $i \mapsto i + n/2 \mod n$, and
in such a circumstance there are two ways to contract the external
state into it for a given colour order. The factor of $\frac12$
absorbs this, assuming $V^s$ is symmetrised accordingly.)

Given what we know about tree-level gluon scattering, we can equate
this with the known Parke--Taylor form of the MHV amplitude to put
\begin{multline}
  \label{eq:mhvym-x1-onshell}
  \frac4{g^2} \int dx^0 \: V^s_{1\cdots n} \left( \frac{ig}{\sqrt2}
  \right)^n \exp\{i(\check 1 + \cdots \check n)x^0\} \tr (T^{a_1}
  \cdots
  T^{a_n}) \\
  = ig^{n-2} (2\pi)^2 \delta^4(p_1 + \cdots + p_n) \frac{\langle 1\:s
    \rangle^4} {\langle 1\:2 \rangle \cdots \langle n-1,n \rangle
    \langle n\:1 \rangle} \tr (T^{a_1} \cdots T^{a_n}).
\end{multline}
Of course, this just tells us that the vertex is the MHV amplitude
on-shell, specifically that
\begin{equation}
  \begin{split}
    \label{eq:mhvym-V}
    V^s(1\cdots n) &= (-i \sqrt 2)^{n-4} \frac{\langle 1\:s
      \rangle^4}{\langle 1\:2 \rangle \cdots \langle n-1,n \rangle
      \langle
      n\:1 \rangle} \\
    &= (-i)^n \frac{\hat2 \cdots \hat n \: (1\:s)^4}{\hat 1 \hat s^2
      (1\:2) \cdots (n-1,n) (n\:1)},
  \end{split}
\end{equation}
using \eqref{eq:lc-spinorbrackets}. It remains for us to show that
this is also true off-shell as per the CSW prescription, and to do so
we follow the argument given by Mansfield in \cite{Mansfield:2005yd}.
Now since we constructed $V^s_{1\cdots n}$ to be independent of $x^0$,
carrying out the integration on the LHS above gives
\begin{multline}
  (2\pi) \delta(\check 1 + \cdots + \check n) (-i\sqrt2)^{4-n}
  V^s_{1\cdots n} = \\ (2\pi) \delta(\check 1 + \cdots + \check n) \:
  (2\pi)^3 \delta^3(\vec p_1 + \cdots + \vec p_n) \frac{\langle 1\:s
    \rangle^4} {\langle 1\:2 \rangle \cdots \langle n-1,n \rangle
    \langle n\:1 \rangle}.
\end{multline}
We would like to simply cancel the $\delta(\check 1 + \cdots + \check
n)$ on either side here and claim \eqref{eq:mhvym-V} applies
off-shell, but we cannot do this: we must demonstrate the absence of
terms in the action that vanish on the support of the $\delta$
function if such a claim is to have any merit.

This will be so if the vertices are holomorphic, in the sense that
they contain no $\bar\partial$ derivatives (or equivalently $\bar p$
momentum factors). First, for $n>3$, any term that vanishes on the
support of this $\delta$ function must depend on a subset of the $\bar
p_i$ when all the momenta are on shell, since $\check p = p \bar p
/\hat p$. The $n=3$ case is exceptional, since
\[
\check 1 + \check 2 + \check 3 = \frac{\lvert(1\:2)\rvert^2}{\hat1
  \hat 2 (\hat1 + \hat2)}.
\]
As we are dealing with real momenta, $(1\:2)$ must vanish on the
support of the $\delta$ function, even though it is independent of any
$\bar p_i$. Nevertheless, we can (and will) check by computing the
relevant terms using the explicit form of the transformation that
\eqref{eq:mhvym-V} still holds off shell for the case of $n=3$.

Now, we will show that the terms of the MHV lagrangian are holomorphic
explicitly in section \ref{sec:mhvym-transf} when we solve the
transformation, but it is worthwhile noting here that Mansfield showed
that this was so in a rather elegant manner by considering the action
of the homogeneous part of gauge transformation
\begin{equation}
  \label{eq:holomorph-trans-1}
  \delta \gA = [\gA, \theta(\bar z)],
  \quad
  \delta \bar\gA = [\bar\gA, \theta(\bar z)]
\end{equation}
for an infinitesimal algebra-valued function $\theta$ of $\bar z$
\emph{alone} on the canonical transformation \eqref{eq:mhvym-transf}: under
the above shift, its LHS transforms as
\[
\delta(L^{-+}+L^\mpp) = \int_\Sigma d^3\vec x\: [\gA, \bar\partial
\theta(\bar z)] (\partial \bar\gA + \hat\partial^{-1} [\gA,
\hat\partial \bar\gA]).
\]
However, the same shift in $L^{-+}+L^\mpp$ can be achieved by just
shifting $\bar\partial \gA$ by
\[
\delta \bar\partial \gA = [\gA, \bar\partial \theta(\bar z)]
\]
and changing nothing else, so varying $\bar\partial \gA$ has the same
effect on the canonical transformation as
\eqref{eq:holomorph-trans-1}. However, the vertices of the MHV
lagrangian are generated by $L^\mmp + L^\mmpp$ and this is invariant
under \eqref{eq:holomorph-trans-1}. Thus the MHV lagrangian can
contain no $\bar\partial$ dependence, and hence for $n>3$ the vertices
contain no terms that generically vanish on mass shell.

We will test this for up to five gluons in section
\ref{sec:mhvym-examples} by explicitly computing the
relevant terms of the MHV lagrangian.


\section{Explicit form of the transformation}
\label{sec:mhvym-transf}

In this section, we will solve for the power series expansions of
$\gA$ and $\bar\gA$ to all orders in $\gB$ and $\bar\gB$. To do this,
we derive recurrence relations from the definition of the transformation
\eqref{eq:mhvym-transf} and \eqref{eq:mhvym-canon-mom}, which follows
from it being canonical.

\subsection{$\gA$ series to all orders}
Taking \eqref{eq:mhvym-transf-2}, we eliminate $\hat\partial \bar\gB$
using \eqref{eq:mhvym-canon-mom} and, dropping the $\hat\partial
\bar\gA$ factor, arrive at the following momentum space equation:
\begin{equation}
\label{eq:mhvym-A-eqn}
\Oomega_1 \gA_1 + i\int_{23} \Ozeta_3 [\gA_2,\gA_3] \delta(\vec p_1 -
\vec p_2 - \vec p_3) = \int_{\vec p} \Oomega_{\vec p} \gB^b_{\vec p}
\frac{\delta \gA_1}{\delta \gB^b_{\vec p}}.
\end{equation}
Here, we use the momentum-space analogues of the $\Oomega$ and
$\Ozeta$ operators, defined as
\[
\Oomega_{\vec p} := \frac{p \bar p}{\hat p}
\quad\text{and}\quad
\Ozeta_{\vec p} := \frac{\bar p}{\hat p}.
\]

As noted earlier, we postulate a solution for $\gA$ as a power series in
$\gB$. In momentum space, we write this in the form
\begin{equation}
\label{eq:A-series}
\gA_1 = \sum_{n=2}^\infty \int_{2\cdots n}
\Upsilon(12\cdots n) \gB_{\bar2} \cdots \gB_{\bar n}
(2\pi)^3 \delta({\textstyle \sum_{i=1}^n \vec p_i}).
\end{equation}
Our task now is to obtain the $\Upsilon$ coefficients. Note that we
have absorbed the various arrangements of $\gB$ strings
in the terms of \eqref{eq:A-series} 
into $\Upsilon$ coefficient. Indeed, brief inspection of \eqref{eq:mhvym-A-eqn}
shows the the various strings that emerge this way may be expressed as
nested commutators of the $\gB_i$, thereby ensuring the
tracelessness of both sides of \eqref{eq:A-series}. By
cyclicity of the trace, this condition leads us to the result that
\begin{equation}
\label{eq:Upsilon-dwi}
  \Upsilon(123 \cdots n)
+ \Upsilon(134 \cdots n2)
+ \cdots + \Upsilon(1,n,2,\dots,n\!-\!2,n\!-\!1) = 0,
\end{equation}
\ie\ an off-shell dual Ward identity.

At the lowest order, one can inspect \eqref{eq:mhvym-A-eqn} and
conclude that for non-vanishing momenta, $\gA_1 = \gB_1$ (\ie\
$\Upsilon(12) = 1$). To do this we had to cancel
$\omega_1$ on either side of \eqref{eq:mhvym-A-eqn}, so we cannot rule out
$\gA$ also having terms with $\delta(p)$ and/or $\delta(\bar p)$
support.
To deal this, we notice that there is some residual gauge freedom left
in the light-cone Yang--Mills action
\eqref{eq:lcym-action}, which we obtained
by gauge-fixing $\hat\gA = 0$ and
integrating out $\check\gA$. Thus, gauge motions of the form $\delta\gA =
[\gD,\theta]$ where $\theta$ does \emph{not} depend on $x^{\bar 0}$
are not fixed by this. (Should $\theta$ depend on $x^0$, the form of
the transformation would be complicated when we integrate out
$\check\gA$.) Let us study the case where $\theta$ is a function of
just $z$, \ie\ holomorphic gauge transformations,
under which the terms of \eqref{eq:lcym-action} transform as
\newcommand{\trint}{\tr \int_\Sigma d^3\vec x \:}
\begin{align}
\delta L^{-+} &= \hphantom{-} \trint \bar\gA
[\partial\theta,\bar\partial \gA], \\
\delta L^{\mpp} &= \hphantom{-} \trint \bar\gA [\bar\partial \gA, \partial\theta], \\
\delta L^{\mmp} &=  -\trint\{
[\bar\gA, \hat\partial \gA] \partial\hat\partial^{-1} [\bar\gA,
\theta]
- [[\hat\partial\gA,\bar\gA],\theta] \partial\hat\partial^{-1} \bar\gA \} \\
\begin{split}
\delta L^{\mmpp} &= -\trint\{
[\bar\gA, \hat\partial \gA] \hat\partial^{-2} [\partial\theta,
\hat\partial \bar\gA] \\
&\hphantom{=-\trint\{} - [[\hat\partial\gA, \bar\gA], \theta]
\hat\partial^{-2} [\gA,\hat\partial \bar\gA] 
- [[\hat\partial \bar\gA, \gA], \theta] \hat\partial^{-2}
[\bar\gA, \hat\partial\gA]\}
\end{split}
\end{align}
(Notice that this requires us to interpret $\bar\partial
\hat\partial^{-1} \theta(z) = \hat\partial^{-1} \bar\partial \theta(z)
= 0$, in line with the Mandelstam--Leibbrandt prescription
\cite{Mandelstam:1982cb, Leibbrandt:1983pj}.
Similarly, it follows that the lagrangian
is also invariant when $\theta$ is a function of just $\bar
z$.)
Significantly, we notice from the above that the LHS of
\eqref{eq:mhvym-transf} is invariant under holomorphic gauge
transformations; this would imply that $\gB$ and $\bar\gB$ are
also invariant. Now under such a transformation $\gA \rightarrow \gA
+ \partial \theta(z) + [\gA,\theta(z)]$. If we apply this
to the leading order term of \eqref{eq:A-series} and take the 
Fourier transform, we obtain
\[
\gA(p,\bar p)  -i p \: \theta(p) \: (2\pi) \delta(\bar p)
 + \int \frac{dk}{2\pi} \: [\gA(p-k,\bar p),\theta(k)]
= \gB(p, \bar p).
\]
Thus we will interpret the form of \eqref{eq:A-series} as tantamount to further
gauge fixing.

That aside, we
substitute \eqref{eq:A-series} into \eqref{eq:mhvym-A-eqn}, remove
the momentum-conserving $\delta$ functions and carefully relabel to
yield the recurrence relation
\begin{equation}
\label{eq:Upsilon-rr}
\Upsilon(1\cdots n) = \frac{-i}{\omega_1 + \cdots + \omega_n}
\sum^{n-1}_{j=2}\left(\zeta_{j+1,n}-\zeta_{2j}\right)
\Upsilon(- ,2,\cdots,j)\Upsilon(- ,j+1,\cdots,n).
\end{equation}
Here, we have made use of the following new notation: first, an
argument ``$-$'' in $\Upsilon$ (and in other related coefficients that
we will encounter later in this work) stands for minus the sum of the
remaining arguments, as per conservation of momentum. Secondly, we
have $\zeta_{jk} := \zeta(P_{jk})$ where $P_{jk} := \sum_{i=j}^k \vec
p_i$. We will use a similar short-hand for $\omega$ below. 
By iterating
\eqref{eq:Upsilon-rr}, we obtain the following coefficients:
\begin{align}
\label{Upsilon3}
\Upsilon(123) &=
   {-i} \frac{\zeta_3 - \zeta_2}
         {\omega_{1} + \omega_2 + \omega_3}, \\
\label{Upsilon4}
\Upsilon(1234) &= \frac 1 {\omega_{1} + \omega_2 + \omega_3 +
\omega_4} \left[
    \frac{(\zeta_4 - \zeta_3) (\zeta_{34} - \zeta_2)}
         {\omega_{34} - \omega_3 - \omega_4}
  + \frac{(\zeta_{4} - \zeta_{23})(\zeta_3 - \zeta_2) }
         {\omega_{23} - \omega_2 - \omega_3}
\right]
\end{align}
At first glance, these seem to not be holomorphic. However, if we use
conservation of momentum to express these in terms of independent
momenta and simplify, we obtain the following very compact
expressions:
\begin{align*}
\Upsilon(123)   &= { i} \frac{\hat1}{(2\: 3)}, \\
\Upsilon(1234)  &= \frac{\hat1\hat3}{(2\:3)(3\:4)},\\
\Upsilon(12345) &= {-i} \frac{\hat1\hat3\hat4}{(2\:3)(3\:4)(4\:5)},
\end{align*}
whence we conjecture
\begin{equation}
\label{eq:Upsilon-coeff}
\Upsilon(1\cdots n) = (-i)^n {\hat1\hat3\hat4\cdots
\widehat{n-1}\over(2\:3)(3\:4)\cdots(n\!-\!1, n)},
\end{equation}
for $n>3$. 

We prove this by induction on $n$. Clearly, the above expressions for
the lower order coefficients provide the initial step. For the
inductive step, we substitute \eqref{eq:Upsilon-coeff} into the RHS of
\eqref{eq:Upsilon-rr} and pull out as many $j$ independent factors as
possible to obtain
\[
 -\frac{\Upsilon(1\cdots n)}{\hat1 (\omega_1 + \cdots + \omega_n)}
\sum_{j=2}^{n-1} \frac{(j,j+1)}{\hat\jmath \: \widehat{j+1}}
\{ P_{2j} \: P_{j+1,n} \}
\]
To evaluate this sum, we expand $(j,j+1)$ so the sum becomes
\[
\sum_{j=3}^n \frac{\tilde j}{\hat j} \{ P_{2,j-1} \: P_{jn} \}
- \sum_{j=2}^{n-1} \frac{\tilde j}{\hat j} \{ P_{2j} \: P_{j+1,n} \}
= \sum_{j=2}^n \frac{\tilde j}{\hat j}(\{ P_{2,j-1} \: P_{jn}\} - \{
P_{2j} \: P_{j+1,n} \})
\]
where the end cases are dealt with by defining $P_{ij} := \vec p_i +
\cdots + \vec p_n + \vec p_1 + \cdots + \vec p_j$ when $j < i$, hence
$P_{i,i-1} = 0$. Upon substituting $P_{jn} = -\vec p_1 - P_{2,j-1}$
and similarly for $P_{j+1,n}$,
the sum collapses to $-\hat 1 (\omega_1 + \cdots + \omega_n)$, 
and the claim \eqref{eq:Upsilon-coeff} is proved.

\subsection{$\bar\gA$ expansion to all orders}
\label{ssec:mhvym-Abar-series}
Differentiating \eqref{eq:A-series} with respect to $\gB$ and
substituting the inverse into \eqref{eq:mhvym-canon-mom} implies a
$\Sigma$ momentum space expansion for $\bar\gA$ of the form
\begin{equation}
\label{eq:Abar-series}
\bar\gA_{\bar 1} = \sum_{m=2}^\infty \sum_{s=2}^m
\int_{2\cdots m} \frac{\hat s}{\hat 1} \Xi^{s-1}(\bar 1 2\cdots m) \:
\gB_{\bar 2} \cdots \gB_{\overline{s-1}} \bar\gB_{\bar s}
\gB_{\overline{s+1}} \cdots \gB_{\bar m}
\: (2\pi)^3 \delta(\vec p_1 - {\textstyle \sum_{i=2}^m \vec p_i})  
\end{equation}
where the superscript on $\Xi$ indicates the relative position of
$\bar\gB$ in the string, not the momentum argument. To extract a
recurrence relation, we use the fact that
\begin{equation}
\label{eq:mhvym-qp-QP}
\tr \int_\Sigma d^3\vec x \: \check\partial \gA \hat\partial \bar\gA
= \tr \int_\Sigma d^3\vec x \: \check\partial \gB \hat\partial \bar\gB,
\end{equation}
which follows from the properties of a canonical transform, specifically
\eqref{eq:qp-QP}. Since all the
fields have the same $x^0$ dependence, and none of the $\Upsilon$
coefficients depend on $x^0$,
\begin{equation}
\check\partial \gA_1 = \sum_{n=2}^\infty \sum_{r=2}^n \int_{2\cdots m}
\Upsilon(1\cdots n) \gB_{\bar 2} \cdots \check\partial \gB_{\bar r}
\cdots \gB_{\bar n}
\: (2\pi)^3 \delta^3({\textstyle \sum_{i=1}^m \vec p_i}),
\end{equation}
which we substitute into the LHS of \eqref{eq:mhvym-qp-QP} along with
\eqref{eq:Abar-series}. By considering each order in powers of $\gB$,
using the cyclic property of the trace to move $\check\partial \gB$ to
the front of each string,
matching up the position of $\bar\gB$ in the strings on both sides,
and then carefully relabelling the momentum
arguments, we arrive at
\begin{multline}
\label{eq:Xi-rr}
\Xi^l(1\cdots n) = - \sum_{r=2}^{n+1-l} \sum_{m=\max(r,3)}^{r+l-1}
\Upsilon(\overbrace{-,n\!-\!r\!+\!3,\cdots,m\!-\!r\!+\!1}^{(m)}) \times \\
\Xi^{l+r-m}(\underbrace{-,m\!-\!r\!+\!2,\cdots,n\!-\!r\!+\!2}_{(n-m+2)})
\end{multline}
(the braces' labels indicate the number of arguments they
enclose). The momentum indices on the RHS should be interpreted
cyclically (\ie\ modulo $n$). Note that in the case where the upper limit of
the inner sum is less than the lower limit, the sum should be taken as
zero (specifically, this happens when $r=2$ and $l=1$).

To obtain concrete expressions for the $\Xi$ coefficients from
\eqref{eq:Xi-rr}, we begin by noting that $\Xi(12)=1$, as required by
the fact that the leading term in the expansion of $\bar\gA$ is $\bar\gB$. 
We can now iterate \eqref{eq:Xi-rr} directly to compute the first few
non-trivial coefficients:
\begin{align}
\Xi^1(123) &= -\Upsilon(231) = -\frac{\hat2}{\hat1} \Upsilon(123),
\label{eq:Xi1123} \\
\Xi^2(123) &= -\Upsilon(312) = -\frac{\hat3}{\hat1} \Upsilon(123).
\label{eq:Xi2123}
\end{align}
Continuing, we have
\begin{equation*}
\begin{split}
\Xi^1(1234) &= -\Upsilon(2+3,4,1) \Xi^1(1+4,2,3) - \Upsilon(2341) \\
 &= -\frac{\hat2 (\hat2 + \hat3)}{(4\:1) (2\:3)} - \frac{\hat2
   \hat4}{(3\:4) (4\:1)} \\
 &= - \frac{\hat2 (\hat2 + \hat3) (3\:4) + \hat2 \hat4 (2\:3)}{(4\:1)
   (2\:3) (3\:4)} \\
&= -\frac{\hat2 \hat3}{(2\:3) (3\:4)}
 = -\frac{\hat2}{\hat1} \Upsilon(1234)
\end{split}
\end{equation*}
for which we have used the Bianchi-like identity \eqref{eq:bianchi},
and similarly,
\begin{align*}
\begin{split}
\Xi^2(1234) &= -\Upsilon(3\!+\!4,1,2)\,\Xi^1(1\!+\!2,3,4)
                -\Upsilon(2\!+\!3,4,1)\,\Xi^2(1\!+\!4,2,3)-\Upsilon(3412)
                \\
&= -\frac{\hat3}{\hat1} \Upsilon(1234),
\end{split} \\
\begin{split}
\Xi^3(1234) &=
-\Upsilon(3\!+\!4,1,2)\,\Xi^2(1\!+\!2,3,4)-\Upsilon(4123) \\
& = -\frac{\hat4}{\hat1} \Upsilon(1234). 
\end{split}
\end{align*}
From this, we claim that
\begin{equation}
\label{eq:Xi-coeff}
\Xi^{s-1}(1\cdots n) = -\frac{\hat s}{\hat 1} \Upsilon(1 \cdots n)
\end{equation}
for $n\ge2$ and $2\le s\le n$.

The proof of this claim follows by induction on $n$. The low-order
coefficients above clearly supply the initial step. The inductive part
can be proved by substituting \eqref{eq:Xi-coeff} into the RHS of
\eqref{eq:Xi-rr}. We pull out the $r$ and $m$ independent
factors to obtain
\begin{multline*}
(-i)^n \: \widehat{l+1} \: \frac{\hat1 \cdots \hat n}{(1\:2) \cdots
  (n\!-\!1,n) (n\:1)} \times \\
\sum_{r=2}^{n+1-l} \sum_{m=\max(r,3)}^{r+l-1}
\frac{(m\!-\!r\!+\!1, m\!-\!r\!+\!2) (n\!-\!r\!+\!2, n\!-\!r\!+\!3) \:
  {\hat P}_{n-r+3, m-r+1}}
{\widehat{m\!-\!r\!+\!1} \quad \widehat{m\!-\!r\!+\!2}
\quad \widehat{n\!-\!r\!+\!2}
\quad \widehat{n\!-\!r\!+\!3}},
\end{multline*}
so the proof follows if we can show that the nested sum here equals
\begin{equation}
\label{eq:Xi-cancels}
-\frac{(1\:2)(n\:1)}{\hat1 \hat 2 \hat n}
\end{equation}
and so the RHS equals the LHS as given by \eqref{eq:Xi-coeff}.
First, we evaluate the inner sum. By performing the change of
variables $j=m-r+1$, this becomes one of the form
\begin{equation}
\sum_{j=a}^b \hat P_{ij} \left(\frac{\widetilde{j+1}}{\widehat{j+1}}  - \frac{\tilde\jmath}{\hat\jmath}\right)
= \hat P_{ib} \frac{\widetilde{b+1}}{\widehat{b+1}} - \hat P_{ia}
\frac{\tilde a}{\hat a} - \tilde P_{a+1,b},
\label{eq:telescope-sum-1}
\end{equation}
so we are left with
\begin{equation}
\label{eq:Xi-rr-outer}
\sum_{r=2}^{n+1-l}
\frac{(n\!-\!r\!+\!2, n\!-\!r\!+\!3)}
{\widehat{n\!-\!r\!+\!2} \quad \widehat{n\!-\!r\!+\!3}} \left\{
-{\tilde P}_{q+2,l} + \frac{\widetilde{l+1}}{\widehat{l+1}}{\hat
P}_{n-r+3,l} - \frac{\widetilde{q+1}}{\widehat{q+1}} {\hat
P}_{n-r+3, q+1}\right\}
\end{equation}
where $q=1$ if $r=2$, otherwise $q=0$. We note that the expression in
the braces vanishes when $r=2$ and $l=1$ given the wrap-around
interpretation of $P_{ij}$. This is consistent with our
statement earlier that this quantity is taken to vanish the lower
limit of the inner sum exceeds the upper limit.

We now evaluate the remaining sum \eqref{eq:Xi-rr-outer}. First, we will treat the end case
$l=n-1$ separately: here, the sum has only one term, $r=2$, and
\eqref{eq:Xi-rr-outer} becomes
\[
\frac{(n\:1)}{\hat n \hat1} \left\{ - \tilde P_{3,n-1} + \frac{\tilde
    n}{\hat n} \hat P_{1,n-1} - \frac{\tilde 2}{\hat 2} \hat P_{12}
\right\}
= \frac{(n\:1)}{\hat n \hat1} \left\{ \tilde1 + \tilde2 - \frac{\tilde
    2}{\hat 2} (\hat1 + \hat2) \right\}
\]
which is clearly equal to \eqref{eq:Xi-cancels}. For the remaining
$l<n-1$, we obtain a telescoping sum from the first term plus
 two sums of the form \eqref{eq:telescope-sum-1}
from the remaining terms
(treating the $r=2$
 term separately as necessary) to obtain
\begin{multline*}
{\tilde
P}_{1,l+1}\left({\tilde1\over\hat1}-{\widetilde{l+1}\over\widehat{l+1}}\right)
+{\tilde n\over\hat
n}\left({\hat1\tilde2\over\hat2}-\tilde1\right)+{\tilde1\over\hat1}\left(-{\tilde
P}_{3l}+{\widetilde{l+1}\over\widehat{l+1}}{\hat
P}_{1l}-\tilde2-{\hat1\tilde2\over\hat2}\right)\\
+{\widetilde{l+1}\over\widehat{l+1}}\left({\tilde
P}_{2l}+\widetilde{l+1}+{\tilde1\over\hat1}{\hat
P}_{l+2,1}\right).
\end{multline*}
Applying conservation of momentum, this collapses to
\[
\frac{\tilde 1^2}{\hat1} + \frac{\tilde n}{\hat n} \left( \frac{\hat1
    \tilde2}{\hat2} - \tilde1 \right) - \frac{\tilde1 \tilde2}{\hat2}
= - \frac{(n\:1) (1\:2)}{\hat1 \hat2 \hat n},
\]
and thus the assertion of \eqref{eq:Xi-coeff} is proved.

We note here that we have proved explicitly that $\Upsilon$ and $\Xi$,
and hence the vertices formed by substituting their parent series
\eqref{eq:A-series} and \eqref{eq:Abar-series}
into \eqref{eq:lcym-mmp} and \eqref{eq:lcym-mmpp} are holomorphic, in
line with Mansfield's analytical argument. If follows that these
vertices give the CSW continuation of the Parke--Taylor MHV amplitudes
when taken off-shell.


\section{Examples}
\label{sec:mhvym-examples}

Let us now verify that the field transformation as described above
results in vertices proportional to the Parke--Taylor amplitudes. We
will do this by computing the interaction terms in the MHV lagrangian
for up to five gluons.

\subsection{Three-gluon vertex}
\label{ssec:mhvym-3g}
Since $\gA$ ($\bar\gA$) is $\gB$ ($\bar\gB$) to the lowest order, the
three-gluon vertex is simply the vertex \eqref{eq:lcym-mmp}:
\[
V^2(123) = i \frac{\hat3}{\hat1\hat2} (1\:2).
\]
Conversely, \eqref{eq:mhvym-V} says it is
\[
i \frac{\hat3}{\hat1\hat2} \frac{(1\:2)^3}{(2\:3)(3\:1)}
\]
and conservation of momentum implies that
$(1\:2) = (2\:3) = (3\:1)$, so the two expressions are equal and $V^2$
is of the Parke--Taylor form.

\subsection{Four-gluon vertices}
\label{ssec:mhvym-4g}
The four-gluon MHV vertices are the first to receive contributions from the
next-to-leading order terms in \eqref{eq:A-series} and
\eqref{eq:Abar-series}, and so are more interesting tests of the technique.
Both receive contributions from $L^\mmp$ of \eqref{eq:lcym-mmp}, and
\eqref{eq:lcym-mmpp}, which is
\begin{multline}
\label{eq:lcym-mmpp-mom}
L^\mmpp = \int_{1234}\Biggl\{
  W^2(1234)
  \tr(\bar\gA_{\bar1}\bar\gA_{\bar2}\gA_{\bar3}\gA_{\bar4}) \\ +
  \frac12 W^3(1234)
  \tr(\bar\gA_{\bar1}\gA_{\bar2}\bar\gA_{\bar3}\gA_{\bar4})
\Biggr\} \:(2\pi)^3 \delta^3({\textstyle \sum_{i=1}^4 \vec p_i}),
\end{multline}
written in $\Sigma$ momentum space, where the coefficients
\[
W^2(1234) = -\frac{\hat1\hat3+\hat2\hat4}{(\hat1+\hat4)^2}
\quad\text{and}\quad
W^3(1234) = \frac{\hat1\hat4+\hat2\hat3}{(\hat1+\hat2)^2}
+ \frac{\hat1\hat2+\hat3\hat4}{(\hat1+\hat4)^2}
\]
after symmetrization.

First, consider $A(1^- 2^- 3^+ 4^+)$, which is proportional to $V^2(1234)$ by
\eqref{eq:mhvym-V}. It comprises four terms in turn from substituting for:
\begin{enumerate}
\item the first $\bar\gA$ in \eqref{eq:lcym-mmp} with
  the ${\cal O}(\gB \bar\gB)$ term in \eqref{eq:Abar-series};
\item the second $\bar\gA$ in \eqref{eq:lcym-mmp} with
  the ${\cal O}(\bar\gB \gB)$ term in \eqref{eq:Abar-series};
\item the $\gA$ in  \eqref{eq:lcym-mmp} with
  the ${\cal O}(\gB^2)$ term in \eqref{eq:A-series}; and
\item the fields in \eqref{eq:lcym-mmpp-mom} at the trivial leading order.
\end{enumerate}
This gives
\begin{equation}
\label{eq:V21234}
V^2(1234) = \frac{\hat1}{\hat5}V^2(523)\Xi^2(\bar541) +
\frac{\hat2}{\hat5}V^2(154)\Xi^1(\bar523) +V^2(125)\Upsilon(\bar534) +
W^2(1234),
\end{equation}
where $\vec p_5$ is determined by momentum conservation
in each term (thus \eg\ in the first term $\vec p_5=\vec p_1 + \vec p_4$).
To compare
this formula to the expected result \eqref{eq:mhvym-V},
we must write both as unique
functions of independent momenta. We eliminate $\vec p_4$ using
conservation of momentum, and simplifying by partial fractions
with the help of computer algebra, both \eqref{eq:mhvym-V} and
\eqref{eq:V21234} give:
\[
\frac{(\hat 1 + \hat 2)^2 (\hat 1 \tilde 2 - \hat 2 \tilde 1)} {\hat 1
  \hat 2 \: [(\hat 1 + \hat 2) \tilde 3 - \hat 3 \tilde 1 - \hat 3
  \tilde 2]} -\frac{\hat 2 (\hat 1 + \hat 2 + \hat 3)(\hat 1 \tilde 2
  - \hat 2 \tilde 1)} {\hat 1 (\hat 2 + \hat 3) (\hat 2 \tilde 3 -
  \hat 3 \tilde 2)} -\frac{\hat 1 \hat 3 (\hat 1 \tilde 2 - \hat 2
  \tilde 1)} {\hat 2 (\hat 2 + \hat 3) (\hat 1 \tilde 2 + \hat 1
  \tilde 3 - \hat 2 \tilde 1 - \hat 3 \tilde 1) }.
\]
Thus we conclude that
\[
V^2(1234) = \frac{\langle1\:2\rangle^3}
{\langle2\:3\rangle\langle3\:4\rangle\langle4\:1\rangle}.
\]

We treat $A(1^- 2^+ 3^- 4^+)$ similarly, and after symmetrization,
\begin{multline}
  V^3(1234) = \hatfrac15 \: V^2(352) \Xi^2(\bar 541) +
  \hatfrac 35 \: V^2(512) \Xi^1(\bar 534) \\
  + \hatfrac 35 \: V^2(154)\, \Xi^2(\bar 523) +
  \hatfrac 15 \: V^2(534) \Xi^1(\bar 512) + W^3(1234),
\end{multline}
and it is straightforward to confirm as above that this agrees with
\eqref{eq:mhvym-V}:
\[
V^3(1234)  = \frac{\hat2\hat4}{\hat1\hat3}
\frac{(1\:3)^4}{(1\:2)(2\:3)(3\:4)(4\:1)}.
\]

\subsection{Five-gluon vertices}
\label{ssec:mhvym-5g}
Finally, we calculate the coefficients of the five-gluon terms in the
MHV lagrangian, $V^2(12345)$ and $V^3(12345)$. This calculation
involves substituting up to the first three terms in
the expansions \eqref{eq:A-series} and \eqref{eq:Abar-series} into
both original vertices
\eqref{eq:lcym-mmp} and \eqref{eq:lcym-mmpp}. We find that
\begin{align*}
\begin{split}
  V^2(12345) &=
  \hatfrac 36 \: V^2(612) \,\Xi^1(\bar 6345)
  + \hatfrac 17 \: V^2(726) \Upsilon(\bar 634) \,\Xi^2(\bar 751)
  + V^2(126) \Upsilon(\bar 6345)\\
  &\quad + \hatfrac 26 \hatfrac 17 \: V^2(764) \,\Xi^1(\bar 623) \,\Xi^2(\bar 751)
  + \hatfrac 26 \: V^2(167) \,\Xi^1(\bar 623) \Upsilon(\bar 745)\\
  &\quad + \hatfrac 26 \: V^2(165) \,\Xi^1(\bar 6234)
  + W^2(1236) \Upsilon(\bar 645)
  + W^2(1265) \Upsilon(\bar 634)\\
  &\quad + \hatfrac 26 \: W^2(1645) \,\Xi^1(\bar 623)
  + \hatfrac 16 \: W^2(6234) \,\Xi^2(\bar 651),\end{split}\\
\begin{split}
  V^3(12345) &=
  \hatfrac 36 \: V^2(612) \,\Xi^1(\bar 6345)
  + \hatfrac 16 \: V^2(634) \,\Xi^2(\bar 6512)
  + \hatfrac 16 \: V^2(637) \,\Xi^1(\bar 612) \Upsilon(\bar 745)\\
  &\quad + \hatfrac 16 \hatfrac 37 \: V^2(675) \,\Xi^1(\bar 612) \,\Xi^1(\bar 734)
  + \hatfrac 17 \hatfrac 36 \: V^2(672) \,\Xi^1(\bar 634) \,\Xi^2(\bar
  751)\\
  &\quad + \hatfrac 16 \hatfrac 37 \: V^2(674) \,\Xi^2(\bar 651) \,\Xi^2(\bar 723)
  + \hatfrac 36 \: V^2(167) \,\Xi^2(\bar 623) \Upsilon(\bar 745)\\
  &\quad + \hatfrac 36 \: V^2(165) \,\Xi^2(\bar 6234)
  + \hatfrac 16 \: V^2(362) \,\Xi^3(\bar 6451)
  + \hatfrac 36 \: W^2(1645) \,\Xi^2(\bar 623)\\
  &\quad + \hatfrac 16 \: W^2(6345) \,\Xi^1(\bar 612)
  + \hatfrac 36 \: W^3(1265) \,\Xi^2(\bar 634)
  + \hatfrac 16 \: W^3(3462) \,\Xi^2(\bar 651)\\
  &\quad + W^3(1236) \Upsilon(\bar 645) \end{split}
\end{align*}
(where, like before, indices 6 and 7 label momenta that are uniquely
determined in terms of the first five by momentum
conservation). Again, we eliminate $\vec p_5$ in favour of the first four
momenta and doing likewise for the corresponding right hand sides of
\eqref{eq:mhvym-V}, we find the expressions agree, and thus confirm that
\begin{align*}
  V^2(12345) &= -i\sqrt2 \frac{\langle 1\:2 \rangle^3} {\langle 2\:3
    \rangle \langle 3\:4 \rangle \langle 4\:5 \rangle \langle 5\:1
    \rangle}, \\
  V^3(12345) &= -i\sqrt2 \frac{\langle 1\:3 \rangle^4} {\langle 1\:2
    \rangle \langle 2\:3 \rangle \langle 3\:4 \rangle \langle 4\:5
    \rangle \langle 5\:1 \rangle}.
\end{align*}

\section{Summary and discussion}
\label{sec:mhvym-conclusion}

In this chapter, we have reviewed the construction of ref.\
\cite{Mansfield:2005yd} in which a lagrangian that produces the CSW
rules of ref.\ \cite{Cachazo:2004kj} at tree-level is obtained by means of a
canonical transformation of the gauge field variables $\gA$ and
$\bar\gA$ of light-cone gauge Yang--Mills theory. This transformation
absorbs the kinetic term and \mhvbar\ vertex into the kinetic term of
the theory in terms of the new variables $\gB$ and $\bar\gB$
according to
\[
L^{-+}[\gA, \bar\gA] + L^\mpp[\gA, \bar\gA] = L^{-+}[\gB, \bar\gB],
\]
and the interaction part of the MHV lagrangian comes from expressing
$L^\mmp + L^\mmpp$ in terms of $\gB$ and $\bar\gB$.

We then solved for the explicit form of $\gA$ and $\bar\gA$ as power
series in $\gB$ and $\bar\gB$ and found that the coefficients take a
particularly simple form.  We summarise these results below:
\begin{align*}
  \gA_1 &= \sum_{n=2}^\infty \int_{2\cdots n} \Upsilon(1\cdots n)
  \gB_{\bar 2} \cdots \gB_{\bar n} \: (2\pi)^3 \delta^3({ \textstyle
    \sum_{i=1}^n \vec p_i
  }),\\
  \bar\gA_1 &= \sum_{m=2}^\infty \sum_{s=2}^m \int_{2\cdots m}
  \frac{\hat s^2}{\hat1^2} \Upsilon(1\cdots m) \gB_{\bar2} \cdots
  \bar\gB_{\bar s} \cdots \gB_{\bar m} \: (2\pi)^3 \delta({\textstyle
    \sum_{i=1}^m \vec p_i}),
\end{align*}
where the coefficient
\[
\Upsilon(1\cdots n) = (-i)^n \frac{\hat1 \hat3 \cdots \widehat{n-1}}
{(2\:3) \cdots (n-1, n)}.
\]
We note that this transformation is local in light-cone time, but
non-local in the quantisation surface.  The validity of this was
tested by computing the terms of the MHV lagrangian for up to five
gluons and showing that the vertices are the Parke--Taylor MHV
amplitudes, continued off-shell by the CSW prescription. Essential to
this is the holomorphic nature of the transformation, and hence the
vertices, which is demonstrated explicitly. The algebraic
manipulations used in this process were greatly aided by the use of
\eqref{eq:lc-spinorbrackets} to express the spinor invariants in terms
of light-cone co-ordinates.

The significance of this work lies in the field-theoretic
understanding it gives to the CSW construction, which here is seen to
be underpinned by light-cone quantisation. Furthermore, since they
have now been shown to have a Lagrangian formulation, we can proceed
to apply well understood quantum field theory techniques. Of
particular relevance to the phenomenological programme is the
calculation of quantum corrections, and hence the regularisation of
the MHV lagrangian: we simply take our favourite regulator, and
observe the consequences unfold as we apply the transformation This
will be considered further in the next chapter, where we will perform
such an analysis for dimensional regularisation; unfortunately, we
will see that this comes at the expense of the simplicity of formalism
that we have in four dimensions.

\subsection{The Case of the Missing Amplitudes}
\label{ssec:mhvym-conclusion-missing}

We hinted at the end of section \ref{sec:background-closing}
the CSW rules were in
a sense `incomplete', in that it was impossible using vertices with an
MHV helicity content alone to construct loop amplitudes with up to
one gluon of negative helicity and an arbitrary number of positive
helicity. This is a particular annoyance since in pure Yang--Mills,
amplitudes like $(\fourplus)$ at one loop are non-vanishing \viz\
\eqref{eq:fourplus-oneloop}. Even more
fundamentally, the \mhvbar\ amplitude is also missing. This takes the
value
\begin{equation}
  A(1^-,2^+,3^+)
  = i g \frac{\phantom{{}^3}[2\:3]^3}{[3\:1][1\:2]}.
  \label{eq:missing-mhvbar}
\end{equation}
For this to be non-vanishing in the on-shell limit, we need to work
with complex momenta or a space-time with a $(2,2)$ signature metric;
in either circumstance, $\tilde\lambda$ is not necessarily the complex
conjugate of $\lambda$, so $[i\:j]$ does not have to vanish as $p_i
\cdot p_j \propto \langle i\:j \rangle [i\:j] \rightarrow 0$.

But all is not lost: the next chapter addresses this issue by showing
that under certain circumstances the transformation `evades' the
$S$-matrix equivalence theorem, bringing in contributions that allow
us to recover precisely these missing pieces.

\begin{subappendices}

\section{Canonical transformations}
\label{sec:canon}

The aim of this appendix is not to give a complete treatment of the
subject of canonical transformations
(which may be found in the literature, \eg\
\cite{Goldstein:2002aa}), but rather develop the points and
results relevant to the main text.

A canonical transformation is a map from one set of phase space
co-ordinates $\vec q = (q^1,\dots,q^n)$ and $\vec p = (p_1,\dots,p_n)$
(the conjugate momenta) for a system with hamiltonian $H$
onto another set $\vec Q$, $\vec P$ with hamiltonian $K$ such that
this map preserves the \emph{form} of Hamilton's equations of motion,
\ie
\begin{equation}
  \label{eq:hamilton-eom}
  \begin{pmatrix} \dot{\vec q} \\ \dot{\vec p} \end{pmatrix}
  = \begin{pmatrix} 0 & +1 \\ -1 & 0 \end{pmatrix}
  \begin{pmatrix} \partial H / \partial \vec q \\ \partial H / \partial
    \vec p \end{pmatrix}
  \quad\longleftrightarrow\quad
  \begin{pmatrix} \dot{\vec Q} \\ \dot{\vec P} \end{pmatrix}
  = \begin{pmatrix} 0 & +1 \\ -1 & 0 \end{pmatrix}
  \begin{pmatrix} \partial K / \partial \vec Q \\ \partial K / \partial
    \vec P \end{pmatrix}.
\end{equation}
There are a number of equivalent ways to approach the analysis of
canonical transformations.  First, let us consider the generating
function approach.  Hamilton's equations of motion
\eqref{eq:hamilton-eom} may be obtained from the variational
principle: under an arbitrary variation $\delta$ of the phase space
co-ordinates,
the action in both
variables must be stationary, \ie
\begin{align*}
  \delta \int_a^b (\dot q^i p_i - H) \: dt &= 0,\\
  \delta \int_a^b (\dot Q^i P_i - K) \: dt &= 0.
\end{align*}
where a
dot denotes a total derivative with respect to time, $t$.  Now in
order that both of these are satisfied, we must have
\begin{equation}
  \label{eq:canonical-hamilt}
  \lambda(\dot q^i p_i - H) = \dot Q^i P_i - K + \frac{dF}{dt}.
\end{equation}
Here, $F$ may depend on $t$ and any of the phase space co-ordinates
(in which it must have continuous second derivatives).  We can add a
total time derivative of it on the RHS above since the variation is
taken to vanish at $t=a$ and $t=b$, and so it does not affect the
variational integrals.  We can then obtain expressions relating the
two sets of phase space co-ordinates
in terms of partial derivatives of $F$, hence
it is referred to as the \emph{generating function}.

Let us study the particular case  where
\begin{equation}
\label{eq:canon-ourform}
F(\vec q, \vec P) = f^i(\vec q)P_i - Q^i P_i,
\end{equation}
where the $Q^i$ on the RHS may be regarded as functions of $\vec q$
and $\vec P$.\footnote{This is a particular case of the `Type 2' canonical
transform in chapter 9 of \cite{Goldstein:2002aa}.} Then
\[
\frac{dF}{dt} = \dot q^i \frac{\partial f^j}{\partial q^i} P_j +
f^i(\vec q) \dot P_i - \dot Q^i P_i - Q^i \dot P_i,
\]
which we plug into the RHS of \eqref{eq:canonical-hamilt} (setting
$\lambda=1$ by scaling the co-ordinates in a straightforward manner
if necessary). Since we may regard $\vec q$ and $\vec P$ as
independent variables, this equation is only satisfied if
\begin{align}
  H &= K, \notag \\
  Q^i &= f^i(\vec q), \label{eq:canon-coord} \\
  p_i &= \frac{\partial f^j}{\partial q^i} P_j = \frac{\partial
    Q^j}{\partial q^i} P_j. \label{eq:canon-mom}
\end{align}
Under these conditions, $\dot F$ vanishes and one immediate
consequence from \eqref{eq:canonical-hamilt} is that
\begin{equation}
  \label{eq:qp-QP}
  \dot q^i p_i = \dot Q^i P_i.
\end{equation}

Let us now generalise to \emph{contact transformations},
time-independent transformations of the phase space co-ordinates (both
co-ordinates and momenta); clearly, the transformation described above
is an example of this. The hamiltonian is invariant under such a
transformation, \ie\ $H = K$, which can be shown by extending the
generating function approach to other forms of $F$
using Legendre transforms, or alternatively
by considering the hamiltonian as a scalar function on the phase space
manifold. Taking this geometric viewpoint, we can deduce some important
conditions that canonical transformations must satisfy in general.  If
we apply the chain rule to the first equation of
\eqref{eq:hamilton-eom}, we have
\[
\begin{split}
  \begin{pmatrix} \dot{\vec q} \\ \dot{\vec p} \end{pmatrix} &=
  \frac{\partial(\vec Q, \vec P)}{\partial(\vec q, \vec
    p)} \begin{pmatrix} \dot{\vec Q} \\ \dot{\vec P} \end{pmatrix} \\
  &= \frac{\partial(\vec Q, \vec P)}{\partial(\vec q, \vec p)}
  \begin{pmatrix} & +1 \\ -1 & \end{pmatrix} \begin{pmatrix} \partial
    K / \partial \vec Q \\ \partial K / \partial
    \vec P \end{pmatrix} \\
  &= \frac{\partial(\vec Q, \vec P)}{\partial(\vec q, \vec p)}
  \begin{pmatrix} & +1 \\ -1 & \end{pmatrix} \frac{\partial(\vec Q,
    \vec P)}{\partial(\vec q, \vec p)}^{\rm
    T} \begin{pmatrix} \partial H / \partial \vec q \\ \partial H
    / \partial \vec p \end{pmatrix},
\end{split}
\]
where the Jacobian matrix
\[
\frac{\partial(\vec Q, \vec P)}{\partial(\vec q, \vec p)} \equiv
\begin{pmatrix} \partial\vec Q/\partial\vec q & \partial\vec
  Q/\partial\vec p \\ \partial\vec P/\partial\vec q & \partial\vec
  P/\partial \vec p \end{pmatrix}.
\]
Comparing with the first equation in \eqref{eq:hamilton-eom}, we infer
that a canonical transform must satisfy
\begin{equation}
  \label{eq:symplectic-invariant}
  \begin{pmatrix} & +1 \\ -1 & \end{pmatrix} =
  \frac{\partial(\vec Q, \vec P)}{\partial(\vec q, \vec p)} 
  \begin{pmatrix} & +1 \\ -1 & \end{pmatrix} \frac{\partial(\vec Q, \vec
    P)}{\partial(\vec q, \vec p)}^{\rm T}.
\end{equation}
This leads us immediately to an important result for canonical
transformations: a theorem, due to Liouville, that states the phase
space volume element is invariant under a canonical transformation. In
particular,
\[
d\vec Q \wedge d\vec P = \left| \det \frac{\partial(\vec Q, \vec
    P)}{\partial(\vec q, \vec p)} \right| d\vec q \wedge d\vec p,
\]
but taking the determinant of both sides of
\eqref{eq:symplectic-invariant} tells us that
\[
\det \frac{\partial(\vec Q, \vec P)}{\partial(\vec q, \vec p)} = \pm
1,
\]
and thus the claim is proved.

The important point of this appendix is that we have shown that an
arbitrary change of the canonical co-ordinates given by
\eqref{eq:canon-coord} induces a canonical transformation given that the
momenta transform as \eqref{eq:canon-mom}, and that this
transformation
preserves the kinetic term
\eqref{eq:qp-QP} and the phase space measure. It should be
noted here that these results hold also for mixtures of c-number and
Grassman-valued dynamical variables, provided the order of the factors
is preserved. When generalising to field theory, of course the partial
derivatives are promoted to functional derivatives, and the sums over
co-ordinate indices to integrals.


\section{The $S$-matrix equivalence theorem}
\label{sec:equivalencetheorem}

The $S$-matrix equivalence theorem \cite{Itzykson:1980rh} states that
while an invertible change of field variables modifies correlation
functions, it does not affect the $S$-matrix elements derived thereof,
up to a wavefunction renormalisation.

Let us illustrate this concretely
using a toy scalar model. We compute
correlation functions from the model's partition function
\begin{equation}
\label{eq:equiv-Z-1}
{\cal Z}[j] = \int {\cal D}\phi \:
     \exp i\left\{ I[\phi] + \int d^4x \: j\phi \right\}
\end{equation}
for some action $I[\phi]$ by
taking functional derivatives with respect to $j(x)$ at $j=0$.
Then the LSZ theorem gives us the $S$-matrix elements as
\begin{equation}
\label{eq:equiv-LSZ}
\begin{split}
\langle \vec q_1 \cdots \vec q_m \rvert S \lvert \vec p_1 \cdots \vec
p_n \rangle = Z_2^{-(m+n)/2} &\prod_{i=1}^m \int d^4x_i \: e^{-ip_i
\cdot x_i} \Delta^{-1}(x_i) \frac{\delta}{i\delta j(x_i)} \\
\times &\prod_{j=1}^n \int d^4y_j \: e^{+iq_j
\cdot y_j} \Delta^{-1}(y_j) \frac{\delta}{i\delta j(y_j)} \:
\left. \frac{{\cal Z}[j]}{{\cal Z}_0} \right|_{j=0}
\end{split}
\end{equation}
where $\Delta^{-1}$ is the kinetic operator,
$Z_2$ is the field strength renormalisation, and ${\cal Z}_0$ is
the free partition function.

Now let us write the action in terms of a new field
variable $\phi'$ given implicitly by the invertible transformation $\phi =
\phi' + F[\phi',\partial_\mu
\phi',\partial_\mu\partial_\nu\phi',\dots]$, where $F$
is a \emph{regular} function(al) of $\phi'$ and its derivatives.  (If
this change of variables has a
non-unit jacobian, we can ignore it for the purposes of this
discussion.) Then \eqref{eq:equiv-Z-1} becomes
\begin{equation}
\label{eq:equiv-Z-2}
{\cal Z}[j] = \int {\cal D}\phi' \:
     \exp i\left\{ I[\phi' + F] + \int d^4x \: j(\phi' + F) \right\}.
\end{equation}
If we define a new action
\[
I'[\phi'] := I[\phi' + F]
\]
then the equivalence theorem tells us that we can just as well use
\begin{equation}
\label{eq:equiv-Z-3}
{\cal Z}'[j] = \int {\cal D}\phi' \:
     \exp i\left\{ I'[\phi'] + \int d^4x \: j\phi' \right\}
\end{equation}
in \eqref{eq:equiv-LSZ} as the original ${\cal Z}[j]$.

Upon taking functional  derivatives with respect to $j$, we see that
additional terms of
$\phi'^2$ and higher powers hidden in $F$
are pulled down from the exponential: an
insertion of $\phi'^n(x)$ will connect to $n$ propagators, whose momenta
sum to that associated with $x$ by the Fourier transform. Unlike when
$n=1$, these propagators will not, in general, cancel the inverse propagator
from LSZ reduction and thus vanish in the on-shell limit. Hence, we can
truncate the source term in \eqref{eq:equiv-Z-2} to $\int d^4x\:j\phi'$. At the quantum level,
self-energy-like terms can be made from insertions of $\phi'^n(x)$, but these
will alter scattering amplitudes by at most a wavefunction renormalisation.

\end{subappendices}

\chapter{Equivalence Theorem Evasion and Dimensional Regularisation}
\label{cha:etv}

In the previous chapter, we saw that by applying a canonical change of
variables to light-cone Yang--Mills theory, we obtained the Canonical
MHV Lagrangian: one with Parke--Taylor MHV vertices and Feynman rules
that follow the CSW rules.  In our closing comments in section
\ref{sec:mhvym-conclusion}, we noted a number of shortcomings. First,
there are the so-called `missing' amplitudes: using the vertex content
alone, it is impossible to construct certain non-vanishing objects,
significantly the one-loop amplitudes with at most one
negative-helicity gluon (something we also mentioned in our review in
section \ref{sec:background-closing}). It was also noted that the
\mhvbar\ amplitude at tree-level, which does not vanish for complex
momenta or in $(2,2)$-signature space-times, does not appear in the
theory.  Secondly, we note that the CSW rules don't appear to suggest
in a natural manner how one should go about imposing a regularisation
structure, which would be required for a complete treatment of the
theory at the quantum level. Can our lagrangian formulation be
extended to provide this? In this chapter, we will begin to address
these issues.

This chapter may be considered to be divided into two segments. The
first, in section \ref{sec:etv-completion}, introduces the notion of
\emph{completion vertices}: contributions to the $S$-matrix that arise
from the canonical transformation itself. We use these to recover the
tree-level $(\mpp)$ amplitude by a mechanism that evades $S$-matrix
equivalence theorem.  We then consider their role in tree-level
amplitudes in general.

The second segment (sections \ref{sec:etv-dimreg} and
\ref{sec:etv-oneloop}) works towards understanding the construction of
the missing one-loop all-$+$ amplitudes in pure Yang--Mills. We begin
by applying the canonical transformation to a light-cone Yang--Mills theory
which has been extended to $D$ dimensions in a particular manner,
which we use to derive $D$-dimensional completion coefficients and
associated completion vertices.  Equivalence theorem evading
contributions are used to construct the one-loop $(\fourplus)$
amplitude, and we analyse their cuts to demonstrate that the quadruple
cut of the propagators coincides with that of the known one-loop
$(\fourplus)$ amplitude.  Then we demonstrate that [for a particular
momentum routing topology] that the equivalence theorem evading
contributions sum up precisely to the amplitude one would have
obtained using light-cone Yang--Mills theory.
 
Finally, we draw our conclusions on this chapter in section
\ref{sec:etv-conclusion}.

This work was published in \cite{Ettle:2007qc}.

\section{MHV completion vertices and evading the equivalence theorem}
\label{sec:etv-completion}

Let us re-visit \eqref{eq:mhvym-transf}, which defined the canonical
transformation by absorbing the \mhvbar\ vertex into the kinetic part
of the theory in terms of $\gB$ and $\bar\gB$. Written in quantisation
surface momentum space, the \mhvbar\ vertex is
\begin{equation}
L^\mpp = \tr \int_{123} \bar V^2(123) \: \gA_{\bar 1} \gA_{\bar 2}
\bar\gA_{\bar 3} \: (2\pi)^3 \delta^3(\vec p_1 + \vec p_2 + \vec p_3)
\end{equation}
where
\begin{equation}
\label{eq:V2bar123}
\bar V^2(123) = i \frac{\hat3}{\hat1 \hat2} \{1\:2\}.
\end{equation}
Now had we written \eqref{eq:mhvym-A-eqn} retaining $\bar V^2$ rather
than writing it out explicitly, we would have arrived at the following
recurrence relation for $\Upsilon$:
\begin{equation}
\label{eq:etv-Upsilon-V-rr}
\Upsilon(1\cdots n) = -\frac1{\hat1(\sum_{i=1}^n \omega_i)}
\sum_{j=2}^{n-1} \bar V^2(P_{2j}, P_{j+1,n}, 1)
\Upsilon(-,2,\dots,j) \Upsilon(-,j+1,\dots,n),
\end{equation}
whence
\begin{equation}
\label{eq:Upsilon-V}
\Upsilon(123) = -\frac{\bar V^2(231)}{\hat1 (\omega_1 + \omega_2 +
  \omega_3)}
= \frac{2\bar V^2(231)}{\hat1 (P_1 + P_2 + P_3)}.
\end{equation}
Here, $P_i := p_i^2/\hat \imath$, and the second line follows using
\eqref{eq:sum-omega}.

Now clearly this can be performed for any choice of $\bar
V^2$. Furthermore, the derivation of the series for $\bar\gA$
in section \ref{ssec:mhvym-Abar-series} does not
involve $\bar V^2$, so the recurrence relation
\eqref{eq:Xi-rr} still holds and taken together with
\eqref{eq:etv-Upsilon-V-rr}, we have a recipe for
computing the series expansions of $\gA$ and $\bar\gA$.
This is an important indicator as to how
amplitudes involving the `missing' \mhvbar\ vertex $\bar V^2$ are recovered, and
to demonstrate that the recovery mechanism is generic in any theory
where a canonical transformation is used to absorb a three-point
vertex into a kinetic term.

\subsection{Defining completion vertices}
\label{ssec:etv-completion-def}
At the end of section \ref{ssec:mhvym-gaugefix}, we commented on how
scattering amplitudes can be formed by applying the LSZ reduction to
correlation functions of $\gA$ and $\bar\gA$ fields.
Let us write such a (momentum space) correlation function somewhat
schematically as
\[
\langle \cdots \gA(p) \cdots \bar \gA(q) \cdots \rangle.
\]
However, it is now the $\gB$ fields which propagate in the theory.
Therefore, we
must regard $\gA$ and $\bar\gA$ above as functions of $\gB$ and
$\bar\gB$ and make the replacements from the series; again, schematically
we can write this as (neglecting the field normalisation factors and
integral symbols for clarity)
\begin{equation}
\label{eq:Wick}
\langle
    \cdots
    \biggl(
        \sum_n \Upsilon_{p2\cdots n}\gB_{\bar 2}\cdots \gB_{\bar n}
    \biggr)\cdots\biggl(
       - \sum_{n,s} \frac{\hat s}{\hat q} \Xi^{s-1}_{q2\cdots n}
            \gB_{\bar 2}\cdots\bar \gB_{\bar s}\cdots \gB_{\bar n}
    \biggr)\cdots
\rangle.
\end{equation}
Order-by-order, we take Wick contractions between the $\gB$ field's
operators using its propagator. This naturally lends itself to a
Feynman graph representation where we have vertices for the
coefficients of the series \eqref{eq:A-series} and
\eqref{eq:Abar-series}, and they are shown in
fig.~\ref{fig:etv-completionverts}. We refer to these vertices as `MHV
completion vertices' and graphs built from them as `MHV completion
graphs' since they allow the construction of amplitudes otherwise
absent from the theory.  Each vertex, which we have drawn here with an
empty circle, corresponds to the insertion of an $\gA$ or $\bar\gA$
operator, labelled diagrammatically by the sign on the curly line. Of
course, $\gB$ propagators attach only to the straight lines.

\begin{figure}[t]
  \begin{align*}
    \begin{matrix}\begin{picture}(80,20)
        \SetOffset(30,10)
        \Gluon(-14,0)(-2,0){2}{2}
        \Line(0,0)(30,0)
        \BCirc(0,0){2}
        \Text(-16,0)[cr]{$\pm$}
        \Text(33,0)[cl]{$\pm$}
      \end{picture}\end{matrix}&=1
    \\
    \begin{matrix}\begin{picture}(80,52)
        \SetOffset(30,24)
        \Gluon(-14,0)(-2,0){2}{2}
        \Line(0,0)(30,15)
        \Line(0,0)(30,-15)
        \BCirc(0,0){2}
        \DashCArc(0,0)(16.7705,-25,25){1}
        \Text(-16,0)[cr]{$1^+$}
        \Text(33,15)[cl]{$2^+$}
        \Text(33,-15)[cl]{$n^+$}
      \end{picture}\end{matrix}&=\Upsilon(1\cdots n)
    \\
    \begin{matrix}\begin{picture}(80,52)
        \SetOffset(30,24)
        \Gluon(-14,0)(-2,0){2}{2}
        \Line(0,0)(30,15)
        \Line(0,0)(30,-15)
        \Line(0,0)(33.5410,0)
        \DashCArc(0,0)(16.7705,-25,-5){1}
        \DashCArc(0,0)(16.7705,5,25){1}
        \BCirc(0,0){2}
        \Text(-16,0)[cr]{$1^-$}
        \Text(33,15)[cl]{$2^+$}
        \Text(36,0)[cl]{$s^-$}
        \Text(33,-15)[cl]{$n^+$}
      \end{picture}\end{matrix}&=-\dfrac{\hat s}{\hat
      1}\Xi^{s-1}(1\cdots n)
  \end{align*}
  \caption{The MHV completion vertices: graphical representations of
    the $\Upsilon$ and $\Xi$ coefficients of the series expansion of
    $\gA$ and $\bar \gA$.  The wavy lines with a $+$($-$) denote
    insertions of $\gA$($\bar\gA$) operators in correlation functions;
    $\gB$ and $\bar\gB$ attach to the straight lines. All momenta are
    outgoing. }
  \label{fig:etv-completionverts}
\end{figure}

Note that the figure shows the vertices appropriate for
$\gA$, $\gB$, etc. so that the normalisation factors of
\eqref{eq:gauge-conv} can be omitted for clarity. If one wishes to
work with canonical normalisation, as we will do in the rest of this
chapter, we make the replacements
\begin{align}
\frac{4i}{g^2} V^s(1\cdots n) &\rightarrow
    \frac{4i}{g^2} \left(-\frac{ig}{\sqrt 2}\right)^{n}
    V^s(1\cdots n),  \label{eq:norm-V} \\
\Upsilon(1\cdots n) &\rightarrow
    \left(-\frac{ig}{\sqrt 2}\right)^{n-2} \Upsilon(1\cdots n), \label{eq:norm-Upsilon} \\
\Xi^s(1\cdots n) &\rightarrow
    \left(-\frac{ig}{\sqrt 2}\right)^{n-2} \Xi^s(1\cdots n), \label{eq:norm-Xi}
\end{align}
which for $V^s$ also includes the normalisation of the action.

\subsection{Equivalence theorem evasion: the tree-level $(\mpp)$ amplitude}
\label{ssec:treelevel-mpp}
\begin{figure}[t]
  \centering \subfigure{
    \begin{picture}(100,80) \SetOffset(50,27) \Gluon(0,27)(0,40){2}{2}
      \Gluon(23.3827,-13.5)(34.6410,-20){2}{2}
      \Gluon(-23.3827,-13.5)(-34.6410,-20){2}{2} \Line(0,25)(21.6506,-12.5)
      \Line(0,25)(-21.6506,-12.5) \BCirc(0,25){2} \BCirc(21.6506,-12.5){2}
      \BCirc(-21.6506,-12.5){2} \Text(0,43)[bc]{$1^+$}
      \Text(35.6410,-19)[tl]{$2^-$} \Text(-35.6410,-19)[tr]{$3^-$}
      \Text(-4,24)[tr]{$+$} \Text(4,24)[tl]{$+$} \Text(-19.0,-8.5)[br]{$-$}
      \Text(19.0,-8.5)[bl]{$-$}
    \end{picture}
    \label{fig:feyn-ppm-1}
  }\quad \subfigure{
    \begin{picture}(100,80) \SetOffset(50,27) \Gluon(0,27)(0,40){2}{2}
      \Gluon(23.3827,-13.5)(34.6410,-20){2}{2}
      \Gluon(-23.3827,-13.5)(-34.6410,-20){2}{2} \Line(0,25)(21.6506,-12.5)
      \Line(21.6506,-12.5)(-21.6506,-12.5) \BCirc(0,25){2}
      \BCirc(21.6506,-12.5){2} \BCirc(-21.6506,-12.5){2}
      \Text(4,24)[tl]{$+$} \Text(19.0,-8.5)[bl]{$-$}
      \Text(-19,-13.5)[tl]{$-$} \Text(19,-13.5)[tr]{$+$}
    \end{picture}
    \label{fig:feyn-ppm-2}
  }\quad \subfigure{
    \begin{picture}(100,100) \SetOffset(50,27) \Gluon(0,27)(0,40){2}{2}
      \Gluon(23.3827,-13.5)(34.6410,-20){2}{2}
      \Gluon(-23.3827,-13.5)(-34.6410,-20){2}{2}
      \Line(0,25)(-21.6506,-12.5) \Line(21.6506,-12.5)(-21.6506,-12.5)
      \BCirc(0,25){2} \BCirc(21.6506,-12.5){2} \BCirc(-21.6506,-12.5){2}
      \Text(-4,24)[tr]{$+$} \Text(-19.0,-8.5)[br]{$-$}
      \Text(-19,-13.5)[tl]{$+$} \Text(19,-13.5)[tr]{$-$}
    \end{picture}
    \label{fig:feyn-ppm-3}
  }
  \caption{Contributions to the tree-level $(\mpp)$ amplitude, before
    applying LSZ reduction.}
  \label{fig:etv-completion-mpp}
\end{figure}
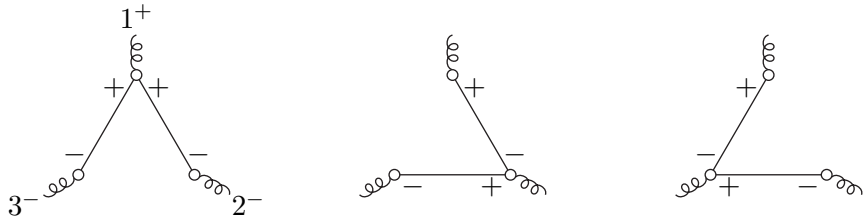

Let us now use the completion vertices to recover the tree-level
$(\mpp)$ amplitude.
By \eqref{eq:mhvym-lsz}, this is obtained
by amputating the $\langle A \bar A \bar A
\rangle$ correlation function. The diagrams for this are
shown in fig.~\ref{fig:etv-completion-mpp},
from which we see 
we need to take the limit
as $p_1^2, p_2^2, p_3^2 \rightarrow 0$ of
\begin{equation}
\begin{split}
&  -i p_1^2 p_2^2 p_3^2 \times \left(-\frac{ig}{\sqrt 2}\right)
    \times (-1)^3 \times \Biggl\{
        \frac1{p_2^2} \frac1{p_3^2} \Upsilon(123)
        - \frac1{p_3^2} \frac1{p_1^2} \frac{\hat 1}{\hat 2} \Xi^2(231)
        - \frac1{p_1^2} \frac1{p_2^2} \frac{\hat 1}{\hat 3} \Xi^1(312)
        \Biggr\}  \\
    &=
       \frac{ig}{\sqrt 2} \frac{\hat 1^2}{(2\:3)} \left(
           \frac{p_1^2}{\hat 1} + \frac{p_2^2}{\hat 2} + \frac{p_3^2}{\hat 3}
       \right).
\end{split}
\label{eq:feyn-ppm-raw}
\end{equation}
In the first line, the leading factor of $-i$ comes from an
un-cancelled inverse propagator, the second from the restoration of
the canonical normalisation using \eqref{eq:gauge-conv}, and the third
is the gluon polarisation factor. Three-particle kinematics mean that
this does not vanish in the null limit: using \eqref{eq:sum-bilinears}
cancels the factor of $(2\:3)$ in the denominator but leaves a
$\{1\:3\}=-\{2\:3\}$ in the numerator, giving the \mhvbar\
amplitude. However, we can see this in a more direct manner if we use
the first equalities of \eqref{eq:Xi1123} and \eqref{eq:Xi2123} to
substitute for $\Xi^1$ and $\Xi^2$ in favour of
$\Upsilon$,\footnote{We note here that in \eqref{eq:Xi1123} and
  \eqref{eq:Xi2123}, and the following three formulae for the
  four-point $\Xi$ coefficients, the first equality in each is derived
  directly from the recurrence relation \eqref{eq:Xi-rr}, and as such
  is independent of the exact form of $\bar V^2$. The subsequent
  equalities depend on knowing the form of $\Upsilon$ and hence $\bar
  V^2$.}  then use \eqref{eq:Upsilon-V} and notice that for
\eqref{eq:V2bar123}
\[
\bar V^2(231) = \frac{\hat1^2}{\hat2^2} \bar V^2(312)
 = \frac{\hat1^2}{\hat3^2} \bar V^2(123),
\]
the first line of \eqref{eq:feyn-ppm-raw} simplifies immediately to
\[
g \sqrt 2 \bar V^2(231),
\]
which is precisely the \mhvbar\ amplitude we sought. Notice that this
occurred \emph{before} taking the on-shell limit, and without any
reference to the specific form of $\bar V^2$. Evaluating $\bar V^2$
and using \eqref{eq:lc-spinorbrackets}, we arrive at the result
\begin{equation}
A(1^-,2^+,3^+)
= g \sqrt 2 \bar V^2(231)
= i g \sqrt 2 \frac{\hat 1}{\hat 2 \hat 3} \{ 2\:3 \}
= i g \frac{\phantom{{}^3}[2\:3]^3}{[3\:1][1\:2]},
\label{eq:feyn-ppm-raw-end}
\end{equation}
which is precisely the Parke--Taylor
 \mhvbar\ amplitude given in \eqref{eq:missing-mhvbar}

\subsubsection*{The origin of equivalence theorem evasion}
\label{sssec:etviol-origin}

It would appear here that the $S$-matrix equivalence theorem, as
described in section \ref{sec:equivalencetheorem} has been violated,
since recovering a non-vanishing amplitude from a theory in which it
is ostensibly absent is certainly not a wave-function renormalisation. 
Clearly, what has happened is that the transformation itself does not
completely satisfy the hypothesis of the theorem, in particular
regarding locality.

The presence of momenta in the denominators of $\Upsilon$, or
equivalently inverse derivatives, shows that the transformation is
non-local on the quantisation surface. However, it is local in
light-cone time; moreover, it contains no $x^0$ dependence, hence the
absence of any terms in $\check p$. As such, one would not expect it
to generate the factor of $1/p^2$ required to cancel the inverse
propagator from the LSZ reduction to allow the graph to survive the
on-shell limit. It is therefore somewhat remarkable that there exist
conditions under which the theorem is circumvented: what we have seen
here is the terms in the first factor of the recurrence relation
\eqref{eq:etv-Upsilon-V-rr} collecting together so that the factor
$\sum_j  p^2_j/\hat j$ is formed (by 
\eqref{eq:sum-omega}) despite such terms being independent of
$\check p$.

\section{Higher order tree-level amplitudes}
\label{sec:etv-completiontree}

\begin{figure}[h]
  \centering \subfigure[]{
    \begin{picture}(100,100) \SetOffset(50,50)
      \Line(-17.6777,0)(17.6777,0) \Line(35.3553,30.6186)(17.6777,0)
      \Line(17.6777,0)(35.3553,-30.6186)
      \Gluon(-26.5164,-15.3093)(-17.6777,0){2}{3}
      \Line(-17.6777,0)(-35.3553,30.6186) \BCirc(-17.6777,0){2}
      \Vertex(17.6777,0){2} \Text(-28,-17)[tr]{$4^-$}
      \Text(-37,32)[br]{$1^-$} \Text(37,32)[bl]{$2^-$}
      \Text(37,-32)[tl]{$3^+$} \Text(-15.6,3)[bl]{$+$}
      \Text(15.6,-3)[tr]{$-$}
    \end{picture}
    \label{fig:etv-completion-mhv-a}
  } \quad \subfigure[]{
    \begin{picture}(100,100) \SetOffset(50,50)
      \Line(-17.6777,0)(17.6777,0) \Line(35.3553,30.6186)(17.6777,0)
      \Line(17.6777,0)(35.3553,-30.6186)
      \Gluon(-26.5164,15.3093)(-17.6777,0){2}{3}
      \Line(-17.6777,0)(-35.3553,-30.6186) \BCirc(-17.6777,0){2}
      \Vertex(17.6777,0){2} \Text(-28,17)[br]{$1^+$}
      \Text(-37,-32)[tr]{$4^+$} \Text(37,32)[bl]{$2^-$}
      \Text(37,-32)[tl]{$3^+$} \Text(-15.6,3)[bl]{$+$}
      \Text(15.6,-3)[tr]{$-$}
    \end{picture}
  } \quad \subfigure[]{
    \begin{picture}(100,100) \SetOffset(50,50)
      \Line(-17.6777,0)(17.6777,0)
      \Line(-35.3553,-30.6186)(-17.6777,0)
      \Line(17.6777,0)(35.3553,-30.6186)
      \Gluon(26.5164,15.3093)(17.6777,0){2}{3}
      \Line(-17.6777,0)(-35.3553,30.6186) \BCirc(17.6777,0){2}
      \Vertex(-17.6777,0){2} \Text(28,17)[bl]{$2^+$}
      \Text(-37,-32)[tr]{$4^+$} \Text(37,-32)[tl]{$3^+$}
      \Text(-37,32)[br]{$1^-$} \Text(-15.6,5)[bl]{$-$}
      \Text(15.6,-5)[tr]{$+$}
    \end{picture}
  }
  \\
  \subfigure[]{
    \begin{picture}(100,100) \SetOffset(50,50)
      \Line(-17.6777,0)(17.6777,0)
      \Line(-35.3553,-30.6186)(-17.6777,0)
      \Line(17.6777,0)(35.3553,30.6186)
      \Gluon(26.5164,-15.3093)(17.6777,0){2}{3}
      \Line(-17.6777,0)(-35.3553,30.6186) \BCirc(17.6777,0){2}
      \Vertex(-17.6777,0){2} \Text(28,-17)[tl]{$3^-$}
      \Text(-37,-32)[tr]{$4^+$} \Text(37,32)[bl]{$2^-$}
      \Text(-37,32)[br]{$1^-$} \Text(-15.6,3)[bl]{$-$}
      \Text(15.6,-3)[tr]{$+$}
    \end{picture}
  } \quad \subfigure[]{
    \begin{picture}(100,100) \SetOffset(50,50)
      \Line(0,-17.6777)(0,17.6777)
      \Line(-30.6186,-35.3553)(0,-17.6777)
      \Line(0,17.6777)(30.6186,35.3553)
      \Gluon(15.3093,-26.5164)(0,-17.6777){2}{3}
      \Line(0,17.6777)(-30.6186,35.3553) \BCirc(0,-17.6777){2}
      \Vertex(0,17.6777){2} \Text(17,-28)[tl]{$3^-$}
      \Text(-32,-37)[tr]{$4^+$} \Text(32,37)[bl]{$2^-$}
      \Text(-32,37)[br]{$1^-$} \Text(3,-15.6)[bl]{$-$}
      \Text(-3,15.6)[tr]{$+$}
    \end{picture}
    \label{fig:etv-completion-mhv-e}
  } \quad \subfigure[]{
    \begin{picture}(100,100) \SetOffset(50,50)
      \Line(0,-17.6777)(0,17.6777)
      \Line(30.6186,-35.3553)(0,-17.6777)
      \Line(0,17.6777)(30.6186,35.3553)
      \Gluon(-15.3093,-26.5164)(0,-17.6777){2}{3}
      \Line(0,17.6777)(-30.6186,35.3553) \BCirc(0,-17.6777){2}
      \Vertex(0,17.6777){2} \Text(-17,-28)[tr]{$4^-$}
      \Text(32,-37)[tl]{$3^+$} \Text(-32,37)[br]{$1^-$}
      \Text(32,37)[bl]{$2^-$} \Text(3,-15.6)[bl]{$-$}
      \Text(-3,15.6)[tr]{$+$}
    \end{picture}
    \label{fig:etv-completion-mhv-f}
  }
  \caption{Graphs containing completion vertices that contribute to
    the amplitude $A(1^-2^-3^+4^+)$. These annihilated in the on-shell
    limit of the LSZ reduction since the momentum in the propagator is
    generically off-shell, and the completion vertices cannot provide
    the $1/p_i^2$ pole required.}
  \label{fig:etv-completion-mhv}
\end{figure}
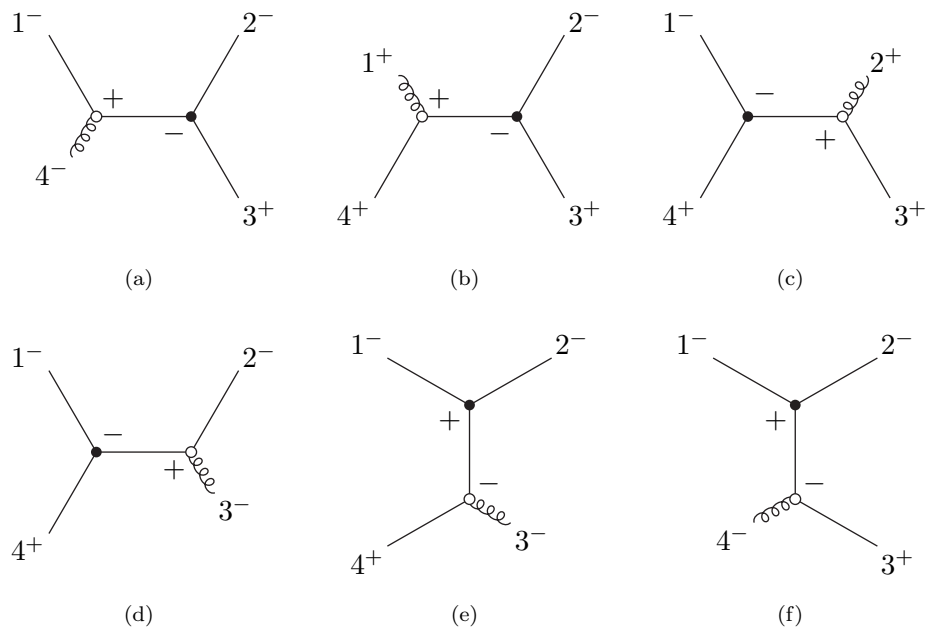

The proof given in section
\ref{ssec:mhvym-amps-verts} that the vertices of the Canonical MHV
Lagrangian were the Parke--Taylor amplitudes continued off shell by
the CSW prescription relied on the tree-level amplitudes and vertices
agreeing (up to polarisation) when on shell. One might begin to worry
at this point that the presence of these
completion vertices could contribute to MHV amplitudes and spoil this.
Indeed, it is possible to construct tree diagrams involving completion
vertices that contribute to MHV (and higher) helicity configurations.
For example, fig.~\ref{fig:etv-completion-mhv}
shows the completion vertex contributions to
$A(1^-2^-3^+4^+)$. Consider the first, fig.~\ref{fig:etv-completion-mhv-a}:
the contribution of this graph is proportional to
\begin{multline*}
\hat1 P_4 \Xi^1(4,1,2+3) \frac1{(p_1 + p_4)^2} V^2(4+1,2,3) \\
= -\hat 1 P_4 \frac{2 \bar V^2(2+3,4,1)}{\hat1(P_1 + P_{2+3} + P_4)}
\frac1{(p_1 + p_4)^2} V^2(4+1,2,3),
\end{multline*}
For generic momenta, the momentum $p_1+p_4$ which runs through the
gluon propagator will not be null,
and as such the expression above cannot produce the $1/p_4^2$ pole
required for it to survive the on-shell limit. By considering the
other diagrams similarly, we see that they also make no contribution
in the on-shell limit.

More generally this happens because the helicity content of MHV and
N${}^r$MHV ($r\ge1$) amplitudes require that any completion
vertex/vertices is/are attached to a propagator(s) which at tree-level will
carry non-null momenta.  By considering the explicit form of
$\Upsilon$, or moreover by inspecting \eqref{eq:etv-Upsilon-V-rr}, it
is clear that the completion vertices cannot generate the required
poles to ensure their survival of the LSZ reduction.  Thus, MHV
completion vertices make no contribution to the non-vanishing
tree-level on-shell
amplitudes for four or more gluons, and the proof at the end of
section \ref{sec:mhvym-struct} 
still holds.

What of the $A(1^-2^+\cdots n^+)$ tree-level amplitude for $n\ge4$? Its helicity
configuration permits contributions of the form of a \emph{single}
completion vertex connected to no internal propagators, which we might
expect to survive. The completion vertex contribution to this
amplitude is proportional to the on-shell limit of
\[
\hat1 P_1 \Upsilon(1\cdots n)
- \hat 1 \sum_{j=2}^n P_j \Xi^{n-j+1}(j,\dots,n,1,\dots,j-1).
\]
Now we can see from \eqref{eq:Xi-rr} that all terms but one in a $\Xi$, namely a
lone $\Upsilon$ with the same number of arguments as the $\Xi$,
consist of products of $\Upsilon$s with generically off-shell momentum
arguments. We therefore discard these terms for the same reasons as
above (they cannot generate the poles required for survival of LSZ),
and make the replacement
$\Xi^r(1\cdots n) \rightarrow \Upsilon(r+1,\dots,n,1,\dots,r)$
to obtain
\begin{equation}
\label{eq:etv-lim-1}
\hat1 (P_1 + \cdots + P_n) \: \Upsilon(1\cdots n).
\end{equation}
However, for $n\ge4$ we know that all other contributions to
this amplitude vanish, as does the amplitude itself, so
\eqref{eq:etv-lim-1} must also vanishes on shell. It is easy to see
that it does simply by inspection using the explicit form of
$\Upsilon(1234\cdots n)$ in \eqref{eq:Upsilon-coeff}, which has only
co-linear singularities.

\subsection{Tree-level off-shell reconstruction}
\label{ssec:etv-completion-mmpp}
Although completion vertices have no contribution to on-shell
tree-level amplitudes, we expect them to contribute off shell to
recover the underlying light-cone Yang--Mills theory.  That this is so
is strongly hinted at by the recurrence relations
\eqref{eq:etv-Upsilon-V-rr} and \eqref{eq:Xi-rr} which express the
$\Upsilon$ and $\Xi$ vertices as sums of terms proportional to $\bar
V^2$ vertices in a manner that consistently matches up the external
helicities (after amputation). So, although the individual graphs of
fig.~\ref{fig:etv-completion-mhv} do not correspond to LCYM graphs due
to the denominator in the leading factor on the RHS of
\eqref{eq:etv-Upsilon-V-rr}, we might expect that when summed and
added to the contributions from the graphs constructed with MHV
vertices, the inverse propagators supplied by the LSZ reduction cancel
the denominators to leave the LCYM graphs, before taking the on-shell limit.

\subsubsection*{An example: tree-level $A(1^-2^-3^+4^+)$}
Let us
demonstrate this in the case of $A(1^-2^-3^+4^+)$.  By a fairly
straightforward calculation, we can see that by taking the
sum of the graphs of fig.~\ref{fig:etv-completion-mhv}, multiplying by
the inverse propagators, and then adding the
contribution from $V^2(1234)$ (the $(--++)$ vertex from the Canonical
MHV Lagrangian shown in \eqref{eq:V21234}), we reconstruct
\emph{algebraically} the off-shell amplitude as computed with LCYM
Feynman rules. It is instructive to illustrate at least in part
how this works, and since the topology of the momentum routing
between the vertices
on the
MHV lagrangian side must match that of the LCYM graphs we seek to
recover, we can
consider just the $(1,2)$ channel, \ie\ the contribution from
figs.~\ref{fig:etv-completion-mhv-e} and
\ref{fig:etv-completion-mhv-f} and from the third term in
\eqref{eq:V21234} (its first two terms clearly have the topology of
the $(1,4)$ channel, and the last term is just the $(--++)$ vertex
from light-cone Yang--Mills theory).  Restoring the normalisation of
$A$, these terms are, respectively
\begin{equation*}
\begin{split}
&ig^2 V^2(1,2,3+4) \Upsilon(1+2,3,4) \\
&+ ig^2 V^2(1,2,3+4) \frac 1{(p_1+p_2)^2} \left( -\frac{\hat1 +
    \hat2}{\hat3} \right) \Xi^2(3,4,1+2) \: \hat3 P_3 \\
&+ ig^2 V^2(1,2,3+4) \frac 1{(p_1+p_2)^2} \left( -\frac{\hat1 +
    \hat2}{\hat4} \right) \Xi^1(4,1+2,3) \: \hat4 P_4 \\
&= 2ig^2 \frac1{(p_1 + p_2)^2} V^2(1,2,3+4) \bar V^2(3,4,1+2) \Biggl\{
\frac{P_{1+2}}{P_{1+2}+P_3+P_4} \\
&\quad +\frac{P_3}{P_{1+2}+P_3+P_4} + \frac{P_4}{P_{1+2}+P_3+P_4} \Biggr\},
\end{split}
\end{equation*}
where we have used the first equality of \eqref{eq:Xi1123} and
\eqref{eq:Xi2123} to evaluate $\Xi^1$ and $\Xi^2$.
This is of course trivially equal to
\[
\begin{matrix}
\begin{picture}(70,55)
\SetOffset(35,28)
\Gluon(0,-10)(0,10){2}{4}
\Gluon(0,10)(20,20){-2}{4}
\Gluon(0,10)(-20,20){-2}{4}
\Gluon(0,-10)(-20,-20){2}{4}
\Gluon(0,-10)(20,-20){2}{4}
\Vertex(0,-10){1}
\Vertex(0,10){1}
\Text(-22,20)[cr]{$1^-$}
\Text(22,20)[cl]{$2^+$}
\Text(22,-20)[cl]{$3^-$}
\Text(-22,-20)[cr]{$4^-$}
\Text(3,9)[tl]{$+$}
\Text(-3,-9)[br]{$-$}
\end{picture}
\end{matrix}
= 2g^2 V^2(1,2,3+4) \frac i{(p_1 + p_2)^2} \bar V^2(3,4,1+2).
\]
We shall see in section \ref{sec:etv-oneloop} that the same features are
responsible for the recovery of the all-$+$ one-loop amplitudes.

\section{The $D$-Dimensional Canonical MHV Lagrangian}
\label{sec:etv-dimreg}

The treatment of quantum corrections to amplitudes in the canonical
MHV lagrangian formalism will require that the theory be regulated,
and we will do so by dimensional regularisation. It turns out that we
can then apply the canonical transformation procedure essentially as
before, save for the fact that pieces outside four dimensions result
in much richer structure and hence more complicated MHV rules.

\subsection{Light-cone Yang-Mills in $D$ dimensions}
\label{ssec:lc-Dd}

We write the co-ordinates in $D=4-2\epsilon$ dimensions as:
\begin{align*}
  x^0 &= \tfrac1{\sqrt 2}(t-x^{D-1}), &
  z^I &= \tfrac1{\sqrt 2}(x^{2I-1}+ix^{2I}),\\
  x^{\bar 0} &= \tfrac1{\sqrt 2}(t+x^{D-1}), & \bar z^I &=
  \tfrac1{\sqrt 2}(x^{2I-1}-ix^{2I}),
\end{align*}
where the index $I$ runs over the $\tfrac12(2-2\epsilon)$ pairs of
transverse directions.  In these co-ordinates, the metric takes block
diagonal form with non-zero components $g_{0\bar0}=g_{\bar0 0}=1$,
$g_{z^I\bar z^J}=g_{\bar z^I z^J}=-\delta_{IJ}$. Again, we introduce a
more compact notation for the components of 1-forms and momenta, for
which we write $(p_0, p_{\bar 0}, p_{z^I}, p_{\bar z^J}) \equiv
(\check{p}, \hat p, p_I, \bar p_I)$, with $(\check{n}, \hat n, n_I,
\bar n_I)$ for momenta labelled by a number.

The reason we make this choice of basis is that it will lead us again
to a lagrangian with an MHV structure and thus inherit some of the
simplicity of MHV rules in four space-time dimensions, for example the
tree-level properties that the first non-vanishing vertices are MHV
vertices, that NMHV
amplitudes are constructed by joining precisely two such vertices
together by the propagator and so on.

In these co-ordinates, the invariant becomes
\begin{equation}
  A \cdot B = \check{A}\,\hat B + \hat A\,\check{B} - A_I \bar B_I
  - \bar A_I B_I,
  \label{eq:paralc-invariant}
\end{equation}
where we have assumed the summation convention that a repeated capital
Roman index in a product is summed over $1,\dots,1-\epsilon$. The
bilinears of \eqref{eq:lc-bilinears} become
\begin{equation}
  \label{eq:lc-d-bilinears}
  (1\:2)_I := \hat 1 2_I - \hat 2 1_I, \quad
  \{1\:2\}_I := \hat 1 \bar 2_I - \hat 2 \bar 1_I.
\end{equation}
They amount to our $D$ dimensional generalisation of the familiar
spinor brackets \eqref{eq:lc-spinorbrackets}.  Scalar products between
these will often be shortened to
\[
(1\:2)\mdot\{1\:2\} \equiv (1\:2)_I \{1\:2\}_I,
\]
where the dot is obviously redundant when the bilinears are purely
four-dimensional.

The Yang-Mills action is written as before in Minkowski co-ordinates
as
\begin{equation}
  S=\frac 1{2g^2} \int d^Dx\:\tr \gF^{\mu\nu}\gF_{\mu\nu}.
  \label{eq:ym-action-Dd}
\end{equation}
The field-strength tensor and group generators are defined as before
in \eqref{eq:gauge-conv}. The quantisation procedure is similar to
that in four dimensions. It takes place on surfaces $\Sigma$ with
normal $\mu = (1,0,\dots,0,1) / \sqrt 2$ (\ie\ of constant $x^0$) in
Minkowski co-ordinates, subject to the axial gauge condition
$\mu~\cdot~\gA = \hat\gA = 0$. We apply the gauge condition and
integrate $\check\gA$ out of the lagrangian by the same procedure as
in section \ref{sec:mhvym-lightcone}, extended to accommodate the
additional degrees of freedom, and are left with a $D$-dimensional
light-cone action in the form \eqref{eq:lcym-action}, where now
\newcommand{\trintx}{\tr\int_\Sigma d^{D-1}\vec{x}\:}
\begin{align}
  L^{-+} &= \phantom{-} \trintx \gA_I (\check\partial\hat\partial
  - \partial_J\bar\partial_J) \bar \gA_I,
  \label{eq:hyperlc-mp}
  \\
  L^{-++} &= - \trintx ( \bar\partial_I \gA_J [\hat\partial^{-1}
  \gA_I, \hat\partial \bar \gA_J] + \bar\partial_I \bar \gA_J
  [\hat\partial^{-1} \gA_I, \hat\partial \gA_J] ),
  \label{eq:hyperlc-mpp}
  \\
  L^{--+} &= - \trintx ( \partial_I \gA_J [\hat\partial^{-1} \bar
  \gA_I, \hat\partial \bar \gA_J] + \partial_I \bar \gA_J
  [\hat\partial^{-1} \bar \gA_I, \hat\partial \gA_J] ),
  \label{eq:hyperlc-mmp}
  \\
  \begin{split}
    L^{--++} &= - \trintx \biggl( \hphantom{+\,} \frac 14
    [\hat\partial \gA_I,\bar \gA_I] \:\hat\partial^{-2}\:
    [\hat\partial \gA_J,\bar \gA_J]  \\
    & \hphantom{=-\trintx\biggl(} + \frac 12 [\hat\partial \gA_I,\bar
    \gA_I] \:\hat\partial^{-2}\:
    [\hat\partial \bar \gA_J, \gA_J] \\
    & \hphantom{=-\trintx\biggl(} - \frac 14 [\hat\partial \bar \gA_I,
    \gA_I] \:\hat\partial^{-2}\:
    [\hat\partial \bar \gA_J, \gA_J] \\
    & \hphantom{=-\trintx\biggl(} - \frac 14 [ \gA_I, \gA_J ] [ \bar
    \gA_I, \bar \gA_J ] - \frac 14 [ \gA_I, \bar \gA_J ] [ \bar \gA_I,
    \gA_J ] \biggr).
  \end{split}
  \label{eq:hyperlc-mmpp}
\end{align}
It may be shown with integration by parts that these expressions
reduce in four dimensions to \eqref{eq:lcym-mp}--\eqref{eq:lcym-mmpp}.

\subsection{The transformation}
\label{ssec:mhvl-trans}

We will now specify the change of field variables from $\gA$ and
$\bar\gA$ to $\gB$ and $\bar\gB$.  From \eqref{eq:hyperlc-mp}, we see
that the momentum conjugate to $\gA_I$ is
$\Pi_I(x)=-\hat\partial\bar\gA_I(x)$, and as such
\begin{equation}
  {\cal D}\gA\:{\cal D}\Pi \equiv \prod_{x,I} d\gA_I(x)\:d\Pi_I(x)
  \label{eq:pathint-measure}
\end{equation}
is proportional (up to a constant) to the path integral measure ${\cal
  D}\gA\:{\cal D}\bar\gA$; therefore under a canonical field
transformation the jacobian will be unity.  Again we choose $\gA$ to
be a functional of $\gB$ alone, and by \eqref{eq:canon-mom} and
\eqref{eq:qp-QP},
\begin{gather}
  \hat\partial \bar\gA^a_I(x^0,\vec x) = \int_\Sigma d^{D-1}\vec y \:
  \frac{\delta\gB^b_J(x^0,\vec y)}{\delta\gA^a_I(x^0,\vec x)}
  \hat\partial \bar\gB^b_J(x^0,\vec y),
  \label{eq:momentum-transform} \\
  \tr \int_\Sigma d^{D-1}\vec
  x\:\check\partial\gA_I\,\hat\partial\bar\gA_I = \tr \int_\Sigma
  d^{D-1}\vec x\:\check\partial\gB_I\,\hat\partial\bar\gB_I.
  \label{eq:second-canon-invar}
\end{gather}

Again, working in momentum space on the quantisation surface, we
express $\gA$ as a series in $\gB$, but this time the series
coefficients carry extra indices for the new transverse directions:
\begin{equation}
  \label{eq:hypermft-A-series}
  \gA_{I_1}(\vec p_1) = \sum_{n=2}^{\infty} \int_{2\cdots n}
  \Upsilon_{I_1\cdots I_n}(1\cdots n) \:
  \gB_{I_2}(-\vec p_2) \cdots \gB_{I_n}(-\vec p_n)
\end{equation}
where $\Upsilon_{IJ}(12) = \delta(\vec p_1+\vec p_2) \delta_{IJ}$.
The integral short-hand used here is
\[
\int_{1\cdots n} = \prod_{k=1}^n \frac1{(2\pi)^{3-2\epsilon}} \int
d\hat k \prod_{I=1}^{1-\epsilon} dk_I d\bar k_I
\]
and for later use we introduce the $\delta$-function stripped form of
a coefficient, given (as the first factor on the right-hand side) by
\begin{equation}
  \Upsilon_{I_1\cdots I_n}(1\cdots n) =
  \Upsilon(1^{I_1}\cdots n^{I_n}) \:
  (2\pi)^{3-2\epsilon} \delta^{3-2\epsilon}(\vec p_1 + \cdots + \vec p_n)
\end{equation}
and similarly for the other vertices $\Xi$, $V$ and $W$, defined
below. They should only be considered to be defined when the sum of
their momentum arguments is $0$. Repeated transverse indices in the
superscripts are also subject to the summation convention. For
convenience, we will often also subsume the index into the momentum
label when the association is obvious (e.g.\ $\Upsilon(1^{I_1}\cdots
n^{I_n}) \rightarrow \Upsilon(1\cdots n)$ above).

The canonical transformation removes the $(\mpp)$ terms from the
lagrangian by absorbing them into the kinetic term for $\gB$:
\begin{equation}
  L^{-+}[\gA,\bar\gA] + L^{-++}[\gA,\bar\gA] = L^{-+}[\gB,\bar\gB].
  \label{eq:AB-implicit}
\end{equation}
Briefly delving into momentum space on the quantisation surface, it is
seen that the term on the right-hand side of \eqref{eq:AB-implicit}
supplies the tree-level propagator
\begin{equation}
  \langle \gB_I \bar\gB_J \rangle = -\frac{ig^2}{2p^2}\delta_{IJ}.
\end{equation}
Similarly, from the quantisation surface Fourier transform of
\eqref{eq:hyperlc-mpp}, expanding the commutator and re-labelling
leads us to
\begin{equation}
  L^{-++} = \tr\int_{123}
  \bar V^2_{IJK}(123) \: \gA_I(\bar1) \gA_J(\bar2) \bar\gA_K(\bar3)
  \label{eq:hyperlc-mom-mpp}
\end{equation}
where
\begin{equation}
  \bar V^2(1^I2^J3^K) = -i \left(
    \frac{\{3\:1\}_J \delta_{KI}}{\hat 2}
    + \frac{\{2\:3\}_I \delta_{JK}}{\hat 1}
  \right).
  \label{eq:Vbar2}
\end{equation}
It obviously follows from \eqref{eq:AB-implicit} and the light-cone
lagrangian that this is the $D$ dimensional equivalent of the $(\ppm)$
\mhvbar\ vertex for the $\gA$ field, and this is reflected in our
choice of notation.

The remaining pieces of the lagrangian, \eqref{eq:hyperlc-mmp} and
\eqref{eq:hyperlc-mmpp}, form MHV vertices in $4-2\epsilon$
dimensions, as explained in the next section.  To obtain the
$\Upsilon$ coefficients, we take the explicit expression of
\eqref{eq:AB-implicit} and use \eqref{eq:momentum-transform} and
\eqref{eq:second-canon-invar}, as before, to further reduce it to
\begin{equation}
  \left\{ \frac{\partial \mdot \bar\partial}{\hat \partial} \gA_I
    - [\bar\partial_J \gA_I, \hat\partial^{-1} \gA_J]
    - \frac{\bar\partial_J}{\hat\partial} [\hat\partial^{-1} \gA_J,
    \hat\partial \gA_I] \right\} (\vec x)
  = \int_\Sigma d^3\vec y \:
  \frac{\delta\gA_I(\vec x)}{\delta\gB^b_J(\vec y)}
  \left(\frac{\partial \mdot \bar\partial}
    {\hat \partial}\right)_{\vec y}
  \gB^b_J(\vec y).
  \label{eq:AB-explicit}
\end{equation}
By again transforming to momentum space and substituting the series
expansion for $\gA$ into both sides of \eqref{eq:AB-explicit} above,
carefully rearranging the fields, and comparing terms order-by-order
in $\gB$, we extract successive $\Upsilon$ coefficients. At
$\mathcal{O}(\gB^2)$, one finds
\begin{align}
  \label{eq:upsilon-3}
  \Upsilon(1^I2^J3^K) &= \frac{i}{(2\:3)\mdot\{2\:3\}}
  (\hat 2 \{2\:3\}_K \delta_{IJ} + \hat 3 \{2\:3\}_J \delta_{KI}) \\
  &= -\frac{1}{\hat{1}} \frac{\bar{V}^2(2^J 3^K 1^I)} {(\Omega_1 +
    \Omega_2 + \Omega_3)}
  \notag \\
  &= \frac{2}{\hat 1} \frac{\bar{V}^2(2^J 3^K 1^I)} {p_1^2/\hat 1 +
    p_2^2/\hat 2 + p_3^2/\hat 3}
  \label{eq:Upsilon-V2-dd}
\end{align}
and for compactness we have defined
\[
\Omega_p := \frac{p_I \bar p_I}{\hat p}.
\]
Again, substituting the series ansatze for $\gA$ in the definition of the
transformation and working order-by-order in $\gB$
leads to the recurrence relation
\begin{equation}
  \Upsilon(1\cdots n) = -
  \frac{1}{\hat1 \sum_{i=1}^n \Omega_i} \sum_{j=2}^{n-1}
  \bar V^2(P^A_{2j},P^B_{j+1,n},1)
  \Upsilon(-^A,2,\dots,j)
  \Upsilon(-^B,j+1,\dots,n),
  \label{eq:upsilon-rr-x}
\end{equation}
a particularly useful case of which is
\begin{equation}
  \begin{split}
    \Upsilon(1234) = \frac{1}{\hat 1 \sum_{i=1}^4 \Omega_i} \Biggl\{&
    \bar V^2(2, \bar 5^A, 1) \frac{1}{\hat 5(\Omega_5 +
      \Omega_3 + \Omega_4)} \bar V^2(3, 4, 5^A) \\
    + & \bar V^2(\bar 5^A, 4, 1) \frac{1}{\hat 5 (\Omega_5 + \Omega_2
      + \Omega_3)} \bar V^2(2, 3, 5^A)\Biggr\}.
  \end{split}
  \label{eq:upsilon1234}
\end{equation}
Note that here (and throughout) $\vec p_5$ is a dummy momentum with scope
limited to each term, and that its value should be taken to be the
negative of the sum of the other arguments that accompany it in each term.

Differentiating \eqref{eq:hypermft-A-series} with respect to $\gB$ and
inserting the inverse into \eqref{eq:momentum-transform} suggests a
series expansion for $\bar \gA$ of the form
\begin{equation}
  \label{eq:hypermft-Abar-series}
  \bar\gA_{I_1}(-\vec p_1) = \sum_{n=2}^{\infty} \sum_{s=2}^n \int_{2\cdots n}
  \frac{\hat s}{\hat 1} \Xi^{s-1}_{I_1\cdots I_n}(\bar12\cdots n) \:
  \gB_{I_2}(-\vec p_2)
  \cdots \bar\gB_{I_s}(-\vec p_s) \cdots
  \gB_{I_n}(-\vec p_n).
\end{equation}
Now, inserting \eqref{eq:hypermft-A-series} and
\eqref{eq:hypermft-Abar-series} into \eqref{eq:second-canon-invar},
and then comparing coefficients order-by-order in $\gB$, we may obtain
expressions for the $\Xi$ coefficients in terms of other $\Xi$s and
$\Upsilon$s of lower order. The results
\begin{equation}
  \label{eq:xi-3pt}
  \Xi^1(1^I2^J3^K) = - \Upsilon(2^J3^K1^I)
  \quad\text{and}\quad
  \Xi^2(1^I2^J3^K) = - \Upsilon(3^K1^I2^J)
\end{equation}
will be of particular relevance to the forthcoming. By careful
examination of the expansion of \eqref{eq:second-canon-invar} at the
$(n-1)^{\rm th}$ order in $\gB$, one finds that this recursion
relation
\begin{equation}
  \begin{split}
    \Xi^s(1\cdots n) = - \sum_{r=\max(2,4-s)}^{n+1-s}
    \sum_{m=\max(r,3)}^{r+s-1}
    &   \Upsilon(-^A,3-r,\dots,m+1-r) \\
    & \times \Xi^{r+s-m}(-^A,m+2-r,\dots,2-r).
  \end{split}
  \label{eq:xi-rr}
\end{equation}
holds, given that $\Xi(1^I2^J)=\delta_{IJ}$.

As in section \ref{ssec:etv-completion-def}, we introduce a
diagrammatic representation of $\Upsilon$ and $\Xi$ in the form of
$D$-dimensional MHV completion vertices, the first few of which are
shown in fig.~\ref{fig:etv-completionverts}. We note that the process
of deriving $\Upsilon$ and $\Xi$ in $4-2\epsilon$ dimensions differs
only from that in four dimensions by the presence of extra transverse
indices, which are seen to each ride alongside (and can therefore be
built into) a momentum index.  It is therefore not surprising that the
relationship between $\Upsilon$ and $\Xi$ is, from this point of view,
identical to that in four dimensions.

\begin{figure}[h]
  \begin{align*}
    \begin{matrix}\begin{picture}(110,20)
        \SetOffset(45,10)
        \Gluon(-14,0)(-2,0){2}{2}
        \Line(0,0)(30,0)
        \BCirc(0,0){2}
        \Text(-16,0)[cr]{${}^I,\pm$}
        \Text(33,0)[cl]{${}^J,\pm$}
      \end{picture}\end{matrix}&=\delta_{IJ}
    \\
    \begin{matrix}\begin{picture}(110,52)
        \SetOffset(45,24)
        \Gluon(-14,0)(-2,0){2}{2}
        \Line(0,0)(30,15)
        \Line(0,0)(30,-15)
        \BCirc(0,0){2}
        \DashCArc(0,0)(16.7705,-25,25){1}
        \Text(-16,0)[cr]{$1^{I_1},+$}
        \Text(33,15)[cl]{$2^{I_2},+$}
        \Text(33,-15)[cl]{$n^{I_n},+$}
      \end{picture}\end{matrix}&=\Upsilon(1^{I_1}\cdots n^{I_n})
    \\
    \begin{matrix}\begin{picture}(110,52)
        \SetOffset(45,24)
        \Gluon(-14,0)(-2,0){2}{2}
        \Line(0,0)(30,15)
        \Line(0,0)(30,-15)
        \Line(0,0)(33.5410,0)
        \DashCArc(0,0)(16.7705,-25,-5){1}
        \DashCArc(0,0)(16.7705,5,25){1}
        \BCirc(0,0){2}
        \Text(-16,0)[cr]{$1^{I_1},-$}
        \Text(33,15)[cl]{$2^{I_2},+$}
        \Text(36,0)[cl]{$s^{I_s},-$}
        \Text(33,-15)[cl]{$n^{I_n},+$}
      \end{picture}\end{matrix}&=-\dfrac{\hat s}{\hat
      1}\Xi^{s-1}(1^{I_1}\cdots n^{I_n})
  \end{align*}
  \caption{Completion vertices for the $D$-dimensional Canonical MHV Lagrangian. }
  \label{fig:etv-completionverts-dd}
\end{figure}

\subsection{$4-2\epsilon$-dimensional MHV vertices}
\label{ssec:hypermhv}

We will now extract the $4-2\epsilon$-dimensional generalisations of
the three and four gluon MHV vertices. The interaction part of the
lagrangian takes the same form as \eqref{eq:mhvym-Vseries} except that
the vertices carry polarisation indices to contract into the
corresponding $\gB$s and $\bar\gB$s:
\[
\frac12 \sum_{s=2}^n \int_{1\cdots n} V^s_{I_1\cdots I_n}(1\cdots n)
\tr [ \bar\gB_{I_1}(-\vec p_1) \gB_{I_2}(-\vec p_2) \cdots
\bar\gB_{I_s}(-\vec p_s) \cdots \gB_{I_n}(-\vec p_n) ]
\]
The Feynman rule for a particular $\gB$ vertex is thus
$4iV^s(1^{I_1}\cdots n^{I_n})/g^2$ (\eqref{eq:norm-V} applies in the
canonical normalisation), and this follows from its definition as the
sum of all contractions of external lines into the term in the action
with the matching colour factor, while accounting for the cyclic
symmetry of the trace.

The three-point MHV vertex follows trivially from $L^\mmp$ using the
leading order terms in the series for $\gA$ and $\bar\gA$. In
quantisation surface momentum space, \eqref{eq:hyperlc-mmp} reads
\begin{equation}
  L^{--+} = \tr\int_{123}
  V^2_{IJK}(123) \: \bar\gA_I(\bar1) \bar\gA_J(\bar2) \gA_K(\bar3)
  \label{eq:hyperlc-mom-mmp}
\end{equation}
where
\begin{equation}
  V^2(1^I2^J3^K) = -i \left(
    \frac{(3\:1)_J \delta_{KI}}{\hat 2}
    + \frac{(2\:3)_I \delta_{JK}}{\hat 1}
  \right).
\end{equation}
Since $\gA=\gB$ and $\bar\gA=\bar\gB$ to leading order, upon
substituting \eqref{eq:hypermft-A-series} and
\eqref{eq:hypermft-Abar-series} into \eqref{eq:hyperlc-mom-mmp}, we
immediately see that $V^2(1^I2^J3^K)$ is the
$\bar\gB_I(1)\bar\gB_J(2)\gB_K(3)$ colour-ordered vertex.

We note that the $\bar\gB\bar\gB\gB\gB$ and $\bar\gB\gB\bar\gB\gB$
colour-ordered vertices receive contributions from $L^\mmp[\gA]$ and
$L^\mmpp[\gA]$. Upon writing the latter in momentum-space, we have
\begin{equation}
  \begin{split}
    L^\mmpp = \tr\int_{1234} \bigl\{ & W^2_{IJKL}(1234) \:
    \bar\gA_I(\bar1) \bar\gA_J(\bar2) \gA_K(\bar3) \gA_L(\bar4) \\
    & + W^3_{IJKL}(1234) \: \bar\gA_I(\bar1) \gA_J(\bar2)
    \bar\gA_K(\bar3) \gA_L(\bar4) \bigr\}
  \end{split}
  \label{eq:hyperlc-mom-mmpp}
\end{equation}
where
\begin{align}
  W^2(1^I2^J3^K4^L) &= \delta_{IK}\delta_{JL} + \delta_{IL}\delta_{JK}
  \frac{\hat 1 \hat 2 + \hat 3 \hat 4}{(\hat 1 + \hat 4)^2}, \\
  W^3(1^I2^J3^K4^L) &= \frac12 \left( \delta_{IL}\delta_{JK}
    \frac{\hat 1 \hat 2 + \hat 3 \hat 4}{(\hat 1 + \hat 4)^2} +
    \delta_{IJ}\delta_{KL} \frac{\hat 1 \hat 4 + \hat 2 \hat 3}{(\hat
      1 + \hat 2)^2} \right).
\end{align}
We substitute \eqref{eq:hypermft-A-series} and
\eqref{eq:hypermft-Abar-series} into $L^\mmp[\gA] + L^\mmpp[\gA]$, and
collect the terms of each colour (trace) order of
$\mathcal{O}(\gB^4)$.  Contracting external lines in a colour-ordered
manner into these terms, we have
\begin{equation}
  \begin{split}
    V^2(1234) = \phantom{+} & \frac{\hat 1}{\hat 5} V^2(5^A23)
    \Xi^2(\bar 5^A41)
    +   \frac{\hat 2}{\hat 5} V^2(15^A4) \Xi^1(\bar 5^A23) \\
    + & V^2(125^A) \Upsilon(\bar 5^A34) + W^2(1234)
  \end{split}
  \label{eq:vtx-4-2}
\end{equation}
for the $\bar\gB_I(1)\bar\gB_J(2)\gB_K(3)\gB_L(4)$ vertex, and
\begin{equation}
  \begin{split}
    V^3(1234) = \phantom{+} & \frac{\hat 1}{\hat 5} V^2(5^A34)
    \Xi^1(\bar 5^A12)
    +   \frac{\hat 3}{\hat 5} V^2(15^A4) \Xi^2(\bar 5^A23) \\
    + & \frac{\hat 3}{\hat 5} V^2(5^A12) \Xi^2(\bar 5^A34)
    +   \frac{\hat 1}{\hat 5} V^2(35^A2) \Xi^1(\bar 5^A41) \\
    + & 2W^3(1234)
  \end{split}
  \label{eq:vtx-4-3}
\end{equation}
for the $\bar\gB_I(1)\gB_J(2)\bar\gB_K(3)\gB_L(4)$ vertex.

That these expressions reduce in four dimensions should be obvious by
comparing the forms of \eqref{eq:vtx-4-2} and \eqref{eq:vtx-4-3} to
their four-dimensional analogs in section \ref{ssec:mhvym-4g} and
noting the reduction of the individual factors.

It is worthwhile noting here that unlike in the four-dimensional case,
the vertices \eqref{eq:vtx-4-2} and \eqref{eq:vtx-4-3} contain terms
which vanish on shell. Furthermore, the transformation coefficients
and the resulting vertices are no longer holomorphic (owing to the
scalar product in the denominator of \eqref{eq:upsilon-3} preventing
cancellation of the antiholomorphic bilinear, something possible in
four dimensions) or have a simple form. The CSW rules are an
inherently four-dimensional construction, and nor do we have any known
$D$-dimensional generalisation of the Parke--Taylor amplitudes with
which to compare. As such, we simply take \eqref{eq:vtx-4-2},
\eqref{eq:vtx-4-3}, and higher vertices computed using this programme
as the definitions of the $D$-dimensional MHV vertices.

\section{The one-loop $(\fourplus)$ amplitude}
\label{sec:etv-oneloop}

It is not possible to construct a one-loop $(\fourplus)$ amplitude
using only the $\gB$ vertices of the canonical MHV lagrangian.
Nevertheless, we know it is non-vanishing and given by
\eqref{eq:fourplus-oneloop}. We will see that it arises
(as it indeed must) from
equivalence theorem evading pieces, constructed from the MHV
completion vertices of fig.~\ref{fig:etv-completionverts-dd}.
In all, we can construct four classes of graphs for this contribution:
boxes, triangles, two classes of bubbles (corresponding to the two
possible arrangements of external lines on either side of the loop),
and the tadpoles.

In the next three subsections, we will consider the
quadruple cut of these diagrams. We will restrict ourselves to
analysing the cuts that arise from the singularities provided by the
propagators, which we refer to as \emph{standard} cuts.  From general
considerations \cite{Eden:1966aa} we expect other \emph{non-standard}
cuts arising from the singular denominators in the vertices. This is
true both of the $D$ dimensional version we have here and the four
dimensional Parke--Taylor forms \eqref{eq:parke-taylor}, but from the earlier
derivation it is clear that this singular behaviour is restricted to
the quantisation surface (they have no dependence on $\check
p$). These cuts therefore depend on the orientation of the
quantisation surface, \ie\ $\mu$, and are thus gauge artifacts which
should all cancel out in any complete on-shell amplitude.


\subsection{Off-shell quadruple cut}
\label{ssec:pppp-4cut}

\begin{figure}[h]
  \centering \subfigure{
    \begin{picture}(120,120) \SetOffset(60,60)
      \Gluon(0,32)(0,45){2}{2} \Gluon(0,-32)(0,-45){2}{2}
      \Gluon(32,0)(45,0){2}{2} \Gluon(-32,0)(-45,0){2}{2}
      \ArrowLine(0,30)(30,0) \ArrowLine(0,-30)(-30,0)
      \ArrowLine(30,0)(0,-30) \ArrowLine(-30,0)(0,30) \BCirc(0,30){2}
      \BCirc(30,0){2} \BCirc(0,-30){2} \BCirc(-30,0){2}
      \Text(0,47)[bc]{$1^-$} \Text(47,0)[lc]{$2^-$}
      \Text(0,-47)[tc]{$3^-$} \Text(-47,0)[rc]{$4^-$}
      \Text(17,17)[bl]{$q_1$} \Text(17,-17)[tl]{$q_2$}
      \Text(-17,-17)[tr]{$q_3$} \Text(-17,17)[br]{$q_4$}
      \Text(4,28)[cl]{$-$} \Text(-4,28)[cr]{$+$} \Text(4,-28)[cl]{$+$}
      \Text(-4,-28)[cr]{$-$} \Text(25,4)[bl]{$+$}
      \Text(25,-4)[tl]{$-$} \Text(-25,4)[br]{$-$}
      \Text(-25,-4)[tr]{$+$}
    \end{picture}
    \label{fig:etv-mhv-box-clock}
  }\quad \subfigure{
    \begin{picture}(100,80) \SetOffset(60,60) \Gluon(0,32)(0,45){2}{2}
      \Gluon(0,-32)(0,-45){2}{2} \Gluon(32,0)(45,0){2}{2}
      \Gluon(-32,0)(-45,0){2}{2} \Line(0,30)(30,0) \Line(0,-30)(-30,0)
      \Line(30,0)(0,-30) \Line(-30,0)(0,30) \BCirc(0,30){2}
      \BCirc(30,0){2} \BCirc(0,-30){2} \BCirc(-30,0){2}
      \Text(4,28)[cl]{$+$} \Text(-4,28)[cr]{$-$} \Text(4,-28)[cl]{$-$}
      \Text(-4,-28)[cr]{$+$} \Text(25,4)[bl]{$-$}
      \Text(25,-4)[tl]{$+$} \Text(-25,4)[br]{$+$}
      \Text(-25,-4)[tr]{$-$}
    \end{picture}
    \label{fig:etv-mhv-box-anticlock}
  }
  \caption{Box contributions to the one-loop $(\fourplus)$ amplitude.
    All external momenta are taken as outgoing. }
  \label{fig:etv-mhv-box}
\end{figure}
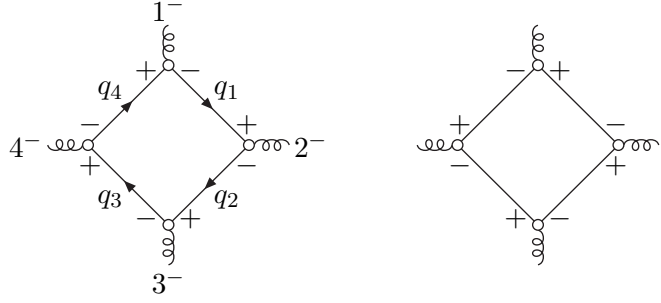

The $(\fourplus)$
amplitude is obtained by amputating $\langle \bar A \bar A \bar A
\bar A \rangle$, the box diagram for which is shown in
fig.~\ref{fig:etv-mhv-box}.  This gives a contribution of

\begin{equation}
  \begin{split}
    \label{eq:boxintegral}
    A^{\text{box}}(1^+,2^+,3^+,4^+) = &\lim_{p_1^2, p_2^2, p_3^2,
      p_4^2 \rightarrow 0} \frac 14 g^4 \: \frac{p_1^2 p_2^2 p_3^2
      p_4^2}{\hat 1 \hat 2 \hat 3 \hat 4} \int \frac{d^D q}{(2\pi)^D}
    \frac{16}{q_1^2 q_2^2 q_3^2 q_4^2} \frac{1}{\Sigma_1 \Sigma_2
      \Sigma_3 \Sigma_4}
    \times \\
    \bigl\{ & \bar V^2(-q_4^D, 1, q_1^A) \bar V^2(-q_1^A, 2, q_2^B)
    \bar V^2(-q_2^B, 3, q_3^C) \bar V^2(-q_3^C, 4, q_4^D) \\
    & + \bar V^2(1, q_1^A, -q_4^D) \bar V^2(2, q_2^B, -q_1^A) \bar
    V^2(3, q_3^C, -q_2^B) \bar V^2(4, q_4^D, -q_3^C) \bigr\},
  \end{split}
\end{equation}
where we have already used \eqref{eq:Upsilon-V2-dd} and
\eqref{eq:xi-3pt}, the internal momenta are defined as $q_i=q-P_{1i}$,
and we define the short-hand
\begin{equation}
  \label{eq:Sigma-sh}
  \Sigma_j := \frac{q_j^2}{\hat q_j} - \frac{q_{j-1}^2}{\hat q_{j-1}}
  + \frac{p_j^2}{\hat p_j}
\end{equation}
(indices interpreted cyclically). Note that the external momenta $p_1,
\dots, p_4$ are in four dimensions and thus their transverse indices
have all been set to one.

Before going on to compute the quadruple standard cut of the box, as
an aside we will show that its double and triple standard cuts vanish
for on-shell external momenta. Consider cutting any three internal
lines of fig.~\ref{fig:etv-mhv-box}, keeping the remaining internal line
strictly off-shell.  Without loss of generality, we choose these
internal lines to be $q_1$, $q_2$ and $q_3$.  In order that the
amplitude survives LSZ reduction, the correlator must generate a
singularity in $p_1^2 p_2^2 p_3^2 p_4^2$.  Clearly, the $\Sigma_2$ and
$\Sigma_3$ denominators provide a singularity $p_2^2 p_3^2$ once the
internal lines are cut. We now claim that the triple cut vanishes as
follows: the required singularity in $p_1^2p_4^2$ must come from the
denominators in the remaining tree graph connecting $p_1$ and
$p_4$. The relevant factors from \eqref{eq:boxintegral} are the $q_4$
propagator, and the $\Sigma_1$ and $\Sigma_4$ denominators, \ie\
\[
\frac1{q_4^2} \left(\frac{q_1^2}{\hat q_1} - \frac{q_4^2}{\hat q_4} +
  \frac{p_1^2}{\hat p_1}\right)^{-1} \left(\frac{q_4^2}{\hat q_4} -
  \frac{q_3^2}{\hat q_3} + \frac{p_4^2}{\hat p_4}\right)^{-1}.
\]
Upon setting $q_1^2$ and $q_3^2$ to zero and discarding (the
non-vanishing) factors of momenta that appear on the numerator, we
arrive at
\[
\frac1{q_4^2} \frac 1{\hat q_4\,p_1^2 - \hat 1 \,q_4^2} \frac 1{\hat 4
  \,q_4^2 + \hat q_4\,p_4^2}.
\]
The above factors clearly cannot cancel $p_1^2 p_4^2$ so long as
$q_4^2 \ne 0$. Hence this cut vanishes as we take all the external
momenta on shell. By similar consideration, one can also see that both
possible double cuts of this graph also vanish.

Now, we will compute the standard quadruple cut. This is obtained by
putting all four internal lines on shell \cite{Eden:1966aa,
  Britto:2004nc, Brandhuber:2005jw}.  The external momenta are kept
off shell momentarily. We see that the $\Sigma_i$ reduce to
$p_i^2/\hat i$ factors upon cutting, producing poles which thus cancel
the factors of $p^2_i$ from LSZ reduction and the $1/\hat i$
factors. The remaining terms have a finite
non-vanishing\footnote{Recall that four-cut solutions are
  non-vanishing because they use complex external and internal momenta
  \cite{Eden:1966aa, Britto:2004nc, Brandhuber:2005jw}.} on-shell limit
and it is already clear that they are exactly what we obtain from the
four-cut box contribution using the light cone Yang--Mills action
\eqref{eq:lcym-action}.

For the purposes of demonstrating that precisely this contribution
arises from the four-cut MHV completion box graph within the present
formalism we do not need to go any further. However, let us show how
this contribution can be straightforwardly computed within the
formalism we have developed here.

\subsection{Explicit evaluation of box quadruple cut}
We begin with \eqref{eq:boxintegral}, and substitute for $\bar V^2$
using \eqref{eq:Vbar2}.  For the moment, the external momenta $p_i$
are four-dimensional and off-shell, whereas the loop momentum $q
\equiv q_4$ in the integral is $D$-dimensional. To compute the cut, we
replace the four propagators with
\[
\delta^{+4}(q_1) \delta^{+4}(q_2) \delta^{+4}(q_3) \delta^{+4}(q_4),
\]
and by splitting the integral over momentum space as in section
\ref{ssec:background-loops-cutcon}, these $\delta$ functions enforce
the four constraints $q^2_i=0$. This in turn fixes the
four-dimensional part of $q$ to a discrete set of solutions (in fact
two) in term of the remaining, orthogonal $-2\epsilon$ components
$\nu$. It is now safe to take the $p_i$ on-shell and, assuming
solution to the constraints exist, we are left with
\begin{equation}
  \label{eq:cutintegral}
  8 g^4 (1-\epsilon) \frac{1}{\hat 1 \hat 2 \hat 3 \hat 4}
  \int \frac{d\nu^{-2\epsilon}}{(2\pi)^{-2\epsilon}}
  \{q\:1\} \{q\!-\!1,2\} \{q\!+\!4,3\} \{q\:4\}.
\end{equation}
Some comments are in order here. First, note
that the $\left\{ \cdots \right\}$ bilinears above have their
index set to $1$, but this has been dropped for clarity.  Now by their $q$
dependence, the bilinears above are functions of $\nu$, but we can say
more: since $q$ can only
contract with either itself or the four-dimensional external momenta,
we see that these solutions can in fact only depend on
\begin{equation}
  \label{eq:mu-q}
  \nu^2 = 2(q_I \bar
  q_I - \tilde q \bar q).
\end{equation}
We would also like to point out the factor of $(1-\epsilon)$ comes
from dimensional regularisation of the gauge field degrees of freedom
(\cf\ the four-dimensional helicity scheme \cite{Bern:1991aq}, which
extends only the loop momentum to $D$ dimensions and therefore lacks
this factor).

What remains is to obtain expressions for the bilinears in terms of
$\nu^2$.  In the following, since the momenta are complex, $\{q\:1\}$
is not related by complex conjugation to $(q\:1)$.
First, consider
$(q\:1) \mdot \{q\:1\}$.  Since, $q_1^2 = q^2 = p^2_1 = 0$,
\eqref{eq:null-p-dot-q}, or \eqref{eq:sum-bilinears}, implies that
this vanishes. Splitting away the four-dimensional part and using
\eqref{eq:mu-q} gives
\begin{align}
  \label{eq:bilinear-mu-1}
  (q\:1)\{q\:1\} + \hat 1^2 \nu^2/2 &= 0, \\
  \intertext{and similarly}
  \label{eq:bilinear-mu-2}
  (q\!-\!1,2)\{q\!-\!1,2\} + \hat 2^2 \nu^2/2 &= 0 \quad\text{and}\quad \\
  \label{eq:bilinear-mu-3}
  (q\:4)\{q\:4\} + \hat 4^2 \nu^2/2 &= 0.
\end{align}
We eliminate $\nu^2$ between \eqref{eq:bilinear-mu-1}, and
\eqref{eq:bilinear-mu-3}, and then use \eqref{eq:bianchi} to eliminate
$(q\:4)$ and its conjugate.  This gives
\begin{equation*}
  \hat q + \hat4 \frac{(q\:1)}{(1\:4)} + \hat4 \frac{\{q\:1\}}{\{1\:4\}}
  = 0.
\end{equation*}
Similarly, eliminating $\nu^2$ between \eqref{eq:bilinear-mu-1} and
\eqref{eq:bilinear-mu-2} leads to
\begin{equation}
  \label{eq:qhat-1-2}
  \hat q\!-\!\hat1 + \hat2 \frac{(q\:1)}{(1\:2)} + \hat2
  \frac{\{q\:1\}}{\{1\:2\}} = 0.
\end{equation}
Subtracting, and using \eqref{eq:bilinear-mu-1} to eliminate
$\{q\:1\}$ yields the quadratic equation
\begin{equation}
  \label{eq:q1-quadratic}
  \alpha (q\:1)^2 + \hat 1 (q\:1) - \bar\alpha \frac{\hat 1^2 \nu^2}{2} = 0
\end{equation}
where
\begin{equation}
  \label{eq:bilinear-alphas}
  \alpha = \frac{\hat 4}{ (1\:4) } - \frac{\hat 2}{ (1\:2) },\quad
  \bar\alpha = \frac{\hat 4}{\{1\:4\}} - \frac{\hat 2}{\{1\:2\}}.
\end{equation}
This has solutions
\begin{equation}
  \label{eq:q1-solutions}
  (q\:1) = -\frac{\hat 1}{2\alpha} (1 \pm \beta),\quad
  \{q\:1\} = -\frac{\hat 1}{2\bar\alpha} (1 \mp \beta),\quad
  \beta = \sqrt{1 + 2\alpha\bar\alpha \nu^2}.
\end{equation}
Next, the Bianchi-like identity \eqref{eq:bianchi} gives
\begin{equation}
  \hat 1\{q\!-\!1,2\} = \hat 2\{q\:1\} + (\hat q - \hat 1)\{1\:2\}.
  \label{eq:q12-bianchi}
\end{equation}
Using \eqref{eq:qhat-1-2} and \eqref{eq:q1-solutions} gives
\begin{equation}
  \label{eq:q12-solutions}
  \{q\!-\!1,2\} = \frac{\hat 2}{2\alpha}\frac{\{1\:2\}}{(1\:2)}(1 \pm \beta).
\end{equation}
Similarly, we find
\begin{equation}
  \label{eq:q4-solutions}
  \{q\:4\} = \frac{\hat 4}{2\alpha}\frac{\{1\:4\}}{(1\:4)}(1 \pm \beta).
\end{equation}
To obtain the final bilinear $\{q\!+\!4, 3\}$, we use
\eqref{eq:bianchi} twice to obtain it in terms of $\{q\:1\}$ and
$\{q\:4\}$.  \eqref{eq:sum-bilinears} is then applied to eliminate a
quotient of $(\cdots)$ bilinears present in one of the terms in favour
of conjugate bilinears, giving
\begin{equation}
  \label{eq:q43-solutions}
  \{q\!+\!4,3\} = \frac 12
  \frac{\{2\:3\}\{3\:4\}}{\{2\:4\}} (1 \pm \beta).
\end{equation}

Assembling the product of the $\{\cdots\}$ bilinears from
\eqref{eq:q1-solutions}, \eqref{eq:q12-solutions},
\eqref{eq:q4-solutions} and \eqref{eq:q43-solutions}, we have
\begin{equation}
  \{q\:1\} \{q\!-\!1,2\} \{q\!+\!4,3\} \{q\:4\}
  = - \tfrac 14 \hat 1^2 \hat 2 \hat 4 \:\nu^4\:
  \frac{\{2\:3\}\{3\:4\}}{(1\:2)(4\:1)}
  = - \tfrac 14 \hat 1 \hat 2 \hat 3 \hat 4 \:\nu^4\:
  \frac{\{1\:2\}\{3\:4\}}{(1\:2)(3\:4)}
\end{equation}
for either of the solutions \eqref{eq:q1-solutions}, where in the
second assertion we have used the fact that the right-hand side of
\eqref{eq:sum-bilinears} is zero for null $p_j$. Using this in
\eqref{eq:cutintegral} and reinstating the propagators, we arrive at
\begin{equation}
  \label{eq:cutintegral-lifted}
  2(1-\epsilon) g^4 \frac{\{1\:2\}\{3\:4\}}{(1\:2)(3\:4)}
  \int \frac{d^4q\:d^{-2\epsilon}\nu}{(2\pi)^D}
  \frac{\nu^4}{q_1^2 q_2^2 q_3^2 q_4^2}.
\end{equation}
Thus we conclude that \eqref{eq:boxintegral} has precisely the
quadruple cut of the $4-2\epsilon$-dimensional box function $K_4$, as
expected of this amplitude \cite{Bern:1993mq, Bern:1993sx,
  Mahlon:1993si, Bern:1993qk, Bern:1995db}.

\subsection{Triangle, bubble and tadpole contributions}
\label{ssec:pppp-trianglebubble}

Typical triangle, bubble and tadpole contributions to the one-loop
$(\fourplus)$ amplitude with internal helicities running from $-$ to
$+$ in a clockwise sense\footnote{The ``sense'' of internal helicity
  orientation is always defined here as propagating from $-$ to $+$.}
are shown in figs.~\ref{fig:mhv-triangles}, \ref{fig:mhv-bubbles} and
\ref{fig:mhv-tadpoles}.
\begin{figure}[h]
  \centering \subfigure{
    \begin{picture}(120,120)(0,0) \ArrowLine(90,30)(30,90)
      \ArrowLine(90,90)(90,30) \ArrowLine(30,90)(90,90)
      \ArrowLine(30,90)(30,30) \Gluon(10,10)(30,30){2}{3}
      \Gluon(110,10)(90,30){2}{3} \Gluon(110,110)(90,90){2}{3}
      \Gluon(10,110)(30,90){2}{3} \BCirc(30,30){2} \BCirc(90,30){2}
      \BCirc(90,90){2} \BCirc(30,90){2} \Text(15,10)[tr]{$4^-$}
      \Text(110,10)[tl]{$3^-$} \Text(15,110)[br]{$1^-$}
      \Text(103,111)[bl]{$2^-$} \Text(21,60)[cc]{$p_4$}
      \Text(60,100)[cc]{$q_1$} \Text(100,60)[cc]{$q_2$}
      \Text(58,50)[cc]{$q_3$} \Text(22,35)[cc]{$-$}
      \Text(22,83)[cc]{$+$} \Text(38,97)[cc]{$-$}
      \Text(82,97)[cc]{$+$} \Text(98,35)[cc]{$+$}
      \Text(98,83)[cc]{$-$} \Text(80,30)[cc]{$-$}
      \Text(37,75)[cc]{$+$}
    \end{picture}
  }\quad \subfigure{
    \begin{picture}(120,120)(0,0) \ArrowLine(90,30)(30,30)
      \ArrowLine(90,90)(90,30) \ArrowLine(30,30)(90,90)
      \ArrowLine(30,30)(30,90) \Gluon(10,10)(30,30){2}{3}
      \Gluon(110,10)(90,30){2}{3} \Gluon(110,110)(90,90){2}{3}
      \Gluon(10,110)(30,90){2}{3} \BCirc(30,30){2} \BCirc(90,30){2}
      \BCirc(90,90){2} \BCirc(30,90){2} \Text(15,10)[tr]{$4^-$}
      \Text(110,10)[tl]{$3^-$} \Text(15,110)[br]{$1^-$}
      \Text(103,111)[bl]{$2^-$} \Text(21,60)[cc]{$p_1$}
      \Text(60,20)[cc]{$q_3$} \Text(100,60)[cc]{$q_2$}
      \Text(60,65)[br]{$q_1$} \Text(22,35)[cc]{$+$}
      \Text(22,83)[cc]{$-$} \Text(38,45)[cc]{$-$}
      \Text(82,90)[cc]{$+$} \Text(98,35)[cc]{$+$}
      \Text(98,83)[cc]{$-$} \Text(80,24)[cc]{$-$}
      \Text(37,24)[cc]{$+$}
    \end{picture}
  }
  \caption{One-loop MHV completion triangle graphs for the
    $(\fourplus)$ amplitude. Note that the propagator carrying an
    external momentum is attached to the $\Xi$ vertex differently in
    each case.}
  \label{fig:mhv-triangles}
\end{figure}
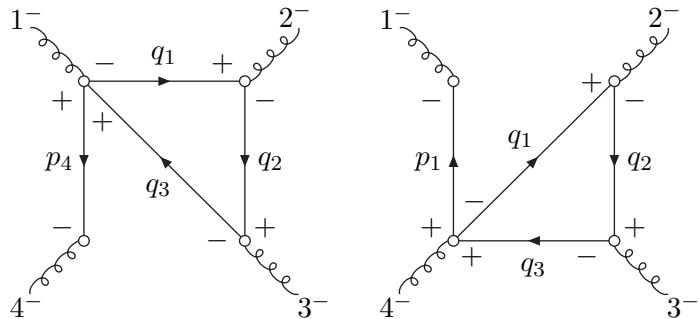
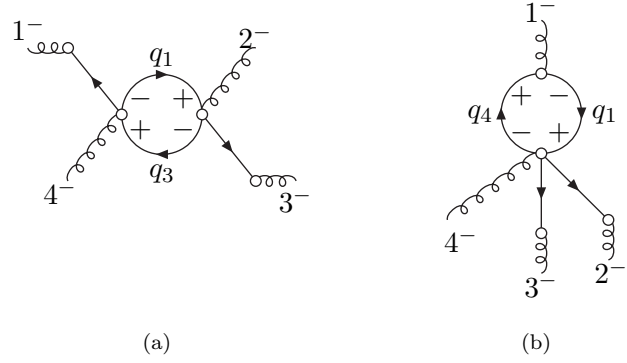
\begin{figure}
  \centering \subfigure[]{
    \begin{picture}(120,120)(0,0) \SetOffset(60,75)
      \ArrowLine(-15,0)(-35,25) \Gluon(-35,-25)(-15,0){2}{4}
      \ArrowArcn(0,0)(15,0,180) \ArrowArcn(0,0)(15,180,0)
      \ArrowLine(15,0)(35,-25) \Gluon(15,0)(35,25){2}{4}
      \Gluon(-35,25)(-50,25){2}{2} \Gluon(35,-25)(50,-25){2}{2}
      \BCirc(-15,0){2} \BCirc(15,0){2} \BCirc(-35,25){2}
      \BCirc(35,-25){2} \Text(35,25)[cb]{$2^-$}
      \Text(50,-28)[ct]{$3^-$} \Text(-38,-25)[ct]{$4^-$}
      \Text(-50,28)[cb]{$1^-$} \Text(0,18)[cb]{$q_1$}
      \Text(0,-19)[ct]{$q_3$} \Text(8,6)[cc]{$+$} \Text(8,-6)[cc]{$-$}
      \Text(-8,6)[cc]{$-$} \Text(-8,-6)[cc]{$+$}
    \end{picture}
    \label{fig:mhv-bubbles-a}
  } \quad \subfigure[]{
    \begin{picture}(120,120)(0,0) \SetOffset(25,0)
      \Gluon(35,110)(35,90){2}{2} \ArrowArcn(35,75)(15,90,270)
      \ArrowArcn(35,75)(15,270,90) \Gluon(35,60)(0,35){-2}{5}
      \ArrowLine(35,60)(35,30) \Gluon(35,30)(35,15){2}{2}
      \ArrowLine(35,60)(60,35) \Gluon(60,35)(60,20){2}{2}
      \BCirc(35,90){2} \BCirc(35,60){2} \BCirc(35,30){2}
      \BCirc(60,35){2} \Text(35,110)[bc]{$1^-$}
      \Text(55,20)[tl]{$2^-$} \Text(35,15)[tc]{$3^-$}
      \Text(5,32)[tc]{$4^-$} \Text(17,75)[cr]{$q_4$}
      \Text(54,75)[cl]{$q_1$} \Text(32,82)[cr]{$+$}
      \Text(46,82)[cr]{$-$} \Text(32,68)[cr]{$-$}
      \Text(46,68)[cr]{$+$}
    \end{picture}
    \label{fig:mhv-bubbles-b}
  }
  \caption{MHV completion bubble graphs. In (a) we show a $2|2$
    bubble.  There are three other graphs like it (up to shifting the
    external momentum labels once by $i\rightarrow i+1$) obtained by
    swapping the external momentum propagators between gluons $1$ and
    $4$, and between gluons $2$ with $3$. The $3|1$ bubble is shown in
    (b); there are two additional graphs (up to rotations of the
    labels), in this case obtained by associating the curly line
    attached to the five-point $\Xi$ with gluon $2$ or $3$ instead of
    $4$.}
  \label{fig:mhv-bubbles}
\end{figure}
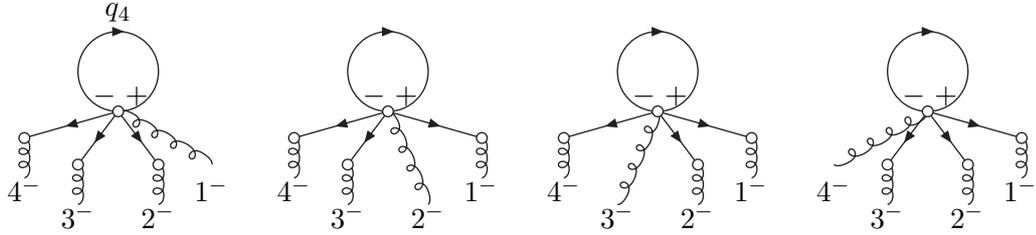
\begin{figure}[h]
  \centering \subfigure{
    \begin{picture}(90,87)(0,0) \Text(45,79)[bc]{$q_4$}
      \SetOffset(0,-5) \ArrowArcn(45,65)(15,269.9999,270.0001)
      \Gluon(45,50)(80,30){2}{4} \Gluon(10,40)(10,25){2}{2}
      \ArrowLine(45,50)(10,40) \BCirc(10,40){2}
      \Text(10,25)[tc]{$4^-$} \ArrowLine(45,50)(30,30)
      \Gluon(30,30)(30,15){2}{2} \ArrowLine(45,50)(60,30)
      \Gluon(60,30)(60,15){2}{2} \BCirc(60,30){2} \BCirc(30,30){2}
      \BCirc(45,50){2} \Text(30,15)[tc]{$3^-$} \Text(60,15)[tc]{$2^-$}
      \Text(80,25)[tc]{$1^-$} \Text(40,56)[cc]{$-$}
      \Text(52,56)[cc]{$+$}
    \end{picture}
  } \subfigure{
    \begin{picture}(90,87)(0,0) \SetOffset(0,-5)
      \ArrowArcn(45,65)(15,269.9999,270.0001)
      \Gluon(10,40)(10,25){2}{2} \ArrowLine(45,50)(10,40)
      \BCirc(10,40){2} \Text(10,25)[tc]{$4^-$}
      \ArrowLine(45,50)(80,40) \ArrowLine(45,50)(30,30)
      \Gluon(80,40)(80,25){2}{2} \Gluon(30,30)(30,15){2}{2}
      \Gluon(45,50)(60,15){2}{4} \BCirc(80,40){2} \BCirc(30,30){2}
      \BCirc(45,50){2} \Text(30,15)[tc]{$3^-$} \Text(60,15)[tc]{$2^-$}
      \Text(80,25)[tc]{$1^-$} \Text(40,56)[cc]{$-$}
      \Text(52,56)[cc]{$+$}
    \end{picture}
  } \subfigure{
    \begin{picture}(90,87)(0,0) \SetOffset(0,-5)
      \ArrowArcn(45,65)(15,269.9999,270.0001)
      \Gluon(10,40)(10,25){2}{2} \ArrowLine(45,50)(10,40)
      \BCirc(10,40){2} \Text(10,25)[tc]{$4^-$}
      \ArrowLine(45,50)(80,40) \ArrowLine(45,50)(60,30)
      \Gluon(80,40)(80,25){2}{2} \Gluon(60,30)(60,15){2}{2}
      \Gluon(45,50)(30,15){2}{4} \BCirc(80,40){2} \BCirc(60,30){2}
      \BCirc(45,50){2} \Text(30,15)[tc]{$3^-$} \Text(60,15)[tc]{$2^-$}
      \Text(80,25)[tc]{$1^-$} \Text(40,56)[cc]{$-$}
      \Text(52,56)[cc]{$+$}
    \end{picture}
  } \subfigure{
    \begin{picture}(90,87)(0,0) \SetOffset(0,-5)
      \ArrowArcn(45,65)(15,269.9999,270.0001)
      \Gluon(45,50)(10,30){2}{4} \Text(10,25)[tc]{$4^-$}
      \ArrowLine(45,50)(80,40) \ArrowLine(45,50)(30,30)
      \ArrowLine(45,50)(60,30) \Gluon(80,40)(80,25){2}{2}
      \Gluon(30,30)(30,15){2}{2} \Gluon(60,30)(60,15){2}{2}
      \BCirc(80,40){2} \BCirc(30,30){2} \BCirc(60,30){2}
      \BCirc(45,50){2} \Text(30,15)[tc]{$3^-$} \Text(60,15)[tc]{$2^-$}
      \Text(80,25)[tc]{$1^-$} \Text(40,56)[cc]{$-$}
      \Text(52,56)[cc]{$+$}
    \end{picture}
  }
  \caption{Tadpole MHV completion graphs. Notice that the coupling of
    the six-point $\Xi$ to the $\bar\gA$ field (denoted by the curly
    line) is associated with a different gluon in each case.}
  \label{fig:mhv-tadpoles}
\end{figure}

Despite appearances, these diagrams do have quadruple cuts as a result
of the singularities in the vertices. We therefore have to consider
also cutting the vertices. Let us first consider the quadruple cut of
the triangle graph. We can restrict our analysis to the graphs
fig.~\ref{fig:mhv-triangles}. Their contribution to the $(\fourplus)$
amplitude is
\begin{equation}
  \label{eq:pppp-tri-Xi}
  \begin{split}
    \lim_{p_1^2, p_2^2, p_3^2, p_4^2 \rightarrow 0} \frac 14 g^4 \:
    \frac{p_1^2 p_2^2 p_3^2 p_4^2}{\hat 1 \hat 2 \hat 3 \hat 4} \int
    &\frac{d^Dq}{(2\pi)^D}
    \frac{\hat q_1 \hat q_2 \hat q_3}{q_1^2 q_2^2 q_3^2} \times \\
    \biggl\{ & - \Xi^1(1, q_1^A, -q_3^C, 4) \Xi^1(2, q_2^B, -q_1^A)
    \Xi^1(3, q_3^C, -q_2^B) \frac{\hat 4}{p_4^2} \\
    & - \Xi^2(4, 1, q_1^A, -q_3^C) \Xi^2(2, q_2^B, -q_1^A) \Xi^2(3,
    q_3^C, -q_2^B) \frac{\hat 1}{p_1^2} \biggr\}.
  \end{split}
\end{equation}
Using the recurrence relations \eqref{eq:upsilon-rr-x} and
\eqref{eq:xi-rr} to evaluate the $\Xi$s, we can re-write this as
\begin{equation}
  \label{eq:pppp-tri-Vbar2}
  \begin{split}
    \lim_{p_1^2, p_2^2, p_3^2, p_4^2 \rightarrow 0} \frac 14 g^4 \:
    \frac{p_1^2 p_2^2 p_3^2 p_4^2}{\hat 1 \hat 2 \hat 3 \hat 4} \int
    \frac{d^Dq}{(2\pi)^D} \frac{16}{q_1^2 q_2^2 q_3^2} \Bigg\{
    \frac{X}{\hat q_4 \Sigma_1 \Sigma_2 \Sigma_3 \Sigma_4 (\Sigma_1 +
      \Sigma_4)} \left( \frac{\Sigma_1 \hat 4}{p_4^2}- \frac{\Sigma_4
        \hat 1}{p_1^2}
    \right)  \\
    + \frac{Y}{(\hat 1 + \hat 4) \Sigma'_{1+4} \Sigma_2 \Sigma_3}
    \left( \frac 1 {\Sigma_{1+4}} - \frac{1}{\Sigma_1 + \Sigma_4}
    \right) \left( \frac{\hat 1}{p_1^2} + \frac{\hat 4}{p_4^2} \right)
    \Bigg\}
  \end{split}
\end{equation}
with
\begin{align}
  X =& \bar V^2(-q_4^D, 1, q_1^A) \bar V^2(-q_1^A, 2, q_2^B) \bar
  V^2(-q_2^B, 3, q_3^C) \bar V^2(-q_3^C, 4, q_4^D),
  \label{eq:pppp-tri-X} \\
  Y =& \bar V^2(4, 1, 2+3^D) \bar V^2(-q_3^C, 1+4^D, q_1^A) \bar
  V^2(-q_1^A, 2, q_2^B) \bar V^2(-q_2^B, 3, q_3^C),
  \label{eq:pppp-tri-Y}
\end{align}
$q_4 := q_3 - p_4$ flowing ``through'' the vertex attached to $p_1$
(or $p_4$) as part of a box-like momentum-flow topology (see section
\ref{ssec:pppp-lc-recon} below for a further investigation of this),
and we define the following extensions of $\Sigma_i$ as
\begin{align}
  \Sigma_{1+4} &:= \frac{q_1^2}{\hat q_1} - \frac{q_3^2}{\hat q_3}
  + \frac{(p_1 + p_4)^2}{\hat 1 + \hat 4}, \\
  \Sigma'_{1+4} &:= \frac{p_1^2}{\hat 1} + \frac{p_4^2}{\hat 4} -
  \frac{(p_1 + p_4)^2}{\hat 1 + \hat 4}.
\end{align}
Note that one can write down expressions for the analogues of $X$ and
$Y$ from graphs with internal helicities of an anti-clockwise
sense. (It is easy to check for the case at hand (where $\bar V^2$ is
the three-point \mhvbar\ vertex), that these are the same as in the
clockwise scenario.)

Now, recall that we are only studying the standard cuts. We will
extract such a quadruple cut contribution here, by keeping the
external momenta off the mass shell, and look for any terms containing
$1/q^2_4$ in addition to the three propagators already appearing in
\eqref{eq:pppp-tri-Vbar2}. Clearly, by inspection of the $\Sigma_i$
factors in \eqref{eq:pppp-tri-Vbar2}, no such $1/q^2_4$ are
generated. Indeed it is impossible to generate such terms from the
vertices since the singularity in $1/q^2_4$ is not restricted to the
quantisation surface. Although the inverse $\Sigma_i$s and
$\Sigma_{1+4}$ appear in \eqref{eq:Sigma-sh} and above to yield
singularities that look superficially similar to those from
propagators, by \eqref{eq:sum-omega} these terms do not contain
$\check q$ components and thus their singularities lie entirely within
the quantisation surface.

A similar analysis of the graphs of figs.~\ref{fig:mhv-bubbles} and
\ref{fig:mhv-tadpoles} leads one quickly to the same conclusion: they
have no contribution to the quadruple cut for off-shell external
momenta, because in this region none of the denominators from their
$\Xi$ vertices form the necessary propagators. Hence, we see that for
the one-loop $(\fourplus)$ amplitude in the Canonical MHV Lagrangian's
formalism, if we keep the external momenta off shell until after the
cuts, only the box graph of fig.~\ref{fig:etv-mhv-box} contributes to the
quadruple cut.

\subsection{Light-cone Yang--Mills reconstructions}
\label{ssec:pppp-lc-recon}

As with section \ref{ssec:etv-completion-mmpp},
expressions \eqref{eq:boxintegral}, \eqref{eq:pppp-tri-X} and
\eqref{eq:pppp-tri-Y} expose the relationship between
between the MHV completion graphs of
figs.~\ref{fig:etv-mhv-box}--\ref{fig:mhv-tadpoles} and the Feynman
graphs one would use to compute the same amplitude in conventional
perturbative LCYM.  We already see parallels of the latter in the
topology of the linking amongst $\bar V^2$ \mhvbar\
vertices.\footnote{Recall that these are the same as the $\ppm$
  vertices in light-cone Yang--Mills.}

\begin{figure}[h]
  \centering \subfigure[]{
    \begin{picture}(80,80)
      \Line(0,0)(20,20) \Line(80,0)(60,20) \Line(20,20)(60,20)
      \Line(0,80)(20,60) \Line(80,80)(60,60) \Line(20,60)(60,60)
      \Line(20,20)(20,60) \Line(60,20)(60,60)
      \Text(0,8)[cc]{$+$} \Text(0,72)[cc]{$+$}
      \Text(80,72)[cc]{$+$} \Text(80,8)[cc]{$+$}
      \Text(14,26)[cc]{$-$} \Text(66,26)[cc]{$+$}
      \Text(14,53)[cc]{$+$} \Text(66,53)[cc]{$-$}
      \Text(27,14)[cc]{$+$} \Text(53,14)[cc]{$-$}
      \Text(27,66)[cc]{$-$} \Text(53,66)[cc]{$+$}
    \end{picture}
\label{fig:pppp-lc-box-tri-bub-a}
  }\qquad \subfigure[]{
    \begin{picture}(103,50)
      \Line(0,0)(20,25) \Line(0,50)(20,25)
      \Line(20,25)(50,25) 
      \Line(50,25)(80,0) \Line(50,25)(80,50)
      \Line(80,0)(80,50)
      \Line(80,0)(105,0) \Line(80,50)(105,50)
      \Text(0,8)[cc]{$+$} \Text(0,42)[cc]{$+$}
      \Text(103,7)[cc]{$+$} \Text(103,43)[cc]{$+$}
      \Text(25,18)[cc]{$-$} \Text(45,18)[cc]{$+$}
      \Text(55,35)[cc]{$-$} \Text(55,12)[cc]{$+$}
      \Text(70,47)[cc]{$+$} \Text(70,2)[cc]{$-$}
      \Text(86,7)[cc]{$+$} \Text(86,43)[cc]{$-$}
    \end{picture}
    \label{fig:pppp-lc-box-tri-bub-b}
  }\qquad \subfigure[]{
    \begin{picture}(120,50)
      \Line(0,0)(20,25) \Line(0,50)(20,25)
      \Line(20,25)(100,25)
      \BCirc(60,25){15}
      \Line(100,25)(120,0) \Line(100,25)(120,50)
      \Text(0,8)[cc]{$+$} \Text(0,42)[cc]{$+$}
      \Text(120,8)[cc]{$+$} \Text(120,42)[cc]{$+$}
      \Text(26,20)[cc]{$-$} \Text(39,20)[cc]{$+$}
      \Text(81,20)[cc]{$+$} \Text(95,20)[cc]{$-$}
      \Text(52,30)[cc]{$-$} \Text(68,30)[cc]{$+$}
      \Text(52,19)[cc]{$+$} \Text(68,19)[cc]{$-$}
    \end{picture}
    \label{fig:pppp-lc-box-tri-bub-c}
  }\\
  \subfigure[]{
    \begin{picture}(140,50) \Line(0,0)(20,25) \Line(0,50)(20,25)
      \Line(20,25)(110,25) \Line(50,0)(50,25) \Line(110,25)(140,25)
      \BCirc(95,25){15} \Text(0,8)[cc]{$+$} \Text(0,42)[cc]{$+$}
      \Text(48,0)[br]{$+$} \Text(78,27)[br]{$+$} \Text(21,27)[bl]{$-$}
      \Text(48,27)[br]{$+$} \Text(52,27)[bl]{$-$}
      \Text(140,27)[br]{$+$} \Text(82,26)[bl]{$-$}
      \Text(82,24)[tl]{$+$} \Text(108,26)[br]{$+$}
      \Text(108,24)[tr]{$-$}
    \end{picture}
    \label{fig:pppp-lc-box-tri-bub-d}
  }\qquad \subfigure[]{
    \begin{picture}(140,50) \Line(0,0)(20,25) \Line(0,50)(20,25)
      \Line(20,25)(110,25) \Line(50,0)(50,25) \Line(80,25)(80,50)
      \BCirc(125,25){15} \Text(0,8)[cc]{$+$} \Text(0,42)[cc]{$+$}
      \Text(48,0)[br]{$+$} \Text(78,50)[tr]{$+$} \Text(21,27)[bl]{$-$}
      \Text(48,27)[br]{$+$} \Text(52,27)[bl]{$-$}
      \Text(78,23)[tr]{$+$} \Text(82,23)[tl]{$-$}
      \Text(108,27)[br]{$+$} \Text(112,26)[bl]{$-$}
      \Text(112,24)[tl]{$+$}
    \end{picture}
    \label{fig:pppp-lc-box-tri-bub-e}
  }
  \caption{Typical LCYM Feynman graphs that contribute to the
    $(\fourplus)$ amplitude. Shown topologies are the (a) box, (b)
    triangle, (c) bubble, (d) typical external leg correction, and (e)
    one of the tadpoles (there is another tadpole not shown here with
    the loop attached to the central leg of the tree instead).}
  \label{fig:pppp-lc-box-tri-bub}
\end{figure}
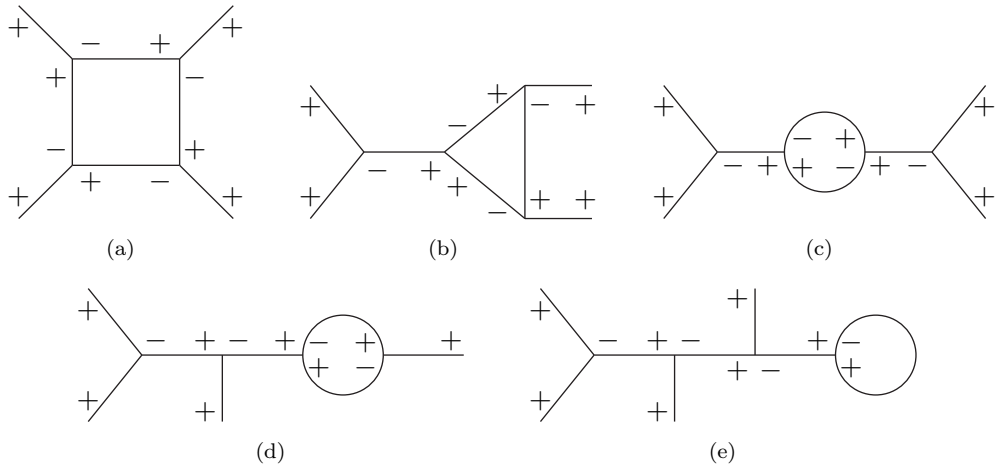

Starting with expression \eqref{eq:boxintegral} for the box graph, it
is immediately apparent that the momentum routing through the $\bar
V^2$s yields the two box-like topologies in LCYM: one with the
internal helicities arranged in a clockwise sense (from the first term
in the curly braces, shown in fig.~\ref{fig:pppp-lc-box-tri-bub-a}),
and another with an anticlockwise arrangement.

The triangle MHV completion graphs of fig.~\ref{fig:mhv-triangles}
reveal a mixture of topologies in the momentum routing. Naturally, one
would expect the triangle diagram of
fig.~\ref{fig:pppp-lc-box-tri-bub-b}, and indeed this arises from the
factor $Y$ in \eqref{eq:pppp-tri-Vbar2}. The MHV completion triangle graph
also has terms with factors of $X$, which we know is nothing but one
of the vertex configurations found in \eqref{eq:boxintegral} of box
topology, \ie\ fig.~\ref{fig:pppp-lc-box-tri-bub-a}. (As we have seen
however, in this case the fourth propagator is missing.)

The bubble and tadpole graphs can be processed in a similar manner:
the graph of fig.~\ref{fig:mhv-bubbles-a} contains the topology of, and
therefore contributes to the reconstruction of, the LCYM self-energy
correction graph in fig.~\ref{fig:pppp-lc-box-tri-bub-c}; similarly
fig.~\ref{fig:mhv-bubbles-b} contributes to the reconstruction of the
external leg corrections (an example of which is seen in
fig.~\ref{fig:pppp-lc-box-tri-bub-d}). The MHV completion tadpole
graphs in fig.~\ref{fig:mhv-tadpoles} contain terms of topology of the
LCYM tadpoles of fig.~\ref{fig:pppp-lc-box-tri-bub-e} (these are
ill-defined but we can take them to vanish in dimensional
regularisation just as we would for the LCYM tadpole), as well as
contributing pieces with the self-energy and external leg correction
topologies. Additionally, both MHV completion bubbles and tadpoles
contribute to the reconstruction of box and triangle LCYM graphs.

\subsubsection*{Full reconstruction of the light-cone Yang--Mills box
  contribution}

We saw in section \ref{ssec:etv-completion-mmpp}
for the case of the $(\mmpp)$ tree-level amplitude that although
individual completion vertex graphs are identical to LCYM graphs, the
\emph{sum} of all completion and MHV graphs with a given momentum-routing
topology did recover the off-shell LCYM graphs of that topology.
In this section, we will demonstrate this at the loop level for the
$(\fourplus)$ amplitude. Again, rather than do this for all graphs displayed in
fig.~\ref{fig:pppp-lc-box-tri-bub}, we will concentrate on the
diagrams of box topology with the internal helicity configuration as
displayed in fig.~\ref{fig:pppp-lc-box-tri-bub-a} (note the sense of
the signs on the internal lines). This corresponds to
terms containing the factor $X$ of \eqref{eq:pppp-tri-X}.

The MHV completion box graph of fig.~\ref{fig:etv-mhv-box} provides a
contribution to the amplitude, given in \eqref{eq:boxintegral} of
\begin{equation}
  \label{eq:mhvbox-X}
  4 g^4 \:
  \int \frac{d^D q}{(2\pi)^D}
  \frac{XC}{q_1^2 q_2^2 q_3^2 q_4^2}
\end{equation}
This is clearly of the box momentum-routing topology, as we remind the
reader that $X$ is simply the product of \mhvbar\ vertices in a box
configuration. The factor $C$ is
\begin{equation}\label{eq:boxbox}
  C_{\text{box}} =  \frac{P_1P_2P_3P_4}{\Sigma_1 \Sigma_2 \Sigma_3
    \Sigma_4}.
\end{equation}
We can repeat this analysis for the two triangle configurations in
fig.~\ref{fig:mhv-triangles},
whose contribution is given in \eqref{eq:pppp-tri-Vbar2}.
We multiply the integrand by
$q_4^2/q_4^2$ and extract the relevant $C$ factor to obtain
\begin{equation}
  \label{eq:trianglebox}
  C_{\text{triangle}} = \frac{P_2P_3}{\Sigma_2\Sigma_3}
  \frac{Q_4}{\Sigma_4+\Sigma_1}\left(\frac{P_1}{\Sigma_4}-
    \frac{P_4}{\Sigma_1}\right),
\end{equation}
where we have introduced the short-hand $Q_i:= q^2_i/\hat q_i$.
By a similar procedure, the $C$ factors for the bubble and tadpole
graphs are obtained. The algebra for this is fairly straightforward,
and we simply state the results. For the four possible bubble
configurations of  fig.~\ref{fig:mhv-bubbles-a} (see the caption for
the description of these configurations),
\begin{equation}\label{eq:22bubblebox}
  C_{2|2} =
  \frac{Q_2}{\Sigma_2+\Sigma_3}\left(\frac{P_3}{\Sigma_2}-\frac{P_2}{\Sigma_3}\right)
  \frac{Q_4}{\Sigma_4+\Sigma_1}\left(\frac{P_1}{\Sigma_4}-\frac{P_4}{\Sigma_1}\right).
\end{equation}
The three bubble configurations of fig.~\ref{fig:mhv-bubbles-b} have
\begin{equation}\label{eq:31bubblebox}
  C_{3|1} = \frac{P_1}{\Sigma_1}\frac{Q_2Q_3}{\Sigma_2}\left\{
    \frac{P_2}{(\Sigma_3+\Sigma_4)\Sigma_4}
    -\frac{P_2+P_3}{(\Sigma_2+\Sigma_3)\Sigma_4}
    +\frac{P_2+P_3+P_4}{(\Sigma_2+\Sigma_3)(\Sigma_2+\Sigma_3+\Sigma_4)}
  \right\},
\end{equation}
and finally the tadpoles of fig.~\ref{fig:mhv-tadpoles} yield
\begin{equation}
  \begin{split}
    C_{\text{tadpole}} = \frac{Q_1Q_2Q_3}{\Sigma_1}\Biggl\{
    \frac1{(\Sigma_1+\Sigma_2)(\Sigma_1+\Sigma_2+\Sigma_3)}
    -\frac{P_1}{(\Sigma_2+\Sigma_3+\Sigma_4)(\Sigma_3+\Sigma_4)\Sigma_4}&\\
    +\frac{P_1+P_2}{(\Sigma_1+\Sigma_2)(\Sigma_3+\Sigma_4)\Sigma_4}
    -\frac{P_1+P_2+P_3}{(\Sigma_1+\Sigma_2)(\Sigma_1+\Sigma_2+\Sigma_3)\Sigma_4}
    & \Biggr\}
  \end{split}
  \label{eq:tadpolebox}
\end{equation}

We must now account for additional contributions that arise from the
images of the graphs under cyclic permutations of the external
momenta. These permutations can be effected in the expressions
\eqref{eq:boxbox}--\eqref{eq:tadpolebox} by permuting \emph{all} the
momentum labels. Now, the box graph is invariant under these cycles so provides
only one contribution; the triangles, $3|1$ bubbles and tadpole graphs
all provide three extra contributions obtained from the three label shifts
$i \rightarrow i+1$ in \eqref{eq:trianglebox}, \eqref{eq:31bubblebox} and
\eqref{eq:tadpolebox}; the $2|2$ bubbles provide one extra contribution
from the shift $i \rightarrow i+2$ acting on \eqref{eq:22bubblebox}.

Adding these contributions and simplifying leads to, after some
considerable but straightforward algebra, the satisfying result that
$C=1$, \ie\ the sum of all the MHV completion graphs contributing to
the box topology is nothing other than what we would have obtained
using LCYM Feynman rules. Note that this happened \emph{before} taking
on-shell limits or performing any integration; recovering the missing
amplitude is a purely algebraic processes.

\subsubsection*{Taking the external momenta on-shell first}

In section \ref{ssec:pppp-4cut}, we reproduced the correct standard
quadruple cut contribution to the one-loop $(\fourplus)$ amplitude by
starting with off-shell external momenta, then cutting the internal
lines, and finally taking the external momenta on shell. Now it
follows that if the cut is well-defined (and it should be, given the
correspondence between MHV completion and LCYM graphs is algebraic),
it should not matter in which order we take these limits.

Inspecting \eqref{eq:boxbox} and
\eqref{eq:22bubblebox}--\eqref{eq:tadpolebox}, it is clear that the
order in which these limits are taken will `switch off' different
contributions. For instance, it is clear to see that upon cutting all
the internal lines by letting $q_i^2$ and hence $Q_i \rightarrow 0$,
only the box
\eqref{eq:boxbox} will survive. Conversely, we see that when we take
the external momenta on shell first, $P_i \rightarrow 0$ so only the tadpole contribution
\begin{equation}
  \label{eq:tadpoleonshell}
  C_\text{tadpole}' =
\frac{Q_1 Q_2 Q_3}{\Sigma_1} \frac1{(\Sigma_1 + \Sigma_2) (\Sigma_1 +
  \Sigma_2 + \Sigma_3)}
=
\frac{Q_1Q_2Q_3}{(Q_1-Q_4)(Q_2-Q_4)(Q_3-Q_4)}.
\end{equation}
and its three cyclic permutations survive. When one sums over these
permutations, one finds that $C=1$ so the correct quadruple cut is
again obtained. Choosing other, more mixed limits would produce more
complicated combinations of terms surviving from the various $C$
coefficients.

As an interesting aside, the reason that the tadpole is the sole
survivor in this limit may be
understood from the structure of the $\Upsilon$ vertices: the tadpole
provides the only configuration in which the correct poles are
generated to cancel the inverse propagators from the LSZ reduction.
Consider the tadpole graphs shown in
fig.~\ref{fig:mhv-tadpoles}: their contribution is
\begin{multline}
\label{eq:etv-tadpole-Xis}
\frac14 g^4 \int \frac{d^Dq_4}{(2\pi)^D}
\: \frac{\hat q_4}{q_4^2} \Biggl\{ -P_1 \Xi^4(1,2,3,4,q_4^A,-q_4^A)
-P_2 \Xi^3(2,3,4,q_4^A,-q_4^A,1) \\
-P_3 \Xi^2(3,4,q_4^A,-q_4^A,1,2)
- P_4\Xi^1(4,q_4^A,-q_4^A,1,2,3) \Biggr\}.
\end{multline}
Inspecting \eqref{eq:xi-rr}, we see that they all contain  the term
$-\Upsilon(q_4^A, -q_4^A,1,2,3,4)$. Since we are concerned only with
the box topology, we can ignore the other terms in \eqref{eq:xi-rr}
since they correspond to the topologies of tree decorations to
triangles, bubbles and tadpoles. A similar argument simplifies the
extraction of the box topology terms from $\Upsilon$: to extract the
terms in $\Upsilon(q_n^A, -q^A,1,\dots,n)$, where $q_n = q$, with an
$n$-gonal topology, we keep only the terms in \eqref{eq:upsilon-rr-x}
with $q$ in each term, so in this situation we can make the replacement
\[
\Upsilon(q^A_n,-q^A,1,\dots,n) \rightarrow - \frac{\bar
  V^2(-q^B_{n-1},n,q^A_n) \Upsilon(q^B_{n-1},-q^A,1,\dots,n-1)} {\hat
  q_n(\Omega_{q_n}-\Omega_q+ \sum_{i=1}^n\Omega_i)}.
\]
Iterating this for $n=4$ gives
\[ \begin{split}
\Upsilon(q^A_4, -q^A,1,2,3,4) &= -\frac{\bar V^2(-q_3^B,4,q^A)
  \Upsilon(q_3^B, -q^A,1,2,3)}{\hat q_4 \sum_{i=1}^4 \Omega_i} \\
&= \cdots \\
&=\frac{16 X}{\hat q_1 \hat q_2 \hat q_3 \hat q_4} 
\frac1{\sum_{i=1}^4 P_i}  \frac1{\sum_{i=1}^4 P_i +Q_3 - Q_4} \\
&\quad \hphantom{\frac{16 X}{\hat q_1 \hat q_2 \hat q_3 \hat q_4}} \times
 \frac1{\sum_{i=1}^4 P_i +Q_2 - Q_4}  \frac1{\sum_{i=1}^4 P_i +Q_1 - Q_4}.
\end{split} \]
using \eqref{eq:sum-omega}.

Making these replacements in \eqref{eq:etv-tadpole-Xis}, reinstating
the propagators by multiplying the integrand by $q_i^2/q_i^2$
($i=1,2,3$) and taking the limit $P_i \rightarrow 0$, it is easy to
see that \eqref{eq:tadpoleonshell} and hence \eqref{eq:tadpolebox} are recovered directly.
\section{Conclusion}
\label{sec:etv-conclusion}

In this chapter, we have seen that the amplitudes that cannot be built
from the CSW rules arise in the Canonical MHV Lagrangian framework as
a result of terms in the field transformation that evade the
$S$-matrix equivalence theorem in the LSZ reduction. 
In detail, we found that the series coefficients of the field
transformation, $\Upsilon$ and $\Xi$, furnish a set of `completion
vertices' that supplant insertions of $A$ and $\bar A$, respectively,
in correlation functions with products of $B$ and $\bar B$ fields.

To demonstrate this, we recovered the three-gluon \mhvbar\ amplitude
(which is non-vanishing in $(2,2)$ signature or for complex momenta),
computed in light-cone Yang--Mills theory using the $\bar V^2$ vertex
eliminated from the lagrangian by the transformation. This is
constructed by the LSZ reduction of the correlation function $\langle
A\bar A\bar A \rangle$. Now not only does this have the correct pole
structure to survive the on-shell limit, it recovers the off-shell
\mhvbar\ vertex $\bar V^2$ algebraically.

For the treatment of amplitudes at the quantum
level, we applied dimensional regularisation to the light-cone
Yang--Mills lagrangian and used the field transformation to obtain
$D$-dimensional versions of the MHV and completion vertices. We
augmented the light-cone co-ordinates in such a way that the ideas of
positive and negative `helicity' are preserved (exact in the
four-dimensional limit). It also allows a clean separation of the two
`helicities' into canonical co-ordinates and momenta, as before.
The result is clearly an MHV lagrangian, insomuch as its vertices
contain just two fields of negative `helicity' and one or more
positive, joined together with a helicity-flipping propagator.
The field transformation that results has a similar structure to the
four-dimensional case, but unlike it lacks the simple, holomorphic
expressions for the vertices and transformation coefficients.

Using this technology, we to constructed the one-loop
$(\fourplus)$ amplitude, whose corresponding correlation function
consists only of completion vertices.
We first studied its generalised unitarity cuts, finding that when the
external momenta are kept off-shell, only the box graphs of
fig.~\ref{fig:etv-mhv-box} contribute to the cut. 
Furthermore, only the quadruple cut of this MHV completion graph
is non-vanishing, 
and we computed it directly and demonstrated that it reproduces
precisely the quadruple cut of the $D$-dimensional massless box
function $K_4$. 

It was then seen that, like at the tree-level, the sum of the MHV completion
diagrams reduces to the LCYM expression for the amplitude, even before
integrating the loop momentum or taking any on-shell limits.
That the loop amplitude is recovered in such a straightforward way is
evidence contrary to the twistor-space inspired
suggestion in \cite{Cachazo:2004zb} that
the $(\fourplus)$ and $(-+++)$ amplitudes come from new local vertices. 

The precise circumstances under which the completion vertices must be
used is a subject for further investigation. We have seen that they
are not required for on-shell tree amplitudes. We presume that 
they may also not be required for amplitudes where only certain legs
are off shell. Nevertheless, they are required for the complete
construction of off-shell tree amplitudes, and on shell at the loop
level due to the existence of regions of propagator phase space that
give rise to poles that cancel inverse propagators from the LSZ
reduction.

In keeping with the field theory spirit which motivated this work, one
might also consider the possibility of directly evaluating individual
MHV completion graphs. However, defining the integrals poses a
technical challenge when it comes to dealing with the unusual,
gauge-dependent singularity structures hidden in the completion
vertices. Conversely, given that we have seen that off-shell LCYM is
recovered algebraically before any integration takes place, one might
question the wisdom of this; again, without knowing the conditions
under which completion vertices must be used, it is hard to see
whether this would provide any computational advantage over LCYM.

\begin{subappendices}
\section{Light-cone vector identities}
\label{sec:lightcone-vectors}
The appendix gives some of the identities particular to vectors in
$D$-dimensional light-cone co-ordinates. Some of these appear in a different
form in \cite{Chakrabarti:2005ny}. First, for any two $D$-vectors $p$ and $q$,
\begin{equation}
\label{eq:dotbilinears}
(p\:q)\mdot\{p\:q\} = -\tfrac12 (\hat p\,q - \hat q\,p)^2
\end{equation}
from which it is clear that for null $p$, $q$,
\begin{equation}
\label{eq:null-p-dot-q}
(p\:q)\mdot\{p\:q\} = \hat p\,\hat q\: p\cdot q.
\end{equation}
The Bianchi-like identity
\begin{equation}
\label{eq:bianchi}
\hat i (j\:k) + \hat j (k\:i) + \hat k (i\:j) = 0
\end{equation}
holds also under replacement of the hat with any transverse component $i_I$ or
$\bar i_I$, and replacement of the bilinear with its adjoint.

For a set of momenta $\{p_j\}$ that sum to zero,
\begin{equation}
\label{eq:sum-bilinears}
\sum_j\frac{(p\:j)\mdot\{j\:q\}}{\hat j} =
    \frac{\hat p\,\hat q}{2} \sum_j \frac{p^2_j}{\hat j}
\end{equation}
for any $p$ and $q$. This is the $D$-dimensional, off-shell generalisation of
the spinor identity $\sum_j \langle p\;j \rangle [j\:q] = \langle p \rvert
(\sum_j \lvert j \rangle [ j \rvert) \lvert q ] = 0$. In four 
dimensions the above identity looks the same except that the 
dot product is simply multiplication. Also,
\begin{equation}
\label{eq:sum-omega}
\sum_j \Omega_j = -\frac 12 \sum_j \frac{p^2_j}{\hat j}.
\end{equation}
(In four dimensions the left hand side has $\omega_j$ in place of
$\Omega_j$.)

\end{subappendices}

\chapter{The Canonical MHV Lagrangian for Massless QCD}
\label{cha:mhvqcd}

In this chapter, we will extend the work of chapter \ref{cha:mhvym}
and develop the ideas sketched by Mansfield in section 3 of
\cite{Mansfield:2005yd} to construct an MHV lagrangian for massless
QCD. The structure of this chapter is as follows.
In section \ref{sec:deriv-lcqcd}, we start with the manifestly
Lorentz-covariant action for massless QCD, and then fix it to the
light-cone gauge, and integrate out the non-dynamical degrees of
freedom.
Then, in section \ref{sec:transform-qcd}, we specify the field
transformation that eliminates the \mhvbar-like vertices from the
action. We establish what form this transformation will take, and
argue that the lagrangian that results will have an infinite series of
terms with MHV helicity content, and that these continue off-shell by
the CSW prescription as in \cite{Wu:2004jxa}.
We solve explicitly for this transformation as a perturbative series.
Next, in section \ref{sec:examples-qcd}, we demonstrate explicitly
that this lagrangian does indeed contain vertices corresponding to the
known expressions for MHV amplitudes containing quarks for the cases
of: two quarks and two gluons, four quarks, and two quarks and three
gluons in the $(1_{\rm q}^+ 2^+ 3^- 4^+ 5_{\bar{\rm q}}^-)$
configuration.
Finally, we draw conclusions on this chapter in section
\ref{sec:conclusion-qcd}.

This work was published in \cite{Ettle:2008ey}.


\section{The light-cone action for massless QCD}
\label{sec:deriv-lcqcd}

Let us begin with the action for a massless QCD theory with ${\rm
  SU}(N_{\rm C})$ gauge symmetry. Its action is
\begin{equation}
  \label{eq:qcd-action}
  S_\text{QCD} = \frac 1{2g^2} \int d^4x \: 2ig^2 \: \bar\psi
  \: \slashed \gD \: \psi
  + \frac 1{2g^2} \int d^4x \: \tr \gF^{\mu\nu} \gF_{\mu\nu}.
\end{equation}
Here, we will use the chiral Weyl representation of the Dirac matrices
\begin{equation*}
  \gamma^\mu = \begin{pmatrix} 0 & \sigma^\mu \\
    \bar\sigma^\mu & 0 \end{pmatrix},
\end{equation*}
and the spinors
\begin{equation*}
  \psi = (\alpha^+, \beta^+, \beta^-, \alpha^-)^{\rm T}
  \quad\text{and}\quad
  \bar\psi = (\bar\beta^+, \bar\alpha^+, \bar\alpha^-, \bar\beta^-)
\end{equation*}
are the quark field and its conjugate.  They have the canonical
normalisation and are in the fundamental representation of ${\rm
  SU}(N_{\rm C})$. Note that the superscripts $\pm$ in the components
denote the physical helicity for \emph{outgoing} particles, as we
shall later see; that $\bar\alpha^+ = (\alpha^-)^*$ should be
understood, and similarly for $\beta$. $\gA$, $\gD$, $\gF_{\mu\nu}$
and the gauge group generator matrices are defined as in the pure
Yang-Mills case of chapter \ref{cha:mhvym}, given in
\eqref{eq:gauge-conv}.

As before, we quantise the theory on surfaces $\Sigma$ of constant
$x^0$, \ie\ those with normal $\mu = (1,0,0,1)/\sqrt 2$ in Minkowski
co-ordinates, and fix to the same axial gauge $\mu \cdot \gA = \hat\gA
= 0$, for which the Faddeev-Popov ghosts are completely decoupled.  We
can set this gauge condition immediately (and discard the infinite,
field-independent factor the Faddeev-Popov procedure produces).  With
this condition in force, the Dirac term in \eqref{eq:qcd-action} may
be expressed as
\begin{equation*}
  \label{eq:lcqcd-quarks-gf1}
  \begin{split}
    \frac 1{2g^2} \int & 2g^2 \: i \bigl\{ \bar\varphi \:
    (\partial\cdot\sigma -\gA\bar\sigma -\bar\gA\sigma) \: \varphi +
    \omega \:(\partial\cdot\bar\sigma + \gA \sigma +
    \bar\gA\bar\sigma) \bar\omega \bigr\} \\
    &+ \tr\ \bigl\{ 2g^2\: i (\bar\varphi\:T^a\hat\sigma\:\varphi +
    \omega \:T^a\check\sigma \bar\omega)T^a \:\check\gA \bigr\},
  \end{split}
\end{equation*}
where for compactness we have split $\psi$, $\bar\psi$ into Weyl spinors:
\begin{align*}
  \bar\omega^{\dot\alpha} &= \begin{pmatrix} \alpha^+ \\
    \beta^+ \end{pmatrix},
\\
\varphi_\alpha &= \begin{pmatrix} \beta^-
    \\ \alpha^- \end{pmatrix},
\\
  \omega^\alpha &= (\bar\alpha^-, \bar\beta^-),
\\
  \bar\varphi_{\dot\alpha} &= (\bar\beta^+, \bar\alpha^+).
\end{align*}
(Note that the meaning of $\sigma$ and $\bar\sigma$ as either a
4-vector, or light-cone co-ordinate component thereof, should be clear
from the context.)  That done, the lagrangian is quadratic in
$\check\gA$ and we can integrate it out. In section
\ref{sec:mhvym-lightcone}, we did this by finding the coefficient of
$\check\gA$ under the trace in \eqref{eq:ym-action-1}, calling it $K_{\rm YM}$, and
replaced all terms in the action containing $\check\gA$ with
\begin{equation*}
  -\frac 1{2g^2} \int \tr 
  \tfrac 18 K_{\rm YM} \hat\partial^{-2} K_{\rm YM}.
\end{equation*}
This time, we must put $K_{\rm YM} \rightarrow K_{\rm YM} + K_\psi$,
where
\begin{equation*}
  K_\psi = 2g^2 \: i
  (\bar\varphi\:T^a\hat\sigma\:\varphi + \omega \:T^a\check\sigma
  \bar\omega)T^a
\end{equation*}
is the coefficient of the Dirac term linear in $\check\gA$ under the
$\frac12 g^{-2} \int \tr$. Hence, all terms in \eqref{eq:qcd-action}
containing $\check\gA$ should be replaced with
\begin{equation}
  \label{eq:Acheck-replacement}
  \frac 1{2g^2}\int \tr\ \bigl\{  -\tfrac 18 K_\psi
  \hat\partial^{-2} K_\psi
  -\tfrac 14 K_\psi \hat\partial^{-2} K_{\text{YM}}
  -\tfrac 18 K_{\text{YM}} \hat\partial^{-2} K_{\text{YM}} \bigr\}.
\end{equation}
The first term here is a four-fermion effective vertex arising from
the integrated-out gauge degree of freedom; similarly, the second term
accounts for the interaction with unphysical gluon states.
The last term produces contributions that result in the same
gluonic terms as before (\cf\
\eqref{eq:lcym-mpp}--\eqref{eq:lcym-mmpp}).  If we evaluate
\eqref{eq:Acheck-replacement}, integrating by parts as necessary, and
use it to substitute for the $\check\gA$ terms in
\eqref{eq:qcd-action}, we arrive at the action
\begin{equation}
  \label{eq:lcqcd-pre-action}
  \begin{split}
    S_{\text{LCQCD}} = \int &\frac 4{g^2}({\cal
      L}^{-+}+{\cal L}^{\mpp}+{\cal L}^{--+}+{\cal L}^{--++}) \\
    & + {\cal L}_{\bar\psi\psi} + {\cal L}_{\bar\psi\gA\psi} + {\cal
      L}_{\bar\psi\bar\gA\psi} + {\cal L}_{\bar\psi\gA\bar\gA\psi} +
    {\cal L}_{\bar\psi\psi\bar\psi\psi}
  \end{split}
\end{equation}
where
\begin{align}
  \label{eq:quark-kinetic}
  {\cal L}_{\bar\psi\psi} &= i \left\{ \bar\varphi
    \: \partial\cdot\sigma \: \varphi + \omega
    \: \partial\cdot\bar\sigma
    \: \bar\omega \right\}, \\
  {\cal L}_{\bar\psi\gA\psi} &= i \left\{ \bar\varphi \left[ (
      \hat\sigma \bar\partial\hat\partial^{-1} - \bar \sigma ) \gA
    \right] \varphi + \omega \left[ ( \check\sigma
      \bar\partial\hat\partial^{-1} + \bar \sigma ) \gA \right]
    \bar\omega \right\}, \nonumber \\
  {\cal L}_{\bar\psi\bar\gA\psi} &= i \left\{ \bar\varphi \left[ (
      \hat\sigma
      \partial\hat\partial^{-1} - \sigma ) \bar\gA \right] \varphi +
    \omega \left[ ( \check\sigma \bar\partial\hat\partial^{-1} +
      \sigma ) \bar\gA \right] \bar\omega \right\}, \nonumber \\
  {\cal L}_{\bar\psi\gA\bar\gA\psi} &= i \left\{ \bar\varphi \:
    \hat\partial^{-2}( [\hat\partial
    \gA,\bar\gA]+[\hat\partial\bar\gA,\gA] )\hat\sigma\: \varphi +
    \omega\: \hat\partial^{-2}( [\hat\partial
    \gA,\bar\gA]+[\hat\partial\bar\gA,\gA] )\check\sigma\: \bar\omega
  \right\}, \nonumber \\
  {\cal L}_{\bar\psi\psi\bar\psi\psi} &= \tfrac 12 g^2 j^a
  \hat\partial^{-2} j^a, \quad j^a = \bar\varphi \: T^a \hat\sigma \:
  \varphi + \omega \: T^a \check\sigma \: \bar\omega. \nonumber
\end{align}

Let us study the quark kinetic terms \eqref{eq:quark-kinetic}
above. Written out, they are
\begin{multline*}
  {\cal L}_{\bar\psi\psi} = i \sqrt 2
    \bigl\{ \bar\beta^+ \hat\partial \beta^- + \bar\beta^+ \partial
    \alpha^- + \bar\alpha^+ \bar\partial \beta^- + \bar\alpha^+
    \check\partial
    \alpha^- \\
    + \bar\alpha^- \check\partial \alpha^+ - \bar\alpha^- \partial
    \beta^+ - \bar\beta^- \bar\partial \alpha^+ + \bar\beta^-
    \hat\partial \beta^+ \bigr\}
\end{multline*}
from which it is clear that $\beta^\pm$ and $\bar\beta^\pm$ are
non-dynamical with respect to $\check\partial$.  Since the terms in
\eqref{eq:lcqcd-pre-action} that couple only non-dynamical fields
contain no other fields, evaluating their path integral amounts to
computing the determinant of a non-field-dependent object
(specifically $\hat\partial$).  Therefore this integration can be
carried out just as well by replacing the non-dynamical fields
according to their classical equations of motion. These are
\begin{align*}
  \hat\partial \beta^- &= - \gD \alpha^-, \\
  \hat\partial \bar\beta^+ &= \bar\alpha^+ \bar\gA - \bar\partial
  \bar\alpha^+, \\
  \hat\partial \beta^+ &= \bar\gD \alpha^+, \\
  \hat\partial \bar\beta^- &= \partial \bar\alpha^- - \bar\alpha^-
  \gA.
\end{align*}
Substituting back into \eqref{eq:lcqcd-pre-action}, we finally arrive
at the following gauge-fixed action that features only dynamical
components:
\begin{equation}
  \label{eq:lcqcd-action}
  \begin{split}
    S_\text{LCQCD} = \frac 4{g^2}\int dx^0 ( &L^{-+} + L^\mpp + L^\mmp
    + L^{--++} + \\
    &L^{\bar\psi\psi} + L^{\bar\psi+\psi} + L^{\bar\psi-\psi} +
    L^{\bar\psi+-\psi} + L^{\bar\psi\psi\bar\psi\psi}),
  \end{split}
\end{equation}
where $L^{-+}$, $L^\mpp$, $L^\mmp$ and $L^{--++}$ form the same
Yang-Mills sector of the theory from chapter \ref{cha:mhvym}, set out
in eqs.\ \eqref{eq:lcym-mpp}--\eqref{eq:lcym-mmpp}, and the new
terms involving the fermions are \newcommand\fintl{\frac{i g^2}{\sqrt
    8}\int_\Sigma d^3{\bf x}\: \Bigl\{}
\begin{align}
  \label{eq:lcqcd-qbarq}
  L^{\bar\psi \psi} &= \phantom{-} \fintl \bar\alpha^+ (\check\partial
  - \omega) \alpha^- + \bar\alpha^-
  (\check\partial - \omega) \alpha^+ \Bigr\}, \\
  \label{eq:lcqcd-qbarpq}
  \begin{split}
    L^{\bar\psi+\psi} &= - \fintl \bar\alpha^+ \:
    \bar\partial\hat\partial^{-1} (\gA \, \alpha^-) + \bar\alpha^-\gA
    \:
    \bar\partial\hat\partial^{-1} \alpha^+ \\
    &\hphantom{=-\fintl} - \bar\alpha^+(\bar\partial \hat\partial^{-1}
    \gA) \alpha^- - \bar\alpha^- (\bar\partial \hat\partial^{-1} \gA)
    \alpha^+ \Bigr\},
  \end{split} \\
  \label{eq:lcqcd-qbarmq}
  \begin{split}
    L^{\bar\psi-\psi} &= - \fintl \bar\alpha^+
    \bar\gA\:\partial\hat\partial^{-1}\alpha^- +
    \bar\alpha^-\partial\hat\partial^{-1}(\bar\gA \, \alpha^+) \\
    &\hphantom{=-\fintl} - \bar\alpha^+(\partial \hat\partial^{-1}
    \bar\gA) \alpha^- - \bar\alpha^- (\partial \hat\partial^{-1}
    \bar\gA) \alpha^+ \Bigr\},
  \end{split} \\
  \label{eq:lcqcd-qbarpmq}
  \begin{split}
    L^{\bar\psi+-\psi} &= - \fintl \bar\alpha^+ \bar\gA
    \hat\partial^{-1} (\gA \alpha^-)
    + \bar\alpha^- \gA \hat\partial^{-1} (\bar\gA \alpha^+) \\
    &\hphantom{=-\fintl} + \bar\alpha^+ \hat\partial^{-2}
    (\hat\partial \gA \bar\gA - \gA \hat\partial \bar\gA) \alpha^- +
    \bar\alpha^+ \hat\partial^{-2} (\hat\partial \bar\gA \gA
    - \bar\gA \hat\partial \gA) \alpha^- \\
    &\hphantom{=-\fintl} + \bar\alpha^- \hat\partial^{-2}
    (\hat\partial \gA \bar\gA - \gA \hat\partial \bar\gA) \alpha^+ +
    \bar\alpha^- \hat\partial^{-2} (\hat\partial \bar\gA \gA - \bar\gA
    \hat\partial
    \gA) \alpha^+ \Bigr\}, \end{split} \\
  \label{eq:lcqcd-qbarqqbarq}
  L^{\bar\psi\psi\bar\psi\psi} &= \frac{g^4}{16} \int_\Sigma d^3{\bf
    x}\: j^a \hat\partial^{-2} j^a, \quad j^a = \sqrt 2 (\bar\alpha^+
  T^a \alpha^- + \bar\alpha^- T^a \alpha^+).
\end{align}


\section{The MHV QCD field transformation}
\label{sec:transform-qcd}

Let us now construct the field transformation that results in a MHV
lagrangian for massless QCD. We label the new algebra-valued gauge
fields $\gB$ and $\bar\gB$ as before, and the new fundamental
representation fermions $\xi^+$, $\bar\xi^-$, $\xi^-$ and $\bar\xi^+$;
their Lorentz transformation properties are the same as those of the
old fields with similar embellishments.  We remove terms in the light-cone
lagrangian with a $(\mpp)$ helicity structure by absorbing them into
the kinetic terms of the new fields as follows:
\begin{equation}
  \label{eq:transf-qcd}
  L^{-+}[\gA, \bar\gA] + L^{-++}[\gA, \bar\gA] +
  L^{\bar\psi\psi}[\alpha^\pm,\bar\alpha^\pm] +
  L^{\bar\psi+\psi}[\gA,\alpha^\pm,\bar\alpha^\pm] =
  L^{-+}[\gB, \bar\gB]
  + L^{\bar\psi\psi}[\xi^\pm,\bar\xi^\pm].
\end{equation}
We remark that this appears sensible, since the theory formed by the
truncation on the LHS of \eqref{eq:transf-qcd} is classically free.
The remaining terms in the lagrangian, \eqref{eq:lcym-mmp},
\eqref{eq:lcym-mmpp} and
\eqref{eq:lcqcd-qbarmq}--\eqref{eq:lcqcd-qbarqqbarq} form the MHV
vertices, as we will show in the next section.

\subsection{Form of the transformation and MHV lagrangian}
\label{ssec:form-transform-qcd}

First, let us establish the general form of the field transformation
and the resulting lagrangian.  We note that the canonical
$(\text{co-ordinate},\text{momentum})$ pairs of the system
\eqref{eq:lcqcd-action} are
\begin{equation*}
  \label{eq:old-coordmom-qcd}
  (\gA, -\hat\partial\bar\gA),\quad
  \left(\alpha^+, -\frac{ig^2}{\sqrt 8} \bar\alpha^- \right)
  \quad\text{and}\quad
  \left(\alpha^-, -\frac{ig^2}{\sqrt 8} \bar\alpha^+ \right),
\end{equation*}
and likewise for the new fields (by replacing $\gA \rightarrow \gB$
and $\alpha \rightarrow \xi$ above).  We have defined the momenta with
respect to the lagrangian \emph{under the integral} in
\eqref{eq:lcqcd-action}. The path integral measure
\begin{equation*}
  {\cal D}\gA {\cal D}\bar\gA \:
  {\cal D}\alpha^+ {\cal D}\bar\alpha^- \:
  {\cal D}\alpha^- {\cal D}\bar\alpha^+,
\end{equation*}
is therefore the phase-space measure (up to an irrelevant constant),
and it will be preserved if the transformation is canonical. This, and
our demands on the helicity content of the resulting lagrangian,
restrict the form of the transformation as follows.

Again, we choose a canonical transformation of the form generated by
\eqref{eq:canon-ourform}, $\gA$ will be a functional of $\gB$
alone. Since the transformation takes place entirely on $\Sigma$ (\ie\
no explicit $x^0$ dependence), and preserves the form of the kinetic
part of the lagrangian, the new $(\text{co-ordinate},\text{momentum})$
pairs are
\begin{equation*}
  \label{eq:new-coordmom-qcd}
  (\gB, -\hat\partial\bar\gB),\quad
  \left(\xi^+, -\frac{ig^2}{\sqrt 8} \bar\xi^- \right)
  \quad\text{and}\quad
  \left(\xi^-, -\frac{ig^2}{\sqrt 8} \bar\xi^+ \right).
\end{equation*}
We remind the reader that they satisfy \eqref{eq:canon-mom}, which in
this case implies that
\begin{equation}
  \label{eq:Abar-canon-qcd}
  \hat\partial \bar\gA^a(\vec x) = \int_\Sigma d^3\vec y \left\{
    \frac{\delta\gB^b(\vec y)}{\delta\gA^a(\vec x)}
    \hat\partial \bar\gB^b(\vec y)
    - \frac{ig^2}{\sqrt 8}\left(
      \bar\xi^-(\vec y)
      \frac{\delta \xi^+(\vec y)}{\delta\gA^a(\vec x)}
      +  \bar\xi^+(\vec y) \frac{\delta \xi^-(\vec y)}{\delta\gA^a(\vec x)}
    \right)
  \right\}.
\end{equation}
(Note that we take all derivatives with respect to Grassman variables
as acting from the left. Also the order of the fermion co-ordinate and
momentum factors above is the opposite of that of
\eqref{eq:canon-mom}, and as such these terms pick up an extra factor
of $-1$.)

By charge conservation, and the requirement that this will be a
canonical transformation that results in a lagrangian whose vertices
have MHV helicity content, tells us that the fermion co-ordinate
transformation takes the form
\begin{equation}
  \label{eq:fqnew-in-old}
  \xi^\pm(\vec x) = \int_\Sigma d^3\vec y \: R^\mp[\gA](\vec x, \vec y)
  \: \alpha^\pm(\vec y).
\end{equation}
The superscript of $R^\pm$ refers to the \emph{chirality} of the Weyl
spinor from which the field components originate: $+$ for
right-handed, $-$ for left-handed.  $R^\pm$ is a matrix-valued
functional of $\gA$. Putting additional factors of $\bar\gA$ into the
RHS of \eqref{eq:fqnew-in-old} would result in terms in the resulting
lagrangian with more than two fields of negative helicity; likewise
with extra quark fields, since charge conservation requires these to
be added in $(+-)$ helicity pairs.  Conversely, the behaviour of the
canonical momenta must be
\begin{equation}
  \label{eq:fpold-in-new}
  \bar\alpha^\pm(\vec x) = \int_\Sigma d^3\vec y \:
  \bar\xi^\pm(\vec y) \: R^\pm[\gA](\vec y, \vec x),
\end{equation}
It will also be useful to define the inverse transformations
\begin{equation}
  \label{eq:fqold-in-new}
  \alpha^\pm(\vec x) = \int_\Sigma d^3\vec y \: S^\mp[\gA](\vec x, \vec y)
  \: \xi^\pm(\vec y)
\end{equation}
as well.

At this point we can immediately read off the propagators for the new
fields from \eqref{eq:transf-qcd} as
\begin{equation}
  \label{eq:propagators-qcd}
  \langle \gB \bar\gB \rangle = - \frac{ig^2}{2p^2}
  \quad\text{and}\quad
  \langle \xi^- \bar\xi^+ \rangle = \langle \xi^+ \bar\xi^- \rangle = i \sqrt 2 \frac{\hat p}{p^2}.
\end{equation}
By using \eqref{eq:gauge-conv}, one obtains the canonically normalised
propagator $\langle B \bar B \rangle = i/p^2$, and indeed in practical
calculations with the MHV lagrangian it is often more convenient to
absorb powers of this factor into the lagrangian's vertices and
transformation series coefficients, as was done in chapter \ref{cha:etv}. For
the purposes of this chapter, however, we will account for these factors
at the end of the calculations we present in the next section.

If we now assume solutions for $R$ and $S$ as infinite series in
$\gA$, it is not hard to see that, upon substitution into the
remaining terms of the light-cone QCD lagrangian, this choice of
transformation gives a set of terms with no more than two fields of
negative helicity ($\bar\gB$, $\xi^-$ and $\bar\xi^-$) in each, but an
increasing number of $\gB$ fields as shown in table
\ref{tbl:mhvqcd-content}. (The number of positive-helicity quark
fields present is, of course, strictly constrained by charge
conservation.)  The sum of these terms has precisely the helicity and
colour structure required to be identified as the interaction part of
a MHV lagrangian, and we claim here that the Feynman rules of its
tree-level perturbation theory follow the CSW rules. We will address
the proof of this in section \ref{ssec:mhvqcd-transf-mhvproof}.

\begin{table}[h]
  \centering\begin{tabular}{ll}
    \toprule
    \textbf{LCQCD term} & \textbf{New field content} \\
    \midrule
    $L^\mmp$ & $\bar\gB\bar\gB\gB\cdots$,
    \quad $\bar\xi\xi\bar\gB\gB\cdots$,
    \quad $\bar\xi\xi\bar\xi\xi\gB\cdots$ \\
    $L^{--++}$ & $\bar\gB\bar\gB\gB\gB\cdots$,
    \quad $\bar\xi\xi\bar\gB\gB\cdots$,\quad
    $\bar\xi\xi\bar\xi\xi\gB\gB\cdots$ \\
    \midrule
    $L^{\bar\psi-\psi}$ & $\bar\xi\xi\bar\gB$,\quad
    $\bar\xi\xi\bar\gB\gB\cdots$,\quad $\bar\xi\xi\bar\xi\xi\gB\cdots$ \\
    $L^{\bar\psi+-\psi}$ & $\bar\xi\xi\bar\gB\gB\cdots$,\quad
    $\bar\xi\xi\bar\xi\xi\gB\cdots$ \\
    $L^{\bar\psi\psi\bar\psi\psi}$ & $\bar\xi\xi\bar\xi\xi$,\quad
    $\bar\xi\xi\bar\xi\xi\gB\cdots$ \\ 
    \bottomrule
  \end{tabular}
  \caption{The contents of the new vertices provided by our choice of
    field transformation. The new fermion fields, $\xi$, always occur
    in bilinear pairs and as such $\bar\xi\xi$ is the sum of a term
    containing exactly one $-$ helicity quark, and another term with
    one $-$ helicity antiquark. An ellipsis $\cdots$ denotes an
    infinite series wherein the field to its immediate left is
    repeated.}
  \label{tbl:mhvqcd-content}
\end{table}

We can deduce from the forms of the light-cone Yang--Mills lagrangian
\eqref{eq:lcqcd-qbarq}--\eqref{eq:lcqcd-qbarqqbarq} and the
transformation that terms containing a single quark-antiquark pair are
of the form
\begin{equation}
  \label{eq:mhvqcd-2q-form}
  \int_{1\cdots n}
  \sum_{n=3}^\infty \sum_{j=2}^{n-1} \sum_\pm
  \bar\xi^\pm_{\bar 1} \gB_{\bar 2} \cdots \bar\gB_{\bar\jmath} \cdots
  \gB_{\overline{n-1}} \xi^\mp_{\bar n} \: V^{j;\pm}_{2{\rm q}\:1\cdots n}.
\end{equation}
Clearly this is coincident with the colour structure of
\eqref{eq:colourstruct-2q}. Similarly, for two quark-antiquark pairs,
we expect a contribution to the lagrangian of the form
\begin{multline}
  \label{eq:mhvqcd-4q-form}
  \int_{1\cdots n} \sum_{n=4}^\infty \sum_{j=3}^{n-1}
  \sum_{(h_1,h_j)\in {\cal H}} \biggl\{ \bar\xi^{h_1}_{\bar1}
  \gB_{\bar 2} \cdots \gB_{\overline{j-2}} \xi^{-h_j}_{\overline{j-1}}
  \bar\xi^{h_j}_{\bar\jmath} \gB_{\overline{j+1}} \cdots
  \gB_{\overline{n-1}} \xi^{-h_1}_{\bar n} \:
  V^{j;h_1h_j}_{4{\rm q}\:1\cdots n}
  \\
  + \frac1{\NC} \bar\xi^{h_1}_{\bar1}
  \gB_{\bar 2} \cdots \gB_{\overline{j-2}} \xi^{-h_1}_{\bar n}
  \bar\xi^{h_j}_{\bar\jmath} \gB_{\overline{j+1}} \cdots 
  \gB_{\overline{n-1}} \xi^{-h_j}_{\overline{j-1}} \:
  V^{j;h_1h_j}_{4{\rm q}(1)\:1\cdots n}
  \biggr\} {\cal S}(h_1,h_j).
\end{multline}
Here, ${\cal H}=\{(+,+),(+,-),(-,-)\}$, and
\[
{\cal S}(h_1,h_j) = \begin{cases}
1/2 & (h_1 = h_j)\\
1 & (\text{otherwise})
\end{cases}
\]
is a symmetry factor whose function is similar to the factor of $1/2$
in \eqref{eq:mhvym-Vseries} (\ie\ it absorbs potential over-counting
issues due to two possible external contractions leading to the same
colour structure). We interpret any $\gB$-string to be the identity above
when it exceeds the bounds at either end of the ellipsis.
This clearly corresponds to
the structure of \eqref{eq:colourstruct-4q}.\footnote{Since
 much of the literature works with ${\rm U}(\NC)$ gauge symmetry,
many of the two-quark-pair amplitudes given there only correspond to
the ${\cal O}(\NC{}^0)$ piece, \ie\ partial amplitudes distilled from
$V^{j;h_1h_j}_{4{\rm q}\:1\cdots n}$.} Both terms have the same
relative sign due to the swapping of the fermion fields induced by the
${\rm SU}(\NC)$ Fierz rearrangement.

We will refer to these forms later in section
\ref{sec:examples-qcd} when we construct explicit examples of
$V_{2{\rm q}}$ and $V_{4{\rm q}}$ for a few low-order cases. Note that
we will also use the same notation as \eqref{eq:V-delta} to factor out $\delta$
functions.

\subsection{Solution to the transformation}
\label{ssec:series-solns}

Extracting explicit solutions proceeds much like in the pure gluon
case. We will obtain the old field variables as perturbative series
expansions in the gauge field $\gB$.

\subsubsection*{$\gA$ and quark transformation $R^\pm$}
\label{ssec:R-series}

Let us begin with \eqref{eq:transf-qcd}. We write it out explicitly,
making use of \eqref{eq:Abar-canon-qcd}, \eqref{eq:fqnew-in-old} and
\eqref{eq:fpold-in-new} to substitute for $\hat\partial\bar\gA$,
$\bar\alpha^\pm$ and $\xi^\pm$ respectively.  For clarity, we show
below only the left-handed chiral fermion components:
\begin{multline}
  \label{eq:transf-qcd-1}
  \int_{\vec x \vec y} \bigl\{ \Oomega \gA + [\gA, \Ozeta \gA]
  \bigr\}^a(\vec x) \frac{\delta \gB^b(\vec y)}{\delta \gA^a(\vec x)}
  \hat\partial \bar\gB^b(\vec y) - \frac{ig^2}{\sqrt 8} \int_{\vec x
    \vec y \vec z} \bigl\{ \Oomega \gA + [\gA, \Ozeta \gA ]
  \bigr\}^a(\vec z) \times \\ \Biggl\{ \bar\xi^-(\vec x) \frac{\delta
    R^-(\vec x, \vec y)}{\delta \gA^a(\vec z)} \alpha^+(\vec y) +
  \text{r.h.} \Biggr\} + \frac{ig^2}{\sqrt 8} \int_{\vec x \vec y}
  \Bigr[ \bar\xi^-(\vec x) \bigl\{ R^-(\vec x,\vec y)[\Ozeta \gA(\vec
  y)] \\ + \Oomega_y R^-(\vec x, \vec y) - \Ozeta_y [R^-(\vec x, \vec
  y) \gA(\vec y)] \bigr\} \alpha^+(\vec y) + \text{r.h.} \Bigr] \\ =
  \int_{\vec x} \Oomega \gB^a \hat\partial\bar\gB^a -
  \frac{ig^2}{\sqrt 8} \int_{\vec x \vec y} \bigl\{ \bar\xi^-(\vec x)
  \Oomega_x R^-(\vec x, \vec y) \alpha^+(\vec y) + \text{r.h.}
  \bigr\}.
\end{multline}
We have adopted the convenient short-hand $\int_{\vec x \vec y \cdots}
:= \int_\Sigma d^3\vec x \: d^3 \vec y \cdots$, and defined the
differential operator $\zeta = \bar\partial / \hat\partial$.  Now
recall from chapter \ref{cha:mhvym} we expressed $\gA$ as a series in
$\gB$, given in eqs.\ \eqref{eq:A-series} and
\eqref{eq:Upsilon-coeff}.  This was obtained by solving
\eqref{eq:mhvym-A-eqn}, reproduced here in configuration space:
\begin{equation*}
  \int_{\vec x} \bigl\{ \Oomega \gA + [\gA, \Ozeta \gA]
  \bigr\}^a(\vec x) \frac{\delta \gB^b(\vec y)}{\delta \gA^a(\vec
    x)}  = \Oomega \gB^b(\vec y),
\end{equation*}
so by substituting for $\gA$ with \eqref{eq:A-series}, we can
eliminate the first terms from either side of
\eqref{eq:transf-qcd-1}. Furthermore, we can use
\eqref{eq:mhvym-A-eqn} on the LHS of \eqref{eq:transf-qcd-1} to trade
the $\delta R^-/\delta \gA$ for $\delta R^-/\delta \gB$, and we arrive
at
\begin{multline}
  \label{eq:transf-gauge-3-lh}
  \int_{\vec x \vec y \vec z} \bar\xi^-(\vec x) \left\{ (\Oomega_x +
    \Oomega_y) R^-(\vec x, \vec y) - [\Oomega \gB^a(\vec z)]
    \frac{\delta R^-(\vec x, \vec y)}{\delta\gB^a(\vec z)} \right\}
  \alpha^+(\vec y) \\
  = \int_{\vec x \vec y \vec z} \bar\xi^-(\vec x) \left\{ \Ozeta_y
    [R^-(\vec x, \vec y) \gA(\vec y)] - R^-(\vec x, \vec y) [\Ozeta
    \gA(\vec y)] \right\} \alpha^+(\vec y).
\end{multline}
The same procedure yields a similar equation for the right-handed
sector:
\begin{multline}
  \label{eq:transf-gauge-3-rh}
  \int_{\vec x \vec y \vec z} \bar\xi^+(\vec x) \left\{ ( \Oomega_x +
    \Oomega_y ) R^+(\vec x, \vec y) - [ \Oomega \gB^a(\vec z) ]
    \frac{\delta R^+(\vec x, \vec y)}{\delta\gB^a(\vec z)} \right\}
  \alpha^-(\vec y) \\
  = \int_{\vec x \vec y \vec z} \bar\xi^+ (\vec x) \left\{ [\Ozeta_y
    R^+(\vec x, \vec y)] \gA(\vec y) - R^+(\vec x, \vec y) [ \Ozeta
    \gA(\vec y) ] \right\} \alpha^-(\vec y).
\end{multline}
As the quark fields are arbitrary, equations
\eqref{eq:transf-gauge-3-lh} and \eqref{eq:transf-gauge-3-rh}
determine the solutions for $R^\pm$ in terms of $\gB$. We switch to
momentum space on $\Sigma$, and postulate a series solution of the
form
\begin{equation*}
  R^\pm(12) = (2\pi)^3 \delta^3(\vec p_1 + \vec p_2) + \sum_{n=3}^\infty
  \int_{3\cdots n} R^\pm (12;3\cdots n) \gB_{\bar 3} \cdots
  \gB_{\bar n} \: (2\pi)^3 \delta^3({\textstyle \sum_{i=1}^n \vec p_i})
\end{equation*}
Here, momenta $\vec p_1$ and $\vec p_2$ are associated with
the Fourier transforms
of $\vec x$ and $\vec y$, respectively, in \eqref{eq:fqnew-in-old}.
For future purposes, it will often be convenient to absorb the first
term above into the sum by defining $R^\pm(12;) = 1$.  As the quark
fields are arbitrary, we can drop them and the integrals over $\vec x$
and $\vec y$ in \eqref{eq:transf-gauge-3-lh} and
\eqref{eq:transf-gauge-3-rh} and what is left determines the solutions
for $R^\pm$ in terms of $\gB$.

Writing equations \eqref{eq:transf-gauge-3-lh} and
\eqref{eq:transf-gauge-3-rh} in momentum space and using
\eqref{eq:A-series} to substitute for $\gA$ leads to the following two
recurrence relations:
\begin{multline}
  \label{eq:Rl-rr}
  R^-(12;3\cdots n) = \\ \frac{-i}{\Oomega_1 + \cdots + \Oomega_n}
  \sum_{j=2}^{n-1} \frac{\{2, P_{j+1,n}\}}{\hat2 \, \hat P_{j+1,n}} \:
  R^-(1,2\!+\!P_{j+1,n};3\dots j) \Upsilon(-,j\!+\!1, \cdots, n)
\end{multline}
and
\begin{multline}
  \label{eq:Rr-rr}
  R^+(12;3\cdots n) = \\\frac{-i}{\Oomega_1 + \cdots + \Oomega_n}
  \sum_{j=2}^{n-1} \frac{\{2, P_{j+1,n}\}}{(\hat2 \!+\! \hat
    P_{j+1,n}) \, \hat P_{j+1,n}} \: R^+(1,2\!+\!P_{j+1,n};3\dots j)
  \Upsilon(-,j\!+\!1, \cdots, n),
\end{multline}
where, as before, we define the momentum space analogue of the
$\Oomega$ operator as $\omega_p := p\bar p/\hat p$,
$P_{ij}:=\sum_{k=i}^j p_k$, and $-$ as a momentum argument denotes the
negative of the sum of the other arguments. We notice immediately from
the above that if we put
\begin{equation}
  \label{eq:Rl-coeff}
  R^-(12;3\cdots n) = -\frac{\hat1}{\hat2} R^+(12;3\cdots n)
\end{equation}
into \eqref{eq:Rl-rr}, we recover \eqref{eq:Rr-rr}; note that this
only fixes the numerator $\hat 1$ above, whereas the sign and the
denominator follow by noting that the lowest-order coefficients
$R^\pm(12;) = 1$ are defined for conserved momentum (\ie\ at $\vec p_1=
-\vec p_2$). Thus, we need only solve for $R^+$.

Now one could obtain (and prove) a form for the $R^+$ coefficients by
direct iteration of \eqref{eq:Rr-rr} and induction on $n$, but in fact
it turns out that the recurrence relation \eqref{eq:Rr-rr} is nothing
other than a re-labelling of the $\Upsilon$ recurrence relation
\eqref{eq:Upsilon-rr}, reproduced below:
\[
\Upsilon(1\cdots n) = \frac{-i}{\Oomega_1 + \cdots + \Oomega_n}
\sum_{j=2}^{n-1} \left( \frac{\bar P_{j+1,n}}{\hat P_{j+1,n}} -
  \frac{\bar P_{2j}}{\hat P_{2j}} \right) \Upsilon(-,2,\cdots,j)
\Upsilon(-,j\!+\!1,\cdots,n).
\]
The solution to this was proved to be \eqref{eq:Upsilon-coeff}, so if
we now put
\begin{equation}
  \label{eq:Rr-coeff}
  R^+(12;3\cdots n) = \Upsilon(213\cdots n)
  = (-i)^n \frac{\hat2 \hat3 \cdots \widehat{n-1}}{(1\:3) (3\:4)
\cdots (n-1,n)} 
\end{equation}
into \eqref{eq:Rr-rr} (and swap momenta $1$ and $2$), we arrive at
\eqref{eq:Upsilon-rr}, and \eqref{eq:Rr-coeff} is thereby proved.

\subsubsection*{Quark inverse transformation $S^\pm$}
\label{sssec:S-soln}

The inverse fermion transformation, $S^\pm$, may be obtained from
$R^\pm$ in an order-by-order manner. Let us begin by writing it as
\begin{equation*}
  S^\pm(12) = \sum_{n=2}^\infty \int_{3\cdots n}
  S^\pm (12;3\cdots n) \gB_{\bar 3} \cdots
  \gB_{\bar n} \: (2\pi)^3 \delta^3({\textstyle \sum_{i=1}^n \vec p_i}),
\end{equation*}
where momenta $\vec p_1$ and $\vec p_2$ correspond to the Fourier transforms of
$\vec x$ and $\vec y$ in \eqref{eq:fqold-in-new}, and we have absorbed
the ${\cal O}({\cal B}^0)$ (\ie\ $n=2$) term into the sum by defining
$S^\pm(12;) = 1$. The $S^\pm$ coefficients satisfy the recurrence
relations
\begin{equation}
  \label{eq:Srl-rr}
  S^\pm(12;3\cdots n) =
  - \sum_{j=2}^{n-1} S^\pm(1,-;3 \dots j) R^\pm(-,2;j\!+\!1, \cdots, n).
\end{equation}
Now it is clear from \eqref{eq:Rl-coeff} that
\begin{equation}
  \label{eq:Sr-coeff}
  S^+(12;3\cdots n) = -\frac{\hat2}{\hat1} S^-(12;3\cdots n)
\end{equation}
where again the overall normalisation is fixed by the lowest order
coefficient.

Direct iteration of \eqref{eq:Srl-rr} gives the first few non-trivial
$S^-$:
\begin{align}
  \label{eq:Sl-1}
  S^-(12;3) &= i \frac{\hat1}{(3\:2)}, \\
  \label{eq:Sl-2}
  S^-(12;34) &= \frac{\hat1\hat4}{(3\:4)(4\:2)},
\end{align}
and so-on, from which we claim that
\begin{equation}
  \label{eq:Sl-coeff}
  S^-(12;3 \cdots n) = (-i)^n
  \frac{\hat1 \hat 4 \cdots \hat n}{(3\:4) \cdots
    (n-1,n)(n\:2)}
  = \Upsilon(13\cdots n2)
\end{equation}
where in the case of $S^-(12;3)$ only the first factor in the
numerator and the last factor in the denominator are retained. The
proof is by induction on $n$. Expressions \eqref{eq:Sl-1} and
\eqref{eq:Sl-2} provide the initial steps. For the inductive step, we
use \eqref{eq:Sl-coeff} to substitute for $S^-$, so the RHS of
\eqref{eq:Srl-rr} becomes
\begin{equation*}
  \begin{split} 
    & (-i)^n \frac{\hat 1 \hat 4 \cdots \widehat{n-1}}{(3\:4) \cdots
      (n\!-\!1,n)} \left\{ \frac{\hat 3}{(3,2\!+\!P_{3n})} +
      \sum_{j=3}^{n-1} \frac{(j,j\!+\!1)}{(j,2\!+\!P_{j,n})}
      \frac{\hat 2+\hat P_{j+1,n}}{(j\!+\!1,2\!+\!P_{j+1,n})} \right\}
    \\
    &= (-i)^n \frac{\hat1 \hat4 \cdots \widehat{n-1}}{(3\:4) \cdots
      (n\!-\!1,n)} \Biggl( x_3 + \sum_{j=3}^{n-1} y_j\Biggr)
  \end{split}
\end{equation*}
where
\begin{equation*}
  y_j = \frac{(j,j\!+\!1)}{(j,2\!+\!P_{jn})} \frac{\hat
    2 + \hat P_{j+1,n}}{(j\!+\!1,2\!+\!P_{j+1,n})}
  \quad\text{and}\quad
  x_j = \frac{\hat\jmath}{(j,2\!+\!P_{jn})}.
\end{equation*}
Now notice that $x_j + y_j = x_{j+1}$, so the sum on the RHS of
\eqref{eq:Srl-rr} collapses to
\begin{equation*}
  (-i)^n \frac{\hat1 \hat4 \cdots \widehat{n-1}}{(3\:4)
    \cdots (n\!-\!1, n)} x_n = S^-(12;3\cdots n),
\end{equation*}
and the proof is complete.\footnote{We could, of course, have obtained
  $S^\pm$ in a manner similar to $R^\pm$ by deriving an explicit
  recurrence relation and then mapping it to that for $\Upsilon$, but
  we believe this to be a quicker route.}

Let us now summarise the quark field transformations we have found
here:
\begin{align}
  \label{eq:alphabar-series}
  \bar\alpha^\pm_1 &=\bar\xi^\pm_1 + \sum_{n=3}^\infty \int_{2\cdots
    n} \left\{\begin{matrix} 1 \\ -\hat1/\hat2
    \end{matrix} \right\} \Upsilon(1\cdots n) \: \bar\xi^\pm_{\bar 2}
  \gB_{\bar 3} \cdots \gB_{\bar n}\: (2\pi)^3 \delta^3({ \textstyle
    \sum_{i=1}^n \vec p_i}), \\
  \label{eq:alpha-series}
  \alpha^\pm_1 &= \xi^\pm_1 + \sum_{n=3}^\infty \int_{2\cdots n}
  \left\{\begin{matrix} 1 \\ -\hat2/\hat1
    \end{matrix} \right\} \Upsilon(1\cdots n) \: \gB_{\bar 2} \cdots
  \gB_{\overline{n-1}} \xi^\pm_{\bar n}\: (2\pi)^3 \delta^3({
    \textstyle \sum_{i=1}^n \vec p_i}),
\end{align}
where the stacked expressions in braces take their value in accordance
with the upper or lower choice of sign.

\subsubsection*{$\bar\gA$ transformation}
\label{sssec:Abar-transf}

Finally, we obtain an expression for $\bar\gA$.  Owing to the form of
the canonical transformation, condition \eqref{eq:qp-QP} still holds
and gives
\begin{equation}
  \label{eq:qdotp-qcd}
  \begin{split}
    \int_{\vec x} \biggl\{ \tr \check\partial \gA \hat\partial \bar\gA
    &- \frac{ig^2}{\sqrt8}\int d^3 \vec x (\bar\alpha^- \check\partial
    \alpha^+ + \bar\alpha^+ \check\partial \alpha^-) \biggr\}
    \\
    &= \int_{\vec x} \biggl\{ \tr \check\partial \gB \hat\partial
    \bar\gB - \frac{ig^2}{\sqrt8}\int d^3 \vec x (\bar\xi^-
    \check\partial \xi^+ + \bar\xi^+ \check\partial \xi^-) \biggr\}.
  \end{split}
\end{equation}
Now consider the functional form of $\hat\partial \bar\gA$ as given by
\eqref{eq:Abar-canon-qcd}. We can split into two pieces,
\begin{equation}
  \label{eq:Abar-qcd-split}
  \hat\partial \bar\gA = \hat\partial \bar\gA^0 + \hat\partial
  \bar\gA^{\rm F},
\end{equation}
where the first term depends only on $\gB$ and $\bar\gB$, and the
second contains the fermion dependence. If we substitute this into
\eqref{eq:qdotp-qcd}, we find the terms
\begin{equation}
  \label{eq:qdotp-qcd-gauge}
  \tr \int_{\vec x} \check\partial \gA \hat\partial \bar\gA^0 
  \quad\text{and}\quad
  \tr \int_{\vec x} \check\partial \gB \hat\partial \bar\gB.
\end{equation}
on the left- and right-hand sides, respectively.  Now, these are the
left- and right-hand side respectively of \eqref{eq:mhvym-qp-QP},
which we solved in chapter \ref{cha:mhvym} to obtain $\bar\gA$ in the
pure Yang--Mills case. Thus, we can consistently identify
$\hat\partial \bar\gA^0$ with the first term in
\eqref{eq:Abar-canon-qcd} and eliminate \eqref{eq:qdotp-qcd-gauge}
from \eqref{eq:qdotp-qcd} by using the same solution, but this time
for $\bar\gA^0$. In quantisation surface momentum space,
\begin{equation}
  \bar\gA^0_{\bar 1} =
  \sum_{m=2}^\infty \sum_{s=2}^m \int_{2\cdots m}
  \frac{\hat s^2}{\hat1^2} \Upsilon(\bar 12\cdots m) \:
  \gB_{\bar2} \cdots \bar\gB_{\bar s}
  \cdots \gB_{\bar m} \: (2\pi)^3 \delta^3(\vec p_1 - {\textstyle
    \sum_{i=2}^m \vec p_i}).
  \label{eq:Abar0-series}
\end{equation}
From this, we immediately see that the pure-gauge MHV lagrangian of
chapter \ref{cha:mhvym} is recovered via the terms that $\bar\gA^0$
contributes to $\bar\gA$ when used in $L^\mmp$ and $L^{--++}$.

Next, we turn to $\gA^{\rm F}$. From \eqref{eq:Abar-canon-qcd} and
\eqref{eq:Abar-qcd-split}, we know that in momentum-space
$\bar\gA^{\rm F}$ is given by
\begin{equation}
  \hat 1 \bar\gA_{\bar 1}^{\rm F} = - \frac{g^2}{\sqrt8} (F_{\bar1}^- +
  F_{\bar1}^+),  \quad
  F_{\bar1}^\pm = \int_{\alpha\beta\gamma} \bar\xi^\pm_{\bar\alpha}
  \frac{\delta R^\pm(\pmb\alpha\pmb\gamma)}{\delta \gA^a_1}
  S^\pm(\bar{\pmb\gamma}\pmb\beta) \xi^\mp_{\bar\beta} \: T^a,
  \label{eq:Abar-qcd-writtenout}
\end{equation}
where we have substituted with \eqref{eq:fqnew-in-old} and then
\eqref{eq:fqold-in-new}. We could, in principle, use the known
expressions for $R^\pm$ and $S^\pm$ to evaluate this, but this is a
laborious process. Instead, we will solve for it outright, as follows.

First, let us study the series expansion for $F^\pm$. Consider the
\emph{form} of the objects between $\bar\xi$ and $\xi$ in
\eqref{eq:Abar-qcd-writtenout}: since $\gA$ depends only on $\gB$, we
see that it must be expressable in the form
\begin{equation}
  \sum_{n=2}^\infty \sum_{j=1}^n \int_{2\cdots n} (\text{coeff})_{\bar
    12\cdots n} \times \gB_{\bar 2} \cdots \gB_{\overline{j-1}} \:
  \delta^3(\vec p_1 + \vec p_j) T^a \: \gB_{\overline{j+1}} \cdots
  \gB_{\bar n}.
\end{equation}
Thus,
\begin{multline}
  \label{eq:AbarF-series}
  F^\pm_{\bar1} = - \sum_{n=3}^\infty \sum_{s=2}^{n-1} \int_{2\cdots
    n} \Delta^{\pm(s)}(\bar1 2 \cdots n) \Biggl \{ \gB_{\bar 2} \cdots
  \gB_{\overline{s-1}} \xi^\mp_{\bar s} \bar\xi^\pm_{\overline{s+1}}
  \gB_{\overline{s+2}} \cdots \gB_{\bar n} \\
  + \frac1{N_{\rm C}} ( \bar\xi^\pm_{\overline{s+1}}
  \gB_{\overline{s+2}} \cdots \gB_{\bar n} \gB_{\bar 2} \cdots
  \gB_{\overline{s-1}} \xi^\mp_{\bar s}) \Biggr\} (2\pi)^3
  \delta^3(\vec p_1 - {\textstyle \sum_{i=2}^n \vec p_i})
\end{multline}
where we have used the ${\rm SU}(N_{\rm C})$ Fierz identity
\eqref{eq:sun-fierz} to evaluate the sum over the gauge
generators. Note the sign change due to the Grassman nature of
$\xi^\pm$ and $\bar\xi^\pm$, and that the ${\cal O}(1/N_{\rm C})$ term
is proportional to the identity matrix. Our goal now is to compute
$\Delta^{\pm(s)}(1\cdots n)$, where $n \ge 3$ and $2 \le s \le n-1$.

Next, we return to \eqref{eq:qdotp-qcd}.  With
\eqref{eq:qdotp-qcd-gauge} eliminated, its momentum-space
representation is left with
\begin{equation*}
  -\tr \int_1 \check\partial \gA_1 \hat 1 \bar\gA^{\rm F}_{\bar 1}
  +\frac{g^2}{\sqrt8} \int_1(\bar{\alpha}^-_1 \check \partial \alpha^+_{\bar 1}
  +\bar{\alpha}^+_1 \check \partial \alpha^-_{\bar 1}) =
  \frac{g^2}{\sqrt8}\int_1 (\bar{\xi}^-_1 \check \partial \xi^+_{\bar 1}+
  \bar{\xi}^+_1 \check \partial \xi^-_{\bar 1} ).
\end{equation*}
We substitute for $\alpha^\pm$ and $\bar\alpha^\pm$ using the
momentum-space representations of \eqref{eq:fqold-in-new} and
\eqref{eq:fpold-in-new}, and this becomes
\begin{equation}
  \label{eq:qdotp-qcd-ferm-1}
  \tr \int_1 \check\partial \gA_1 (F^-_{\bar 1} + F^+_{\bar 1}) =
  -\int_{\alpha\beta\gamma} \bigl\{ \bar\xi^-_{\bar\alpha}
  R^-(\pmb\alpha\pmb\gamma) (\check\partial S^-(\bar{\pmb\gamma}\pmb\beta)) \xi^+_{\bar\beta}
  + \bar\xi^+_{\bar\alpha}
  R^+(\pmb\alpha\pmb\gamma) (\check\partial S^+(\bar{\pmb\gamma}\pmb\beta)) \xi^-_{\bar\beta}
  \bigr\}.
\end{equation}
We can split this into left- and right-handed chiral pieces and solve
for $F^+$ and $F^-$ separately.  Let us first study the terms on the
RHS of \eqref{eq:qdotp-qcd-ferm-1}.
\begin{equation}
  \begin{split}
    \label{eq:qdotp-qcd-ferm-rhs}
    & - \int_{\alpha\beta\gamma} \bar\xi^\pm_{\bar\alpha}
    R^\pm(\pmb\alpha\pmb\gamma) [\check\partial
    S^\pm(\bar{\pmb\gamma}\pmb\beta)] \xi^\mp_{\bar\beta} \\
    &= \tr \int_{\alpha\beta\gamma} \xi^\mp_{\bar\beta}
    \bar\xi^\pm_{\bar\alpha} R^\pm(\pmb\alpha\pmb\gamma)
    [\check\partial
    S^\pm(\bar{\pmb\gamma}\pmb\beta)] \\
    &= \tr \sum_{n=3}^\infty \sum_{s=2}^{n-1} \sum_{l=s+1}^n
    \int_{1\cdots n} R^\pm(s\!+\!1,-;s\!+\!2,\dots,l)
    S^\pm(-,s;l\!+\!1,\dots,n,1,\dots,s\!-\!1) \times \\
    &\hphantom{= \tr \sum_{n=3}^\infty \sum_{s=2}^{n-1}
      \sum_{l=s+1}^n} (\gB)_{n,s} \times (2\pi)^3 \delta({\textstyle
      \sum_{i=1}^n \vec p_i}),
  \end{split}
\end{equation}
where we use the short-hand
\begin{equation*}
  (\gB)_{n,s} \equiv
  \check\partial \gB_{\bar 1} \gB_{\bar 2} \cdots \gB_{\overline{s-1}}
  \xi^\mp_{\bar s} \bar\xi^\pm_{\overline{s+1}} \gB_{\overline{s+2}}
  \cdots \gB_{\bar n}.
\end{equation*}

The LHS of \eqref{eq:qdotp-qcd-ferm-1} is
\begin{multline*}
  \tr \int_1 \check\partial \gA_1 F^\pm_{\bar 1} = -\tr \int_1
  \sum_{i=2}^\infty\sum_{j=2}^i \sum_{k=3}^\infty\sum_{l=2}^{k-1}
  \int_{2\cdots i} \int_{2'\cdots k'}
  \Upsilon(12\cdots i) \Delta^{\pm(s)}(\bar12'\cdots k') \\
  \times \check\partial \gB_{\bar\jmath} \cdots \gB_{\bar \imath}
  \gB_{\bar2'} \cdots \gB_{\overline{l-1}'} \xi^\mp_{\bar l'}
  \bar\xi^\pm_{\overline{l+1}'} \gB_{\overline{l+2}'} \cdots \gB_{\bar
    k'} \gB_{\bar 2} \cdots \gB_{\overline{j-1}} \\
  \times (2\pi)^6 \delta({\textstyle \sum_{m=1}^i \vec p_m})
  \delta(\vec p_1 - {\textstyle \sum_{n=2}^k \vec p_n}),
\end{multline*}
summed over the $\pm$ superscript.  Notice that the ${\cal O}(1/N_{\rm
  C})$ piece has vanished since it is proportional to the identity
matrix, and $\check\partial \gA$ is traceless.  After carefully
relabelling the momenta, starting with $j \rightarrow 1$, this becomes
\begin{multline*}
  -\tr \sum_{n=3}^\infty \sum_{s=2}^{n-1} \sum_{m=2}^{n-1}
  \sum_{p=\max(0,m-s)}^{\min(m-2,n-s-1)}
  \Upsilon(-,n-p+1,\dots,n,1,\dots,m-p-1) \\
  \times \Delta^{\pm(s+p-m+2)}(-,m-p,\dots,n-p) (\gB)_{n,s}.
\end{multline*}
Now fix $n$ and $s$ above. The following change of variables
\begin{equation*}
  q=n-p,\quad r=m-p-1
\end{equation*}
allows us to write the coefficient of $(\gB)_{n,s}$ as
\begin{multline}
  - \Delta^{\pm(s)}(1\cdots n) - \sum_{r=2}^{s-1}
  \Upsilon(-,1,\dots,r)
  \Delta^{\pm(s-r+1)}(-,r+1,\dots,n) \\
  - \sum_{q=s+1}^{n-1} \sum_{r=1}^{s-1} \Upsilon(-,q+1,\dots, n, 1,
  \dots, r) \Delta^{\pm(s-r+1)}(-, r+1, \dots, q).
  \label{eq:qdotp-qcd-ferm-lhs}
\end{multline}
We can now equate the coefficients of $(\gB)_{n,s}$ from
\eqref{eq:qdotp-qcd-ferm-rhs} and \eqref{eq:qdotp-qcd-ferm-lhs} to
obtain the following recurrence relation for $\Delta^\pm$
\footnote{Note that in the interest of clarity concerning the origin
  of certain terms in the forthcoming, we have \emph{not} substituted
  for $R^\pm$ and $S^\pm$ in terms of $\Upsilon$.}:
\begin{equation}
  \begin{split}
    \Delta^{\pm(s)}(1\cdots n) =& - \sum_{q=s+1}^n R^\pm(s+1,-;
    \underbrace{s+2, \dots, q}_{(q-s-1)}) S^\pm(-, s;
    \underbrace{q+1,\dots, n, 1, \dots,
      s-1}_{(n+s-q+1)}) \\
    & - \sum_{r=2}^{s-1} \Upsilon(\underbrace{-,1,\dots,r}_{(r+1)})
    \Delta^{\pm(s-r+1)}(\underbrace{-,r+1,\dots,n}_{(n-r+1)}) \\
    & - \sum_{q=s+1}^{n-1} \sum_{r=1}^{s-1}
    \Upsilon(\underbrace{-,q+1,\dots, n, 1, \dots, r}_{(n-q+r+1)})
    \Delta^{\pm(s-r+1)}(\underbrace{-, r+1, \dots, q}_{(q-r+1)}).
  \end{split}
  \label{eq:Delta-rr}
\end{equation}
The numbers below the underbraces denote the number of arguments they
enclose. In cases where the upper limit of a sum is less than the
lower limit --- such as when $s=2$ in the sum over $r$ in the second
term, or when $s=n-1$ in the sum over $q$ in the third --- the sum is
taken to vanish.

Now if we notice that, using \eqref{eq:Rl-coeff} and
\eqref{eq:Sr-coeff},
\begin{equation*}
  \begin{split}
    &R^+(s+1,-; s+2, \dots, q) S^+(-,s; q+1,\dots,
    n, 1, \dots, s-1) \\
    &= \left(-\frac{\hat P_{s+1,q}}{\widehat{s+1}} \right)
    \left(-\frac{\hat s}{\hat P_{q+1,n}+\hat P_{1s}} \right) \\
    &\quad \times R^-(s+1,-; s+2, \dots, q) S^-(-,s; q+1,\dots,
    n, 1, \dots, s-1) \\
    &= -\frac{\hat s}{\widehat{s+1}} R^-(s+1,-; s+2, \dots, q)
    S^-(-,s; q+1,\dots, n, 1, \dots, s-1).
  \end{split}
\end{equation*}
(since $P_{1s} + P_{q+1,n} = -P_{s+1,q}$ on account of conservation of
momentum) and place this into \eqref{eq:Delta-rr} for $\Delta^+$, it
is easy to see that putting
\begin{equation}
  \label{eq:Delta+-coeff}
  \Delta^{+(s)}(1\cdots n) = -\frac{\hat s}{\widehat{s+1}}
  \Delta^{-(s)}(1\cdots n)
\end{equation}
leads back to \eqref{eq:Delta-rr} for $\Delta^-$. Hence, we will solve
for just $\Delta^-$ and use the above relationship to obtain
$\Delta^+$.

Let us compute by hand first few $\Delta^-$ coefficients. First,
\begin{equation*}
  \Delta^{-(2)}(123) = -S^-(32;1) = -i \frac{\hat 3}{(2\:3)}.
\end{equation*}
Next,
\begin{equation*}
  \begin{split}
    \Delta^{-(2)}(1234) &= -\Upsilon(-,4,1)\Delta^{-(2)}(-,2,3)
    -S^-(32;41) -R^-(3,-;4) S^-(-,2;1)
    \\
    &= -\frac{\hat3^2}{(2\:3)(3\:4)}.
  \end{split}
\end{equation*}
Similarly, we can obtain next few coefficients:
\begin{align*}
  \Delta^{-(3)}(1234) &= - \frac{\hat3 \hat4}{(2\:3)(3\:4)}, \\
  \Delta^{-(2)}(12345) &= i \frac{\hat3^2\hat4}{(2\:3)(3\:4)(4\:5)}, \\
  \Delta^{-(3)}(12345) &= i \frac{\hat3\hat4^2}{(2\:3)(3\:4)(4\:5)},  \\
  \Delta^{-(4)}(12345) &= i
  \frac{\hat3\hat4\hat5}{(2\:3)(3\:4)(4\:5)},
\end{align*}
from which we conjecture
\begin{equation}
  \label{eq:Delta--coeff}
  \Delta^{-(s)}(1\cdots n) = -(-i)^n \frac{\widehat{s+1} \: \hat3 \hat4
    \cdots \widehat{n-1}}{(2\:3) (3\:4) \cdots (n-1,n)}
  = -\frac{\widehat{s+1}}{\hat1} \Upsilon(1\cdots n)
\end{equation}
This can be proved by induction on $n$. The foregoing calculations of
the lowest-order coefficients obviously furnish the initial step. For
the inductive part, we substitute \eqref{eq:Delta--coeff} into the
recurrence relation. For each term on the RHS of \eqref{eq:Delta-rr},
we can pull out a factor of
\[
\frac{\hat 1 \cdots \hat n}{(1\:2)\cdots(n\:1)}
\]
leaving telescoping sums of the form \eqref{eq:telescope-sum-1}.  (In
the second term of \eqref{eq:Delta-rr}, two such sums are nested.)
One might worry about the cases discussed below \eqref{eq:Delta-rr}
where certain terms are taken to vanish.  It turns out that we can
handle these cases consistently by understanding the sum $P_{ij} = p_i
+ p_{i+1} + \cdots + p_n + p_1 + \cdots + p_j$ when $j<i$. When this
is so and we evaluate the sums using \eqref{eq:telescope-sum-1}, we
find that they vanish in these particular conditions because of terms
of the form $P_{i,i-1}=0$.  To complete the proof, one simply
evaluates the sums and does the algebra while applying the
conservation of momentum. This results in an expression on the RHS of
\eqref{eq:Delta-rr} equal to the given for $\Delta^{-(s)}(1\cdots n)$
in \eqref{eq:Delta--coeff}.

To finish this section, we state the series expansion of $\bar\gA$,
assembled from \eqref{eq:Abar-qcd-split}, \eqref{eq:Abar0-series} and
\eqref{eq:AbarF-series}, and using \eqref{eq:Delta+-coeff} and
\eqref{eq:Delta--coeff}:
\begin{multline*}
  \bar\gA_1 = \sum_{m=2}^\infty \sum_{s=2}^m \int_{2\cdots m}
  \frac{\hat s^2}{\hat1^2} \Upsilon(1 \cdots m) \gB_{\bar2} \cdots
  \bar\gB_{\bar s} \cdots \gB_{\bar m} \: (2\pi)^3 \delta({\textstyle
    \sum_{i=1}^m \vec p_i}) \\
  + \frac{g^2}{\hat1^2 \sqrt 8} \sum_\pm \sum_{n=3}^\infty
  \sum_{s=2}^{n-1} \int_{2\cdots n} \left\{\begin{matrix} -\hat s \\
      \widehat{s+1}
    \end{matrix} \right\} \Upsilon(1 \cdots n) \Biggl\{ \gB_{\bar 2}
  \cdots \gB_{\overline{s-1}} \xi^\mp_{\bar s}
  \bar\xi^\pm_{\overline{s+1}} \gB_{\overline{s+2}} \cdots \gB_{\bar
    n} + \\ \frac1{\NC} \bar\xi^\pm_{\overline{s+1}}
  \gB_{\overline{s+2}} \cdots \gB_{\bar n} \gB_{\bar 2} \cdots
  \gB_{\overline{s-1}} \xi^\mp_{\bar s} \Biggr\} \times (2\pi)^3
  \delta^3({\textstyle \sum_{i=1}^n \vec p_i}).
\end{multline*}
But there is one further simplification we can apply: if we relabel the
fields in the ${\cal O}(1/\NC)$ terms, we can use the dual Ward
identity \eqref{eq:Upsilon-dwi} to evaluate the sum over $s$ in this
term to leave
\begin{multline}
  \label{eq:Abar-loworder-qcd}
  \bar\gA_1 = \sum_{m=2}^\infty \sum_{s=2}^m \int_{2\cdots m}
  \frac{\hat s^2}{\hat1^2} \Upsilon(1 \cdots m) \gB_{\bar2} \cdots
  \bar\gB_{\bar s} \cdots \gB_{\bar m} \: (2\pi)^3 \delta({\textstyle
    \sum_{i=1}^m \vec p_i}) \\
  + \frac{g^2}{\hat1^2 \sqrt 8} \sum_\pm \sum_{n=3}^\infty
  \int_{2\cdots n} \Upsilon(1 \cdots n)
  \Biggl[ \sum_{s=2}^{n-1} \left\{ \begin{matrix} -\hat s \\
      \widehat{s+1} \end{matrix} \right\} \gB_{\bar 2} \cdots
  \gB_{\overline{s-1}} \xi^\mp_{\bar s} \bar\xi^\pm_{\overline{s+1}}
  \gB_{\overline{s+2}} \cdots \gB_{\bar n} \\ 
  + \frac1{\NC} \left\{\begin{matrix} \hat n \\ 
    -\hat 2 \end{matrix}\right\} \bar\xi^\pm_{\bar 2} \gB_{\bar3}
  \cdots \gB_{\overline{n-1}} \xi^\mp_{\bar n} \Biggr] \times
  (2\pi)^3 \delta^3({\textstyle \sum_{i=1}^n \vec p_i}).
\end{multline}

\subsubsection*{Completion vertices}
\label{ssec:mhvqcd-transf-completion}
Naturally, one can define completion vertices for massless QCD by the
same protocol as used in the pure-gauge scenario. From
\eqref{eq:alphabar-series}, \eqref{eq:alpha-series} and
\eqref{eq:Abar-loworder-qcd}, we can write down the rules for the
completion vertices, and the non-trivial ones are shown in
fig.~\ref{fig:mhvqcd-completionverts}. These augment those from the
pure-gauge theory, shown in fig.~\ref{fig:etv-completionverts}.
\begin{figure}[t]
  \begin{align*}
    \begin{matrix}\begin{picture}(90,67)
        \SetOffset(45,42)
        \ArrowLine(0,13)(0,1)
        \DashArrowLine(0,0)(28.9778,-7.7646){2}
        \Line(12.6785,-27.1892)(0,0)
        \Line(0,0)(-28.9778,-7.7646) \BCirc(0,0){2}
        \DashCArc(0,0)(15,-160,-70){1}
        \Text(0,16)[bc]{$1^\pm$} \Text(31,-5)[tl]{$2^\pm$}
        \Text(12,-29)[tl]{$3^+$}
        \Text(-31,-5)[tr]{$n^+$}
      \end{picture}\end{matrix} &= \left\{ \begin{matrix} 1 \\ -\hat1/\hat2 \end{matrix} \right\}
    \Upsilon(1\cdots n)
    \\
    \begin{matrix}\begin{picture}(90,67)
        \SetOffset(45,42)
        \ArrowLine(0,1)(0,13)
        \Line(0,0)(28.9778,-7.7646)
        \Line(0,0)(-12.6785,-27.1892)
        \DashArrowLine(-28.9778,-7.7646)(0,0){2} \BCirc(0,0){2}
        \DashCArc(0,0)(15,-110,-20){1}
        \Text(0,16)[bc]{$1^\pm$} \Text(31,-5)[tl]{$2^+$}
         \Text(0,-29)[tr]{$(n\!-\!1)^+$}
        \Text(-31,-5)[tr]{$n^\pm$}
      \end{picture}\end{matrix} &= \left\{\begin{matrix} 1 \\ -\hat2/\hat1
      \end{matrix} \right\} \Upsilon(1\cdots n)
    \\
    \begin{matrix}\begin{picture}(90,67)
        \SetOffset(45,42)
        \Gluon(0,1)(0,13){2}{2}
        \Line(0,0)(28.9778,-7.7646)
        \DashArrowLine(12.6785,-27.1892)(0,0){2}
        \DashArrowLine(0,0)(-12.6785,-27.1892){2}
        \Line(0,0)(-28.9778,-7.7646) \BCirc(0,0){2}
        \DashCArc(0,0)(15,-55,-20){1}
        \DashCArc(0,0)(15,-160,-125){1}
        \Text(0,16)[bc]{$1^-$} \Text(31,-5)[tl]{$2^+$}
        \Text(12,-29)[tl]{$s^\mp$} \Text(0,-29)[tr]{$(s\!+\!1)^\pm$}
        \Text(-31,-5)[tr]{$n^+$}
      \end{picture}\end{matrix} &= \frac{g^2}{\hat1^2 \sqrt 8}\left\{\begin{matrix} -\hat s  \\ \widehat{s+1}
      \end{matrix} \right\} \Upsilon(1 \cdots n)
  \end{align*}
  \label{fig:mhvqcd-completionverts-c}
  \caption{The non-trivial MHV completion vertices for massless
    QCD. Curly and solid, arrowless lines are as in
  fig.~\ref{fig:etv-completionverts}. Solid lines with arrows
  represent correlation function insertions of $\alpha^\pm$
  ($\bar\alpha^\pm$) when the arrow points outwards (inwards), and
  $\xi^\pm$ ($\bar\xi^\pm$) attach to the dotted lines with the
  arrows pointing inwards (outwards).  The direction of the arrow
  shows charge flow in both cases, and the missing lines are for $+$-helicity gluons.
  (Stacked expressions in braces
  take their value in accordance with the upper or lower choice of
  sign for the fermion helicities.)  }
\label{fig:mhvqcd-completionverts}
\end{figure} (Note that the figure shows the
vertices as appropriate for the normalisation of $\gA$, $\gB$, \etc\
so that the normalisation factors from \eqref{eq:gauge-conv} can be
omitted for clarity.)  For an example of these vertices in action, see
section \ref{ssec:mhvqcd-completion}, where they are used to
reconstruct the `missing' $A(1_{{\rm q}}^+ 2^+ 3_{\bar{\rm q}}^-)$
partial amplitude.

\subsection{An indirect proof of MHV vertices}
\label{ssec:mhvqcd-transf-mhvproof}

We now return to the discussion started at the end of section
\ref{ssec:form-transform-qcd}. There we claimed that the vertices of
the transformed lagrangian are proportional to the MHV amplitudes
continued off-shell by the CSW prescription (in this case by using the
choice \eqref{eq:lc-spinors} for the Weyl spinors). To prove this, let
us first review the proof from the pure gauge case. First, for each
vertex in the Canonical MHV Lagrangian, its unique helicity and colour
structure means that it is the sole contributor to the corresponding
tree-level MHV amplitude \emph{on shell}. This was first stated in
section \ref{ssec:mhvym-amps-verts}, and later in section
\ref{sec:etv-completiontree} we plugged a possible hole in this claim
by showing that for on-shell tree-level amplitudes, diagrams
constructed using completion vertices were annihilated in the LSZ
reduction procedure, and so did not contribute.  Then to show that the
vertex was also valid off shell, we argued that due to the
holomorphicity of the vertices, they could contain no terms vanishing
on the support of the on-shell condition.

It is straightforward to extend this argument to the QCD theory
studied in this chapter. First, it is quite clear that the vertex
coupling a $-$-helicity gluon to a quark-antiquark pair in the MHV
lagrangian is the same as that of light-cone gauge QCD, \ie\
\eqref{eq:lcqcd-qbarmq}.  The helicity configuration precludes any
contribution from completion vertices. For higher order vertices, we
can see again from the form of the completion vertices shown
fig.~\ref{fig:mhvqcd-completionverts} that the diagrams through which they
contribute terms in MHV amplitudes do not survive LSZ reduction for
generic momenta (for the same reason as in the pure-gauge case; notice
that they are all proportional to $\Upsilon$). Finally, that there are
no terms that vanish on-shell follows since the transformation is
performed on a surface of equal light-cone time and has no $x^0$
dependence, and transformed lagrangian is manifestly holomorphic (\cf\
\eqref{eq:lcqcd-qbarpmq}, \eqref{eq:lcqcd-qbarqqbarq},
\eqref{eq:alphabar-series}, \eqref{eq:alpha-series} and
\eqref{eq:Abar-loworder-qcd}). This completes the proof that the
vertices of the transformed lagrangian are indeed MHV vertices.


\section{Example vertices}
\label{sec:examples-qcd}

We now have all-orders expressions for the `old' fields $\gA$,
$\bar\gA$, $\alpha^\pm$ and $\bar\alpha^\pm$ in terms of their new
counterparts $\gB$, $\bar\gB$, $\xi^\pm$ and $\bar\xi^\pm$.  Let us
now verify explicitly that the vertices obtained by substituting for
the old fields in remaining terms of the LCYM lagrangian,
\eqref{eq:lcym-mmp} and \eqref{eq:lcym-mmpp}, are proportional
off-shell to the known tree-level MHV amplitudes.

\subsection{On external states, vertices and amplitudes}
\label{ssec:mhvqcd-external}

As in section \ref{sec:mhvym-examples}, a partial MHV amplitude is
obtained from the MHV lagrangian by contracting an external state into
the vertex with the relevant helicity and charge content, and summing
if there is more than once contraction which picks out the desired
colour structure.

We must define the polarisation vectors and spinors. As before, the
relevant components of the gluon polarisation vectors are given by
\eqref{eq:gluon-poln} of section \ref{sec:mhvym-examples}.  In
co-ordinates, $E_+ = \bar E_-= -1$ so again by the LSZ theorem, when
an external $+$ ($-$) polarisation state is contracted into a $\gA$
($\bar\gA$) vertex from the lagrangian, it contributes a factor of
\begin{equation*}
  -1 \times -\frac{ig}{\sqrt 2},
\end{equation*}
where the second factor accounts restores the canonical normalisation
of the gauge field from \eqref{eq:gauge-conv}.  For the polarisation
spinors for the massless quarks, we must solve the Dirac equation
$\slashed p \psi = 0$. This has the following positive-energy
solutions for $\psi$:
\begin{equation*}
  \bar u^+(p) \equiv (\bar\varphi(p), 0)
  \quad\text{and}\quad
  \bar u^-(p) \equiv (0, \omega(p)).
\end{equation*}
For the purposes of the LSZ theorem (or, more simply as polarisation
spinors contracted into $\gamma$ matrices for external lines in
Feynman diagrams), these correspond to positive- and negative-helicity
\emph{outgoing} quarks, respectively. Similarly, the negative-energy
solutions are
\begin{equation*}
  v^+(p) \equiv \begin{pmatrix} \bar\omega(p) \\ 0 \end{pmatrix}
  \quad\text{and}\quad
  v^-(p) \equiv \begin{pmatrix} 0 \\ \varphi(p) \end{pmatrix}
\end{equation*}
for \emph{outgoing} antiquarks, where again the sign in the
superscript denotes physical helicity. Here we have used the following
definitions for the Weyl spinors:
\begin{align}
  \label{eq:poln-antiquark-}
  \varphi(p) &= 2^{1/4} \begin{pmatrix}
    - p/\sqrt{\hat p} \\ \sqrt{\hat p} \end{pmatrix}, \\
  \label{eq:poln-antiquark+}
  \bar\omega(p) &= 2^{1/4} \begin{pmatrix} \sqrt{\hat p} \\
    \bar p/\sqrt{\hat p} \end{pmatrix},
  \\
  \label{eq:poln-quark+}
  \bar\varphi(p) &= 2^{1/4} ( {-\bar p/\sqrt{\hat
      p}}, \sqrt{\hat p}), \\
  \label{eq:poln-quark-}
  \omega(p) &= 2^{1/4} ( \sqrt{\hat p}, {p/\sqrt{\hat p}} ),
\end{align}
These are such that the Dirac spinors have the conventional
phenomenologists' normalisation of $u^\dagger(p) u(p) = 2p^t$.

Let us consider the LSZ reduction \emph{before} we remove the
non-dynamical fermionic degrees of freedom. For example, in this
context, an outgoing $+$ helicity quark with momentum $p$ produces a
term
\begin{equation*}
  \bar u^+(p) (-i\slashed p) \langle \cdots \psi \cdots \rangle =
  \bar\varphi(p)_{\dot\alpha} (-i p\cdot\sigma)^{\dot\alpha \beta}
  \langle \cdots \begin{pmatrix} \beta^- \\ \alpha^- \end{pmatrix}_\beta
  \cdots \rangle,
\end{equation*}
where $-i p\cdot\sigma$ is the inverse of the propagator obtained from
\eqref{eq:lcqcd-qbarq}. We can now compute the correlation function
using the partition function generated by the MHV lagrangian. Since
$\varphi_1 \equiv \beta^-$ is replaced by its equation of motion, one
might expect this non-propagating component to complicate
things. Fortunately,
\begin{equation}
  \label{eq:lsz-factor-qcd}
  \bar\varphi(p) (-i p\cdot\sigma) = - \frac i{2^{1/4} \sqrt{\hat p}}
  (0, p^2),
\end{equation}
so it does not arise in the computation. We proceed to replace
$\varphi_2 \equiv \alpha^-$ with its expression in terms of the new
variables, and note that momentum conservation implies only the
leading order term $\xi^-$ survives the on-shell limit for generic
momenta at tree-level. The propagator $\langle \xi^- \bar\xi^+
\rangle$ of \eqref{eq:propagators-qcd} cancels factors in
\eqref{eq:lsz-factor-qcd} to leave a polarisation factor
\begin{equation*}
  2^{1/4}\sqrt{\hat p}.
\end{equation*}
One may show similarly that the same expression applies for the $-$
helicity state, and for the anti-quarks. In summary, we state the
polarisation spinors and the fields in the lagrangian associated with
each outgoing state in table \ref{tbl:field-state-qcd}.

\begin{table}[h]
  \centering\begin{tabular}{lccc}
    \toprule
    \multicolumn{2}{l}{\textbf{State}} &
    \textbf{Polarisation} & \textbf{Field} \\
    \midrule
    \multirow{2}{*}{particle}
    & $+$ & $\bar\varphi(p)$ & $\bar\xi^+$ \\
    & $-$ & $\omega(p)$      & $\bar\xi^-$ \\
    \multirow{2}{*}{antiparticle}
    & $+$ & $\bar\omega(p)$  & $\xi^+$ \\
    & $-$ & $\varphi(p)$     & $\xi^-$ \\
    \bottomrule
  \end{tabular}
  \caption{The polarisation spinor and lagrangian field associated
    with each outgoing quark state.}
  \label{tbl:field-state-qcd}
\end{table}

Let us now frame this in the context of the lagrangian vertices
$V_{2{\rm q}}$ and $V_{4{\rm q}}$, used to express the general form of
the quark-gluon interaction terms in \eqref{eq:mhvqcd-2q-form} and
\eqref{eq:mhvqcd-4q-form}. As we noted before, these contain the only
terms in the MHV QCD lagrangian that contribute to tree-level
MHV quark-gluon
amplitudes, and as in the pure-gluon case of section \ref{sec:mhvym-struct}
we extract partial amplitudes by contracting external states into
these terms in a manner that picks out the desired colour order.  As
with the purely gluonic case, our outgoing states are constructed by
having annihilation operators act to the right on the `out' vacuum
state $\langle 0 \rvert$. We note that when dealing with fermions, we
must take extra care with statistics, which essentially comes down to
a choice of sign convention, or equivalently annihilation operator
ordering.  Now in section \ref{sec:swis}, we obtained SUSY partial
amplitudes using supersymmetric Ward identities by associating them
with $S$-matrix elements where the order of the annihilation operators
placed between the free and physical vacua was the same as that of the
arguments of the partial amplitude. Since we wish to compare the
output of the MHV lagrangian with SWI-derived amplitudes, we will
adopt the same convention when defining the outgoing states.

First, consider the MHV amplitude with one quark-antiquark pair. Its
external state is
\[
\langle 0 \rvert q^\pm_1 A^+_2 \cdots A^-_j \cdots A^+_{n-1} \bar
q^\mp_n.
\]
This contracts into the vertex in \eqref{eq:mhvqcd-2q-form} multiplied
by an external state factor of
\[
(-1) \times \frac{4i}{g^2} \times
\left(-\frac{ig}{\sqrt2}\right)^{n-2} (-1)^{n-2} \times
2^{1/4}\sqrt{\hat1} \times 2^{1/4}\sqrt{\hat n}.
\]
Considering the factors delimited by $\times$ symbols, the first comes
from Fermi statistics; the second from the path integral; the third
from gluon polarisation and normalisation; and the final two from the
external state spinors (see above). Thus, when the partial amplitude
is defined as in \eqref{eq:colourstruct-2q},
\begin{equation}
  \label{eq:A-V-2q}
  A(1_{{\rm q}}^\pm,2^+,\dots, j^-, \dots,(n\!-\!1)^+,n_{\bar{\rm q}}^\mp)
  = 2^{(7-n)/2} \, i^{n+1} \, g^{n-4} \, \sqrt{\hat1 \hat n} \:
  V^{j;\pm}_{2{\rm q}}(1\cdots n).
\end{equation}

With two quark-antiquark pairs, we would contract the external state
\[
\langle 0 \rvert q^{h_1}_1 A^+_2 \cdots A^+_{j-2} \bar q^{-h_j}_{j-1}
\: q^{h_j}_j A_{j+1}^+ \cdots A^+_{n-1} \bar q^{-h_1}_n
\]
into \eqref{eq:mhvqcd-4q-form}, taking care with the fermion
statistics. By construction
${\cal S}(h_1,h_j)$ drops out so we are left with
\begin{multline}
  \label{eq:A-V-4q}
A(1^{h_1}_{{\rm q}}, 2^+, \dots ,(j\!-\!2)^+,
    (j\!-\!1)^{-h_j}_{\bar{\rm q}}; 
   j^{h_j}_{{\rm q}}, (j\!+\!1)^+, \dots, (n\!-\!1)^+,
   n^{-h_1}_{\bar{\rm q}})
\\
= 2^{5-n/2}i^{n+1} g^{n-6} \sqrt{\hat1 \: \hat j \: \widehat{j\!-\!1} \: \hat
  n} \:
V^{j;h_1h_j}_{4{\rm q}}(1\cdots n),
\end{multline}
and similarly for the sub-leading partial amplitudes
\begin{multline*}
A_{(1)}(1^{h_1}_{{\rm q}}, 2^+, \dots ,(j\!-\!2)^+, n^{-h_1}_{\bar{\rm q}}; 
   j^{h_j}_{{\rm q}}, (j\!+\!1)^+, \dots, (n\!-\!1)^+,
   (j\!-\!1)^{-h_j}_{\bar{\rm q}})
\\
= 2^{5-n/2}i^{n+1} g^{n-6} \sqrt{\hat1 \: \hat j \: \widehat{j\!-\!1} \: \hat
  n} \:
V^{j;h_1h_j}_{4{\rm q}(1)}(1\cdots n).
\end{multline*}

\subsection{Two quarks and two gluons}
\label{ssec:two-quarks-two-gluons}
Let us now consider the partial amplitude $A(1_{\rm q}^+ 2^+ 3^-
4_{\bar{\rm q}}^-)$. In the colour structure decomposition of the
$S$-matrix, this arises as the coefficient of $(T^{a_2}
T^{a_3})\ud14$. This amplitude is known to be (see \eqref{eq:csw-ferm-q+gq-})
\begin{equation}
  \label{eq:qqgg-swi}
  -i g^2 \frac{\langle 1\:3 \rangle \langle 4\:3 \rangle^2}
  {\langle 1\:2 \rangle
    \langle 2\:3 \rangle
    \langle 4\:1 \rangle}
\end{equation}
(something which can be obtained most easily using supersymmetric Ward
identities --- see \eqref{eq:swi-qg3q}).  We wish to compare this to
the result of \eqref{eq:A-V-2q}; we just have to compute
$V^{3+}_{2{\rm q}}(1234)$.
%
Note that this amplitude has a single quark-antiquark pair, and as
such we may safely discard any ${\cal O}(1/N_{\rm C})$ terms
that arise in
the analysis.  Were we to retain these terms, we would find they
vanish anyway due to antisymmetries in the gluonic coefficients.

Looking table \ref{tbl:mhvqcd-content}, we see that, based upon field
content, the term we are considering receives contributions from from
$L^{\mmp}$, $L^{\bar\psi+-\psi}$ and $L^{\bar\psi-\psi}$ (in
\eqref{eq:lcym-mmp}, \eqref{eq:lcqcd-qbarpmq} and
\eqref{eq:lcqcd-qbarmq} respectively). Let us consider each in turn.
Written in momentum space, \eqref{eq:lcym-mmp} is
\begin{equation}
  \label{eq:lcqcd-mmp-M}
  L^{\mmp} = i \tr \int_{123} \frac{\hat 3}{\hat 1 \hat 2} (1\:2) \:
  \bar\gA_{\bar 1} \bar\gA_{\bar 2} \gA_{\bar 3},
\end{equation}
where here and in the foregoing, a momentum-conserving $\delta$
function of the sum of all the momenta in the integral measure is
omitted for clarity.  We start by substituting for $\bar\gA$:
the term in $\bar\xi^+_{\bar 1}
\gB_{\bar 2} \bar\gB_{\bar 3} \xi^-_{\bar 4}$ comes from the second
$\bar\gA$ being replaced with the $\bar\xi^+ \xi^-$
term in \eqref{eq:Abar-loworder-qcd}. Relabelling the momenta, the
contribution from \eqref{eq:lcqcd-mmp-M} to the vertex is
\begin{equation*}
  - \frac{g^2}{\sqrt 8} \int_{1234} \frac{\hat 2 \hat 4}{\hat 3
    (\hat 1 + \hat 4)^2} \frac{(3\:2)}{(4\:1)} \: \bar\xi^+_{\bar 1}
  \gB_{\bar 2} \bar\gB_{\bar 3} \xi^-_{\bar 4}.
\end{equation*}

Next, consider the contribution from $L^{\bar\psi+-\psi}$: here, only
the leading order substitutions are needed for the fields involved so
we simply extract the relevant term from the momentum-space
representation of \eqref{eq:lcqcd-qbarpmq}, giving
\begin{equation*}
  \frac{g^2}{\sqrt 8} \int_{1234} \frac{\hat 3 - \hat 2}{(\hat 1 + \hat
    4)^2} \: \bar\xi^+_{\bar 1} \gB_{\bar 2} \bar\gB_{\bar 3} \xi^-_{\bar 4}.
\end{equation*}

Finally $L^{\bar\psi-\psi}$, which in momentum space is
\begin{equation}
  \label{eq:lcqcd-qbarmq-M}
  L^{\bar\psi-\psi} = -\frac{ig^2}{\sqrt 8} \int_{123} \left\{ \left(
      \frac{\tilde 3}{\hat 3} - \frac{\tilde 2}{\hat 2} \right)
    \bar\alpha^+_{\bar 1} \bar\gA_{\bar 2} \alpha^-_{\bar 3} + 
    \left(
      \frac{\tilde2 + \tilde3}{\hat2 + \hat3} - \frac{\tilde 2}
      {\hat 2} \right) \bar\alpha^-_{\bar 1} \bar\gA_{\bar 2} \alpha^+_{\bar 3}
  \right\},
\end{equation}
contributes two terms from the next-to-leading order substitutions for
$\bar\alpha^+$ and $\bar\gA$ from \eqref{eq:fqold-in-new} and
\eqref{eq:Abar-loworder-qcd}, leading to
\begin{equation*}
  \frac{g^2}{\sqrt 8} \int_{1234} \left\{
    \frac{\hat1 + \hat2}{\hat3 \hat4} \frac{(4\:3)}{(1\:2)}
    + \frac{\hat3^2}{\hat4 (\hat1 + \hat4)^2} \frac{(1\:4)}{(2\:3)}
  \right\} \bar\xi^+_{\bar 1} \gB_{\bar 2} \bar\gB_{\bar 3} \xi^-_{\bar 4}.
\end{equation*}
This is clearly of the form \eqref{eq:mhvqcd-2q-form}.

The sum of these coefficients is, accounting for conservation of
momentum, the vertex
\begin{equation*}
  V^{3;+}_{2{\rm q}}(1234) = 
  -\frac{g^2}{\sqrt 8} \frac{\hat 2}{\hat3 \hat4}
  \frac{(1\:3) (4\:3)^2}{(1\:2) (2\:3) (4\:1)}.
\end{equation*}
Plugging this into the RHS of \eqref{eq:A-V-2q} gives us
\begin{equation*}
  -ig^2 \frac{\sqrt{\hat 1}\hat2}{\hat3 \sqrt{\hat4}}
  \frac{(1\:3) (4\:3)^2}{(1\:2) (2\:3) (4\:1)},
\end{equation*}
which may be shown to equal \eqref{eq:qqgg-swi} using
\eqref{eq:lc-spinorbrackets}.

One may also show that the remaining three partial amplitudes bearing
this colour structure can be obtained in a similar manner by
considering the other possible choices of substitutions.  We reproduce
these amplitudes below:
\begin{align*}
  A(1_{\rm q}^- 2^+ 3^- 4_{\bar{\rm q}}^+) &= \hphantom{-} ig^2
  \frac{\hat 2 \sqrt{\hat 4}}{\sqrt{\hat1} \hat 3} \frac{(1\:3)^3}
  {(1\:2)(2\:3)(4\:1)} = \hphantom{-} ig^2\frac{\langle 1\:3
    \rangle^3}{\langle 1\:2 \rangle \langle 2\:3 \rangle \langle 4\:1
    \rangle}, \\
  A(1_{\rm q}^+ 2^- 3^+ 4_{\bar{\rm q}}^-) &= -ig^2 \frac{\sqrt{\hat1}
    \hat3}{\hat2 \sqrt{\hat4}} \frac{(2\:4)^3} {(2\:3)(3\:4)(4\:1)} =
  -ig^2 \frac{\langle 2\:4 \rangle^3}{\langle 2\:3 \rangle \langle
    3\:4 \rangle \langle 4\:1 \rangle}, \\
  A(1_{\rm q}^- 2^- 3^+ 4_{\bar{\rm q}}^+) &= \hphantom{-} ig^2
  \frac{\hat3 \sqrt{\hat4}}{\sqrt{\hat1} \hat2} \frac{(1\:2)^2 (2\:4)}
  {(2\:3)(3\:4)(4\:1)} = \hphantom{-} ig^2 \frac{\langle 1\:2
    \rangle^2 \langle 2\:4 \rangle}{\langle 2\:3 \rangle \langle 3\:4
    \rangle \langle 4\:1 \rangle},
\end{align*}
in agreement with the known expressions.

\subsection{Four quarks}
\label{ssec:four-quarks}
In amplitudes with two or more quark-antiquark pairs, terms ${\cal
  O}(1/\NC)$ and higher contribute at the tree-level through the
sub-leading colour structures.
Thus we must keep track of these terms in the analysis.

Let us first compute the terms in the MHV lagrangian containing just
two quark-antiquark pairs. To that end, we note they receive
contributions from $L^{\bar\psi-\psi}$ and
$L^{\bar\psi\psi\bar\psi\psi}$. In momentum space, these are given by
\eqref{eq:lcqcd-qbarmq-M} and
\begin{equation}
  \label{eq:lcqcd-qbarqqbarq-M}
  \begin{split}
    L^{\bar\psi\psi\bar\psi\psi} = \frac{g^4}8 \int_{1234} \Biggl\{ &
    \left( \frac1{(\hat1 + \hat4)^2} + \frac 1{\NC} \frac1{(\hat1 +
        \hat2)^2} \right) (\bar\alpha^-_{\bar1} \alpha^+_{\bar2}
    \bar\alpha^-_{\bar3} \alpha^+_{\bar4} + \bar\alpha^+_{\bar1}
    \alpha^-_{\bar2} \bar\alpha^+_{\bar3}
    \alpha^-_{\bar4}) \\
    &+ \frac2{(\hat1 + \hat4)^2} \bar\alpha^+_{\bar1} \alpha^+_{\bar2}
    \bar\alpha^-_{\bar3} \alpha^-_{\bar4} + \frac1{\NC} \frac1{(\hat1
      + \hat2)^2} \bar\alpha^-_{\bar1} \alpha^+_{\bar2}
    \bar\alpha^+_{\bar3} \bar\alpha^-_{\bar4} \Biggr\}.
  \end{split}
\end{equation}
We substitute for $\bar\gA$ in \eqref{eq:lcqcd-qbarmq-M} using the
leading order fermion terms in \eqref{eq:Abar-loworder-qcd}
and for the fermions in
\eqref{eq:lcqcd-qbarqqbarq-M}. Summing and symmetrising over the
momentum labels as much as possible leads to the following terms in
the MHV lagrangian:
\begin{multline}
  \label{eq:lmhv-qqqq-M}
  \frac{g^4}8 \int_{1234} \Biggl\{ \frac12 \frac{(2\:4)^2}{\hat2 \hat4
    (1\:4) (3\:2)} \left( \bar\xi^+_{\bar1} \xi^-_{\bar2}
    \bar\xi^+_{\bar3} \xi^-_{\bar4} + \frac1{\NC} \bar\xi^+_{\bar1}
    \xi^-_{\bar4}
    \bar\xi^+_{\bar3} \xi^-_{\bar2} \right) \\
  + \frac12 \frac{(1\:3)^2}{\hat1 \hat3 (1\:4) (3\:2)} \left(
    \bar\xi^-_{\bar1} \xi^+_{\bar2} \bar\xi^-_{\bar3} \xi^+_{\bar4} +
    \frac1{\NC} \bar\xi^-_{\bar1} \xi^+_{\bar4}
    \bar\xi^-_{\bar3}  \xi^+_{\bar2} \right)\\
  -\frac{(3\:4)^2}{\hat3 \hat4 (1\:4) (3\:2)} \left( \bar\xi^+_{\bar1}
    \xi^+_{\bar2} \bar\xi^-_{\bar3} \xi^-_{\bar4} + \frac1{\NC}
    \bar\xi^+_{\bar1} \xi^-_{\bar4} \bar\xi^-_{\bar3} \xi^+_{\bar2}
  \right) \Biggr\}.
\end{multline}
The colour structure in each of these terms is $\di12\di34 -
\di14\di32 / \NC$. Clearly this conforms to \eqref{eq:mhvqcd-4q-form}
with $V^{3;h_1h_3}_{4{\rm q}(1)}(1234) = V^{3;h_1h_3}_{4{\rm
    q}}(1234)$ and
\begin{align}
  \label{eq:V-4q-1}
  V^{3;++}_{4{\rm q}}(1234) &= \hphantom{-} \frac{g^4}8
  \frac{(2\:4)^2}{\hat2 \hat4 (1\:4) (3\:2)}, \\
  \label{eq:V-4q-2}
  V^{3;--}_{4{\rm q}}(1234) &=\hphantom{-} \frac{g^4}8
  \frac{(1\:3)^2}{\hat1 \hat3 (1\:4) (3\:2)}, \\
  \label{eq:V-4q-3}
  V^{3;+-}_{4{\rm q}}(1234) &= - \frac{g^4}8 \frac{(3\:4)^2}{\hat3
    \hat4 (1\:4) (3\:2)}.
\end{align}

We see that there are three independent colour-ordered four-quark partial
amplitudes here: $A(1_{\rm q}^+ 2_{\rm{\bar q}}^- 3_{\rm q}^+
4_{\rm{\bar q}}^-)$, $A(1_{\rm q}^- 2_{\rm{\bar q}}^+ 3_{\rm q}^-
4_{\rm{\bar q}}^+)$ and $A(1_{\rm q}^- 2_{\rm{\bar q}}^- 3_{\rm q}^+
4_{\rm{\bar q}}^+)$. It suffices to check the leading-in-$1/\NC$
amplitudes and we shall check them all in turn.  By plugging
\eqref{eq:V-4q-1} and \eqref{eq:V-4q-2} into \eqref{eq:A-V-4q}, we
obtain
\begin{equation}
  \label{eq:4qpartial1}
  A(1_{\rm q}^+ 2_{\rm{\bar q}}^- 3_{\rm q}^+ 4_{\rm{\bar q}}^-) = ig^2 \sqrt{\frac{\hat1 \hat3}{\hat2 \hat4}}
  \frac{(2\:4)^2}{(1\:4) (3\:2)} = ig^2 \frac{\langle 2\:4
    \rangle^2}{\langle 1\:4 \rangle \langle 3\:2 \rangle},
\end{equation}
and
\begin{equation}
  \label{eq:4qpartial2}
  A(1_{\rm q}^- 2_{\rm{\bar q}}^+ 3_{\rm q}^- 4_{\rm{\bar q}}^+) = ig^2 \sqrt{\frac{\hat2 \hat4}{\hat1 \hat3}}
  \frac{(1\:3)^2}{(1\:4) (3\:2)} = ig^2 \frac{\langle 1\:3
    \rangle^2}{\langle 1\:4 \rangle \langle 3\:2 \rangle},
\end{equation}
respectively. These
are readily seen to be the known results \eqref{eq:csw-ferm-qgqqgq}.
(They may also be checked against a
calculation made \eg\ using the light-cone QCD Feynman rules obtained
from \eqref{eq:lcqcd-qbarmq}--\eqref{eq:lcqcd-qbarqqbarq}, discussed
in section \ref{sec:mhvqcd-4qproblem}. The helicity arrangements here
are such that both partial amplitudes lift to the same colour trace on
the SUSY side, and so they will both contribute to the SUSY partial
amplitude.  As such, we cannot obtain these QCD amplitudes from
supersymmetric Ward identities.)  Finally, putting \eqref{eq:V-4q-3}
into \eqref{eq:A-V-4q} gives
\begin{equation}
  \label{eq:mhvl-partial-qpqbpqmqbm}
  A(1_{\rm q}^+ 2_{\rm{\bar q}}^+ 3_{\rm q}^- 4_{\rm{\bar q}}^-) = -ig^2
  \sqrt{\frac{\hat1 \hat2}{\hat3 \hat4}} \frac{(3\:4)^2}{(1\:4) (3\:2)} =
  -ig^2 \frac{\langle 3\:4
    \rangle^2}{\langle 1\:4 \rangle \langle 3\:2 \rangle}
\end{equation}
in agreement with the known result in
\eqref{eq:csw-ferm-qgqqgq}. (Otherwise it can be quickly checked
either by direct computation or by using SWIs to compute the gluino
amplitudes with the same colour structure; we note that unlike the
previous case, the helicity arrangements constrain the $\text{QCD}
\rightarrow \text{SUSY}$ mapping such that it is one-to-one.)

\subsection{Two quarks and three gluons}
\label{ssec:two-quarks-three-gluons}

Finally, we consider $A(1_{\rm q}^+ 2^+ 3^- 4^+ 5_{\bar{\rm q}}^-)$,
which is given by \eqref{eq:csw-ferm-q+gq-} as
\begin{equation}
  \label{eq:qgggq-swi}
  A(1_{\rm q}^+ 2^+ 3^- 4^+ 5_{\bar{\rm q}}^-) = -ig^3 \frac{\langle
    1\:3 \rangle \langle 3\:5 \rangle^3}{\langle 1\:2 \rangle \langle
    2\:3 \rangle \langle 3\:4 \rangle \langle 4\:5 \rangle \langle 5\:1
    \rangle}.
\end{equation}
This partial amplitude is tied to the $(T^{a_2} T^{a_3} T^{a_4})\ud15$
colour structure.
In the MHV lagrangian, the term we seek is $\bar\xi^+_{\bar 1}
\gB_{\bar 2} \bar\gB_{\bar 3} \gB_{\bar 4} \xi^-_{\bar 5} \:
V^{3+}_{2{\rm q}}(12345)$ and it receives contributions from $L^\mmp$,
$L^{--++}$, $L^{\bar\psi-\psi}$ and $L^{\bar\psi+-\psi}$, so we will
write it as\footnote{Note that we will retain $R^\pm$, $S^\pm$ and
  $\Delta^\pm$ explicitly here and in other forthcoming examples
  (rather than used their expressions in $\Upsilon$) to elucidate the
  origin of each term.}
\begin{equation}
  \label{eq:qgggq-struct}
  \int_{12345} (W^\mmp + W^{--++} + W^{\bar\psi-\psi} + W^{\bar\psi+-\psi})
  \: \bar\xi^+_{\bar 1} \gB_{\bar 2}
  \bar\gB_{\bar 3} \gB_{\bar 4} \xi^-_{\bar 5}.
\end{equation}
Let us consider each of the $W$s in turn. First, we observe from
\eqref{eq:lcqcd-mmp-M} and the structures of
\eqref{eq:Abar-loworder-qcd} that $L^\mmp$ yields four terms with the
structure of \eqref{eq:qgggq-struct} coming from the different
possible choices substitution for $\gA$. We carry this out and
carefully relabel the momenta while accounting for the anticommuting
nature of the fermions to obtain
\newcommand\gintl[1]{\frac{#1g^2}{\sqrt 8}\int_{12345} \Biggl\{}
\begin{multline}
  \label{eq:qgggq-1}
  W^\mmp = - \frac{ig^2}{\sqrt 8} \Biggl\{ \frac{\hat2 \hat3
    (2,3+4)}{(\hat1 + \hat5)^2 (\hat3 + \hat4)^2} \frac{\hat3}{\hat3 +
    \hat4} \Upsilon(-,3,4) \Delta^{+(2)}(-,5,1) \\
  + \frac{\hat3 \hat4 (2+3,4)}{(\hat1 + \hat5)^2 (\hat2 + \hat3)^2}
  \frac{\hat3}{\hat2 + \hat3} \Upsilon(-,2,3) \Delta^{+(2)}(-,5,1) \\
  + \frac{\hat4 (3\:4)}{\hat3 (\hat3 + \hat4)^2}
  \Delta^{+(2)}(-,5,1,2) + \frac{\hat2 (2\:3)}{\hat3 (\hat2 +
    \hat3)^2} \Delta^{+(3)}(-,4,5,1) \Biggr\}.
\end{multline}
We remind the reader here that in the argument lists of $\Upsilon$,
$\Delta$, \etc, $-$ is a placeholder whose value should be taken to be
the negative of the sum of the other momenta passed to that
coefficient.

Next, $L^{--++}$ has terms of the form $\tr (\bar\gA \gA \bar\gA \gA)$
and $\tr (\bar\gA \bar\gA \gA \gA)$, of which only the former
contribute terms to \eqref{eq:qgggq-struct}. In momentum space, this
is
\begin{equation*}
  \tr \int_{1234} \left\{ \frac{\hat2 \hat3}{(\hat3 + \hat4)^2}
    + \frac{\hat3 \hat4}{(\hat2 + \hat3)^2} \right\} \bar\gA_{\bar 1}
  \gA_{\bar 2} \bar\gA_{\bar 3} \gA_{\bar 4}.
\end{equation*}
Substituting for each $\bar\gA$ in turn using the lowest-order terms
in \eqref{eq:AbarF-series} gives
\begin{multline}
  \label{eq:qgggq-2}
  W^{--++} = -\frac{g^2}{\sqrt 8} \Biggl\{ \frac 1{\hat1 + \hat5}
  \left( \frac{\hat2 \hat3}{(\hat3 + \hat4)^2} + \frac{\hat3
      \hat4}{(\hat2 + \hat3)^2} \right) \Delta^{+(2)}(-,5,1) \\ +
  \frac 1{\hat1 + \hat5} \left( \frac{\hat4 (\hat1 + \hat5)}{(\hat3 +
      \hat4)^2} + \frac{\hat2 (\hat1 + \hat5)}{(\hat2 + \hat3)^2}
  \right) \Delta^{+(2)}(-,5,1) \Biggr\}.
\end{multline}

$L^{\bar\psi-\psi}$ contributes four terms in the structure of
\eqref{eq:qgggq-struct}, owing to the fact that we will now also see
terms from the fermion series expansions \eqref{eq:fqold-in-new} and
\eqref{eq:fpold-in-new} when these are substituted into
\eqref{eq:lcqcd-qbarmq-M}. Upon re-arrangement and re-labelling of the
momenta, we arrive at the contribution
\begin{multline}
  \label{eq:qgggq-3}
  W^{\bar\psi-\psi} = -\frac{ig^2}{\sqrt 8} \Biggl\{ \left(
    \frac{\tilde4 + \tilde5}{\hat4 + \hat5} - \frac{\tilde3}{\hat3}
  \right) R^+(1,-;2) S^+(-,5;4) \\
  + \left( \frac{\hat3}{\hat3 + \hat4} \right)^2 \left(
    \frac{\tilde5}{\hat5} - \frac{\tilde3 + \tilde4}{\hat3 + \hat4}
  \right) R^+(1,-;2)  \Upsilon(-,3,4) \\
  + \left( \frac{\hat3}{\hat2 + \hat3 + \hat4} \right)^2 \left(
    \frac{\tilde5}{\hat5} - \frac{\tilde1 + \tilde5}{\hat1 + \hat5}
  \right) \Upsilon(-,2,3,4)
  \\
  + \left( \frac{\hat3}{\hat2 + \hat3} \right)^2 \left( \frac{\tilde4
      + \tilde5}{\hat4 + \hat5} - \frac{\tilde2 + \tilde3}{\hat2 +
      \hat3} \right) \Upsilon(-,2,3) S^+(-,5;4) \Biggr\}.
\end{multline}

Finally, $L^{\bar\psi+-\psi}$, which has momentum-space representation
\begin{multline*}
  L^{\bar\psi+-\psi} = -\frac{g^2}{\sqrt 8} \int_{1234} \Biggl\{
  \left( \frac 1{\hat3 + \hat4} + \frac{\hat2 - \hat3}{(\hat2 +
      \hat3)^2} \right) \bar\alpha^+_{\bar1} \bar\gA_{\bar2}
  \gA_{\bar3}
  \alpha^-_{\bar4} \\
  + \frac{\hat2 - \hat3}{(\hat2 + \hat3)^2} \bar\alpha^+_{\bar1}
  \gA_{\bar2} \bar\gA_{\bar3} \alpha^-_{\bar4} + \text{l.h.\ pieces}
  \Biggr\},
\end{multline*}
contributes four terms to \eqref{eq:qgggq-struct} from substitutions
for $\bar\gA$ and the fermions:
\begin{multline}
  \label{eq:qgggq-4}
  W^{\bar\psi+-\psi} = - \frac{g^2}{\sqrt 8} \Biggl\{ \left( \frac
    1{\hat4 + \hat5} + \frac{\hat3 - \hat4}{(\hat3 + \hat4)^2} \right)
  R^+(1,-;2) + \frac{\hat2 - \hat3}{(\hat2 + \hat3)^2}
  S^+(-,5;4) \\
  + \left( \frac{\hat3}{\hat3 + \hat4} \right)^2 \frac{\hat2 -
    \hat3 - \hat4}{(\hat2 + \hat3 + \hat4)^2}  \Upsilon(-,3,4) \\
  +\left( \frac{\hat3}{\hat2 + \hat3} \right)^2 \left( \frac 1{\hat4 +
      \hat5} + \frac{\hat2 + \hat3 - \hat4}{(\hat2 + \hat3 + \hat4)^2}
  \right) \Upsilon(-,2,3) \Biggr\}.
\end{multline}

We take the sum of \eqref{eq:qgggq-1}, \eqref{eq:qgggq-2},
\eqref{eq:qgggq-3} and \eqref{eq:qgggq-4} to be $V^{3;+}_{2{\rm q}}(12345)$
and when plugged into
\eqref{eq:A-V-2q}, its RHS becomes
\begin{equation*}
  -\frac{ig^3}{\sqrt 2} \frac{\sqrt{\hat1} \hat2 \hat4}{\hat3
    \sqrt{\hat5}}
  \frac{(1\:3) (3\:5)^3}{(1\:2) (2\:3) (3\:4) (4\:5) (5\:1)},
\end{equation*}
which may be shown to equal \eqref{eq:qgggq-swi} using
\eqref{eq:lc-spinorbrackets}.


\subsection{On missing amplitudes}
\label{ssec:mhvqcd-completion}

We learned in chapter \ref{cha:etv} that contributions to the
$S$-matrix from completion vertices (arising from the series expansions
of the transformations themselves) are required to obtain certain
`missing' amplitudes. In particular,
we demonstrated this by calculating $A(1^-2^+3^+)$ at tree-level (which has no
vertex in the MHV lagrangian, and is non-vanishing for complex momenta),
and by showing that $A(1^+2^+3^+4^+)$ at one-loop
matches what would be obtained were one to compute it using
light-cone Yang--Mills theory.

\begin{figure}[t]
  \centering \subfigure{
    \begin{picture}(100,80) \SetOffset(50,27) \Gluon(0,27)(0,40){2}{2}
      \ArrowLine(34.6410,-20)(23.3827,-13.5)
      \ArrowLine(-23.3827,-13.5)(-34.6410,-20)
      \DashArrowLine(21.6506,-12.5)(0,25){1}
      \DashArrowLine(0,25)(-21.6506,-12.5){1} \BCirc(0,25){2}
      \BCirc(21.6506,-12.5){2} \BCirc(-21.6506,-12.5){2}
      \Text(0,43)[bc]{$2^-$} \Text(35.6410,-19)[tl]{$3_{{\rm q}}^+$}
      \Text(-35.6410,-19)[tr]{$1_{\bar{\rm q}}^-$}
      \Text(-4,24)[tr]{$+$} \Text(4,24)[tl]{$-$}
      \Text(-19.0,-8.5)[br]{$-$} \Text(19.0,-8.5)[bl]{$+$}
    \end{picture}
    \label{fig:mhv-completion-qgq-1}
  }\quad \subfigure{
    \begin{picture}(100,80) \SetOffset(50,27) \Gluon(0,27)(0,40){2}{2}
      \ArrowLine(34.6410,-20)(23.3827,-13.5)
      \ArrowLine(-23.3827,-13.5)(-34.6410,-20)
      \Line(21.6506,-12.5)(0,25)
      \DashArrowLine(21.6506,-12.5)(-21.6506,-12.5){1} \BCirc(0,25){2}
      \BCirc(21.6506,-12.5){2} \BCirc(-21.6506,-12.5){2}
      \Text(4,24)[tl]{$-$} \Text(19.0,-8.5)[bl]{$+$}
      \Text(-19,-13.5)[tl]{$-$} \Text(19,-13.5)[tr]{$+$}
    \end{picture}
    \label{fig:mhv-completion-qgq-2}
  }\quad \subfigure{
    \begin{picture}(100,100) \SetOffset(50,27)
      \Gluon(0,27)(0,40){2}{2} \ArrowLine(34.6410,-20)(23.3827,-13.5)
      \ArrowLine(-23.3827,-13.5)(-34.6410,-20)
      \Line(0,25)(-21.6506,-12.5)
      \DashArrowLine(21.6506,-12.5)(-21.6506,-12.5){1} \BCirc(0,25){2}
      \BCirc(21.6506,-12.5){2} \BCirc(-21.6506,-12.5){2}
      \Text(-4,24)[tr]{$-$} \Text(-19.0,-8.5)[br]{$+$}
      \Text(-19,-13.5)[tl]{$-$} \Text(19,-13.5)[tr]{$+$}
    \end{picture}
    \label{fig:mhv-completion-qgq-3}
  }
  \caption{Contributions to the tree-level $A(1_{{\rm q}}^+ 2^+
    3_{\bar{\rm q}}^-)$ amplitude, before applying LSZ reduction. All
    momenta are directed out of the diagrams, arrows indicate colour
    flow.}
  \label{fig:mhvqcd-completion-qgq}
\end{figure}
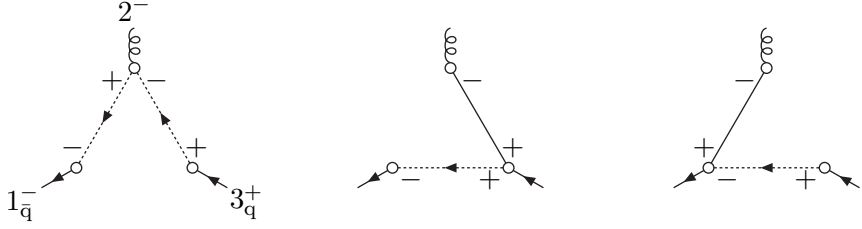

The situation with quarks added is no different: quark-gluon
amplitudes whose construction requires erstwhile \mhvbar-like vertices
are recovered through completion vertices found in
\eqref{eq:fpold-in-new}, \eqref{eq:fqold-in-new}, \eqref{eq:A-series} 
and \eqref{eq:Abar-loworder-qcd}. As an example of this, let us study
the partial amplitude $A(1_{{\rm q}}^+ 2^+ 3_{\bar{\rm q}}^-)$.  Is it
easy to evaluate this from either light-cone QCD, and it is
given by
\begin{equation}
  A(1_{{\rm q}}^+ 2^+ 3_{\bar{\rm q}}^-) = ig \frac{[2\:1]^2}{[3\:1]}.
  \label{eq:qcd-completion}
\end{equation}
The LSZ reduction gives this amplitude as
\begin{equation}
  \label{eq:qcd-completion-1}
  A(1_{{\rm q}}^+ 2^+ 3_{\bar{\rm q}}^-) = \lim_{p_1^2, p_2^2, p_3^2
    \rightarrow 0}
  ip_2^2 \times \frac{-ip_1^2}{2^{1/4}\sqrt1} \times
  \frac{-ip_3^2}{2^{1/4}\sqrt3} \times \langle \alpha^-_1 \, \bar A_2 \,
  \bar\alpha^+_3 \rangle.
\end{equation}
Here, the first
factor contributes the gluon polarisation and inverse propagators; and
the second and third factors are \eqref{eq:lsz-factor-qcd} for the
quarks.  The correlation function may be computed by substituting for
each of the fields involved with their next-to-leading-order
expressions from \eqref{eq:alphabar-series}, \eqref{eq:alpha-series} and
\eqref{eq:Abar-loworder-qcd} (in the case of $\bar\gA$ taking the
right-handed fermionic part), or equivalently by evaluating the
 sum of the three diagrams of
fig.~\ref{fig:mhvqcd-completion-qgq}, constructed from the vertices of
fig.~\ref{fig:mhvqcd-completionverts}.
Thus, \eqref{eq:qcd-completion-1} becomes (accounting for the
normalisation of $\gA$, which simply amounts to using
\eqref{eq:gauge-conv} to substitute for the gauge fields in the series):
\begin{equation*}\begin{split}
    A(1_{{\rm q}}^+ 2^+ 3_{\bar{\rm q}}^-) &= -\frac{i}{\sqrt{2 \:
        \hat1 \hat3}} p_1^2 p_2^2 p_3^2 \Biggl\{ \frac{\hat1}{p_1^2}
    \frac{\hat3}{p_3^2} \frac{ig}{\hat2^2} \hat3 \Upsilon(231) 
    + \frac{\hat1}{p_1^2} \frac1{p_2^2} \: ig \:
    \Upsilon(312)
    + \frac1{p_2^2} \frac{\hat3}{p_3^2} \: ig \: 
    \frac{\hat3}{\hat1} \Upsilon(123) \Biggr\} \\
    &= \frac{ig}{\sqrt2} \sqrt{\hat 1 \hat 3} \frac{\hat3}{(1\:2)}
    \left( \frac{p_1^2}{\hat 1} + \frac{p_2^2}{\hat 2} +
      \frac{p_3^2}{\hat 3} \right) \\
    &= ig \sqrt2 \frac1{\hat2} \sqrt{\frac{\hat3}{\hat1}} \: \{3\:1\}.
  \end{split}\end{equation*}
Note that the first term on the first line acquires an extra factor of
$-1$ since we need to transpose the quark fields to evaluate the
correlation function.
This expression may be shown to equal \eqref{eq:qcd-completion} using
\eqref{eq:lc-spinorbrackets}. (Note that we make use of
\eqref{eq:sum-bilinears} to obtain the final line.)

\section{Conclusion}
\label{sec:conclusion-qcd}

In this chapter, we extended the canonical MHV lagrangian formalism of
\cite{Mansfield:2005yd} and chapters \ref{cha:mhvym} and \ref{cha:etv}
to a full massless QCD theory with ${\rm SU}(N_{\rm C})$ gauge
symmetry.  We started with massless QCD in the light-cone gauge with
the non-dynamical field components integrated out. By applying a
canonical transformation to the field variables, we obtain a
lagrangian incorporating gluon-gluon and quark-gluon interactions
whose vertices are proportional (up to polarisation factors) to the
MHV amplitudes in the literature (obtained, for example, by
supersymmetry). This has been checked explicitly for amplitudes with
two quarks and two gluons, with four quarks, and with two quarks and
three gluons in the $(1_{\rm q}^+ 2^+ 3^- 4^+ 5_{\bar{\rm q}}^-)$
configuration. The field transformations for the fermions
and $\bar\gA$ are summarised in
\eqref{eq:alphabar-series}, \eqref{eq:alpha-series} and
\eqref{eq:Abar-loworder-qcd}.

The MHV QCD lagrangian we have found maintains a certain `backward
compatibility' with the pure-gauge case found in chapter
\ref{cha:mhvym}. The solution for $\gA$ in terms of $\gB$ is the same,
whereas $\bar\gA$ acquires new terms in the new fermion fields brought
on by the requirement that the transformation is canonical. As in the
pure-gauge case, the explicit form of this transformation as a series
expansion has coefficients that have simple, holomorphic expressions
in the momenta.

As we found out in chapter \ref{cha:etv}, the $S$-matrix receives
contributions beyond the vertex content of an MHV lagrangian ---
specifically, the completion vertices that originate in the
transformation itself. These allow us to construct otherwise missing
amplitudes, notably those eliminated by the choice of field
transformation; we demonstrated this for the MHV QCD lagrangian in the
simple case of the otherwise missing $(1^+_{\rm q} 2^+ 3^-_{\bar{\rm
    q}})$ partial amplitude. Similarly, we would expect the completion
vertices to be important for recovery of the full off-shell theory,
and for on-shell loop-level amplitudes.

\begin{subappendices}
\section{On SUSY and
$A(1_{\rm q}^+ 2_{\rm{\bar q}}^- 3_{\rm q}^+
  4_{\rm{\bar q}}^-)$ and $A(1_{\rm q}^- 2_{\rm{\bar q}}^+
3_{\rm q}^-  4_{\rm{\bar q}}^+)$}
\label{sec:mhvqcd-4qproblem}

In the main text, we compared the the expressions given in
\eqref{eq:4qpartial1} and \eqref{eq:4qpartial2} for
the partial amplitudes
$A(1_{\rm q}^+ 2_{\rm{\bar q}}^- 3_{\rm q}^+ 4_{\rm{\bar q}}^-)$
and $A(1_{\rm q}^- 2_{\rm{\bar q}}^+ 3_{\rm q}^- 4_{\rm{\bar q}}^+)$,
respectively,
with those obtained by directly by light-cone gauge QCD.  This is
because mapping these results onto the SUSY theory to check that they
then satisfy the SWIs requires some care, which we alluded to in
section \ref{sec:swis}: these helicity arrangements are such that both
partial amplitudes lift to the same colour trace on the SUSY side, and
so they will both contribute to the SUSY partial amplitude.  As such,
we cannot obtain these QCD amplitudes from supersymmetric Ward
identities.

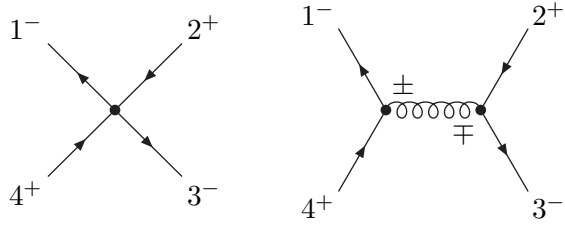
\begin{figure}[t]
  \centering \subfigure{
    \begin{picture}(80,86)
      \SetOffset(40,43)
      \ArrowLine(0,0)(-25,25)
      \ArrowLine(0,0)(25,-25)
      \ArrowLine(25,25)(0,0)
      \ArrowLine(-25,-25)(0,0)
      \Vertex(0,0){2}
      \Text(-27,27)[br]{$1^-$}
      \Text(27,27)[bl]{$2^+$}
      \Text(27,-27)[tl]{$3^-$}
      \Text(-27,-27)[tr]{$4^+$}
    \end{picture}
    \label{fig:mhvqcd-4q-feyn-a}
  }\qquad \subfigure{
    \begin{picture}(100,86)
      \SetOffset(50,43)
      \Gluon(-17.6777,0)(17.6777,0){3}{5}
      \ArrowLine(35.3553,30.6186)(17.6777,0)
      \ArrowLine(17.6777,0)(35.3553,-30.6186)
      \ArrowLine(-35.3553,-30.6186)(-17.6777,0)
      \ArrowLine(-17.6777,0)(-35.3553,30.6186)
      \Vertex(-17.6777,0){2}
      \Vertex(17.6777,0){2}
      \Text(-37,32)[br]{$1^-$}
      \Text(37,32)[bl]{$2^+$}
      \Text(37,-32)[tl]{$3^-$}
      \Text(-37,-32)[tr]{$4^+$}
      \Text(-15.6,5)[bl]{$\pm$}
      \Text(15.6,-5)[tr]{$\mp$}
    \end{picture}
    \label{fig:mhvqcd-4q-feyn-b}
  }
  \caption{Feynman graphs from light-cone gauge QCD that contribute to
    the $\di12 \di34 - \di14 \di32 /{N_{\rm C}}$
    term of the $(1_{\rm q}^- 2_{\rm{\bar q}}^+ 3_{\rm q}^- 4_{\rm{\bar
        q}}^+)$  amplitude.}
  \label{fig:mhvqcd-4q-feyn}
\end{figure}

First, we outline the light-cone QCD calculation used to obtain these
amplitudes. For the case of $A(1_{\rm{q}}^- 2_{\rm{\bar q}}^+
3_{\rm{q}}^- 4_{\rm{\bar q}}^+)$, the contributing diagrams are shown
in fig.~\ref{fig:mhvqcd-4q-feyn}. From these, we pull out the terms
proportional to $\di12 \di34 - \di14 \di32 /{\NC}$.  (There are no
such terms in the $(1,2)$ channel, hence the absence of these graphs
from fig.~\ref{fig:mhvqcd-4q-feyn}.)  The coefficient of this colour
structure found this way is
\[
\frac{2ig^2\sqrt{\hat1\hat2\hat3\hat4}}{(\hat2 + \hat3)^2} \left\{ 1 +
  \frac1{(p_2 + p_3)^2} \left[ \frac{\{1\:4\} (2\:3)}{\hat3\hat4} +
    \frac{(1\:4) \{3\:2\}}{\hat1\hat2} \right] \right\},
\]
and by putting all momenta on shell and simplifying, this expression
can be shown to be equal to \eqref{eq:4qpartial2}.

Now let us see precisely how the correspondence to the SUSY partial
amplitude works in this case.  Note that $A(1_{\rm q}^+ 2_{\rm{\bar
    q}}^- 3_{\rm q}^+ 4_{\rm{\bar q}}^-)$ and $A(4_{\rm q}^-
1_{\rm{\bar q}}^+ 2_{\rm q}^- 3_{\rm{\bar q}}^+) $ (a re-labelling of
\eqref{eq:4qpartial2})
have the same helicity arrangement, and suggest comparison with
$A(\Lambda_1^+ \Lambda_2^- \Lambda_3^+ \Lambda_4^-)$; that this mixes
quark and antiquark is of no concern here as the SUSY amplitude does
not distinguish the two, the gluinos being in the adjoint
representation.  This particular SUSY partial amplitude contains
contributions from gluon exchanges in both the $(1,4)$ and $(1,2)$
channels, and it is associated with the $\tr (T^{a_1} T^{a_2} T^{a_3}
T^{a_4})$ colour structure (shown on the right of
fig.~\ref{fig:thooft-2q}). On the QCD side, $A(1_{\rm q}^+ 2_{\rm{\bar
    q}}^- 3_{\rm q}^+ 4_{\rm{\bar q}}^-)$ is associated with the
$\di12\di34$ leading-order colour structure (see
fig.~\ref{fig:thooft-2q} left) corresponding to a $(1,4)$ gluon
exchange, and $-A(4_{\rm q}^- 1_{\rm{\bar q}}^+ 2_{\rm q}^-
3_{\rm{\bar q}}^+)$ with $\di14\di32$ (see fig.~\ref{fig:thooft-2q}
centre), implying a $(1,2)$ exchange. Since both colour structures
lift to $\tr (T^{a_1} T^{a_2} T^{a_3} T^{a_4})$, we would expect that
the SUSY partial amplitude is the sum of the two QCD partial
amplitudes, \ie:
\begin{equation*}
  \begin{split}
    A(\Lambda_1^+ \Lambda_2^- \Lambda_3^+ \Lambda_4^-) &= A(1_{\rm
      q}^+ 2_{\rm{\bar q}}^- 3_{\rm q}^+ 4_{\rm{\bar q}}^-) - A(4_{\rm
      q}^-
    1_{\rm{\bar q}}^+ 2_{\rm q}^- 3_{\rm{\bar q}}^+) \\
    &= ig^2 \frac{\langle 1\:3 \rangle \langle 2\:4 \rangle^3}{\langle
      1\:2 \rangle \langle 2\:3 \rangle\langle 3\:4 \rangle\langle
      4\:1 \rangle }
    \\
    &= ig^2 \left\{ \frac{\langle 2\:4 \rangle^2}{\langle 1\:4 \rangle
        \langle 3\:2 \rangle} - \frac{\langle 2\:4 \rangle^2}{\langle
        1\:2 \rangle \langle 3\:4 \rangle} \right\},
  \end{split}
\end{equation*}
in agreement with the result from SWIs.
The $-$ sign for the second term on the first line above comes from
Fermi statistics. Note that we have used the
Schouten identity \eqref{eq:schouten}.

\end{subappendices}

\chapter{Discussion}
\label{cha:conclusion}

As noted in the introduction, the isolation of New Physics signatures
and improved measurements of the QCD coupling constant at the LHC
require the theoretical community to have a good understanding of QCD
processes, particularly those involving multiple partons, at leading
and next-to-leading order \cite{Buttar:2006zd}. The past two decades
have seen considerable advancement in technology for computing
amplitudes for multipartonic processes, driven by the observation that
gauge theories appear to have structures much simpler than the
traditional calculation techniques would imply. The Parke--Taylor
formula \eqref{eq:parke-taylor} for the tree-level MHV amplitude is a
remarkably simple expression in light of the number of Feynman
diagrams that would be required to compute it. Subsequent developments
have yielded tree-level computational techniques that provide
polynomial-in-$n$ time algorithms for $n$ gluon amplitudes. In
particular the BCF recursion relations between on-shell amplitudes
have been very useful in pushing forward the set of known analytic
expressions for tree-level amplitudes  (see section
\ref{sec:background-bcf} for background and references to its
applications), as have comparable numerical techniques based upon
Berends--Giele recursion relations \cite{Berends:1987me}. Taking a
more formal approach, the CSW rules (explained in section
\ref{sec:background-csw}) used insight from twistor-string theory to
show tree-level amplitudes could be obtained by sewing Parke--Taylor
MHV amplitudes together with scalar propagators. This superficially
looks like a field theory of a charged scalar with an infinite tower
of vertices of ever-increasing valence, and thus grows no faster than
$n^2$, again a significant improvement over Feynman diagrams.  The
natural question to ask is what field theoretic motivation underpins
this, and how this can be extended to quantum corrections.

Developments at the loop level have not been so straightforward, but
significant progress has been made using a number of techniques. In
particular, the CSW rules have been applied successfully at one loop
for supersymmetric theories \cite{Brandhuber:2004yw, Quigley:2004pw,
  Bedford:2004py} and the cut-constructible parts of pure Yang--Mills
amplitudes \cite{Bedford:2004nh, Brandhuber:2005kd} (reviewed in
section \ref{ssec:background-csw-loop}). Many parts of QCD amplitudes
can be obtained by unitarity in four dimensions \cite{Bern:1994cg,
  Bern:1994zx, Bern:2004bt, Britto:2005ha, Britto:2006sj} (indeed,
\emph{all} of the supersymmetric components are cut-constructible)
leading one to promising ideas such as loop-level on-shell recursion
relations/bootstrapping \cite{Bern:2005hs, Bern:2005cq}, unitarity
\cite{Bern:1995db} and generalised unitarity \cite{Brandhuber:2005jw}
outside four dimensions, and direct computation of the rational parts
from Feynman diagrams \cite{Xiao:2006vr,Su:2006vs,Xiao:2006vt}.

In this thesis, we have seen how one of these modern techniques, the
CSW rules, may be understood from the field theory point of view at
the action level, both for pure Yang--Mills and massless QCD, and admit
the addition of dimensional regularisation structure.

\section{Summary of work undertaken}
\textit{This section is intended as an overview of the research
  described in chapters \ref{cha:mhvym}--\ref{cha:mhvqcd}. For
  detailed discussions we refer the reader to the relevant concluding
  section in each, specifically sections \ref{sec:mhvym-conclusion},
  \ref{sec:etv-conclusion} and \ref{sec:conclusion-qcd}.}

The CSW rules are underwritten by field theory, as shown by Mansfield
\cite{Mansfield:2005yd} and demonstrated explicitly in chapter
\ref{cha:mhvym}. In particular, the CSW rules are obtained by a
canonical transformation of the fields of light-cone gauge Yang--Mills
theory that absorbs the $(\mpp)$ vertex into the kinetic term of the
theory expressed in the new variables. The series solution this
entails expresses the remaining pieces of the LCYM lagrangian as an
infinite tower of terms with an MHV helicity content, each a
Parke--Taylor amplitude continued off shell by the CSW
prescription. Of particular utility in this process was moving from
the spinor formalism to light-cone co-ordinates.

The precise form of this transformation has series coefficients with a
simple, holomorphic form. Chapter \ref{cha:etv} showed how these
provide additional vertices at the level of correlation functions and
hence contribute terms to $S$-matrix elements via the LSZ
reduction. That under certain circumstances these contributions do not
vanish is a peculiarity of the non-local nature of the transformation
that allows it to evade the equivalence theorem.  These vertices
`complete' the CSW rules by recovering the parts of Yang--Mills theory
that required the $(\mpp)$ vertex for their construction, in what
(ultimately) turns out to be an algebraic re-arrangement of the
contributions arising from this eliminated vertex. In particular we
showed this to occur for the one-loop $(\fourplus)$ amplitude.

By exploiting the links we made to Yang--Mills theory, the
transformation can be applied to derive MHV vertices in $D$ dimensions
and hence indicates how to apply dimensional regularisation to MHV
techniques.  Unfortunately, the price we pay for this is the
destruction of the pleasing holomorphic character of the
transformation as well as the simple form of the series coefficients.

In chapter \ref{cha:mhvqcd}, the transformation was extended to
include massless quarks in the fundamental representation. This again
resulted in an MHV-form lagrangian, with off-shell vertices coincident
to on-shell amplitudes and Feynman rules following the CSW
prescription as laid out in \cite{Wu:2004jxa}.

\section{Related developments}
Field transformation techniques have since been applied to a variety
of theories in order to obtain an MHV lagrangian and/or obtain
specific results at the action level. In \cite{Feng:2006yy}, Feng and
Huang use two field transformations (one of which is canonical) to
obtain a MHV lagrangian for ${\cal N}=4$ supersymmetric
Yang--Mills. In ref.\ \cite{Ananth:2007zy}, the authors use a field
transformation to make manifest the KLT relations \cite{Kawai:1985xq}
for the three- and four-graviton vertices at the action level.

Ref.\ \cite{Brandhuber:2006bf} describes a field transformation that
is holomorphic but \emph{not} canonical, giving rise to a non-unit
jacobian in the path integral. The authors argue that this jacobian
gives rise to one-loop amplitudes with at most one gluon of negative
helicity.  In ref.\ \cite{Brandhuber:2007vm} a superset of these
authors address the issues of the missing amplitudes (and the rational
pieces of amplitudes with non-vanishing cuts) by using the
transformation associated with the Canonical MHV Lagrangian of chapter
\ref{cha:mhvym} in conjunction with a four-dimensional `light-cone
world-sheet friendly' regulator of Qiu, Thorn and Chakrabarti
\cite{Thorn:2005ak, Chakrabarti:2005ny, Chakrabarti:2006mb}.  This
regulator violates Lorentz covariance by giving the gluon propagator
a non-vanishing $++$ component. This must be removed by a
counterterm. By applying the field transformation to the fields in
this counterterm, it was
demonstrated that one could recover the one-loop $(\fourplus)$ amplitude, and it
was argued that he all-$+$ amplitude could be recovered similarly.

Boels \etal\ have been pursuing formal developments complementary to
those herein in a series of papers that trace their origins back to
the (ambi)twistor Yang--Mills studies of Mason and Skinner
\cite{Mason:2005kn, Boels:2006ir}. The crux of this idea is that the
twistor-space Yang--Mills action has a larger gauge group than that of
the usual space-time formulation. By making local, linear gauge
transformations on the twistor side that are inaccessible from
space-time, the effect on the theory pushed forward to space-time
is that it
undergoes a non-local, non-linear transformation that recovers the
traditional formulation of Yang--Mills, or one which makes the CSW
rules manifest \cite{Boels:2007qn}. In other words, the MHV lagrangian
arises as a result of a choice of gauge fixing in twistor space.
Twistor actions that lead to space-time MHV lagrangians have been
constructed for pure Yang--Mills, extended to include adjoint scalars
and fundamental representation fermions \cite{Boels:2007qn}, and used
in the background gauge to study renormalisability
\cite{Boels:2007gv}. In \cite{Boels:2007pj,Boels:2007qy}, CSW rules
for a massive scalar are obtained using both the twistor action and
space-time field transformation. The research in \cite{Boels:2008du}
showed explicitly that these two approaches produce identical field
transformations, and initiates a study at the loop-level.

\section{Future work}
It is not yet clear how to decide for a given amplitude and order in
perturbation theory when to use the completion vertices. Ideally one
would like an algorithmic means of making this decision.  We know that
by the validity of the CSW construction, completion vertices do not
contribute to on-shell tree amplitudes in space-times with a Minkowski
signature. Yet we have seen that they are required for the
construction of certain classes of non-vanishing amplitudes.  One might
also guess that keeping subsets of momenta on shell might eliminate
the need for certain subsets of diagrams constructed with completion
vertices.  The fact that only the tadpole MHV completion graph
contributes to the [box topology of the] one-loop $(\fourplus)$
amplitude (when all external momenta are taken on shell from the
beginning) has been used as starting point to an attempt to tackle these
questions in recent research \cite{Fudger:2008aa}. It is hoped that a
study of how the field transformation behaves under BRST
transformation might shed some light on this.

The issue of direct evaluation of \emph{individual} MHV completion
diagrams is, at time of writing, an unresolved challenge. It is not
completely clear at present what pole prescription is required to
correctly define the integrals in the face of the non-standard
singularity structure of the completion vertices.  Of course, one
might question why one would even consider this, given algebraic
reconstruction of LCYM before integration --- all the more reason to
get to grips with the points raised in the previous paragraph.

Whether the MHV lagrangian will lead to a better paradigm for
perturbation theory is uncertain at this time. Even if the ideas
explored herein do not yield any new computational advantage, we
believe it has provided insight into the mechanism underlying
methodologies such as the CSW construction and its progeny. More
generally, the application of the technology underlying the MHV
lagrangian may yet yield insight into possible simplifications hidden
in other field theories.

\bibliographystyle{utphys}
\bibliography{bibliography}

\providecommand{\href}[2]{#2}\begingroup\raggedright\begin{thebibliography}{10%
0}

\bibitem{Peskin:1995ev}
M.~E. Peskin and D.~V. Schroeder, {\em ``An Introduction to Quantum Field
  Theory''}.
\newblock Addison-Wesley, 1995.

\bibitem{DeGrand:2006zz}
T.~DeGrand and C.~E. Detar, {\em {Lattice methods for quantum chromodynamics}}.
\newblock {World Scientific}, 2006.

\bibitem{Gross:1973id}
D.~J. Gross and F.~Wilczek, ``{Ultraviolet behavior of non-abelian gauge
  theories},''
\href{http://dx.doi.org/10.1103/PhysRevLett.30.1343}{{\em Phys. Rev. Lett.}
  {\bf 30} (1973)  1343--1346}.

\bibitem{Politzer:1973fx}
H.~D. Politzer, ``{Reliable perturbative results for strong interactions?},''
\href{http://dx.doi.org/10.1103/PhysRevLett.30.1346}{{\em Phys. Rev. Lett.}
  {\bf 30} (1973)  1346--1349}.

\bibitem{Wilson:1977nj}
K.~G. Wilson, ``{Quantum Chromodynamics on a Lattice}.'' {Presented at Cargese
  Summer Inst., Cargese, France, Jul 12--31, 1976}.

\bibitem{Montvay:1994cy}
I.~Montvay and G.~Munster, {\em {Quantum fields on a lattice}}.
\newblock Cambridge University Press, 1994.

\bibitem{Maldacena:1997re}
J.~M. Maldacena, ``{The large N limit of superconformal field theories and
  supergravity},'' {\em Adv. Theor. Math. Phys.} {\bf 2} (1998)  231--252,
\href{http://arxiv.org/abs/hep-th/9711200}{{\tt arXiv:hep-th/9711200}}.

\bibitem{Sterman:1995fz}
G.~Sterman, ``{Partons, factorization and resummation},''
\href{http://arxiv.org/abs/hep-ph/9606312}{{\tt arXiv:hep-ph/9606312}}.

\bibitem{Tung:2001cv}
W.~K. Tung, ``{Perturbative QCD and the parton structure of the nucleon},'' in
  {\em {At the frontier of particle physics}}, M.~Shifman, ed., vol.~2,
  pp.~887--971.
\newblock 2001.

\bibitem{Yao:2006px}
{\bf Particle Data Group} Collaboration, W.~M. Yao {\em et al.}, ``{Review of
  particle physics},''
\href{http://dx.doi.org/10.1088/0954-3899/33/1/001}{{\em J. Phys.} {\bf G33}
  (2006)  1--1232}.

\bibitem{Martin:1997ns}
S.~P. Martin, ``{A supersymmetry primer},''
\href{http://arxiv.org/abs/hep-ph/9709356}{{\tt arXiv:hep-ph/9709356}}.

\bibitem{Buttar:2006zd}
C.~Buttar {\em et al.}, ``{Les Houches physics at TeV colliders 2005, standard
  model, QCD, EW, and Higgs working group: Summary report},''
\href{http://arxiv.org/abs/hep-ph/0604120}{{\tt arXiv:hep-ph/0604120}}.

\bibitem{Parke:1986gb}
S.~J. Parke and T.~R. Taylor, ``{An Amplitude for $n$ Gluon Scattering},''
\href{http://dx.doi.org/10.1103/PhysRevLett.56.2459}{{\em Phys. Rev. Lett.}
  {\bf 56} (1986)  2459}.

\bibitem{Berends:1988zn}
F.~A. Berends and W.~T. Giele, ``{Multiple Soft Gluon Radiation in Parton
  Processes},''
\href{http://dx.doi.org/10.1016/0550-3213(89)90398-2}{{\em Nucl. Phys.} {\bf
  B313} (1989)  595}.

\bibitem{Cachazo:2004kj}
F.~Cachazo, P.~Svrcek, and E.~Witten, ``{MHV vertices and tree amplitudes in
  gauge theory},'' \href{http://dx.doi.org/10.1088/1126-6708/2004/09/006}{{\em
  JHEP} {\bf 09} (2004)  006},
\href{http://arxiv.org/abs/hep-th/0403047}{{\tt arXiv:hep-th/0403047}}.

\bibitem{Risager:2005vk}
K.~Risager, ``{A direct proof of the CSW rules},'' {\em JHEP} {\bf 12} (2005)
  003,
\href{http://arxiv.org/abs/hep-th/0508206}{{\tt arXiv:hep-th/0508206}}.

\bibitem{Britto:2004ap}
R.~Britto, F.~Cachazo, and B.~Feng, ``{New recursion relations for tree
  amplitudes of gluons},''
  \href{http://dx.doi.org/10.1016/j.nuclphysb.2005.02.030}{{\em Nucl. Phys.}
  {\bf B715} (2005)  499--522},
\href{http://arxiv.org/abs/hep-th/0412308}{{\tt arXiv:hep-th/0412308}}.

\bibitem{Britto:2005fq}
R.~Britto, F.~Cachazo, B.~Feng, and E.~Witten, ``{Direct proof of tree-level
  recursion relation in Yang- Mills theory},''
  \href{http://dx.doi.org/10.1103/PhysRevLett.94.181602}{{\em Phys. Rev. Lett.}
  {\bf 94} (2005)  181602},
\href{http://arxiv.org/abs/hep-th/0501052}{{\tt arXiv:hep-th/0501052}}.

\bibitem{Ozeren:2005mp}
K.~J. Ozeren and W.~J. Stirling, ``{MHV techniques for QED processes},'' {\em
  JHEP} {\bf 11} (2005)  016,
\href{http://arxiv.org/abs/hep-th/0509063}{{\tt arXiv:hep-th/0509063}}.

\bibitem{Luo:2005rx}
M.-x. Luo and C.-k. Wen, ``{Recursion relations for tree amplitudes in super
  gauge theories},''
  \href{http://dx.doi.org/10.1088/1126-6708/2005/03/004}{{\em JHEP} {\bf 03}
  (2005)  004},
\href{http://arxiv.org/abs/hep-th/0501121}{{\tt arXiv:hep-th/0501121}}.

\bibitem{Ozeren:2006ft}
K.~J. Ozeren and W.~J. Stirling, ``{Scattering amplitudes with massive fermions
  using BCFW recursion},'' {\em Eur. Phys. J.} {\bf C48} (2006)  159--168,
\href{http://arxiv.org/abs/hep-ph/0603071}{{\tt arXiv:hep-ph/0603071}}.

\bibitem{Badger:2005zh}
S.~D. Badger, E.~W.~N. Glover, V.~V. Khoze, and P.~Svrcek, ``{Recursion
  relations for gauge theory amplitudes with massive particles},'' {\em JHEP}
  {\bf 07} (2005)  025,
\href{http://arxiv.org/abs/hep-th/0504159}{{\tt arXiv:hep-th/0504159}}.

\bibitem{Badger:2005jv}
S.~D. Badger, E.~W.~N. Glover, and V.~V. Khoze, ``{Recursion relations for
  gauge theory amplitudes with massive vector bosons and fermions},'' {\em
  JHEP} {\bf 01} (2006)  066,
\href{http://arxiv.org/abs/hep-th/0507161}{{\tt arXiv:hep-th/0507161}}.

\bibitem{Bedford:2005yy}
J.~Bedford, A.~Brandhuber, B.~J. Spence, and G.~Travaglini, ``{A recursion
  relation for gravity amplitudes},''
  \href{http://dx.doi.org/10.1016/j.nuclphysb.2005.016}{{\em Nucl. Phys.} {\bf
  B721} (2005)  98--110},
\href{http://arxiv.org/abs/hep-th/0502146}{{\tt arXiv:hep-th/0502146}}.

\bibitem{Cachazo:2005ca}
F.~Cachazo and P.~Svrcek, ``{Tree level recursion relations in general
  relativity},''
\href{http://arxiv.org/abs/hep-th/0502160}{{\tt arXiv:hep-th/0502160}}.

\bibitem{Benincasa:2007qj}
P.~Benincasa, C.~Boucher-Veronneau, and F.~Cachazo, ``{Taming tree amplitudes
  in general relativity},''
  \href{http://dx.doi.org/10.1088/1126-6708/2007/11/057}{{\em JHEP} {\bf 11}
  (2007)  057},
\href{http://arxiv.org/abs/hep-th/0702032}{{\tt arXiv:hep-th/0702032}}.

\bibitem{Georgiou:2004wu}
G.~Georgiou and V.~V. Khoze, ``{Tree amplitudes in gauge theory as scalar MHV
  diagrams},'' {\em JHEP} {\bf 05} (2004)  070,
\href{http://arxiv.org/abs/hep-th/0404072}{{\tt arXiv:hep-th/0404072}}.

\bibitem{Wu:2004jxa}
J.-B. Wu and C.-J. Zhu, ``{MHV vertices and fermionic scattering amplitudes in
  gauge theory with quarks and gluinos},''
  \href{http://dx.doi.org/10.1088/1126-6708/2004/09/063}{{\em JHEP} {\bf 09}
  (2004)  063},
\href{http://arxiv.org/abs/hep-th/0406146}{{\tt arXiv:hep-th/0406146}}.

\bibitem{Cachazo:2004zb}
F.~Cachazo, P.~Svrcek, and E.~Witten, ``{Twistor space structure of one-loop
  amplitudes in gauge theory},''
  \href{http://dx.doi.org/10.1088/1126-6708/2004/10/074}{{\em JHEP} {\bf 10}
  (2004)  074},
\href{http://arxiv.org/abs/hep-th/0406177}{{\tt arXiv:hep-th/0406177}}.

\bibitem{Cachazo:2004by}
F.~Cachazo, P.~Svrcek, and E.~Witten, ``{Gauge theory amplitudes in twistor
  space and holomorphic anomaly},''
  \href{http://dx.doi.org/10.1088/1126-6708/2004/10/077}{{\em JHEP} {\bf 10}
  (2004)  077},
\href{http://arxiv.org/abs/hep-th/0409245}{{\tt arXiv:hep-th/0409245}}.

\bibitem{Cachazo:2004dr}
F.~Cachazo, ``{Holomorphic anomaly of unitarity cuts and one-loop gauge theory
  amplitudes},''
\href{http://arxiv.org/abs/hep-th/0410077}{{\tt arXiv:hep-th/0410077}}.

\bibitem{Britto:2004nj}
R.~Britto, F.~Cachazo, and B.~Feng, ``{Computing one-loop amplitudes from the
  holomorphic anomaly of unitarity cuts},''
  \href{http://dx.doi.org/10.1103/PhysRevD.71.025012}{{\em Phys. Rev.} {\bf
  D71} (2005)  025012},
\href{http://arxiv.org/abs/hep-th/0410179}{{\tt arXiv:hep-th/0410179}}.

\bibitem{Eden:1966aa}
R.~J. Eden, P.~V. Landshoff, D.~I. Olive, and J.~C. Polkinghorne, {\em The
  Analytic $S$-Matrix}.
\newblock Cambridge University Press, 1966.

\bibitem{Bern:1994cg}
Z.~Bern, L.~J. Dixon, D.~C. Dunbar, and D.~A. Kosower, ``{Fusing gauge theory
  tree amplitudes into loop amplitudes},''
  \href{http://dx.doi.org/10.1016/0550-3213(94)00488-Z}{{\em Nucl. Phys.} {\bf
  B435} (1995)  59--101},
\href{http://arxiv.org/abs/hep-ph/9409265}{{\tt arXiv:hep-ph/9409265}}.

\bibitem{Brandhuber:2005jw}
A.~Brandhuber, S.~McNamara, B.~J. Spence, and G.~Travaglini, ``{Loop amplitudes
  in pure Yang-Mills from generalised unitarity},'' {\em JHEP} {\bf 10} (2005)
  011,
\href{http://arxiv.org/abs/hep-th/0506068}{{\tt arXiv:hep-th/0506068}}.

\bibitem{Britto:2004nc}
R.~Britto, F.~Cachazo, and B.~Feng, ``{Generalized unitarity and one-loop
  amplitudes in N = 4 super-Yang-Mills},''
  \href{http://dx.doi.org/10.1016/j.nuclphysb.2005.07.014}{{\em Nucl. Phys.}
  {\bf B725} (2005)  275--305},
\href{http://arxiv.org/abs/hep-th/0412103}{{\tt arXiv:hep-th/0412103}}.

\bibitem{Xiao:2006vr}
Z.~Xiao, G.~Yang, and C.-J. Zhu, ``{The rational part of QCD amplitude. I: The
  general formalism},''
  \href{http://dx.doi.org/10.1016/j.nuclphysb.2006.09.008}{{\em Nucl. Phys.}
  {\bf B758} (2006)  1--34},
\href{http://arxiv.org/abs/hep-ph/0607015}{{\tt arXiv:hep-ph/0607015}}.

\bibitem{Su:2006vs}
X.~Su, Z.~Xiao, G.~Yang, and C.-J. Zhu, ``{The rational part of QCD amplitude.
  II: The five-gluon},''
  \href{http://dx.doi.org/10.1016/j.nuclphysb.2006.09.007}{{\em Nucl. Phys.}
  {\bf B758} (2006)  35--52},
\href{http://arxiv.org/abs/hep-ph/0607016}{{\tt arXiv:hep-ph/0607016}}.

\bibitem{Xiao:2006vt}
Z.~Xiao, G.~Yang, and C.-J. Zhu, ``{The rational part of QCD amplitude. III:
  The six-gluon},''
  \href{http://dx.doi.org/10.1016/j.nuclphysb.2006.09.006}{{\em Nucl. Phys.}
  {\bf B758} (2006)  53--89},
\href{http://arxiv.org/abs/hep-ph/0607017}{{\tt arXiv:hep-ph/0607017}}.

\bibitem{Bern:2005hs}
Z.~Bern, L.~J. Dixon, and D.~A. Kosower, ``{On-shell recurrence relations for
  one-loop QCD amplitudes},''
  \href{http://dx.doi.org/10.1103/PhysRevD.71.105013}{{\em Phys. Rev.} {\bf
  D71} (2005)  105013},
\href{http://arxiv.org/abs/hep-th/0501240}{{\tt arXiv:hep-th/0501240}}.

\bibitem{Bern:2005ji}
Z.~Bern, L.~J. Dixon, and D.~A. Kosower, ``{The last of the finite loop
  amplitudes in QCD},''
  \href{http://dx.doi.org/10.1103/PhysRevD.72.125003}{{\em Phys. Rev.} {\bf
  D72} (2005)  125003},
\href{http://arxiv.org/abs/hep-ph/0505055}{{\tt arXiv:hep-ph/0505055}}.

\bibitem{Bern:2005cq}
Z.~Bern, L.~J. Dixon, and D.~A. Kosower, ``{Bootstrapping multi-parton loop
  amplitudes in QCD},''
  \href{http://dx.doi.org/10.1103/PhysRevD.73.065013}{{\em Phys. Rev.} {\bf
  D73} (2006)  065013},
\href{http://arxiv.org/abs/hep-ph/0507005}{{\tt arXiv:hep-ph/0507005}}.

\bibitem{Brandhuber:2004yw}
A.~Brandhuber, B.~J. Spence, and G.~Travaglini, ``{One-loop gauge theory
  amplitudes in N = 4 super Yang-Mills from MHV vertices},''
  \href{http://dx.doi.org/10.1016/j.nuclphysb.2004.11.023}{{\em Nucl. Phys.}
  {\bf B706} (2005)  150--180},
\href{http://arxiv.org/abs/hep-th/0407214}{{\tt arXiv:hep-th/0407214}}.

\bibitem{Bedford:2004nh}
J.~Bedford, A.~Brandhuber, B.~J. Spence, and G.~Travaglini,
  ``{Non-supersymmetric loop amplitudes and MHV vertices},''
  \href{http://dx.doi.org/10.1016/j.nuclphysb.2005.01.032}{{\em Nucl. Phys.}
  {\bf B712} (2005)  59--85},
\href{http://arxiv.org/abs/hep-th/0412108}{{\tt arXiv:hep-th/0412108}}.

\bibitem{Brandhuber:2005kd}
A.~Brandhuber, B.~Spence, and G.~Travaglini, ``{From trees to loops and
  back},'' {\em JHEP} {\bf 01} (2006)  142,
\href{http://arxiv.org/abs/hep-th/0510253}{{\tt arXiv:hep-th/0510253}}.

\bibitem{Mansfield:2005yd}
P.~Mansfield, ``{The Lagrangian origin of MHV rules},'' {\em JHEP} {\bf 03}
  (2006)  037,
\href{http://arxiv.org/abs/hep-th/0511264}{{\tt arXiv:hep-th/0511264}}.

\bibitem{Gorsky:2005sf}
A.~Gorsky and A.~Rosly, ``{From Yang-Mills Lagrangian to MHV diagrams},'' {\em
  JHEP} {\bf 01} (2006)  101,
\href{http://arxiv.org/abs/hep-th/0510111}{{\tt arXiv:hep-th/0510111}}.

\bibitem{Ettle:2006bw}
J.~H. Ettle and T.~R. Morris, ``{Structure of the MHV-rules Lagrangian},'' {\em
  JHEP} {\bf 08} (2006)  003,
\href{http://arxiv.org/abs/hep-th/0605121}{{\tt arXiv:hep-th/0605121}}.

\bibitem{Ettle:2007qc}
J.~H. Ettle, C.-H. Fu, J.~P. Fudger, P.~R.~W. Mansfield, and T.~R. Morris,
  ``{S-Matrix Equivalence Theorem Evasion and Dimensional Regularisation with
  the Canonical MHV Lagrangian},'' {\em JHEP} {\bf 05} (2007)  011,
\href{http://arxiv.org/abs/hep-th/0703286}{{\tt arXiv:hep-th/0703286}}.

\bibitem{Ettle:2008ey}
J.~H. Ettle, T.~R. Morris, and Z.~Xiao, ``{The MHV QCD Lagrangian},''
\href{http://arxiv.org/abs/0805.0239}{{\tt arXiv:0805.0239 [hep-th]}}.

\bibitem{Weinberg:1996kr}
S.~Weinberg, ``{The quantum theory of fields. Vol. 2: Modern applications},''.
  Cambridge, UK: Univ. Pr. (1996) 489 p.

\bibitem{Kleiss:1988ne}
R.~Kleiss and H.~Kuijf, ``{Multi-gluon cross-sections and five jet production
  at hadron colliders},''
\href{http://dx.doi.org/10.1016/0550-3213(89)90574-9}{{\em Nucl. Phys.} {\bf
  B312} (1989)  616}.

\bibitem{Witten:2003nn}
E.~Witten, ``{Perturbative gauge theory as a string theory in twistor space},''
  \href{http://dx.doi.org/10.1007/s00220-004-1187-3}{{\em Commun. Math. Phys.}
  {\bf 252} (2004)  189--258},
\href{http://arxiv.org/abs/hep-th/0312171}{{\tt arXiv:hep-th/0312171}}.

\bibitem{Mangano:1990by}
M.~L. Mangano and S.~J. Parke, ``{Multiparton amplitudes in gauge theories},''
  \href{http://dx.doi.org/10.1016/0370-1573(91)90091-Y}{{\em Phys. Rept.} {\bf
  200} (1991)  301--367},
\href{http://arxiv.org/abs/hep-th/0509223}{{\tt arXiv:hep-th/0509223}}.

\bibitem{tHooft:1973jz}
G.~'t~Hooft, ``{A planar diagram theory for strong interactions},''
\href{http://dx.doi.org/10.1016/0550-3213(74)90154-0}{{\em Nucl. Phys.} {\bf
  B72} (1974)  461}.

\bibitem{Berends:1987me}
F.~A. Berends and W.~T. Giele, ``{Recursive Calculations for Processes with n
  Gluons},''
\href{http://dx.doi.org/10.1016/0550-3213(88)90442-7}{{\em Nucl. Phys.} {\bf
  B306} (1988)  759}.

\bibitem{Bern:1990ux}
Z.~Bern and D.~A. Kosower, ``{Color decomposition of one loop amplitudes in
  gauge theories},''
\href{http://dx.doi.org/10.1016/0550-3213(91)90567-H}{{\em Nucl. Phys.} {\bf
  B362} (1991)  389--448}.

\bibitem{Bern:1994zx}
Z.~Bern, L.~J. Dixon, D.~C. Dunbar, and D.~A. Kosower, ``{One loop n point
  gauge theory amplitudes, unitarity and collinear limits},''
  \href{http://dx.doi.org/10.1016/0550-3213(94)90179-1}{{\em Nucl. Phys.} {\bf
  B425} (1994)  217--260},
\href{http://arxiv.org/abs/hep-ph/9403226}{{\tt arXiv:hep-ph/9403226}}.

\bibitem{Grisaru:1976vm}
M.~T. Grisaru, H.~N. Pendleton, and P.~van Nieuwenhuizen, ``{Supergravity and
  the S Matrix},''
\href{http://dx.doi.org/10.1103/PhysRevD.15.996}{{\em Phys. Rev.} {\bf D15}
  (1977)  996}.

\bibitem{Dixon:1996wi}
L.~J. Dixon, ``{Calculating scattering amplitudes efficiently},''
\href{http://arxiv.org/abs/hep-ph/9601359}{{\tt arXiv:hep-ph/9601359}}.

\bibitem{Grisaru:1977px}
M.~T. Grisaru and H.~N. Pendleton, ``{Some Properties of Scattering Amplitudes
  in Supersymmetric Theories},''
\href{http://dx.doi.org/10.1016/0550-3213(77)90277-2}{{\em Nucl. Phys.} {\bf
  B124} (1977)  81}.

\bibitem{Berger:2006uc}
C.~F. Berger, ``{Bootstrapping one-loop QCD amplitudes},''
  \href{http://dx.doi.org/10.1063/1.2735149}{{\em AIP Conf. Proc.} {\bf 903}
  (2007)  157--160},
\href{http://arxiv.org/abs/hep-ph/0608027}{{\tt arXiv:hep-ph/0608027}}.

\bibitem{Brown:1952eu}
L.~M. Brown and R.~P. Feynman, ``{Radiative corrections to Compton
  scattering},''
\href{http://dx.doi.org/10.1103/PhysRev.85.231}{{\em Phys. Rev.} {\bf 85}
  (1952)  231--244}.

\bibitem{Passarino:1978jh}
G.~Passarino and M.~J.~G. Veltman, ``{One Loop Corrections for e+ e-
  Annihilation Into mu+ mu- in the Weinberg Model},''
\href{http://dx.doi.org/10.1016/0550-3213(79)90234-7}{{\em Nucl. Phys.} {\bf
  B160} (1979)  151}.

\bibitem{'tHooft:1978xw}
G.~'t~Hooft and M.~J.~G. Veltman, ``{Scalar One Loop Integrals},''
{\em Nucl. Phys.} {\bf B153} (1979)  365--401.

\bibitem{Stuart:1987tt}
R.~G. Stuart, ``{Algebraic reduction of one loop Feynman diagrams to scalar
  integrals},''
\href{http://dx.doi.org/10.1016/0010-4655(88)90202-0}{{\em Comput. Phys.
  Commun.} {\bf 48} (1988)  367--389}.

\bibitem{Stuart:1989de}
R.~G. Stuart and A.~Gongora, ``{Algebraic reduction of one loop Feynman
  diagrams to scalar integrals 2},''
\href{http://dx.doi.org/10.1016/0010-4655(90)90019-W}{{\em Comput. Phys.
  Commun.} {\bf 56} (1990)  337--350}.

\bibitem{vanNeerven:1983vr}
W.~L. van Neerven and J.~A.~M. Vermaseren, ``{Large loop integrals},''
\href{http://dx.doi.org/10.1016/0370-2693(84)90237-5}{{\em Phys. Lett.} {\bf
  B137} (1984)  241}.

\bibitem{Melrose:1965kb}
D.~B. Melrose, ``{Reduction of Feynman diagrams},''
{\em Nuovo Cim.} {\bf 40} (1965)  181--213.

\bibitem{vanOldenborgh:1989wn}
G.~J. van Oldenborgh and J.~A.~M. Vermaseren, ``{New Algorithms for One Loop
  Integrals},''
\href{http://dx.doi.org/10.1007/BF01621031}{{\em Z. Phys.} {\bf C46} (1990)
  425--438}.

\bibitem{vanOldenborgh:1990fy}
G.~J. van Oldenborgh, {\em {One loop calculations with massive particles}}.
\newblock PhD thesis, University of Amsterdam, 1990.
\newblock {RX-1313 (AMSTERDAM)}.

\bibitem{Aeppli:1992aa}
A.~Aeppli.
\newblock PhD thesis, University of Zurich, 1992.

\bibitem{Landau:1959fi}
L.~D. Landau, ``{On analytic properties of vertex parts in quantum field
  theory},''
\href{http://dx.doi.org/10.1016/0029-5582(59)90154-3}{{\em Nucl. Phys.} {\bf
  13} (1959)  181--192}.

\bibitem{Mandelstam:1959bc}
S.~Mandelstam, ``{Analytic properties of transition amplitudes in perturbation
  theory},''
\href{http://dx.doi.org/10.1103/PhysRev.115.1741}{{\em Phys. Rev.} {\bf 115}
  (1959)  1741--1751}.

\bibitem{Cutkosky:1960sp}
R.~E. Cutkosky, ``{Singularities and discontinuities of Feynman amplitudes},''
{\em J. Math. Phys.} {\bf 1} (1960)  429--433.

\bibitem{vanNeerven:1985xr}
W.~L. van Neerven, ``{Dimensional regularization of mass and infrared
  singularities in two loop on-shell vertex functions},''
\href{http://dx.doi.org/10.1016/0550-3213(86)90165-3}{{\em Nucl. Phys.} {\bf
  B268} (1986)  453}.

\bibitem{Bern:1995db}
Z.~Bern and A.~G. Morgan, ``{Massive Loop Amplitudes from Unitarity},''
  \href{http://dx.doi.org/10.1016/0550-3213(96)00078-8}{{\em Nucl. Phys.} {\bf
  B467} (1996)  479--509},
\href{http://arxiv.org/abs/hep-ph/9511336}{{\tt arXiv:hep-ph/9511336}}.

\bibitem{Nair:1988bq}
V.~P. Nair, ``{A current algebra for some gauge theory amplitudes},''
\href{http://dx.doi.org/10.1016/0370-2693(88)91471-2}{{\em Phys. Lett.} {\bf
  B214} (1988)  215}.

\bibitem{Quigley:2004pw}
C.~Quigley and M.~Rozali, ``{One-loop MHV amplitudes in supersymmetric gauge
  theories},'' {\em JHEP} {\bf 01} (2005)  053,
\href{http://arxiv.org/abs/hep-th/0410278}{{\tt arXiv:hep-th/0410278}}.

\bibitem{Bedford:2004py}
J.~Bedford, A.~Brandhuber, B.~J. Spence, and G.~Travaglini, ``{A twistor
  approach to one-loop amplitudes in N = 1 supersymmetric Yang-Mills theory},''
  \href{http://dx.doi.org/10.1016/j.nuclphysb.2004.11.031}{{\em Nucl. Phys.}
  {\bf B706} (2005)  100--126},
\href{http://arxiv.org/abs/hep-th/0410280}{{\tt arXiv:hep-th/0410280}}.

\bibitem{Feynman:1972mt}
R.~P. Feynman, ``{Closed loop and tree diagrams},''. In *J R Klauder, Magic
  Without Magic*, San Francisco 1972, 355-375.

\bibitem{Brandhuber:2006bf}
A.~Brandhuber, B.~Spence, and G.~Travaglini, ``{Amplitudes in pure Yang-Mills
  and MHV diagrams},'' {\em JHEP} {\bf 02} (2007)  088,
\href{http://arxiv.org/abs/hep-th/0612007}{{\tt arXiv:hep-th/0612007}}.

\bibitem{Dinsdale:2006sq}
M.~Dinsdale, M.~Ternick, and S.~Weinzierl, ``{A comparison of efficient methods
  for the computation of Born gluon amplitudes},'' {\em JHEP} {\bf 03} (2006)
  056,
\href{http://arxiv.org/abs/hep-ph/0602204}{{\tt arXiv:hep-ph/0602204}}.

\bibitem{Giele:2008bc}
W.~T. Giele and G.~Zanderighi, ``{On the Numerical Evaluation of One-Loop
  Amplitudes: the Gluonic Case},''
\href{http://arxiv.org/abs/0805.2152}{{\tt arXiv:0805.2152 [hep-ph]}}.

\bibitem{Ellis:2006ss}
R.~K. Ellis, W.~T. Giele, and G.~Zanderighi, ``{The one-loop amplitude for
  six-gluon scattering},'' {\em JHEP} {\bf 05} (2006)  027,
\href{http://arxiv.org/abs/hep-ph/0602185}{{\tt arXiv:hep-ph/0602185}}.

\bibitem{Giele:2008ve}
W.~T. Giele, Z.~Kunszt, and K.~Melnikov, ``{Full one-loop amplitudes from tree
  amplitudes},'' \href{http://dx.doi.org/10.1088/1126-6708/2008/04/049}{{\em
  JHEP} {\bf 04} (2008)  049},
\href{http://arxiv.org/abs/0801.2237}{{\tt arXiv:0801.2237 [hep-ph]}}.

\bibitem{Bern:1993mq}
Z.~Bern, L.~J. Dixon, and D.~A. Kosower, ``{One loop corrections to five gluon
  amplitudes},'' \href{http://dx.doi.org/10.1103/PhysRevLett.70.2677}{{\em
  Phys. Rev. Lett.} {\bf 70} (1993)  2677--2680},
\href{http://arxiv.org/abs/hep-ph/9302280}{{\tt arXiv:hep-ph/9302280}}.

\bibitem{Bidder:2004tx}
S.~J. Bidder, N.~E.~J. Bjerrum-Bohr, L.~J. Dixon, and D.~C. Dunbar, ``{N = 1
  supersymmetric one-loop amplitudes and the holomorphic anomaly of unitarity
  cuts},'' \href{http://dx.doi.org/10.1016/j.physletb.2004.11.073}{{\em Phys.
  Lett.} {\bf B606} (2005)  189--201},
\href{http://arxiv.org/abs/hep-th/0410296}{{\tt arXiv:hep-th/0410296}}.

\bibitem{Britto:2005ha}
R.~Britto, E.~Buchbinder, F.~Cachazo, and B.~Feng, ``{One-loop amplitudes of
  gluons in SQCD},'' \href{http://dx.doi.org/10.1103/PhysRevD.72.065012}{{\em
  Phys. Rev.} {\bf D72} (2005)  065012},
\href{http://arxiv.org/abs/hep-ph/0503132}{{\tt arXiv:hep-ph/0503132}}.

\bibitem{Britto:2006sj}
R.~Britto, B.~Feng, and P.~Mastrolia, ``{The cut-constructible part of QCD
  amplitudes},'' \href{http://dx.doi.org/10.1103/PhysRevD.73.105004}{{\em Phys.
  Rev.} {\bf D73} (2006)  105004},
\href{http://arxiv.org/abs/hep-ph/0602178}{{\tt arXiv:hep-ph/0602178}}.

\bibitem{Bern:2004bt}
Z.~Bern, L.~J. Dixon, and D.~A. Kosower, ``{All next-to-maximally
  helicity-violating one-loop gluon amplitudes in N = 4 super-Yang-Mills
  theory},'' \href{http://dx.doi.org/10.1103/PhysRevD.72.045014}{{\em Phys.
  Rev.} {\bf D72} (2005)  045014},
\href{http://arxiv.org/abs/hep-th/0412210}{{\tt arXiv:hep-th/0412210}}.

\bibitem{Mahlon:1993si}
G.~Mahlon, ``{Multi - gluon helicity amplitudes involving a quark loop},''
  \href{http://dx.doi.org/10.1103/PhysRevD.49.4438}{{\em Phys. Rev.} {\bf D49}
  (1994)  4438--4453},
\href{http://arxiv.org/abs/hep-ph/9312276}{{\tt arXiv:hep-ph/9312276}}.

\bibitem{Berger:2006vq}
C.~F. Berger, Z.~Bern, L.~J. Dixon, D.~Forde, and D.~A. Kosower, ``{All
  one-loop maximally helicity violating gluonic amplitudes in QCD},''
  \href{http://dx.doi.org/10.1103/PhysRevD.75.016006}{{\em Phys. Rev.} {\bf
  D75} (2007)  016006},
\href{http://arxiv.org/abs/hep-ph/0607014}{{\tt arXiv:hep-ph/0607014}}.

\bibitem{Berger:2006ci}
C.~F. Berger, Z.~Bern, L.~J. Dixon, D.~Forde, and D.~A. Kosower,
  ``{Bootstrapping one-loop QCD amplitudes with general helicities},''
  \href{http://dx.doi.org/10.1103/PhysRevD.74.036009}{{\em Phys. Rev.} {\bf
  D74} (2006)  036009},
\href{http://arxiv.org/abs/hep-ph/0604195}{{\tt arXiv:hep-ph/0604195}}.

\bibitem{Thorn:2005ak}
C.~B. Thorn, ``{Notes on one-loop calculations in light-cone gauge},''
\href{http://arxiv.org/abs/hep-th/0507213}{{\tt arXiv:hep-th/0507213}}.

\bibitem{Chalmers:1996rq}
G.~Chalmers and W.~Siegel, ``{The self-dual sector of {QCD} amplitudes},''
  \href{http://dx.doi.org/10.1103/PhysRevD.54.7628}{{\em Phys. Rev.} {\bf D54}
  (1996)  7628--7633},
\href{http://arxiv.org/abs/hep-th/9606061}{{\tt arXiv:hep-th/9606061}}.

\bibitem{Mandelstam:1982cb}
S.~Mandelstam, ``{Light Cone Superspace and the Ultraviolet Finiteness of the
  N=4 Model},''
\href{http://dx.doi.org/10.1016/0550-3213(83)90179-7}{{\em Nucl. Phys.} {\bf
  B213} (1983)  149--168}.

\bibitem{Leibbrandt:1983pj}
G.~Leibbrandt, ``{The Light Cone Gauge in Yang-Mills Theory},''
\href{http://dx.doi.org/10.1103/PhysRevD.29.1699}{{\em Phys. Rev.} {\bf D29}
  (1984)  1699}.

\bibitem{Goldstein:2002aa}
{Goldstein, Herbet and Poole, Charles and Safko, John}, {\em Classical
  Mechanics}.
\newblock Addison-Wesley, third~ed., 2002.

\bibitem{Itzykson:1980rh}
C.~Itzykson and J.~B. Zuber, {\em Quantum Field Theory}.
\newblock McGraw-Hill, 1980.

\bibitem{Bern:1991aq}
Z.~Bern and D.~A. Kosower, ``{The Computation of loop amplitudes in gauge
  theories},''
\href{http://dx.doi.org/10.1016/0550-3213(92)90134-W}{{\em Nucl. Phys.} {\bf
  B379} (1992)  451--561}.

\bibitem{Bern:1993sx}
Z.~Bern, L.~J. Dixon, and D.~A. Kosower, ``{New QCD results from string
  theory},''
\href{http://arxiv.org/abs/hep-th/9311026}{{\tt arXiv:hep-th/9311026}}.

\bibitem{Bern:1993qk}
Z.~Bern, G.~Chalmers, L.~J. Dixon, and D.~A. Kosower, ``{One loop N gluon
  amplitudes with maximal helicity violation via collinear limits},''
  \href{http://dx.doi.org/10.1103/PhysRevLett.72.2134}{{\em Phys. Rev. Lett.}
  {\bf 72} (1994)  2134--2137},
\href{http://arxiv.org/abs/hep-ph/9312333}{{\tt arXiv:hep-ph/9312333}}.

\bibitem{Chakrabarti:2005ny}
D.~Chakrabarti, J.~Qiu, and C.~B. Thorn, ``{Scattering of glue by glue on the
  light-cone worldsheet. I: Helicity non-conserving amplitudes},''
  \href{http://dx.doi.org/10.1103/PhysRevD.72.065022}{{\em Phys. Rev.} {\bf
  D72} (2005)  065022},
\href{http://arxiv.org/abs/hep-th/0507280}{{\tt arXiv:hep-th/0507280}}.

\bibitem{Feng:2006yy}
H.~Feng and Y.-t. Huang, ``{MHV lagrangian for N = 4 super Yang-Mills},''
\href{http://arxiv.org/abs/hep-th/0611164}{{\tt arXiv:hep-th/0611164}}.

\bibitem{Ananth:2007zy}
S.~Ananth and S.~Theisen, ``{KLT relations from the Einstein-Hilbert
  Lagrangian},'' \href{http://dx.doi.org/10.1016/j.physletb.2007.07.003}{{\em
  Phys. Lett.} {\bf B652} (2007)  128--134},
\href{http://arxiv.org/abs/0706.1778}{{\tt arXiv:0706.1778 [hep-th]}}.

\bibitem{Kawai:1985xq}
H.~Kawai, D.~C. Lewellen, and S.~H.~H. Tye, ``{A Relation Between Tree
  Amplitudes of Closed and Open Strings},''
\href{http://dx.doi.org/10.1016/0550-3213(86)90362-7}{{\em Nucl. Phys.} {\bf
  B269} (1986)  1}.

\bibitem{Brandhuber:2007vm}
A.~Brandhuber, B.~Spence, G.~Travaglini, and K.~Zoubos, ``{One-loop MHV Rules
  and Pure Yang-Mills},''
  \href{http://dx.doi.org/10.1088/1126-6708/2007/07/002}{{\em JHEP} {\bf 07}
  (2007)  002},
\href{http://arxiv.org/abs/0704.0245}{{\tt arXiv:0704.0245 [hep-th]}}.

\bibitem{Chakrabarti:2006mb}
D.~Chakrabarti, J.~Qiu, and C.~B. Thorn, ``{Scattering of glue by glue on the
  light-cone worldsheet. II: Helicity conserving amplitudes},''
  \href{http://dx.doi.org/10.1103/PhysRevD.74.045018}{{\em Phys. Rev.} {\bf
  D74} (2006)  045018},
\href{http://arxiv.org/abs/hep-th/0602026}{{\tt arXiv:hep-th/0602026}}.

\bibitem{Mason:2005kn}
L.~J. Mason and D.~Skinner, ``{An ambitwistor Yang-Mills Lagrangian},''
  \href{http://dx.doi.org/10.1016/j.physletb.2006.02.061}{{\em Phys. Lett.}
  {\bf B636} (2006)  60--67},
\href{http://arxiv.org/abs/hep-th/0510262}{{\tt arXiv:hep-th/0510262}}.

\bibitem{Boels:2006ir}
R.~Boels, L.~Mason, and D.~Skinner, ``{Supersymmetric gauge theories in twistor
  space},'' {\em JHEP} {\bf 02} (2007)  014,
\href{http://arxiv.org/abs/hep-th/0604040}{{\tt arXiv:hep-th/0604040}}.

\bibitem{Boels:2007qn}
R.~Boels, L.~Mason, and D.~Skinner, ``{From twistor actions to MHV diagrams},''
  \href{http://dx.doi.org/10.1016/j.physletb.2007.02.058}{{\em Phys. Lett.}
  {\bf B648} (2007)  90--96},
\href{http://arxiv.org/abs/hep-th/0702035}{{\tt arXiv:hep-th/0702035}}.

\bibitem{Boels:2007gv}
R.~Boels, ``{A quantization of twistor Yang-Mills theory through the background
  field method},'' \href{http://dx.doi.org/10.1103/PhysRevD.76.105027}{{\em
  Phys. Rev.} {\bf D76} (2007)  105027},
\href{http://arxiv.org/abs/hep-th/0703080}{{\tt arXiv:hep-th/0703080}}.

\bibitem{Boels:2007pj}
R.~Boels and C.~Schwinn, ``{CSW rules for a massive scalar},''
  \href{http://dx.doi.org/10.1016/j.physletb.2008.02.038}{{\em Phys. Lett.}
  {\bf B662} (2008)  80--86},
\href{http://arxiv.org/abs/0712.3409}{{\tt arXiv:0712.3409 [hep-th]}}.

\bibitem{Boels:2007qy}
R.~Boels, C.~Schwinn, and S.~Weinzierl, ``{Recent developments for multi-leg
  QCD amplitudes with massive particles},''
\href{http://arxiv.org/abs/0712.3506}{{\tt arXiv:0712.3506 [hep-ph]}}.

\bibitem{Boels:2008du}
R.~Boels and C.~Schwinn, ``{CSW rules for massive matter legs and glue
  loops},''
\href{http://arxiv.org/abs/0805.4577}{{\tt arXiv:0805.4577 [hep-th]}}.

\bibitem{Fudger:2008aa}
{Fudger, Jonathan P. and Morris, Tim R.} {(in preparation)}, 2008.

\end{thebibliography}\endgroup


\providecommand{\href}[2]{#2}\begingroup\raggedright\endgroup

\end{document}